\documentclass[fleqn,usenatbib]{mnras}

\usepackage{newtxtext,newtxmath}

\usepackage[T1]{fontenc}

\DeclareRobustCommand{\VAN}[3]{#2}
\let\VANthebibliography\thebibliography
\def\thebibliography{\DeclareRobustCommand{\VAN}[3]{##3}\VANthebibliography}

\usepackage{graphicx}	
\usepackage{amsmath}	

\usepackage[mode=buildnew]{standalone} 

\usepackage{pdflscape}
\usepackage{tikz}
\usetikzlibrary{shapes.geometric, arrows}
\usepackage{multirow}   

\title[Characterisation of \textit{HST} confirmed lenses]{Characterisation of \textit{Herschel}-selected strong lens candidates through \textit{HST} and sub-mm/mm observations}

\author[E. Borsato et al.]{
E. Borsato,$^{1}$\thanks{E-mail: edoardo.borsato.1@phd.unipd.it}
L. Marchetti,$^{2,3,36}$
M. Negrello,$^{4}$
E. M. Corsini,$^{1,5}$
D. Wake,$^{6}$
A. Amvrosiadis,$^{7}$ 
A.~J. Baker,$^{8,9}$ 
\newauthor T.~J.~L.~C. Bakx,$^{10,11}$ 
A. Beelen,$^{12}$
S. Berta,$^{13}$
D.~L. Clements,$^{14}$
A. Cooray,$^{15}$
P. Cox,$^{16}$
H. Dannerbauer,$^{17,18}$ 
\newauthor G. de Zotti,$^{5}$ 
S. Dye,$^{19}$
S.~A. Eales,$^{4}$
A. Enia,$^{20,21}$
D. Farrah,$^{22,23}$
J. Gonzalez-Nuevo,$^{24,25}$
D.~H. Hughes,$^{26}$
\newauthor D. Ismail,$^{27}$
S. Jin,$^{28,29}$ 
A. Lapi,$^{30}$
M.~D. Lehnert,$^{31}$ 
R. Neri,$^{13}$ 
I.~P\'{e}rez-Fournon,$^{17,18}$ 
G. Rodighiero,$^{1,5}$ 
\newauthor D. Scott,$^{32}$  
S. Serjeant,$^{33}$  
F. Stanley,$^{34}$ 
S. Urquhart,$^{33}$  
P. van der Werf,$^{35}$  
M. Vaccari,$^{3,36,37}$ 
L. Wang,$^{38,39}$  
\newauthor C. Yang,$^{34}$
A. Young$^{8}$
\\
\\
\emph{\normalsize Affiliations are listed at the end of the paper}}

\date{Accepted XXX. Received YYY; in original form ZZZ}

\pubyear{2022}

\begin{document}

\def\simlt{\mathrel{\rlap{\lower 3pt\hbox{$\sim$}}\raise 2.0pt\hbox{$<$}}}
\def\simgt{\mathrel{\rlap{\lower 3pt\hbox{$\sim$}}\raise 2.0pt\hbox{$>$}}}

\label{firstpage}
\pagerange{\pageref{firstpage}--\pageref{lastpage}}
\maketitle

\begin{abstract}
We have carried out \textit{HST} snapshot observations at 1.1 $\mu$m of 281 candidate strongly lensed galaxies identified in the wide-area extragalactic surveys conducted with the \textit{Herschel space observatory}. Our candidates comprise systems with flux densities at $500\,\mu$m $S_{500}\geq 80$ mJy.  We model and subtract the surface brightness distribution for 130 systems, where we identify a candidate for the foreground lens candidate. After combining visual inspection, archival high-resolution observations, and lens subtraction, we divide the systems into different classes according to their lensing likelihood. We confirm 65 systems to be lensed. Of these, 30 are new discoveries. We successfully perform lens modelling and source reconstruction on 23 systems, where the foreground lenses are isolated galaxies and the background sources are detected in the \textit{HST} images. All the systems are successfully modelled as a singular isothermal ellipsoid. The Einstein radii of the lenses and the magnifications of the background sources are consistent with previous studies. However, the background source circularised radii (between 0.34 kpc and 1.30 kpc) are $\sim$3 times smaller than the ones measured in the sub-mm/mm for a similarly selected and partially overlapping sample. We compare our lenses with those in the SLACS survey, confirming that our lens-independent selection is more effective at picking up fainter and diffuse galaxies and group lenses. This sample represents the first step towards characterising the near-IR properties and stellar masses of the gravitationally lensed dusty star-forming galaxies.

\end{abstract}




\section{Introduction}

The very bright tail of the number counts of galaxies at sub-millimetre and millimetre wavelengths comprises a mixture of distinct galaxy populations: low-redshift ($z\simlt0.1$) late-type galaxies, flat spectrum radio sources, high-redshift ($z\simgt1$) gravitationally-lensed dusty star-forming galaxies (DSFGs), and hyper-luminous infrared galaxies (HyLIRGs) \citep{Negrello2007, Negrello2010, Negrello2017, Wardlow2013, Vieira2013, Nayyeri2016, Rowan-Robinson2018, Ward2022}.
This distribution allows us to efficiently select strong gravitational lensing systems by combining a flux density cut (e.g., $S_{500}\geq100$\,mJy at 500\,$\mu$m) with shallow optical and radio surveys to remove the contaminants. This approach became effective in the last decade thanks to the wide-area surveys carried out, in particular, by the \textit{Herschel Space Observatory} \citep[][]{Pilbratt2010} and the South Pole Telescope \citep[SPT; e.g.,][]{Vieira2010, Carlstrom2011, Mocanu2013}. 
\citet{Negrello2017} and then \citet{Ward2022} identified 80$+$11 candidate lensed galaxies with $S_{\rm 500}\geq100\,$mJy in the $>600\,$deg$^{2}$ of the \textit{Herschel}-Astrophysical Terahertz Large Area Survey \citep[H-ATLAS;][]{Eales2010}. \citet{Bakx2018} extended the selection down to $S_{\rm 500}=80\ {\rm{mJy}}$ by including DSFGs with photometric redshift $z_{\rm phot}>2$, as derived from the sub-mm colours (the \textit{Herschel} bright sources -- HerBS -- sample). At flux densities $S_{500}<100$\,mJy, the number density of unlensed DSFGs exponentially increases (see \citealt{Negrello2007, Cai2013} for details); therefore, the HerBS sample is expected to contain a mixture of both lensed and unlensed DSFGs, with the latter dominating over the former at $S_{\rm 500\,\mu m}\simlt100\,{\rm mJy}$. \citet{Wardlow2013} identified 42 candidate high-$z$ lensed DSFGs with $S_{\rm 500}\geq80\ {\rm{mJy}}$ in the $95\,{\rm deg}^{2}$ of the \textit{Herschel} Multi-tiered Extragalactic Survey \citep[HerMES,][]{Oliver2012}, while \citet{Nayyeri2016} published a catalogue of 77 candidate lensed DSFGs with $S_{\rm 500}\geq100\,{\rm mJy}$ in the $300\,{\rm deg}^{2}$ of the HerMES Large Mode Survey \citep[HeLMS,][]{Oliver2012} and the $79\,{\rm deg}^{2}$ of the \textit{Herschel} Stripe 82 Survey \citep[HerS;][]{Viero2014}. 
Following a different approach, \citet{Gonzalez-nuevo2012, Gonzalez-nuevo2019} lowered the SPIRE flux density cut ($S_{350}\geq85\, {\rm{mJy}}$ at $350\, {\rm{\mu m}}$ and $S_{250}\geq35\ {\rm{mJy}}$ at $250\, {\rm{\mu m}}$) constraints on both the Herschel-SPIRE colours and the presence of close-by near-infrared (near-IR) sources acting as potential lensing galaxies, increasing the number of strong lens candidates by a factor of $\sim5$. 

Using a similar approach to \citet{Negrello2010} but at mm wavelengths, the SPT collaboration produced a sample of 48 candidate lensed galaxies with deboosted $S_{\rm 1.4}\geq20\,{\rm mJy}$ at 1.4 mm over an area of $2500\,{\rm deg}^{2}$ \citep[][]{Vieira2013, Spilker2016, Everett2020, Reuter2020, Cai2022}.
Lastly, \citet{Canameras2015} applied a combination of flux density cut and colour selection on candidates extracted from the \textit{Planck} Catalogue of Compact Sources, identifying 11 lensed DSFGs. Later, \citet{Harrington2016}, \citet{Berman2022}, and \citet{Kamieneski2023} identified 30 strongly lensed galaxies by cross-matching the \textit{Planck} catalogue, the \textit{WISE} All-Sky Survey, and the \textit{Herschel} surveys. Three and two of these systems were also included in the \citet{Wardlow2013} and \citet{Negrello2017} samples, respectively.

To confirm whether a sub-mm/mm bright galaxy is gravitationally lensed, high-resolution follow-up observations are needed. The multiple images of a background source, which are the distinctive features of lensing, have typical separations of a few arcseconds. They can not be resolved by either {\it Herschel} (FWHM\,$\simeq$\,18,\,24,\,and\,35\,arcsec at 250,\,350,\,and\,500\,$\mu$m, respectively) or SPT (FWHM\,$\simeq$\,1\,arcmin at 1.4 mm). Because lensed DSFGs emit mostly in the far-IR/sub-mm/mm \citep{Negrello2014}, the best way to detect and characterise the multiple images is via high-angular resolution observations at sub-mm/mm wavelengths obtained with interferometers, such as the Atacama Large Millimetre Array (ALMA), the Submillimetre Array (SMA), and the Northern Extended Millimetre Array (NOEMA) \citep[e.g.,][]{Bussmann2013, Amvrosiadis2018}. These data can then be used to reconstruct the intrinsic morphology of the background DSFGs via lens modelling techniques \citep[e.g.,][]{Dye2018, Maresca2022}, which provide crucial information on the spatial distribution of the gas and dust and on the star-formation rate surface density in those galaxies \citep[e.g.,][]{Canameras2017, Yang2019, Sun2021, Jarugula2021, Dye2022}. Interestingly, at these wavelengths, the object acting as a lens, which is usually a massive red-and-dead foreground elliptical galaxy with very low dust content, remains undetected, thus facilitating the source reconstruction.
However, it is not possible to constrain all the physical properties of the lensed DSFGs with the long wavelength data alone. Indeed, one parameter that remains elusive is the background source stellar mass, which is crucial to understand better the evolutionary stage of a galaxy \citep[e.g.,][]{Renzini2009}. Constraining the stellar mass of the background sources requires them to be detected at optical/near-IR wavelengths, where the emission of the foreground lens dominates \citep[e.g.,][]{Negrello2014}. Hence, careful lens photometric modelling and subtraction are needed to reveal the lensed background sources in the high-resolution optical/near-IR images. This analysis represents the first step to constrain the stellar masses by fitting the background source spectral energy distribution \citep[SED,][]{Negrello2014}.

So far, high-resolution near-IR follow-up observations of tens of sub-mm/mm selected candidate lensed galaxies have been obtained in the near-IR with the \textit{Hubble Space Telescope} \citep[\textit{HST}; e.g.,][]{Negrello2014} and with ground-based telescopes exploiting adaptive optics \citep[e.g.,][]{Fu2012, Calanog2014, Messias2014}. These observations are characterised by long integration times (i.e., from tens of minutes to hours) and have been mainly aimed at identifying and studying the background source. Instead, we report on \textit{HST} snapshot observations of a much larger sample of 281 sub-mm bright {\it Herschel}-selected systems. The primary goal of these short observations is to efficiently confirm the gravitational lensing nature of the sub-mm selected systems either by detecting the characteristic lensing features, such as arcs or multiple images, or by identifying massive low-$z$ early-type galaxies (ETGs) located at the position of the sub-mm emission and thus potentially acting as the lens. These observations represent the first step to enable more detailed follow-ups to study the background sources and, at the same time, provide data to study the properties of the lenses.

This paper is organized as it follows. Section~\ref{sec2} presents the full sample of candidates we have followed up with \textit{HST} and details the \textit{HST} snapshot observations. Section~\ref{sec3} discusses the visual classification of the \textit{HST} images. In Section~\ref{sec4}, we focus on the sub-sample of systems that show clear evidence of lensing features in the \textit{HST} images and describe the fitting techniques adopted for modelling the surface brightness of the lenses. Section~\ref{sec5} presents the lens modelling we apply and the results for a subsample of candidates. Section~\ref{sec6} discusses our results, focusing on the comparison with the Sloan Lens ACS Survey (SLACS), while the main conclusions are summarised in Section~\ref{sec7}.
In this paper we adopt the values of $H_0=67.7$ km s$^{-1}$ Mpc$^{-1}$, $\Omega_{0,{\rm m}}=0.31$, and $\Omega_{0,\Lambda}=0.69$ from Planck18 cosmology (see \citealt{PC2020}\footnote{Available in the \texttt{astropy.cosmology} package through \texttt{astropy} see, \citealt{Astropy2013, Astropy2018}).} for details).

\section{Sample, observations, and data reduction}
\label{sec2}

\subsection{Sample selection}
\label{subsec_2.1}
Our sample consists of candidate $z\simgt1$ (gravitationally lensed and unlensed) DSFGs extracted from the {\it Herschel} wide-area extragalactic surveys (i.e., H-ATLAS, HerMES, HeLMS, and HerS) for which we obtained the \textit{HST} snapshot observations described in the following section. Many of our targets are selected from \citet{Bussmann2013}, \citet{Wardlow2013}, \citet{Calanog2014}, \citet{Nayyeri2016}, \citet{Negrello2017}, and \citet{Bakx2018}. Due to the varying quality of {\it Herschel} and ancillary data available at the time of these different works, they could select candidates only in specific sky areas and applied slightly different selection criteria to identify their most reliable candidate lensed system. Nevertheless, these selections generally relied upon a bright flux density cut at 500\,$\mu$m following the approach by \citet{Negrello2010}. In summary, the objects we observed with \textit{HST} were identified as candidate lensing systems by having either the flux density measured at 500 $\mu$m $S_{500}\simgt100\,$mJy (as done in \citealt{Wardlow2013}, \citealt{Nayyeri2016}, \citealt{Negrello2017}) or by having $S_{500}\simgt80\,$mJy and photometric redshifts of the potential background source $z>2$ (as the candidates presented in \citealt{Bakx2018}). Exceptions are the candidates from \cite{Calanog2014} that were selected by means of their bright ($S_{500}\simgt80\,$mJy) SPIRE 500\,$\mu$m flux densities, but also thanks to multiwavelength observations and specific source-extraction techniques applied on deeper {\it Herschel} maps available only for the HerMES fields. 
We refer the reader to \citet{Calanog2014} for more details. As a result, \citet{Calanog2014} candidates include targets with flux densities at 500\,$\mu$m below the flux limit used in the other quoted works.
For the candidates in the H-ATLAS survey, we use the flux densities coming from \citet{Harris2012}, \citet{Negrello2017}, and \citet{Bakx2018}, when available, if not, we adopt the flux densities published in the H-ATLAS Data Release 1 \citep[DR1,][]{Valiante2016} or Data Release 2 \citep[DR2,][]{Maddox2018}. The candidates belonging to the other surveys were covered by the catalogues of \citet{Wardlow2013}, \citet{Bussmann2013, Bussmann2015}, \citet{Calanog2014}, and \citet{Nayyeri2016}.

The goal of our \textit{HST} observations is to observe as many candidates as possible to confirm their nature by means of the higher-resolution \textit{HST} imaging and thus increase the number of confirmed {\it Herschel}-selected lensing systems to allow for better statistical studies. Over the years, we have been awarded \textit{HST} snapshot time to follow up on our targets. However, since the time granted was not enough to observe the full list of {\it Herschel}-selected lensing systems candidates, and because of the filler nature of the snapshot program, which only guarantees a partial completion rate, we had to prioritise those candidates with a higher probability of being lensed to maximise our success rate. This resulted in prioritising the brightest sources at 500\,$\mu$m (i.e., the higher the flux density, the higher the probability for the system to be lensed) or those having other multi-wavelength data and/or redshift information. The success rate of the \textit{HST} observations was entirely determined by the \textit{HST} observing schedule and does not depend on the candidate properties. In the end, only a fraction (290 out of 398) of the full sample has been observed, as detailed in the following section, which is the focus of this paper. Given that this sub-sample represents a random collection of the entire candidate population, we limit our focus on presenting the properties of this sub-sample without drawing any statistical conclusion on the parent sample, which would otherwise be biased by observational constraints that were out of our control.
\begin{figure*}
    \centering
    \includegraphics[width=0.95\textwidth]{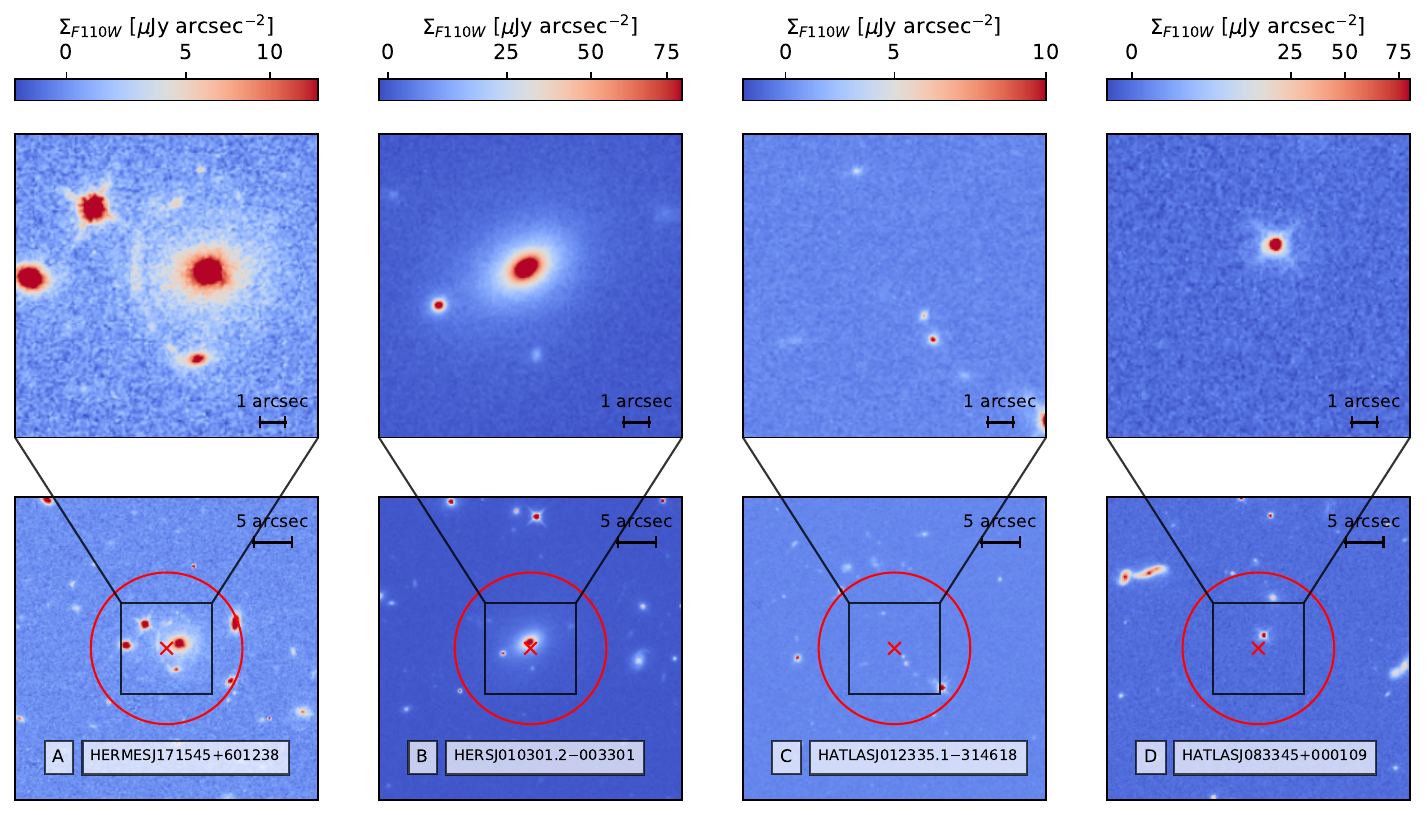}
    \caption{\textit{From left to right panels:} Examples of sample systems assigned to classes A, B, C, and D, respectively. In each column, the bottom panel shows a $20\, {\rm arcsec}\times\,20\, {\rm arcsec}$ cutout of the \textit{HST} image, while the top panel shows a $12\, {\rm arcsec} \times\,12\, {\rm arcsec}$ zoom-in on the central source. The cross marks the position of the {\it Herschel} detection used to identify our targets, while the circle roughly corresponds to the $\sim20\,{\rm arcsec}$ full width at half maximum (FWHM) of the {\it Herschel}/SPIRE beam at $250\ {\rm{\mu m}}$. The images are oriented with N up and E to the left.}
    \label{fig:vis_class}
\end{figure*}

\subsection{\textit{HST} snapshot observations and data reduction}
\label{subsec_2.2}

Our \textit{HST} snapshot follow-up observations were carried out in three different cycles.
\begin{itemize}
    \item Proposal ID 12488 (PI M. Negrello) in Cycle 19: 200 targets were proposed with $S_{500\rm}\simgt85\,$mJy, that were selected from the preliminary source catalogues of HerMES (63 targets) and H-ATLAS (137 targets). At the time, the only {\it Herschel} data available were for four of the final five H-ATLAS fields (the three fields on the Celestial Equator named GAMA09, GAMA12 and GAMA15 and the one close to the North Galactic Pole named NGP), so the targets were spread across the Equatorial and Northern sky. The observations were carried out from October 2011 to August 2013.
    \item Proposal ID 15242 (PI L. Marchetti), in Cycle 25: 200 targets were proposed with $S_{500}\geq100\,$mJy from the full {\it Herschel} coverage, but in particular from HeLMS and HerS \citep[][]{Nayyeri2016} and with $S_{500}\geq80\,$mJy from the H-ATLAS SGP field near the South Galactic Pole. The sample also included some of the targets submitted for observations in the Cycle 19 proposal that had not been observed or for which the \textit{HST} data obtained were corrupted due to issues with the \textit{HST} tracking system. The observations were carried out from October 2017 to June 2018.
    \item Proposal ID 16015 (PI L. Marchetti) in Cycle 26: this was a continuation of the Cycle 25 proposal. The observations were carried out from November 2019 to September 2020.
\end{itemize}
In summary, a total of 290 candidates were observed as part of these three snapshot programmes: 173 from H-ATLAS, 60 from HerMES, 42 from HeLMS, and 15 from HerS. Unfortunately, the data of nine systems are corrupted due to issues with the \textit{HST} tracking system, so the final sample with available \textit{HST} snapshot observations amount to 281 systems. All the \textit{HST} snapshot proposals share the same observing strategy with short observations of $\sim\,$4 or $\sim\,$11 minutes on source performed with the wide-\textit{YJ} filter \textit{F110W} of the Wide Field Camera 3 (WFC3) IR channel \citep{Dressel2022}. The \textit{F110W} filter has a pivot wavelength $\lambda_{\rm p} = 1153.4$ nm and a passband rectangular width of 443.0 nm, corresponding to a wavelength coverage between 883.2 nm and 1412.1 nm. This observing band was chosen to maximise the signal-to-noise ratio (SNR) for short exposures and efficiently cover the stellar emission from the foreground galaxy in the gravitational lensing system, usually an ETG \citep[e.g.,][]{Auger2009A}.  
The IR channel of the WFC3 allows to read out the detector with different samplings while the exposure accumulates. The sampling strategy varies depending on the specific scientific case. Our targets were observed with the \texttt{MULTIACCUM} observing mode using either eight samplings with 10-s sampling intervals linearly spaced between one another (\texttt{SPARS10} sampling sequence) for a total exposure time of 251.75$\,$s, or nine samplings with 25-s sampling intervals linearly spaced between one another (\texttt{SPARS25} sampling sequence) for a total exposure time of 711.74$\,$s. Due to the filler nature of the snapshot programmes, the total exposure time was defined according to the available time slot. Each set of observations is flatfield corrected, background subtracted, cosmic-ray cleaned, distortion corrected, rotated and drizzled. For drizzling, we use a 4-point sub-pixel dithering pattern that results in an output pixel size of $0.064\ {\rm{arcsec}}$, which is roughly half of the original pixel size. The size of the field of view (FOV) is $136\, {\rm arcsec} \times\,123\, {\rm arcsec}$. The reduction steps are performed with the \texttt{AstroDrizzle} package \citep{Hoffmann2021}. Additionally, we perform a first-order astrometric correction by matching the \textit{Gaia} early Data Release 3 sources \citep{Gaia2021} with their nearest \textit{HST} counterparts.

\subsection{Multiwavelength data}
\label{subsec_2.3}

We use archival ALMA, SMA, and NOEMA high-resolution multiwavelength observations for 77 systems and include them in our analysis. For the ALMA data, we download the Stokes I continuum images from the science archive\footnote{\url{https://almascience.eso.org/aq/?result_view=observations}.}. For the SMA data, we either use the reduced continuum images from \citet{Enia2018}, or we download the UV tables from the science archive\footnote{\url{https://lweb.cfa.harvard.edu/cgi-bin/sma/smaarch.pl}.} and we produce the continuum images with the Common Astronomy Software Applications\footnote{Available at \url{https://casa.nrao.edu/}.} (\texttt{CASA}) software. 
HerBS-89a is the only source for which we use NOEMA data. The details of the data reduction are available in \citet[][but see \citealt{Berta2021} for a detailed study of this galaxy]{Neri2020}.
These multiwavelength observations have an angular resolution ranging from $1$ to $0.03$ arcsec for ALMA, $0.5$ arcsec for SMA, and $0.3$ arcsec for NOEMA. More details on the instrumental setup and references will be given in the source-by-source description in Sec.~\ref{sec4.3}. 
In addition to these high-resolution observations, we search both previous literature works and the NASA/IPAC Extragalactic Database\footnote{Available at \url{https://ned.ipac.caltech.edu/}} for the lens candidates and/or potential lenses redshifts. The details on the redshifts and their references are available in the source-by-source description and Table~\ref{tab:z}.

\section{Lens classification}
\label{sec3}
We assign each one of the 281 sources in our sample to a class that describes the likelihood of the source being a strong lensing system:
\begin{itemize}
    \item Class A: the source is confirmed to be strongly lensed 
    because it satisfies one or both of the following criteria: (i) the \textit{HST} image shows a near-IR object located close to the position of the {\it Herschel} detection, while lensing features (such as arcs and/or multiple images) are visible either before or only after the subtraction of the near-IR source (supposedly acting as a lens), (ii) lensing features are clearly detected in high spatial resolution sub-mm/mm ancillary data described in Sec.~\ref{subsec_2.3};
    \item Class B: a single galaxy or a group of galaxies is visible close to the position of the {\it Herschel} detection, but there is no detection of lensing features even after the subtraction of the foreground galaxy. Therefore, this class tags objects that are likely to be a lens, but the background source is probably too faint to be detected in the relatively shallow \textit{HST} data;
    \item Class C : no source is detected in the \textit{HST} image approximately within the $\sim20$ arcsec SPIRE FWHM at 250 $\mu m$ of the position of the {\it Herschel} detection. These cases are likely to be either high-$z$ unlensed DSFGs or lensing events where the lens is either at high redshift or has an intrinsically low luminosity, or both;
    \item Class D: the source is a contaminant, e.g., a dusty star, an unlensed QSO, or a low redshift, sub-mm bright, spiral galaxy (previously misinterpreted as a lens candidate).
\end{itemize}
The adopted classes are similar to those introduced by \citet{Negrello2017}. The main difference is our use of the information provided by the \textit{HST} images after the modelling and subtracting the detected near-IR source. This mainly affects the definition of class A [criterion (i)] and class B.

In order to allocate the 281 objects to the four classes, we proceed by first carrying out a visual inspection of the \textit{HST} data alone, without performing any galaxy subtraction and without relying on any multiwavelength ancillary data.
Based on this simple visual analysis, we are able to assign 25 objects to class A, 105 to class B, 146 to class C, and 5 to class D. Four examples extracted from the A, B, C and D classes are shown in Fig~\ref{fig:vis_class}.

As a second step in the classification, we exploit the available high spatial resolution sub-mm/mm data to look for evidence of multiple images and arcs. In this way, we are able to identify 31 more lensed objects (promoting them to A class) that have been previously classified as B or C and 26 individual unlensed DSFGs or overdensities of unlensed DSFGs. 
In summary, after the use of ancillary data, the number of objects assigned to class A increases from 25 to 56 and the number of systems in class B and C decreases to 76 and 120, respectively. In contrast, that of objects in class D increases from 5 to 29.  

Finally, as described in detail in Sec.~\ref{sec4}, in the \textit{HST} images, we perform, if possible, the modelling of the surface brightness of the foreground galaxies and subtract it from the \textit{HST} image to reveal any lensing feature. In this way, we are able to identify 9 new lensed galaxies that have been previously assigned to class B. 

In conclusion, after the visual inspection of the \textit{HST} images, use of ancillary data, and surface brightness modelling, we are able to assign 65 objects to class A, 67 to class B, 120 to class C, and 29 to class D. The systems classified as A are listed in Table~\ref{tab:source_table}, those classified as D are listed in Table~\ref{tab:unlensed}, while the rest of the sample is listed in Table~\ref{tab:full_candidate_table}.

\begin{landscape}
\begin{table}
\centering
\caption{Properties of the 65 systems classified as A by visual inspection, through multiwavelength follow-up observations, or after the lens subtraction.}
\label{tab:source_table}

{\scriptsize


\begingroup
\setlength{\tabcolsep}{8pt}
\renewcommand{\arraystretch}{0.9}
\begin{tabular}{c l l l c c c c c c c c c l}
\hline
\hline
	 No. & IAU name & Alt. Name & Ref. & RA & Dec & Vis. Class & Multiw. Obs. & Prev. Classification & Ref. & $S_{250}$ & $S_{350}$ & $S_{500}$ & Ref.  \\ 
	   &   &   &   & [h m s] & [d m s] &   &   &   &   & [mJy] & [mJy] & [mJy] &    \\ 
	 (1) & (2) & (3) & (4) & (5) & (6) & (7) & (8) & (9) & (10) & (11) & (12) & (13) & (14)  \\ 
\hline
	 S\_1 & HATLASJ000330.6$-$321136 & HERBS155 & Ba18 & 00:03:31 & $-$32:11:36.00 & A & -- & -- & -- & $59.9\pm5.8$ & $94.2\pm5.8$ & $85.6\pm7.2$ & Ba18  \\ 
	 S\_2 & HATLASJ000912.7$-$300807 & SD.v1.70 & Zh18 & 00:09:13 & $-$30:08:07.00 & A & -- & -- & -- & $352.8\pm5.4$ & $272.6\pm6.1$ & $156.1\pm6.8$ & Ne17  \\ 
	 S\_3 & HELMSJ001353.5$-$060200 & HELMS31 & Na16 & 00:13:54 & $-$06:02:00.00 & A & -- & -- & -- & $178.0\pm7.0$ & $176.0\pm6.0$ & $120.0\pm7.0$ & Na16  \\ 
	 S\_4 & HELMSJ003619.8$+$002420 & HELMS14 & Na16 & 00:36:20 & $+$00:24:20.00 & A & -- & -- & -- & $251.0\pm6.0$ & $247.0\pm6.0$ & $148.0\pm7.0$ & Na16  \\ 
	 S\_5 & HELMSJ005841.2$-$011149 & HELMS23 & Na16 & 00:58:41 & $-$01:11:49.00 & A & -- & -- & -- & $391.0\pm7.0$ & $273.0\pm6.0$ & $126.0\pm8.0$ & Na16  \\ 
	 S\_6 & HERSJ011722.3$+$005624 & HERS10 & Na16 & 01:17:22 & $+$00:56:24.00 & A & -- & -- & -- & $105.0\pm6.0$ & $125.0\pm6.0$ & $117.0\pm7.0$ & Na16  \\ 
	 S\_7 & HERSJ012620.5$+$012950 & HERS5 & Na16 & 01:26:21 & $+$01:29:50.00 & A & -- & -- & -- & $268.0\pm8.0$ & $228.0\pm7.0$ & $133.0\pm9.0$ & Na16  \\ 
	 S\_8 & HERSJ020941.2$+$001558 & 9io9 & Ge15 & 02:09:41 & $+$00:15:58.00 & A & A & Lensed & Ge15 & $826.0\pm7.0$ & $912.0\pm7.0$ & $718.0\pm8.0$ & Na16  \\ 
	   &   &  HERS1 & Na16 &   &   &   &   &   & Hr16  &   &   &   &    \\ 
	   &   &  PJ020941.3 & Hr16  &   &   &   &   &   & Li22  &   &   &   &    \\ 
	   &   & ACT-S J0210$+$0016 & Su17 &   &   &   &   &   &   &   &   &   &    \\ 
	 S\_9 & HERMESJ032637$-$270044 & HECDFS05 & Wa13 & 03:26:36 & $-$27:00:44.00 & A & -- & Lensed & Ca14 & $155.0$ & $131.0$ & $84.0$ & Ca14  \\ 
	 S\_10 & HERMESJ033732$-$295353 & HECDFS02 & Wa13 & 03:37:32 & $-$29:53:53.00 & A & -- & Lensed & Wa13 & $133.0$ & $147.0$ & $122.0$ & Ca14  \\ 
	   &   &   &  &   &   &   &   &   & Ca14  &  &   &   &   \\ 
	 S\_11 & HATLASJ083051$+$013225 & G09v1.97 & Bu13 & 08:30:51 & $+$01:32:24.87 & A & A & Lensed & Bu13 & $248.5\pm7.5$ & $305.3\pm8.1$ & $269.1\pm8.7$ & Ne17  \\ 
	   &   &  HERBS4 & Ba18 &   &   &   &   &   & Ca14  &  &   &   &   \\ 
	   &   &   &  &   &   &   &   &   & Ne17  &  &   &   &   \\ 
	   &   &   &  &   &   &   &   &   & Am18  &  &   &   &    \\ 
      &   &   &  &   &   &   &   &   & En18  &  &   &   &    \\ 
	   &   &   &  &   &   &   &   &   & Ya19  &  &   &   &    \\ 
	   &   &   &  &   &   &   &   &   & Ma22  &  &   &   &    \\ 
	 S\_12 & HERMESJ100144$+$025709 & HCOSMOS01 & Ca14 & 10:01:44 & $+$02:57:08.62 & A & A & Lensed & Ca14 & $86.0\pm6.0$ & $96.0\pm6.0$ & $71.0\pm6.0$ & Bu15  \\ 
  	   &   &   &  &   &   &   &   &   & Bu15  &   &   &   &   \\ 
	 S\_13 & HERMESJ103827$+$581544 & HLock04 & Wa13 & 10:38:27 & $+$58:15:43.60 & A & A & Lensed & Bu13 & $190.0$ & $156.0$ & $101.0$ & Ca14  \\ 
  	   &   &   &  &   &   &   &   &   & Wa13  &   &   &   &    \\ 
      &   &   &  &   &   &   &   &   & Ca14  &   &   &   &    \\ 
	 S\_14 & HERMESJ110016$+$571736 & HLock12 & Ca14 & 11:00:16 & $+$57:17:35.92 & A & -- & Lensed & Ca14 & $224.0$ & $159.0$ & $79.0$ & Ca14  \\ 
	   &   &   &  &   &   &   &   &   & Am18  &   &   &   &    \\ 
	 S\_15 & HATLASJ114638$-$001132 & G12v2.30 & Bu13 & 11:46:38 & $-$00:11:32.00 & A & A & Lensed & Bu13 & $316.0\pm6.6$ & $357.9\pm7.4$ & $291.8\pm7.7$ & Ne17  \\ 
	   &   &  HERBS2 &   &   &   &   &   &   & Om13  &   &   &   &    \\ 
	   &   &    &   &   &   &   &   &   & Ca14  &   &   &   &    \\ 
      &   &    &   &   &   &   &   &   & Ne17  &   &   &   &    \\ 
      &   &    &   &   &   &   &   &   & Am18  &   &   &   &    \\ 
	 S\_16 & HATLASJ125126$+$254928 & HERBS52 & Ba18 & 12:51:26 & $+$25:49:28.31 & A & A & -- & -- & $57.4\pm7.4$ & $96.8\pm8.2$ & $109.4\pm8.8$ & Ne17  \\ 
	 S\_17 & HATLASJ125760$+$224558 & -- & -- & 12:58:00 & $+$22:45:57.82 & A & -- & -- & -- & $272.4\pm7.3$ & $215.0\pm8.1$ & $137.8\pm8.7$ & Ne17  \\ 
	 S\_18 & HATLASJ133008$+$245900 & NBv1.78 & Bu13 & 13:30:08 & $+$24:58:59.70 & A & A & Lensed & Hs12 & $271.2\pm7.2$ & $278.2\pm8.1$ & $203.5\pm8.5$ & Ne17  \\ 
	   &   &  HERBS12 & Ba18 &   &   &   &   &   & Bu13  &   &   &   &    \\
	   &   &          &      &   &   &   &   &   & Ca14  &   &   &   &    \\
      &   &          &      &   &   &   &   &   & Ne17  &   &   &   &    \\
      &   &          &      &   &   &   &   &   & En18  &   &   &   &    \\
	 S\_19 & HATLASJ133846$+$255057 & HERBS29 & Ba18 & 13:38:46 & $+$25:50:56.84 & A & A & Lensed & Ne17 & $159.0\pm7.4$ & $183.1\pm8.2$ & $137.6\pm9.0$ & Ne17  \\ 
  	   &   &   &  &   &   &   &   &   & Am18  &   &   &   &   \\ 
	 S\_20 & HATLASJ142935$-$002837 & G15v2.19 & Ca14 & 14:29:35 & $-$00:28:37.00 & A & A & Lensed & Ca14 & $801.8\pm6.6$ & $438.5\pm7.5$ & $199.8\pm7.7$ & Ne17  \\ 
	   &   & H1429$-$0028  & Me14 &   &   &   &   &   & Me14  &   &   &   &   \\
	   &   &   &  &   &   &   &   &   & Ne17  &   &   &   &    \\ 
	   &   &   &  &   &   &   &   &   & Am18  &   &   &   &    \\ 
	   &   &   &  &   &   &   &   &   & Dy18  &   &   &   &    \\ 
	 S\_21 & HERMESJ171451$+$592634 & HFLS02 & Wa13 & 17:14:51 & $+$59:26:34.12 & A & -- & Lensed & Ca14 & $164.0$ & $148.0$ & $87.0$ & Ca14  \\ 
	 S\_22 & HERMESJ171545$+$601238 & HFLS08 & Ca14 & 17:15:45 & $+$60:12:38.34 & A & -- & Lensed & Ca14 & $86.0$ & $93.0$ & $67.0$ & Ca14  \\ 
	 S\_23 & HATLASJ225844.7$-$295124 & HERBS26 & Ba18 & 22:58:45 & $-$29:51:25.00 & A & -- & -- & -- & $175.4\pm5.2$ & $187.0\pm5.9$ & $142.6\pm7.5$ & Ne17  \\ 
	 S\_24 & HELMSJ232210.3$-$033559 & HELMS19 & Na16 & 23:22:10 & $-$03:35:59.00 & A & -- & -- & -- & $114.0\pm6.0$ & $160.0\pm7.0$ & $134.0\pm8.0$ & Na16  \\ 
	 S\_25 & HATLASJ233037.2$-$331217 & HERBS123 & Ba18 & 23:30:37 & $-$33:12:18.00 & A & -- & -- & -- & $106.2\pm5.9$ & $107.9\pm6.0$ & $90.0\pm7.5$ & Ba18  \\

\hline
\hline
\end{tabular}
\endgroup

\begin{flushleft}
\textit{Notes}: Col.~(1): Source reference number. Col.~(2): IAU name of the {\it Herschel} detection. Cols.~(3) and (4): Alternative name and reference. Cols.~(5) and (6): ICRS RA and Dec coordinate (J2000.0) of the {\it Herschel} detection. Col.~(7): Classification after the visual inspection [criterium (i)]. Col.~(8): Lens classification including the multiwavelength observations. Cols.~(9) and (10): Confirmed lensed system and reference. Col.~(11), (12), (13), and (14): SPIRE flux density at 250 $\mu {\rm m}$, 350 $\mu {\rm m}$, 500 $\mu {\rm m}$ and reference. Following are the abbreviations used for the references:
Al17: \citet{Albareti2017};
Am18: \citet{Amvrosiadis2018};
Ba18: \citet{Bakx2018};
Be21: \citet{Berta2021};
Bo06: \citet{Borys2006};
Bu13: \citet{Bussmann2013};
Bu15: \citet{Bussmann2015}; 
Bu21: \citet{Butler2021};
Ca14: \citet{Calanog2014}; 
Co11: \citet{Cox2011};
Dy18: \citet{Dye2018};
Dy22: \citet{Dye2022};
En18: \citet{Enia2018};
Fa17: \citet{Falgarone2017};
Ga11: \citet{Gavazzi2011};
Ge15: \citet{Geach2015};
Gi23: \citet{Giulietti2023};
Hs12: \citet{Harris2012}; 
Hr16: \citet{Harrington2016};
Ik11: \citet{Ikarashi2011};
Ma19: \citet{Ma2019};
Ma22: \citet{Maresca2022};
Me14: \citet{Messias2014};
Na16: \citet{Nayyeri2016}; 
Na17: \citet{Nayyeri2017}; 
Ne17: \citet{Negrello2017}; 
Om13: \citet{Omont2013}; 
Su17: \citet{Su2017};
Sh21: \citet{Shirley2021};
Ur22: \citet{Urquhart2022};
Va16: \citet{Valiante2016};
Wa13: \citet{Wardlow2013};
Ya17: \citet{Yang2017};
Ya19: \citet{Yang2019}.
\end{flushleft}
}

\end{table}
\end{landscape}

\begin{landscape}
\begin{table}
\centering
\contcaption{}
{\scriptsize
\begingroup
\setlength{\tabcolsep}{8pt}
\renewcommand{\arraystretch}{0.9}
\begin{tabular}{c l l l c c c c c c c c c l}
\hline
\hline
	 No. & IAU name & Alt. name & Ref. & RA & Dec & Vis. class & Multiw. class & Prev. classification & Ref. & $S_{250}$ & $S_{350}$ & $S_{500}$ & Ref.  \\ 
	   &   &   &   & [h m s] & [d m s] &   &   &   &   & [mJy] & [mJy] & [mJy] &    \\ 
	 (1) & (2) & (3) & (4) & (5) & (6) & (7) & (8) & (9) & (10) & (11) & (12) & (13) & (14)  \\ 
\hline

\\
\multicolumn{14}{l}{\textit{Confirmed after the lens subtraction}} \\
\\

	 S\_26 & HELMSJ001626.0$+$042613 & HELMS22 & Na16 & 00:16:26 & $+$04:26:13.00 & B & A & Lensed & Am18 & $130.0\pm15.0$ & $180.0\pm18.0$ & $130.0\pm15.0$ & Zh18  \\
	   &   &   &  &   &   &   &   &   & Dy18  &   &   &   &   \\  
	 S\_27 & HATLASJ002624.8$-$341737 & HERBS22 & Ba18 & 00:26:25 & $-$34:17:38.00 & B & -- & -- & -- & $137.7\pm5.2$ & $185.9\pm5.8$ & $148.8\pm6.8$ & Ne17  \\ 
	 S\_28 & HELMSJ004723.6$+$015751 & HELMS9 & Na16 & 00:47:24 & $+$01:57:51.00 & C & A & Lensed & Am18 & $398.0\pm6.0$ & $320.0\pm6.0$ & $164.0\pm8.0$ & Na16  \\
	   &   &   &  &   &   &   &   &   & Dy18  &   &   &   &   \\   
  S\_29 & HERSJ012041.6$-$002705 & HERS2 & Na16 & 01:20:42 & $-$00:27:05.00 & B & -- & -- & -- & $240\pm6$ & $260\pm6$ & $189\pm7$ & Na16  \\ 
	 S\_30 & HATLASJ085112$+$004934 & -- & -- & 08:51:12 & $+$00:49:33.83 & B & -- & -- & -- & $125.1\pm7.3$ & $118.3\pm8.3$ & $77.7\pm8.8$ & Va16  \\ 
	 S\_31 & HATLASJ085359$+$015537 & G09v1.40 & Bu13 & 08:53:59 & $+$01:55:37.21 & B & A & Lensed & Bu13 & $396.4\pm7.6$ & $367.9\pm8.2$ & $228.2\pm8.9$ & Ne17  \\ 
	   &   &   &  &   &   &   &   &   & Ca14  &   &   &   &   \\
	   &   &   &  &   &   &   &   &   & Am18  &   &   &   &   \\
      &   &   &  &   &   &   &   &   & En18  &   &   &   &   \\
	   &   &   &  &   &   &   &   &   & Bu21  &   &   &   &   \\
	 S\_32 & HERMESJ104549$+$574512 & HLock06 & Wa13 & 10:45:49 & $+$57:45:11.52 & B & -- & Lensed & Ca14 & $136.0$ & $127.0$ & $96.0$ & Ca14  \\ 
	 S\_33 & HERMESJ105551$+$592845 & HLock08 & Wa13 & 10:55:51 & $+$59:28:45.44 & B & -- & -- & -- & $142.0$ & $119.0$ & $84.0$ & Ca14  \\ 
	 S\_34 & HERMESJ105751$+$573026 & HLSW$-$01 & Ga11 & 10:57:51 & $+$57:30:26.42 & B & A & Lensed & Ga11 & $402.0$ & $377.0$ & $249.0$ & Ca14  \\ 
	   &   &  HLock01 & Wa13  &   &   &   &   &   & Wa13  &   &   &   &    \\ 
	   &   &    &    &   &   &   &   &   & Bu13  &   &   &   &    \\
	   &   &    &    &   &   &   &   &   & Ca14  &   &   &   &    \\
	   &   &    &    &   &   &   &   &   & Am18  &   &   &   &    \\
	 S\_35 & HATLASJ132630$+$334410 & NAv1.195 & Bu13 & 13:26:30 & $+$33:44:09.90 & C & A & Lensed & Bu13 & $190.6\pm7.3$ & $281.4\pm8.2$ & $278.5\pm9.0$ & Ne17  \\ 
  	   &   &    &    &   &   &   &   &   & Am18  &   &   &   &    \\
  	   &   &    &    &   &   &   &   &   & En18  &   &   &   &    \\
	 S\_36 & HATLASJ133543$+$300404 & NA.v1.489 & Na17 & 13:35:43 & $+$30:04:03.66 & B & A & Lensed & Na17 & $136.6\pm7.2$ & $145.7\pm8.0$ & $125.0\pm8.5$ & Ne17  \\ 
	   &   &  HERBS35 & Ba18 &   &   &   &   &   &   &   &   &   &   \\ 
  S\_37 & HATLASJ142140$+$000448 & HerBS140 & Ba18 & 14:21:40 & $+$00:04:48.00  & B & --  & --  & --  & $96.8\pm7.2$ & $98.5\pm8.2$ & $87.4\pm8.7$ & Ba18  \\ 
	 S\_38 & HERMESJ142824$+$352620 & HBootes03 & Wa13 & 14:28:24 & $+$35:26:19.54 & B & A & Lensed & Bo06 & $323.0$ & $243.0$ & $139.0$ & Ca14  \\ 
  	   &   &    &    &   &   &   &   &   & Bu13  &   &   &   &    \\  
	 S\_39 & HATLASJ223753.8$-$305828 & HERBS68 & Ba18 & 22:37:54 & $-$30:58:28.00 & B & -- & -- & -- & $139.1\pm4.9$ & $144.9\pm5.1$ & $100.6\pm6.2$ & Ne17  \\ 
	 S\_40 & HATLASJ225250.7$-$313657 & HERBS47 & Ba18 & 22:52:51 & $-$31:36:58.00 & B & -- & -- & -- & $127.4\pm4.2$ & $138.7\pm4.9$ & $111.4\pm5.9$ & Ne17  \\ 
	 S\_41 & HELMSJ233441.0$-$065220 & HELMS1 & Na16 & 23:34:41 & $-$06:52:20.00 & B & A & -- & -- & $431.0\pm6.0$ & $381.0\pm7.0$ & $272.0\pm7.0$ & Na16  \\ 
	 S\_42 & HELMSJ233633.5$-$032119 & HELMS41 & Na16 & 23:36:34 & $-$03:21:19.00 & B & -- & -- & -- & $130.0\pm6.0$ & $131.0\pm6.0$ & $110.0\pm7.0$ & Na16  \\ 

\\
\multicolumn{14}{l}{\textit{Confirmed through sub-mm/mm follow-up}} \\
\\

	 S\_43 & HELMSJ001615.7$+$032435 & HELMS13 & Na16 & 00:16:16 & $+$03:24:35.00 & B & A & Lensed & Am18 & $176.0\pm13.0$ & $210.0\pm15.0$ & $134.0\pm11.0$ & Zh18  \\ 
	   &   &    &    &   &   &   &   &   & Dy18  &   &   &   &    \\
	 S\_44 & HELMSJ002220.9$-$015524 & HELMS29 & Na16 & 00:22:21 & $-$01:55:24.00 & B & A & -- & -- & $66.0\pm6.0$ & $102.0\pm6.0$ & $121.0\pm7.0$ & Na16  \\ 
	 S\_45 & HELMSJ003814.1$-$002252 & HELMS24 & Na16 & 00:38:14 & $-$00:22:52.00 & C & A & -- & -- & $73.35\pm5.55$ & $119.01\pm6.01$ & $122.87\pm6.69$ & Su17  \\ 
	   &   &  ACT-S J0038$-$0022 & Su17 &   &   &   &   &   &   &   &   &   &    \\ 
	 S\_46 & HELMSJ003929.6$+$002426 & HELMS11 & Na16 & 00:39:30 & $+$00:24:26.00 & C & A & -- & -- & $140.0\pm7.0$ & $157.0\pm7.0$ & $154.0\pm8.0$ & Na16  \\ 
	 S\_47 & HELMSJ004714.2$+$032454 & HELMS8 & Na16 & 00:47:14 & $+$03:24:54.00 & C & A & Lensed & Am18 & $312.0\pm6.0$ & $244.0\pm7.0$ & $168.0\pm8.0$ & Na16  \\
	   &   &   &  &   &   &   &   &   & Dy18  &   &   &   &    \\  
	 S\_48 & HELMSJ005159.4$+$062240 & HELMS18 & Na16 & 00:52:00 & $+$06:22:41.00 & B & A & Lensed & Am18 & $163.0\pm13.0$ & $202.0\pm15.0$ & $142.0\pm12.0$ & Zh18  \\ 
	   &   &   &  &   &   &   &   &   & En18  &   &   &   &   \\ 
	   &   &   &  &   &   &   &   &   & Ma22  &   &   &   &    \\ 
	 S\_49 & HATLASJ005724.2$-$273122 & HERBS60 & Ba18 & 00:57:24 & $-$27:31:22.00 & B & A & -- & -- & $73.3\pm5.8$ & $101.2\pm6.1$ & $103.6\pm7.5$ & Ba18  \\ 
	 S\_50 & HERMESJ021831$-$053131 & SXDF1100.001 (Orochi) & Ik11 & 02:18:31 & $-$05:31:31.00 & B & A & Lensed & Bu13 & $78.0\pm7.0$ & $122.0\pm8.0$ & $99.0\pm7.0$ & Bu15  \\ 
	   &   &  HXMM02 & Wa13  &   &   &   &   &   & Bu15  &   &   &   &    \\ 
	 S\_51 & HERMESJ033211$-$270536 & HECDFS04 & Wa13 & 03:32:11 & $-$27:05:36.00 & C & A & Lensed & Bu15 & $56.0\pm6.0$ & $61.0\pm6.0$ & $55.0\pm6.0$ & Bu15  \\ 
	 S\_52 & HERMESJ044154$-$540352 & HADFS01 & Wa13 & 04:41:54 & $-$54:03:52.00 & B & A & Lensed & Bu15 & $76.0\pm6.0$ & $100.0\pm6.0$ & $94.0\pm6.0$ & Bu15  \\ 
	 S\_53 & HATLASJ083932$-$011760 & HERBS105 & Ba18 & 08:39:32 & $-$01:18:00.00 & C & A & -- & -- & $73.8\pm7.4$ & $88.5\pm8.1$ & $93.2\pm8.7$ & Ba18  \\ 
	 S\_54 & HATLASJ091841$+$023048 & G09v1.326 & Ca14 & 09:18:41 & $+$02:30:47.97 & B & A & -- & -- & $125.7\pm7.2$ & $150.7\pm8.2$ & $128.4\pm8.7$ & Ne17  \\ 
	   &   &  HERBS32 & Ba18 &   &   &   &   &   &   &   &   &   &    \\ 
	 S\_55 & HATLASJ113526$-$014606 & G12v2.43 & Bu13 & 11:35:26 & $-$01:46:06.00 & C & A & Lensed & Gi23 & $278.8\pm7.4$ & $282.9\pm8.2$ & $204.0\pm8.6$ & Ne17  \\ 
	   &   &  HERBS10 & Ba18 &   &   &   &   &   &   &   &   &   &   \\

\hline
\hline
\end{tabular}
\endgroup
}

\end{table}
\end{landscape}

\begin{landscape}
\begin{table}
\centering
\contcaption{}
{\scriptsize
\begingroup
\setlength{\tabcolsep}{8pt}
\renewcommand{\arraystretch}{0.9}
\begin{tabular}{c l l l c c c c c c c c c l}
\hline
\hline
	 No. & IAU name & Alt. name & Ref. & RA & Dec & Vis. class & Multiw. obs. & Prev. classification & Ref. & $S_{250}$ & $S_{350}$ & $S_{500}$ & Ref.  \\ 
	   &   &   &   & [h m s] & [d m s] &   &   &   &   & [mJy] & [mJy] & [mJy] &    \\ 
	 (1) & (2) & (3) & (4) & (5) & (6) & (7) & (8) & (9) & (10) & (11) & (12) & (13) & (14)  \\ 
\hline
	 S\_56 & HATLASJ115433.6$+$005042 & HERBS177 & Ba18 & 11:54:34 & $+$00:50:42.30 & B & A & -- & -- & $53.9\pm7.4$ & $85.8\pm8.1$ & $83.9\pm8.6$ & Ba18  \\ 
	 S\_57 & HATLASJ120127.6$-$014043 & HERBS61 & Ba18 & 12:01:28 & $-$01:40:44.00 & B & A & Lensed & En18 & $67.4\pm6.5$ & $112.1\pm7.4$ & $103.9\pm7.7$ & Ne17  \\ 
	 S\_58 & HATLASJ131611$+$281220 & HERBS89 & Ba18 & 13:16:11 & $+$28:12:20.39 & C & A & Lensed & Be21 & $71.8\pm5.7$ & $103.4\pm5.7$ & $95.7\pm7.0$ & Ba18  \\ 
	 S\_59 & HATLASJ134429$+$303036 & NAv1.56 & Bu13 & 13:44:29 & $+$30:30:35.77 & B & A & Lensed & Bu13 & $462.0\pm7.4$ & $465.7\pm8.6$ & $343.3\pm8.7$ & Ne17  \\ 
	   &   &  HERBS1 & Ba18 &   &   &   &   &   & Fa17  &   &   &   &   \\ 
	   &   &    &   &   &   &   &   &   & Am18  &   &   &   &    \\
	   &   &    &   &   &   &   &   &   & En18  &   &   &   &    \\
	 S\_60 & HATLASJ141352$-$000027 & G15v2.235 & Bu13 & 14:13:52 & $-$00:00:27.00 & B & A & Lensed & Bu13 & $188.6\pm7.4$ & $217.0\pm8.1$ & $176.4\pm8.7$ & Ne17  \\ 
	   &   &  HERBS15 & Ba18 &   &   &   &   &   & Am18  &   &   &   &    \\ 
	 S\_61 & HATLASJ142414$+$022304 & ID 141 & Co11 & 14:24:14 & $+$02:23:03.62 & B & A & Lensed & Co11 & $112.2\pm7.3$ & $182.2\pm8.2$ & $193.3\pm8.5$ & Ne17  \\ 
	   &   &  G15v2.779 & Bu13  &   &   &   &   &   & Bu12  &   &   &   &    \\ 
	   &   &  HERBS13 & Ba18 &   &   &   &   &   & Dy18  &   &   &   &   \\ 
           &   &   &  &   &   &   &   &   & En18  &   &   &   &   \\ 
           &   &   &  &   &   &   &   &   & Dy22  &   &   &   &   \\ 
  S\_62 & HERMESJ142826$+$345547 & HBootes02 & Wa13 & 14:28:26 & $+$34:55:47.03 & B & A & Lensed & Wa13 & $159.0$ & $195.0$    & $156.0$ & Ca14  \\ 
  	   &   &    &    &   &   &   &   &   & Bu13  &   &   &   &    \\
  	   &   &    &    &   &   &   &   &   & Ca14  &   &   &   &    \\
	 S\_63 & HATLASJ230815.5$-$343801 & HERBS28 & Ba18 & 23:08:16 & $-$34:38:01.00 & B & A & -- & -- & $79.4\pm5.4$ & $135.4\pm5.7$ & $140.0\pm7.0$ & Ne17  \\ 
	 S\_64 & HELMSJ232439.5$-$043936 & HELMS7 & Na16 & 23:24:40 & $-$04:39:36.00 & B & A & Lensed & Am18 & $214.0\pm7.0$ & $218.0\pm7.0$ & $172.0\pm9.0$ & Na16  \\ 
           &   &   &  &   &   &   &   &   & Ma22  &   &   &   &    \\ 
	 S\_65 & HELMSJ233620.8$-$060828 & HELMS6 & Na16 & 23:36:21 & $-$06:08:28.00 & B & A & -- & -- & $193.0\pm7.0$ & $252.0\pm6.0$ & $202.0\pm8.0$ & Na16  \\ 

\hline
\hline
\end{tabular}
\endgroup
}

\end{table}
\end{landscape}

\begin{table*}
\centering
\caption{Properties of the 26 systems confirmed to be unlensed by multiwavelength follow-up.}
\label{tab:unlensed}
{\scriptsize
\begingroup
\setlength{\tabcolsep}{4pt}
\begin{tabular}{c l l c c c c c c c c c c}
\hline
No. & Name & Alt. Name & Ref. & RA & Dec & Multiw. & Ref. & Nature & Ref. & \textit{HST} Counterpart & z & Ref. \\
& & & & $[{\rm{h\ m\ s}}]$ & $[{\rm{d\ m\ s}}]$ & & & & & & \\
 (1) & (2) & (3) & (4) & (5) & (6) & (7) & (8) & (9) & (10) & (11) & (12) & (13) \\
\hline

S\_72 &  HERMESJ003824$-$433705  & HELAISS02  & Ca14 & 00:38:24 &  $-$43:37:05.00  & ALMA band 7 & Bu13  & multiple & -- & yes & --  & -- \\

S\_86 &  HERMESJ022022$-$015329 & HXMM04 & Wa13 & 02:20:22 & $-$01:53:29.00 & ALMA band 7 &  Bu15 & single & Bu15 & yes & ($0.21\pm0.14$) & Wa13 \\

S\_87 &  HERMESJ022029$-$064846 & HXMM09 & Wa13 & 02:20:29 & $-$06:48:46.00 & ALMA band 7 &  Bu15 & multiple/merger & Bu15 & no & ($0.21\pm0.09$) & Wa13 \\

S\_89 &  HERMESJ022548$-$041750  & HXMM05  & Wa13 & 02:25:48 &  $-$04:17:50.00  & ALMA band 7, 8 & Le19  & single DSFG & Le19 & no & 2.985 & Wa13  \\
 &                      &         &      &  &    & SMA at 340 GHz    & Bu13  &             &    & &             &       \\
                         
S\_90 &  HERMESJ045027$-$524126  & HADFS08  & Ca14 & 04:50:27 &  $-$52:41:26.00  & ALMA band 7 & Bu15 & merger & -- & yes & -- & -- \\

S\_92 &  HATLASJ084933$+$021443  & HerBS8  & Ca14 & 08:49:33 &  $+$02:14:43.14  & ALMA band 3 & 2018.1.01146.S  & protocluster & Iv13 & yes & 2.41 & Bu13  \\
 &                      &         &      &  &    & ALMA band 4, 6, 7 & Iv13  &             &    & &             &       \\
 &                      &         &      &  &    & SMA at 340 GHz & Iv13  &             &    & &            &       \\

S\_119 &  HATLASJ222503.7$-$304847  & HerBS166  & Ba18 & 22:25:04 &  $-$30:48:48.00  & ALMA band 7 & Ma19  & multiple & -- & yes & (4.52$\pm$0.62) & Ba18  \\

S\_140 &  HATLASJ000455.3$-$330811 & HerBS170  & Ba18 & 00:04:55 &  $-$33:08:12.00  & ALMA band 6 & 2018.1.00526.S & multiple & -- & no & (4.24$\pm$0.58) & Ba18  \\ 

S\_151 &  HELMSJ005258.6$+$061319 & HELMS10 & Na16 & 00:52:59 &  $+$06:13:19.00  & ALMA band 7 & 2013.1.00749.S  & multiple & -- & yes & ($3.2\pm0.2$) & Le18 \\ 
                         
S\_156 &  HATLASJ012335.1$-$314618  & HerBS145  & Ba18 & 01:23:35 &  $-$31:46:19.00  & ALMA band 6 & 2018.1.00526.S & single/QSO & -- & yes & 2.73 & Ur22  \\

S\_164 &  HERMESJ021943$-$052433 & HXMM20 & Ca14 & 02:19:43 &  $-$05:24:33.00  & ALMA band 7 & GG19 & protocluster & GG19 & yes & 2.602 & GG19  \\ 

S\_165 &  HERMESJ022206$-$070727 & HXMM23 & Ca14 & 02:22:06 &  $-$07:07:27.00  & ALMA band 7 & Bu15  & single & Bu15 & yes & -- & -- \\ 

S\_167 &  HERMESJ022251$-$032414 & HXMM22 & Ca14 & 02:22:51 & $-$03:24:14.00 & ALMA band 7 &  Bu15 & single & Bu15 & no & -- & -- \\

S\_176 &  HERMESJ043830$-$541832 & HADFS02 & Ca14 & 04:38:30 & $-$54:18:32.00 & ALMA band 7 &  Bu15 & multiple/merger & Bu15 & yes & ($3.40\pm0.70$) & Ag18 \\

S\_177 &  HERMESJ044947$-$525427 & HADFS09 & Ca14 & 04:49:47 &  $-$52:54:27.00  & ALMA band 7 &  Bu15 & multiple & Bu15 & yes & -- & --  \\ 

S\_178 &  HATLASJ083153$+$014014  & --  & -- & 08:31:53 &  $+$01:40:14.43  & ALMA band 3 & Pe20 & quasar & DS21 & yes & 3.9136 & DS21  \\
 &                      &     &    &  &    & ALMA band 5 & 2019.2.00027.S &        &    & &      &       \\
 &                       &     &    &  &    & ALMA band 8 & DS21 &        &    & &      &       \\

S\_188 &  HATLASJ091454$-$010357 & HerBS142 & Ba18 & 09:14:54 & $-$01:03:57.00 & ALMA band 6 &  2018.1.00526.S  & multiple/merger & Bu15 & yes & ($3.22\pm0.44$) & Ba18 \\ 
S\_193 &  HERMESJ100057$+$022014 & HCOSMOS02 & Ca14 & 10:00:57 &  $+$02:20:13.70  & ALMA band 3 &  Wa18, Ch21  & protocluster & GG19 & yes & 2.497 & Ca14 \\ 
 &  &  &  &  &    & ALMA band 4 &  2019.1.00151.S  & & & & & \\ 
 &  &  &  &  &    & ALMA band 6 &  2013.1.00118.S  & & & & & \\ 
 &  &  &  &  &    & ALMA band 6 & Ch21  & & & & & \\
 &  &  &  &  &    & ALMA band 7 &  2016.1.00463.S  & & & & & \\ 
 &  &  &  &  &    & ALMA band 7 & 2016.1.00478.S  & & & & & \\ 

S\_201 &  HATLASJ115521$-$021332 & HERBS179 & Ba18 & 11:55:21 &  $-$02:13:32.00  & ALMA band 6 & 2018.1.00526.S & single & -- & yes & ($4.07\pm$0.56) & Ba18 \\

S\_207 &  HATLASJ121812.8$+$011841  & HerBS83  & Ba18 & 12:18:13 &  $+$01:18:41.67  & ALMA band 6 & 2018.1.00526.S & multiple & -- & no & * & Co23  \\

S\_211 &  HATLASJ122407.4$-$003247 & HerBS161 & Ba18 & 12:24:07 &  $-$00:32:47.00  & ALMA band 6 & 2018.1.00526.S & multiple & -- & yes & ($3.82\pm0.52$) & Ba18 \\

S\_212 &  HATLASJ122459.1$-$005647  & HerBS150  & Ba18 & 12:24:59 &  $-$00:56:47.00  & ALMA band 6 & 2018.1.00526.S & multiple & -- & yes & (4.57$\pm$0.63) & Ba18  \\

S\_240 &  HATLASJ141118$-$010655 & HerBS201 & Ba18 & 14:11:18 &  $-$01:06:55.00  & ALMA band 6 &  2018.1.00526.S  & multiple & -- & no & ($4.00\pm0.55$) & Ba18 \\ 

S\_271 &  HATLASJ232200.0$-$355622 & HerBS118 & Ba18 & 23:22:00 &  $-$35:56:22.00  & ALMA band 6 & 2018.1.00526.S & single & -- & yes & ($3.80\pm0.52$) & Ba18 \\

S\_278 &  HATLASJ083345$+$000109 & HerBS88  & Ba18 & 08:33:45 &  $+$00:01:09.41  & ALMA band 7 & 2013.1.00358.S & single/QSO & -- & yes & 2.530596 & Al17  \\ 

S\_279 &  HATLASJ090613.8$-$010042 & HerBS165 & Ba18 & 09:06:14 &  $-$01:00:43.00  & ALMA band 6 &  2018.1.00526.S  & single/QSO & -- & yes & * & Co23 \\ 

\hline
\hline

\end{tabular}

\endgroup
\begin{flushleft}
\textit{Notes}: Col.~(1): Source reference number. The candidates are ordered first by their lens classification and then by their RA. Col.~(2): IAU name of the {\it Herschel} detection. Cols.~(3) and (4): Alternative name and reference. Cols.~(5) and (6): ICRS RA and Dec coordinate (J2000.0) of the {\it Herschel} detection. Cols.~(7) and (8): Multiwavelength observations and reference. Cols.~(9) and (10): Possible nature of the system and reference. Col.~(11): Presence of near-IR counterparts. Cols.~(12) and (13): Redshift of the source and reference. Following are the abbreviations used for the references that are not already included in Tables~\ref{tab:source_table} and \ref{tab:z}:
Ag18: \citet{Aguirre2018};
Ch21: \citet{Champagne2021};
DS21: \citet{Diaz-Santos2021}; 
GG19: \citet{Gomez-Guijarro2019};
Iv13: \citet{Ivison2013};
Le18: \citet{Lewis2018};
Le19: \citet{Leung2019};
Pe20: \citet{Penney2020};
Wa18: \citet{Wang2018}.

\end{flushleft}
}
\end{table*}

\begin{table}

\caption{Redshift of the candidate lenses and background sources of the confirmed lensing systems.}
\label{tab:z}
    \centering
    {\scriptsize

    \renewcommand{\arraystretch}{0.9}
    \begin{tabular}{c l c l c l}

\hline
   
No. & Name & $z_{\rm l}$ & Ref. & $z_{\rm s}$ & Ref.\\
(1) & (2) & (3) & (4) & (5) & (6) \\
\hline
S\_1 & HATLASJ000330.6$-$321136 & (0.38{\scriptsize $\pm$0.10}) & Sh21 & 3.077  & Ur22\\

S\_2 & HATLASJ000912.7$-$300807 & (0.28{\scriptsize $\pm$0.08}) & Sh21 & (1.19{\scriptsize $\pm$0.10}) & Zh18\\

S\_3 & HELMSJ001353.5$-$060200 & (0.60{\scriptsize $\pm$0.18}) & Na16 & 1.948 & Co23 \\

S\_4 & HELMSJ003619.8$+$002420 & 0.257573 & Na16 & 1.617 & Co23\\

S\_5 & HELMSJ005841.2$-$011149 & (0.38{\scriptsize $\pm$0.08}) & Na16 & 1.498 & Co23\\

S\_6 & HERSJ011722.3$+$005624 & (0.87{\scriptsize $\pm$0.05}) & Na16 & 2.469 & Co23\\

S\_7 & HERSJ012620.5$+$012950 & (0.43{\scriptsize $\pm$0.05}) & Na16 & 1.449 & Co23\\

S\_8 & HERSJ020941.2$+$001558 & 0.201854 & Na16 & 2.55293 & Ge15\\

S\_9 & HERMESJ032637$-$270044 & -- & -- & -- & -- \\

S\_10 & HERMESJ033732$-$295353 & (0.19{\scriptsize $\pm$0.05}) & Sh21 & -- & -- \\

S\_11$_1$ & HATLASJ083051$+$013225$_1$ & 0.626 & Bu13  & 3.6345 & Ya17\\

S\_11$_2$ & HATLASJ083051$+$013225$_2$ & 1.002 & Bu13  & 3.6345 & Ya17 \\

S\_12 & HERMESJ100144$+$025709 & 0.608 & Ca14 & -- & -- \\

S\_13 & HERMESJ103827$+$581544 & 0.591465 & Al17 & -- & -- \\

S\_14 & HERMESJ110016$+$571736 & 0.780518 & Al17 & -- & -- \\

S\_15 & HATLASJ114638$-$001132 & 1.2247 & Bu13 & 3.2596 & Y17\\

S\_16 & HATLASJ125126$+$254928 & (0.62{\scriptsize $\pm$0.10}) & Ne17 & 3.4419 & Ba20\\

S\_17 & HATLASJ125760$+$224558 & 0.555449 & Al17 & (1.53{\scriptsize $\pm$0.30}) & Ne17\\

S\_18 & HATLASJ133008$+$245860 & 0.4276 & Bu13 & 3.112 & Ca14\\

S\_19 & HATLASJ133846$+$255057 & (0.42{\scriptsize $\pm$0.10}) & Ne17 & (2.34{\scriptsize $\pm$0.40})  & Ba18\\ 

S\_20 & HATLASJ142935$-$002837 & 0.21844 & Me14 & 1.0271 & Me14\\

S\_21 & HERMESJ171451$+$592634 & 1.236 & Sh21 & 3.17844 & HC16\\

S\_22 & HERMESJ171545$+$601238 & (0.40{\scriptsize $\pm$0.09}) & Sh21 & 2.264 & Ca14\\

S\_23 & HATLASJ225844.7$-$295124 & (0.69{\scriptsize $\pm$0.21}) & Sh21 & (2.48{\scriptsize $\pm$0.040}) & Ba18\\

S\_24 & HELMSJ232210.3$-$033559 & (0.14{\scriptsize $\pm$0.09}) & Na16 & 4.688 & Co23\\

S\_25$_1$ & HATLASJ233037.2$-$331217$_1$ & (0.66{\scriptsize $\pm$0.17}) & Sh21 & 2.170 & Ur22\\
S\_25$_2$ & HATLASJ233037.2$-$331217$_2$ & (0.66{\scriptsize $\pm$0.17}) & Sh21 & 2.170 & Ur22\\

S\_26 & HELMSJ001626.0$+$042613 & 0.2154 & Am18 & 2.509 & Na16 \\

S\_27 & HATLASJ002624.8$-$341737 & (0.93{\scriptsize$\pm$0.35})  & Wa22 & 3.050 & Ur22 \\

S\_28 & HELMSJ004723.6$+$015751 & 0.3650  & Am18 & 1.441 & Na16 \\

S\_29 & HERSJ012041.6$-$002705  & (0.73{\scriptsize$\pm$0.04}) & Na16 & 2.015 & Co23 \\

S\_30 & HATLASJ085112$+$004934 & (0.66{\scriptsize$\pm$0.32}) & Sh21 & (1.77{\scriptsize$\pm$0.27}) & MG19 \\

S\_31 & HATLASJ085359$+$015537  & (1.16{\scriptsize$\pm$0.22})  & Sh21 & 2.0925 & Ya16 \\

S\_32 & HERMESJ104549$+$574512  & (0.20{\scriptsize$\pm$0.02}) & Wa13 & 2.991 & Wa13 \\

S\_33 & HERMESJ105551$+$592845  & (0.38{\scriptsize$\pm$0.11}) & Wa13 & 1.699$^{\rm a}$ & Wa13 \\

S\_34 & HERMESJ105751$+$573026  &  (0.60{\scriptsize$\pm$0.04}) & Ga11 & 2.9575 & Ga11 \\

S\_35 & HATLASJ132630$+$334410  &  0.7856  & Bu13 & 2.951 & Bu13 \\

S\_36$_{1}$ & HATLASJ133543$+$300404$_{1}$  & 0.9825 & St14 & 2.685 & Ca14 \\
S\_36$_{2}$ & HATLASJ133543$+$300404$_{2}$  & 0.9845 & St14 & 2.685 & Ca14 \\
S\_36$_{3}$ & HATLASJ133543$+$300404$_{3}$  & 0.9815 & St14 & 2.685 & Ca14 \\
S\_36$_{4}$ & HATLASJ133543$+$300404$_{4}$  & 0.9945 & St14 & 2.685 & Ca14 \\

S\_37 & HATLASJ142140$+$000448  & (1.11$\pm$0.41) & Sh21 & 2.781 & Co23 \\ 

S\_38 & HERMESJ142824$+$352620  & 1.034 & Bo06 & 1.325 & Bo06 \\

S\_39 & HATLASJ223753.8$-$305828 & (0.54$\pm$0.14) & Sh21 & (2.13$\pm$0.38) & Wa22 \\

S\_40 & HATLASJ225250.7$-$313657 & (0.69$\pm$0.26) & Sh21 & 2.433 & Ur22 \\

S\_41 & HELMSJ233441.0$-$065220  & -- & -- & 1.905 & Co23 \\

S\_42 & HELMSJ233633.5$-$032119  & -- & -- & 2.335 & Co23 \\

S\_43 & HELMSJ001615.7$+$032435  & 0.663  & Na16 & 2.765 & Na16 \\

S\_44 & HELMSJ002220.9$-$015524  & (0.90$\pm$0.15)  & Sh21 & 5.162 & As16 \\

S\_45 & HELMSJ003814.1$-$002252  & (0.17$\pm$0.08)  & Na16 & 4.984 & Co23 \\

S\_46 & HELMSJ003929.6$+$002426  & (0.72$\pm$0.23) & Sh21 & 2.848 & Co23 \\

S\_47 & HELMSJ004714.2$+$032454  & (0.48$\pm$0.08) & Na16 & 1.19 & Na16 \\

S\_48$_1$ & HELMSJ005159.4$+$062240$_1$  & 0.60266 & Ok21 & 2.392 & Na16 \\
S\_48$_3$ & HELMSJ005159.4$+$062240$_3$  & 0.59852 & Ok21 & 2.392 & Na16 \\

S\_49 & HATLASJ005724.2$-$273122  & (0.89$\pm$0.41) & Wa22 & 3.261 & Ur22 \\

S\_50 & HERMESJ021831$-$053131  & 1.350 & Wa13 & 3.3950 & Wa13 \\

S\_51 & HERMESJ033211$-$270536  & -- & -- & -- & -- \\

S\_52 & HERMESJ044154$-$540352  & -- & -- & -- & -- \\

S\_53 & HATLASJ083932$-$011760  & (0.42$\pm$0.12) & Sh21 & 2.669 & Co23 \\

S\_54$_1$ & HATLASJ091841$+$023048$_1$  & (0.91$\pm$0.32) & Sh21 &  2.5811 & Ha12 \\
S\_54$_2$ & HATLASJ091841$+$023048$_2$  & (0.91$\pm$0.32) & Sh21 &  2.5811 & Ha12 \\

S\_55 & HATLASJ113526$-$014606  & -- & -- & 3.1276 & Ha12 \\

S\_56 & HATLASJ115433.6$+$005042  & (0.52$\pm$0.11) & Sh21 & (3.90$\pm$0.50) & Ba20 \\

S\_57 & HATLASJ120127.6$-$014043  & (0.88$\pm$0.35) & Sh21 & (4.06$\pm$0.38) & MG19 \\

S\_58 & HATLASJ131611$+$281220  & (0.90$\pm$0.13) & Be21 & 2.9497 & Ne20 \\

S\_59 & HATLASJ134429$+$303036  & 0.6721 & Bu13 & 2.3010 & Ha12 \\

S\_60$_1$ & HATLASJ141352$-$000027$_1$  & 0.5478 & Bu13 & 2.4782 & Ha12 \\
S\_60$_3$ & HATLASJ141352$-$000027$_3$  & 0.5494 & Bu13 & 2.4782 & Ha12 \\

S\_61$_1$ & HATLASJ142414$+$022304$_1$  & 0.595 & Bu13 & 4.243 & Co11 \\
S\_61$_2$ & HATLASJ142414$+$022304$_2$  & 0.595 & Bu13 & 4.243 & Co11 \\

S\_62 & HERMESJ142826$+$345547  & 0.414 & Wa13 & 2.804 & Wa13 \\

S\_63 & HATLASJ230815.5$-$343801  & (0.72$\pm$0.21) & Wa22 & (4.03$\pm$0.38) & MG19 \\

S\_64 & HELMSJ232439.5$-$043936  & (0.67$\pm$0.26) & Sh21 & 2.473 & Na16 \\

S\_65 & HELMSJ233620.8$-$060828  & 0.3958 & Na16 & 3.434 & Na16 \\

\hline
\end{tabular}

\begin{flushleft}
\textit{Notes}: Col.~(1): Source reference number. Col.~(2): IAU name of the {\it Herschel} detection. Cols.~(3) and (4): Redshift of the lens candidate and reference. Cols.~(5) and (6): Redshift of the background source and reference. The bracketed values are photometric redshifts. Following are the abbreviations used for the references that are not already included in Table~\ref{tab:source_table}:  
As16: \citet{Asboth2016};
Ba20: \citet{Bakx2020};
Co23: \citet{Cox2023};
HC16: \citet{Hernan-Caballero2016};
MG19: \citet{Manjon-garcia2019};
Ne20: \citet{Neri2020};
Ok21: \citet{Okido2021};
St14: \citet{Stanford2014};
Wa22: \citet{Ward2022};
Ya16: \citet{Yang2016}.
\end{flushleft}}
    
\end{table}

\section{Surface brightness modelling}
\label{sec4}

We group all the systems, for which we are able to identify the foreground lensing galaxies or a suitable candidate according to the following configurations:
\begin{itemize}
    \item systems in which a clearly isolated galaxy is acting as a lens. In this case, we model this single object (type 1);
    \item systems in which two or three galaxies are probably acting as lenses (e.g., they are very close to each other or even blended, and the lensing features are consistent with multiple lenses). In this case, we model all the galaxies at the same time (type 2);
    \item systems in which more than one galaxy is clearly visible in the foreground, but either they are sufficiently small or distant with respect to each other that they can be separately modelled (type 3);
    \item one system which shows no trace of a lens, likely due to a combination of high redshift and obscuration (HATLASJ113526$-$014606; see \citealt{Giulietti2023} for details). This system will not be included in the lens surface-brightness modelling.
\end{itemize}

For the systems confirmed as lensed by multiwavelength data alone, we use the location of the background sources sub-mm emission to identify the foreground lensing galaxies in the \textit{HST} observations.

It is worth noticing that for type 2 and type 3 systems, we can not be sure whether all the galaxies are contributing to the lensing. In most cases, we do not know the redshift of all the galaxies, so we can not rule out the possibility that they are unrelated field objects. 

We still provide the models of all the galaxies of interest to enable future studies once all the redshifts are known. In order to best characterise the galaxies acting as lenses and provide the best-fitting model of their surface brightness, we use either a parametric or a non-parametric approach depending on the lens morphology and configuration of the lensing system. 

For the parametric modelling, we adopt the Galaxy Surface Photometry 2-Dimensional Decomposition algorithm (\texttt{GASP2D}; \citealt{Mendez-abreu2008, Mendez-abreu2017}) to model the type 1 and type 3 lensing systems and the Galaxy Fitting algorithm (\texttt{GALFIT}; \citealt{Peng2002, Peng2010}) for the type 2 foreground lensing systems. \texttt{GASP2D} has the advantage of automatically setting the initial guess values needed to fit the galaxy image, whereas \texttt{GALFIT} allows us to perform the simultaneous fit of multiple systems. We adopt the non-parametric Isophote Fitting algorithm (\texttt{ISOFIT}; \citealt{Ciambur2015}) to model highly-inclined disk lenses for which \texttt{GASP2D} and \texttt{GALFIT} fail. In the following sections, we describe the approaches adopted for the different cases. We point out in the source-by-source descriptions the cases for which we made exceptions to the above prescriptions.

\subsection{Parametric approaches}

Both \texttt{GASP2D} and \texttt{GALFIT} adopt a set of analytical functions to model the light distribution of the galaxy components. They derive the best-fitting values of the structural parameters by comparing the model of the surface brightness distribution with the observed photon counts of the galaxy in each image pixel with an iterative procedure of non-linear least-squares minimization. The model surface brightness is convolved with the point spread function (PSF) measured on the galaxy image to deal with the smearing effects due to the telescope optics and instrumental setup. Each image pixel is weighted according to the inverse of the variance of its total photon counts due to the contribution of both the galaxy and sky, the detector photon-noise limitation and readout noise are also taken into account. 
\texttt{GASP2D} and \texttt{GALFIT} are written in \texttt{IDL}\footnote{\url{https://www.l3harrisgeospatial.com/Software-Technology/IDL}} and {\texttt{C}}\footnote{\url{https://users.obs.carnegiescience.edu/peng/work/galfit/galfit.html}}, respectively. They provide the best-fitting values of the structural parameters of the galaxy components and their formal errors, as well as the model and residual image of the galaxy to be compared to the observed one.

\subsubsection{Residual sky subtraction}
\label{sec4.1.1}

The sky level provided by the image reduction consists of a global estimate across the FOV after automatically masking the luminous sources. This subtraction proves to be unreliable in the analysis of the faintest lensing features since a residual sky level could still be present in the image. Therefore, we measure and subtract the residual sky level as follows.

We start by masking the system as well as the foreground stars, companion and background galaxies, and lensing features using the \texttt{make\char`_sources\char`_mask} task of the \texttt{photutils.background} \texttt{python} package\footnote{\url{https://photutils.readthedocs.io/en/stable/index.html}} \citep{larry_bradley_2021_4624996}. The source segmentation algorithm identifies all the above sources and creates their footprints by selecting the $N_{\rm pixel}$ connected pixels with photon counts over a given threshold $\sigma$. Then, the algorithm enlarges the available footprints with binary dilation \citep[e.g.][]{Nadadur2000} through an $m\, {\rm pixel} \times\, m\, {\rm pixel}$ matrix to build the pixel mask of all the luminous sources in the image. We repeat this process twice to identify both the compact ($N_{\rm pixel}=5$, $m=11$, $\sigma=2$) and extended ($N_{\rm pixel}=10$, $m=51$, $\sigma=2$) sources. In both cases, we reiterate the source identification and footprint dilation of the masked image until no more pixels are added to the mask. The values of $N_{\rm pixel}$, $m$, and $\sigma$ are chosen such that the resulting masks cover most of the sources and, at the same time, leave some background-dominated regions in the image. The final pixel mask is the combination of the footprints built for compact and extended sources, which we visually check and manually edit to remove the spurious sources, such as residual cosmic rays and bad pixels.

We calculate the residual sky level as the biweight location of the unmasked pixels. This is an estimator for the central location of a distribution that is very robust against outliers \citep[e.g.][]{Beers1990}. Then, we subtract it from the image. We derive the standard deviation ${\rm rms}_{\rm bkg}$ of the sky-subtracted image through the median absolute deviation of the unmasked pixels using $\sigma = MAD/\Phi^{-1} (3/4)$, where $MAD$ is the median absolute deviation, $1/\Phi^{-1} (3/4) \approx 1.4826$ is the normal inverse cumulative distribution function evaluated at a probability of $3/4$. The biweight location and $MAD$ are available through \texttt{astropy.stats} package \citep{Astropy2018}. The image of HERSJ012620.5$+$012950 \citep[HERS5, S\_7;][]{Nayyeri2016} with the pixel mask adopted to estimate the residual sky level is shown in Fig.~\ref{fig:all_mask} (left panel) as an example.

\subsubsection{Pixel mask and noise map}
\label{sec4.1.2}

\begin{figure*}
 \includegraphics[width=\textwidth]{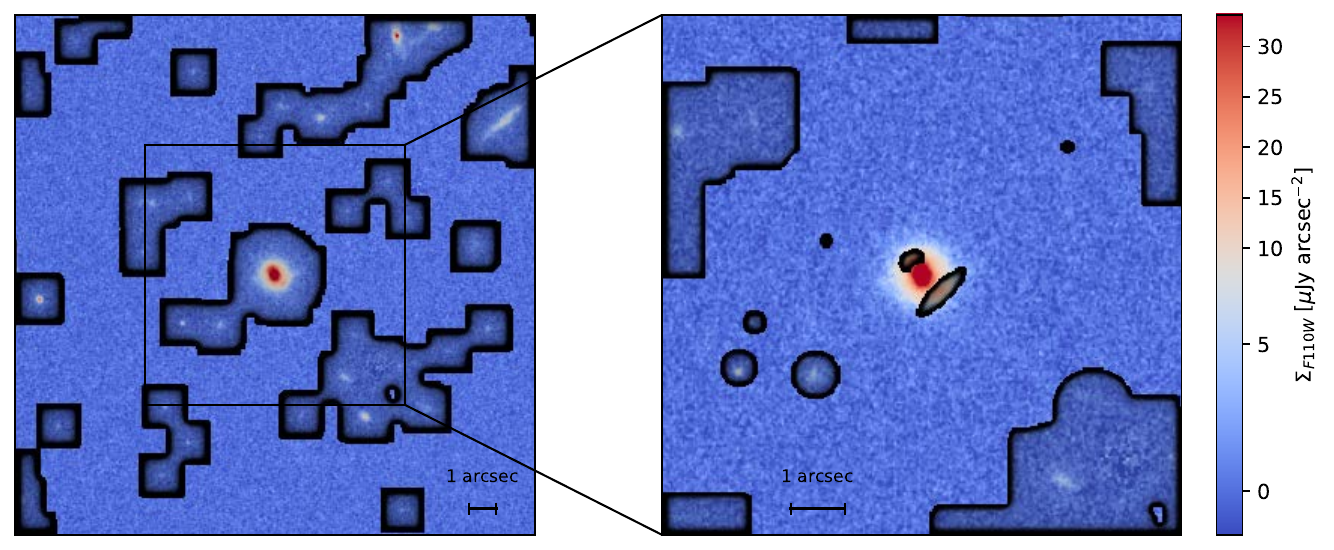}
 \caption{{\em Left panel:\/} Image of S\_7 with the pixel mask (corresponding to the regions with black contours) used to estimate the residual sky level and standard deviation of the background. The highlighted box marks the FOV of the zoom-in shown in the right panel. {\em Right panel:\/} Trimmed sky-subtracted image and pixel mask used to fit the surface brightness distribution of the lensing galaxy of S\_7.}
 \label{fig:all_mask}
\end{figure*}

We trim the sky-subtracted images to reduce the computing time to perform a reliable photometric decomposition. Each galaxy is centred in an FOV ranging from $200\,{\rm pixel} \times\, 200\, {\rm pixel}$ (corresponding to $12.8\,{\rm arcsec} \times\, 12.8\, {\rm arcsec}$) to $400\,{\rm pixel} \times\, 400\, {\rm pixel}$ ($25.6\,{\rm arcsec} \times\, 25.6\, {\rm arcsec}$) to fully cover all the lensing features.
We also trim the pixel mask to the same region. Then, we unmask the pixels corresponding to the source in the central circular region of $\sim 100$-pixel ($6.4\, {\rm arcsec}$) radius. This allows us to set up the final pixel mask, which we use for the parametric fitting of the source surface brightness. The trimmed image and final pixel mask that we use to model the surface brightness distribution of S\_7 are shown in Fig.~\ref{fig:all_mask} (right panel). In addition, we build the noise map of the images by calculating the variance (in units of electrons) of each pixel as $\sigma^2 = ({\rm RON}^2 + \sigma_{\rm sky}^2 + \sigma_{\rm gal}^2)/{\rm GAIN}^2$, where $\sigma_{\rm sky}^2 = I_{\rm sky} \cdot$GAIN and $\sigma_{\rm gal}^2=I_{\rm gal} \cdot$GAIN with $I_{\rm sky}$ and $I_{\rm gal}$ the surface brightness of the sky and lens in ADU, RON is the detector readout noise (in ADU) scaled for the number of samplings performed during the exposure (see \citealt{Dressel2022} for details), and GAIN is the detector gain (in e$^{-}$ ADU$^{-1}$). Since the sky level that was actually subtracted during image reduction is not readily available, we use ${\rm rms}_{\rm bkg}^{2}  = {\rm RON}^2 + \sigma_{\rm sky}^2$. Nevertheless, we estimate the expected variance of the sky from the equation above, verifying that $\sigma_{\rm sky}>{\rm RON}$.

\subsubsection{PSF model}
\label{sec4.1.3}

\begin{figure}
 \includegraphics[width=0.45\textwidth]{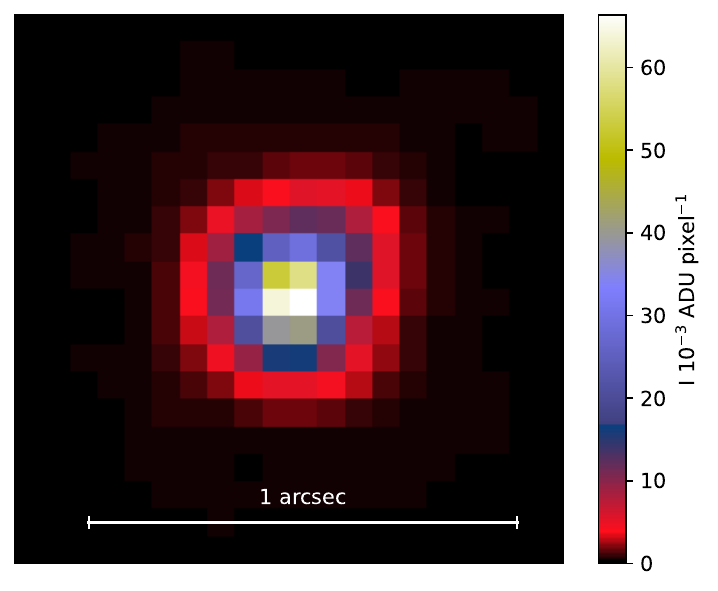}
 \caption{Image of the normalized PSF model adopted to fit the surface brightness distribution of S\_7.}
 \label{fig:PSF}
\end{figure}

For each image, we build a PSF model from the non-saturated point sources with the highest SNR available in the full \textit{WFC3/F110W} FOV. For each point source, we identify the centroid with the \texttt{find\char`_centroids} task of the \texttt{photutils} package and we extract a cutout of $100\,{\rm pixel} \times\,100\, {\rm pixel}$ ($6.4\,{\rm arcsec} \times\,6.4\, {\rm arcsec}$) corresponding to $\sim20\, {\rm FWHM} \times\,20\, {\rm FWHM}$. We get the stacked PSF by summing all the point-source cutouts. Before stacking, we normalise the total flux of each point source to unity after removing spurious sources and surface brightness gradient due to nearby bright and extended sources. The stacked PSF is smoothed using a \texttt{python} two-dimensional Savitsky-Golay filter{\footnote{\url{https://github.com/espdev/sgolay2}}} \citep[e.g.][]{Ratzlaff1989} to remove noise. Then, we mask the smoothed PSF model using image segmentation and replace the background pixels with their mean value. We adopt the resulting stacked, smoothed, and background-subtracted PSF to model the surface brightness distribution of the source. The average number of point sources we use to build the PSFs is 2. We show in Fig.~\ref{fig:PSF} the PSF model adopted for the parametric fit of S\_7.

\begin{table*}
    \centering
    \caption{Input files adopted to perform the surface brightness modelling with the different fitting algorithms used in this work.}
    \label{tab:sbmodelling}
    \begin{tabular}{c c c c c c c}
    \hline
    Algorithm   &   Noise Map   &   PSF Model   &   Isophote Table  &   Parametric Functions  &  Parameter Guesses  & Pixel Mask \\
    \hline
    \texttt{GASP2D}     &  yes &  yes &  yes &  yes &  optional  &  yes \\
    \texttt{GALFIT}     &  yes &  yes &  no &  yes &  yes  &  yes \\
    \texttt{ISOFIT+CMODEL}     &  no &  no &  yes &  no &  no  &  yes \\
     \hline
    \end{tabular}
   
\end{table*}

\subsubsection{Parametric modelling}

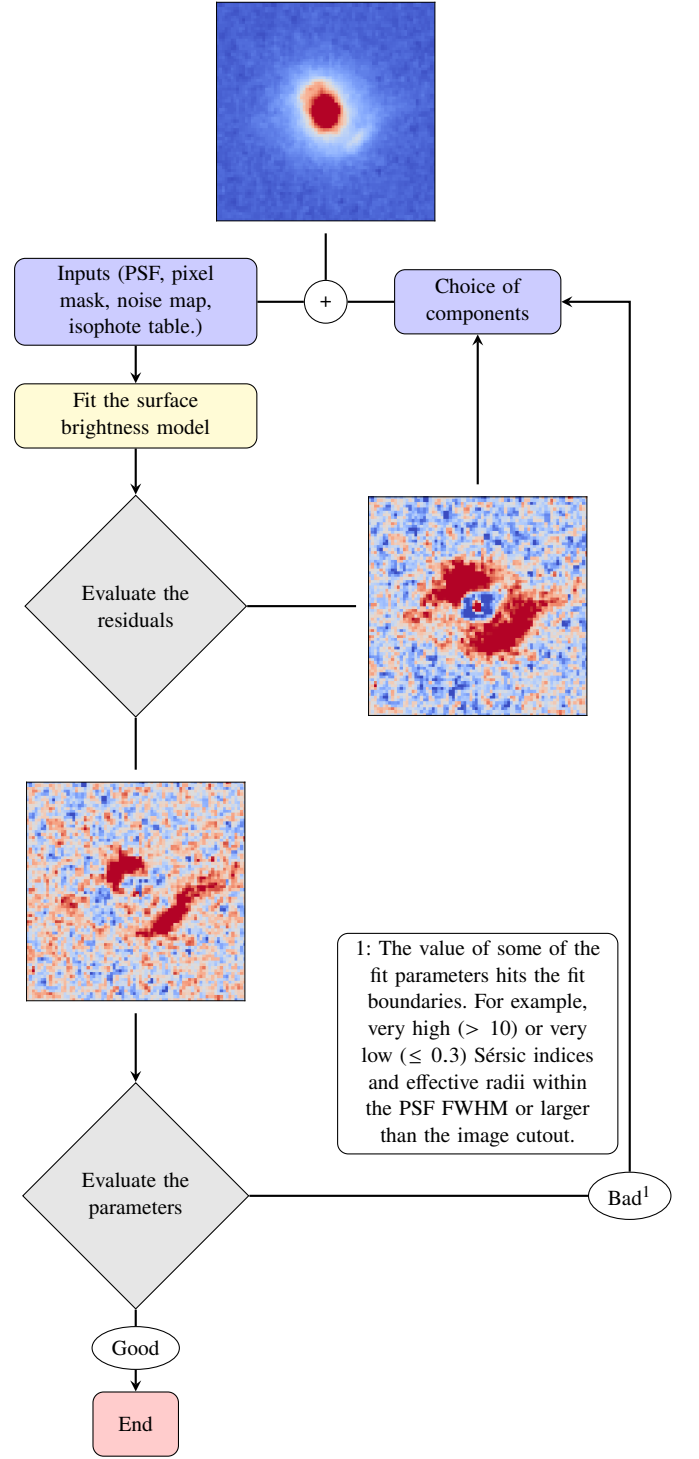
\begin{figure}

\caption{Flowchart describing the steps of the parametric surface-brightness modelling.}
    \label{fig:flowchart1}

\tikzstyle{startstop} = [rectangle, draw, fill=red!20, text centered, rounded corners, minimum height=3em, minimum width=4em]
\tikzstyle{decision} = [diamond, draw, fill=gray!20, text centered, minimum height=3em, minimum width=3em]
\tikzstyle{data} = [rectangle, draw, fill=blue!20, text centered, rounded corners, minimum height=3em, minimum width=4em]
\tikzstyle{procedure} = [rectangle, draw, fill=yellow!20, text centered, rounded corners, minimum height=3em, minimum width=4em]
\tikzstyle{cloud} = [draw, ellipse,fill=white!20, text centered, minimum height=2em, minimum width=2em]

\tikzstyle{arrow} = [thick,->,>=stealth]

\begin{tikzpicture}[node distance=2cm]

\node (start) [xshift=2cm] {\includegraphics[width=3cm]{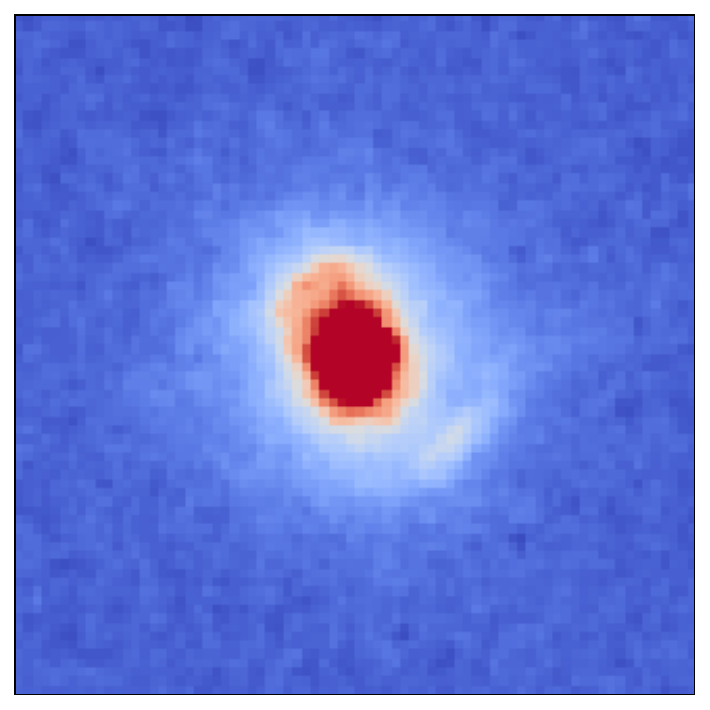}};
\node (05) [cloud, below of=start, node distance=2.5cm] {+};
\node (0) [data, left of=05, text width=3cm, node distance=2.5cm] {Inputs (PSF, pixel \\ mask, noise map, \\ isophote table.)};
\node (1) [data, right of=05, xshift=0cm, text width=2cm] {Choice of components};
\node (2) [procedure, below of=0, node distance=1.5cm, text width=3cm] {Fit the surface brightness model};
\node (3) [decision, below of=2, node distance=2.5cm, text width=2cm] {Evaluate the residuals};
\node (4) [right of=3, xshift=2.5cm] {\includegraphics[width=3cm]{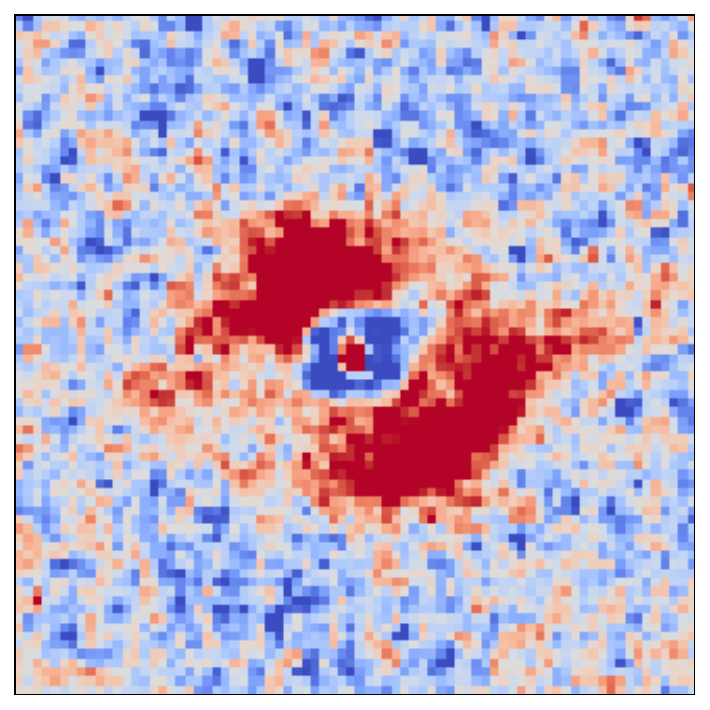}};
\node (5) [below of=3, yshift=-1.75cm] {\includegraphics[width=3cm]{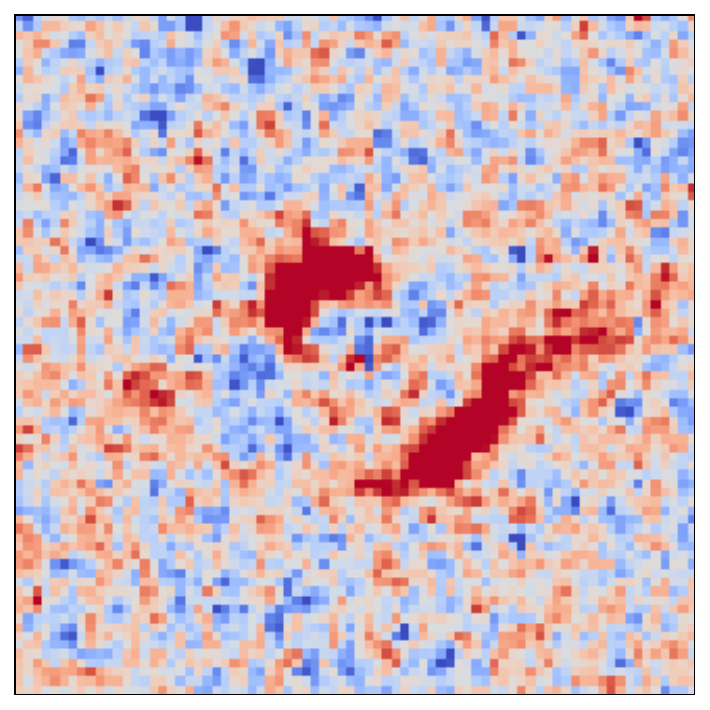}};
\node (6) [decision, below of=5, node distance=2cm, yshift=-2cm, text width=2cm] {Evaluate the parameters};
\node (75) [cloud, right of=6, node distance=6.5cm] {Bad$^1$};
\node (175) [below of=1, node distance=2cm] {};
\node (8) [cloud, below of=6, node distance=2cm] {Good};
\node (end) [startstop, below of=8, node distance=1cm] {End};

\node (comment) [rectangle, draw, fill=white!20, text centered, rounded corners, minimum height=3em, minimum width=4em, left of=75, yshift=2cm, text width=3.5cm] {1: The value of some of the fit parameters hits the fit boundaries. For example, very high ($> 10$) or very low ($\leq 0.3$) Sérsic indices and effective radii within the PSF FWHM or larger than the image cutout.};

\draw [thick,-] (start) -- (05);
\draw [thick,-] (1) -- (05);
\draw [thick,-] (05) -- (0);
\draw [arrow] (0) -- (2);
\draw [arrow] (2) -- (3);
\draw [thick,-] (3) -- (4);
\draw [thick,-] (3) -- (5);
\draw [arrow] (4) -- (1);
\draw [arrow] (5) -- (6);
\draw [thick,-] (6) -- (75);
\draw [arrow] (75) |- (1);
\draw [thick,-] (6) -- (8);
\draw [arrow] (8) -- (end);

\end{tikzpicture}
\end{figure}

We perform the parametric modelling of the systems following the steps reported in the flowchart given in Fig.~\ref{fig:flowchart1} and taking into account the PSF model, pixel mask, and noise map associated with the image.

Both \texttt{GASP2D} and \texttt{GALFIT} algorithms are based on a $\chi^2$ minimization. Thus, it is important to adopt initial guesses for the free parameters as close as possible to their actual values. This ensures that the iteration procedure does not stop on a local minimum of the $\chi^2$ distribution. 
\texttt{GASP2D} automatically sets the guess values by performing a one-dimensional decomposition of the azimuthally-averaged radial profiles of the surface brightness, ellipticity, and position angle of the source, which we measured by fitting ellipses to the isophotes with the \texttt{ISOFIT} task (see Sect.~\ref{isofit} for details) and adopting the final pixel mask. 
For \texttt{GALFIT}, we preliminarily fit a de Vaucouleurs law and adopt for the ellipticity and position angle of the mean values of the azimuthally-averaged radial profiles. The best-fitting values of this preliminary decomposition are adopted as guess values for the actual parametric fitting of the source.

We always start by fitting a single Sérsic component to the source. Then, we visually inspect the residual image to look for spurious sources and/or lensing features. We update the pixel mask to account for them and repeat the fit. If significant residuals are still visible, we add a second component and repeat again the fit. We check the new residual image and iterate the fitting procedure up to a maximum of three Sérsic extended components and one unresolved nuclear component. The additional luminous components are included to remove the residual under/over-subtracted structures at scales similar to the PSF or larger.
As a final step, we double-check the best-fitting values to avoid hitting the boundaries of the allowed ranges for the structural parameters, for which we adopt the ones implemented in \texttt{GASP2D}, which are $[0.3,\, 10]$ for the Sérsic index and $[0.5\, {\rm FWHM_{PSF}},\, +\infty)$ for the effective radius \citep[see,][for details]{Mendez-abreu2008, Mendez-abreu2017}. This issue occurs mostly for the Sérsic index of a few luminous components, which we fix to 0.5, 1, or 4. If the best-fitting effective radius of the single Sérsic fit is smaller than the PSF FWHM or larger than the cutout radius, we add a nuclear point source or enlarge the cutout. When necessary, we mask out the dust patches and lanes as much as possible to recover a reliable model of the surface brightness distribution.

The input files adopted by the two algorithms to perform the surface brightness modelling are listed in Table~\ref{tab:sbmodelling}.

\subsubsection{Error estimate}

The formal errors on the fitted parameters obtained from the $\chi^2$ minimisation procedure usually underestimate the real errors of the structural parameters \citep[e.g.,][]{Mendez-abreu2008}. Therefore, we estimate the errors on the fitted parameters by analysing a sample of mock galaxies we build through a series of Monte Carlo (MC) simulations following \citet{DallaBonta2018}.

To mimic the observational setup, we carry out the MC simulations in two different exposure-time bins, corresponding to 251 s and 712 s, respectively.
For each modelled galaxy (Tables~\ref{tab:models} and \ref{tab:SBmodels_full}), we build at least 70 mock galaxies by randomly choosing each component parameters $p_i$ in the range $\hat{p_i} - 0.3 \hat{p_i} < p_i < \hat{p_i} + 0.3 \hat{p_i}$ where $\hat{p_i}$ is the best-fitting value. We create more than 7800 mock galaxies in total. The image size of each mock source range between $200\,{\rm pixel} \times\,200\, {\rm pixel}$ ($12.8\,{\rm arcsec} \times\,12.8\, {\rm arcsec}$) to $400\,{\rm pixel} \times\,400\, {\rm pixel}$ ($25.6\,{\rm arcsec} \times\, 25.6\, {\rm arcsec}$) to enclose the largest fitted lens of that particular magnitude bin. We convolve the mock galaxies of each magnitude bin with a PSF that was randomly chosen from those produced to fit the systems of that particular magnitude bin. We adopt the same pixel scale, detector gain, and readout noise of the real images. In addition, we add a background level corresponding to the median ${\rm rms}_{\rm bkg}$ measured in the real images of that particular magnitude bin, and we add the photon noise in order to match the SNR of the mock and real images.

We run \texttt{GASP2D} or \texttt{GALFIT} with the appropriate combination of components to analyse the images of the mock galaxies. We then group the modelled galaxies according to their component combination (i.e. Sérsic, de Vaucouleurs, Sérsic-Sérsic, etc.). For each component of each modelled galaxy, we bin the different values of the mock galaxies in bins of one or two-magnitude width centred on the magnitude of the best-fitting model component.
We study the distribution of the relative errors on the effective surface brightness $I_{\rm e}$, effective radius $R_{\rm e}$, and Sérsic index $n$ as $(p_{\rm output}/p_{\rm input} - 1)$ and of the absolute errors on the position angle PA and axial ratio $q$ as $(p_{\rm output} - p_{\rm input})$. All the distributions appear to be nearly Gaussian after removing all the systems for which the fit failed, e.g. when one of the fitted components goes to zero or the model extends significantly more than the cutout.
We measure the biweight location and median absolute deviation of each distribution to detect possible systematic errors and derive the errors on the single parameters, respectively. We do not identify any systematic error, as all biweight location values are consistent with zero.

\subsection{Non-parametric approach}
\label{isofit}
We use \texttt{ISOFIT} to model the surface brightness distribution of the highly-inclined disk foreground lensing galaxies ($i>80^\circ$) for which \texttt{GASP2D} and \texttt{GALFIT} fail to recover the structural parameters. \texttt{ISOFIT} models the galaxy isophotes, taking into account deviations from a perfect elliptical shape. It supersedes the \texttt{IRAF}\footnote{\url{https://github.com/iraf-community/iraf}} task \texttt{ELLIPSE} \citep{Jedrzejewski1987} by sampling the isophotes according to eccentric anomaly rather than azimuthal angle and fitting simultaneously all the higher-order Fourier harmonics. \texttt{ISOFIT} provides a table with the best-fitting parameters of the galaxy isophotes, which can be used to build the model image through the \texttt{IRAF} task \texttt{CMODEL} \citep{Ciambur2015}.
Before fitting the isophotes, we subtract the residual sky level and built the pixel mask, as it has been done for the images of the systems to be fitted with the parametric approach and described in Sections \,\ref{sec4.1.1} and \,\ref{sec4.1.2}.
Since \texttt{ISOFIT} models the observed surface brightness distribution of the galaxy through an isophotal fit and without disentangling the light contribution of its structural components, we do not need to account for the image PSF and noise map. The input files needed by \texttt{ISOFIT} are listed in Table~\ref{tab:sbmodelling}. Unfortunately, we are not able to model HATLASJ142935$-$002837~(G15v2.19; S\_20) due to the low SNR of the image, which is heavily obscured by a complex pattern of dust that is not possible to be successfully masked. For the analysis of S\_20, we refer the reader to \citet{Calanog2014} and \citet{Messias2014}, who managed to model this lens using Keck and ancillary ALMA observations. As such, the following results are limited to the 63 systems we could model.

\subsection{Surface brightness modelling results}
\label{sec4.3}

After subtracting the foreground lens candidates, we check the residuals for lensing morphology, finding 9 new lensed systems. The confirmed lenses increase to 65 systems. For 42 of these, we detect the observed-frame near-IR emission of the background source. As a result, the fraction of confirmed systems increases more than doubled, from $9\%$ to $23\%$. The number of B, C, and D systems becomes 67, 120, and 29, respectively, while their fractions change to $24\%$, $43\%$, and $10\%$. 
\begin{figure}
    \centering
    \includegraphics[width=0.5\textwidth]{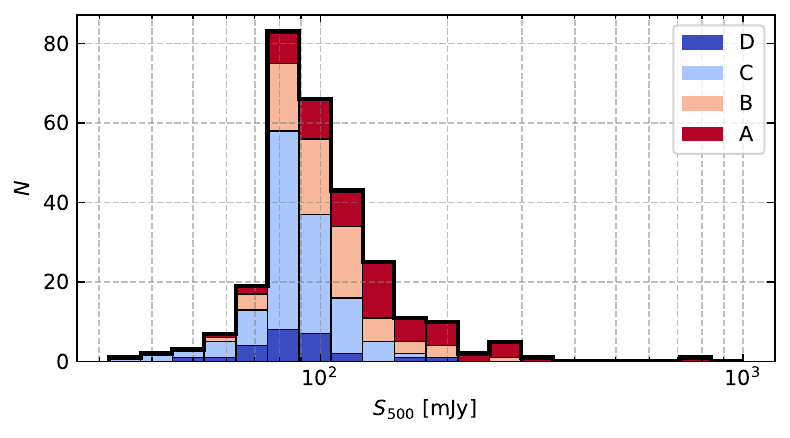}
    \caption{Distribution of the 281 sample systems with \textit{HST} snapshot observations as a function of the $S_{500}$ flux density. The red, pink, light blue and blue histograms refer to the systems classified as A, B, C, and D, respectively.}
    \label{fig:hist_fd}
\end{figure}
Figure\,\ref{fig:hist_fd} shows the distribution of the flux densities at 500 $\mu$m of our sample systems divided according to their classification. Most of the systems have $S_{\rm 500}\geq80\,{\rm mJy}$, except for those from \cite{Calanog2014}. We note that, as a consequence of the preliminary nature of the H-ATLAS catalogues used in \citet{Negrello2014}, several systems end up having $S_{\rm 500}<80\,{\rm mJy}$ in the final release of the H-ATLAS catalogues \citep{Valiante2016, Maddox2018}. We adopt the revised $S_{500}$ values in Fig.~\ref{fig:hist_fd}. We confirm, according to expectations \citep{Negrello2007, Wardlow2013}, that the number of A and B systems increases at increasing flux densities, whereas the number of C and D systems peaks at lower flux densities. Unfortunately, due to the sporadic coverage of both the multiwavelength follow-ups and \textit{HST} snapshots, we can not draw meaningful statistical conclusions on the population of lensed DSFGs.
\begin{figure}
    \centering
    \includegraphics[width=0.5\textwidth]{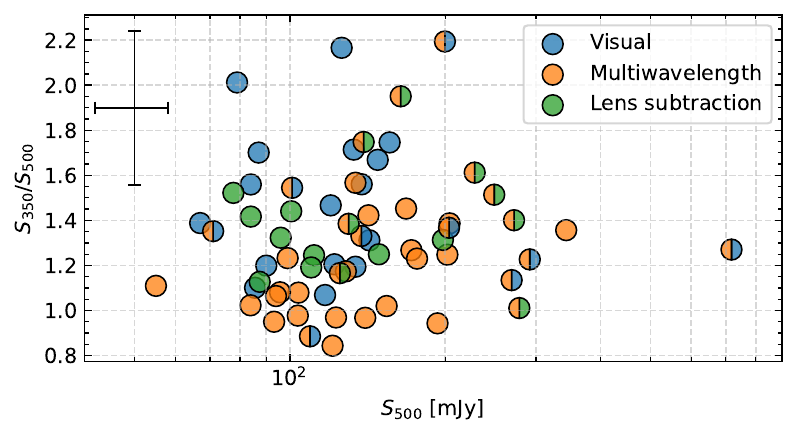}
    \caption{Flux density ratio $S_{350}/S_{500}$ as function of the flux density at 500 $\mu$m for the systems classified as A. The blue, orange, and green colors marks systems are classified according to visual inspection, multiwavelength follow-up observations, and after the lens subtraction, respectively. The cross at the top left corner gives the median measurement error.}
    \label{fig:A_500_350-500}
\end{figure}
In Fig.~\ref{fig:A_500_350-500}, we show the $S_{\rm 350}/S_{\rm 500}$ distribution as a function of the flux density at 500 $\mu$m for the A systems. In this case, the different colors refer to the different classification methods. \textit{HST} confirmed systems, both before and after the lens subtraction, tend to have lower $S_{\rm 350}/S_{\rm 500}$ flux density ratios than those of the systems confirmed by multiwavelength follow-ups. This is consistent with them being at lower redshifts and sampling redder, less obscured, stellar emission.
\begin{figure}
    \centering
    \includegraphics[width=0.5\textwidth]{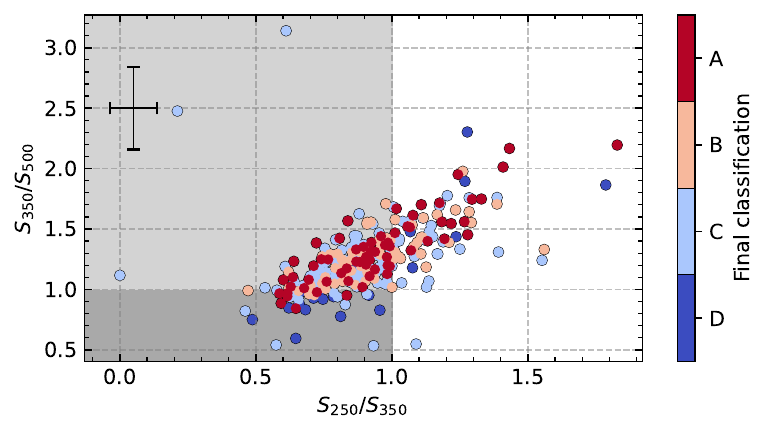}
    \caption{Distribution of the {\it Herschel} sub-mm flux density ratios of the 281 systems with \textit{HST} snapshot observations, which are color-coded according to their class. The light grey area marks the region of the systems with the SED peaking at $350\ \mu{\rm m}$, while the dark grey area indicates the region of the systems with the SED peaking at $500\ \mu{\rm m}$. The cross at the top left corner gives the median measurement error.}
    \label{fig:color_vs_color}
\end{figure}
Most of the systems have SEDs peaking at 350\,$\mu$m, as illustrated in Fig.\,\ref{fig:color_vs_color}. This should be the case for galaxies with bright $S_{500}$ flux densities, which are expected to be located at $1\simlt z \simlt4$ and potentially lensed \citep{Gonzalez-nuevo2012}. We note that the C and D systems tend to spread over a wider range of flux density ratios than the A or B systems. This is in line with them being mostly contaminants with a varying range of intrinsic properties.
We find that 47 of the 64 systems classified as A (73\%) have $S_{500}>100\,{\rm mJy}$, 12 (19\%) have $80<S_{500}<100\,{\rm mJy}$ and 5 (8\%) have $S_{500}<80\,{\rm mJy}$. 

In the following, we analyse the systems confirmed as strong lenses (class A). 
Among the 63 successfully modelled systems, which comprise one or more than one galaxy, we include the surface brightness distribution of 87 galaxies. A high fraction of these galaxies ($\sim 47\%$) show complex surface brightness profiles that often need more than one component (e.g., two or three Sérsic profiles or a combination of Sérsic profiles and PSF). For three systems (S\_47, S\_48, and S\_81) we include spiral arms too. The parameters of the best-fitting models for four representative systems are given in Table~\ref{tab:models}. They are three type 1 lenses, one visually confirmed (S\_12) and one with multiwavelength observations (S\_50). In addition, there is one type 2 system (S\_11) and one type 3 system (S\_8). The remaining systems can be found in Table~\ref{tab:SBmodels_full}. In Fig.~\ref{fig:models}, we show the cutouts of the \textit{HST} images, best-fitting models, \textit{HST} residuals after model subtraction, and SNR maps of the systems listed in Table~\ref{tab:models}, and one system modelled with \texttt{ISOFIT} (S\_5). The remaining systems are shown in Fig.~\ref{fig:models_full}.

In the rest of this section, we summarise the properties of each candidate lens we modelled. Table~\ref{tab:z} reports the references for the redshifts of the lens and source mentioned hereafter. We find that 34 of the systems discussed here have already been studied in the literature. \citet{Bussmann2013, Bussmann2015}, \citet{Enia2018}, \citet{Dye2018}, and \citet{Maresca2022} performed lens modelling and, in some cases, SED fitting of SMA and/or ALMA observations of 30, 29, 12, 6, and 7 sources, respectively, of these 11, 4, 8, 6, 4 sources were target candidates of our observations and are thus included in our analysis. \citet{Giulietti2022} studied the far-IR/radio correlation of 28 sources, 18 of which are included in our sample. \citet{Calanog2014} analysed 11 more systems of our sample (Table~\ref{tab:source_table}), and performed the source reconstruction on the lens-subtracted \textit{HST} images.
\citet{Cox2011}, \citet{Gavazzi2011}, \citet{Bussmann2012}, \citet{Fu2012}, \citet{Messias2014}, \citet{Geach2015}, \citet{Nayyeri2017}, \citet{Butler2021}, \citet{Liu2022}, \citet{Dye2022}, \citet{Giulietti2023} analysed single sources.
%
%
None of these previous works focused on the photometric properties of the lenses or performed a detailed analysis and modelling of their surface brightness. Our results thus represent the best reference for such information. In the following description, we report any relevant information available for each system from these previous works.
\subsubsection{\textit{HST} confirmed lenses}

\textbf{HATLASJ000330.6$-$321136~(HerBS155; S\_1)}: We model this type 1 foreground lens at $z_{\rm l}^{\rm phot}=0.38$ \citep{Shirley2021} with two Sérsic components using \texttt{GASP2D}. The background lensed source is located at $z_{\rm s}^{\rm spec}=3.08$, as measured by the BEARS survey \citep{Urquhart2022}. It forms two regular arcs on the SE and NW sides of the lens.

\textbf{HATLASJ000912.7$-$300807~(SD.v1.70; S\_2)}: We model this type 1 foreground lens at $z_{\rm l}^{\rm phot}=0.28$ \citep{Shirley2021} with three Sérsic components using \texttt{GALFIT} to account for the different location of their centres. The background lensed source is located at $z_{\rm s}^{\rm phot}=1.19$ \citep{Zhang2018}, and it forms a complex and multi-component arc-like main image on the SE side of the lens and a compact irregular secondary image on the NW side.

\textbf{HELMSJ001353.5$-$060200~(HELMS31; S\_3)}: We model this type 1 foreground lens at $z_{\rm l}^{\rm phot}=0.60$ \citep{Nayyeri2016} with a single Sérsic component and an exponential component using \texttt{GASP2D}.

\textbf{HELMSJ003619.8$+$002420~(HELMS14; S\_4)}: We model this type 1 foreground lens at $z_{\rm l}^{\rm spec}=0.26$ \citep{Nayyeri2016} with three Sérsic components using \texttt{GALFIT} to account for the different location of their centres. The background lensed source is located at $z_{\rm s}^{\rm spec}=1.62$ \citep{Cox2023} and forms an almost complete ring ($r\sim2$ arcsec) consisting of a diffuse component and three brighter knots located on the SE, SW, and NW sides of the lens. Two additional faint NE and SW structures are visible close to the ring, but they do not have a clear lensed morphology. 

\textbf{HELMSJ005841.2$-$011149~(HELMS23; S\_5)}: The foreground lens is a type 1 edge-on galaxy at $z_{\rm l}^{\rm phot}=0.38$ \citep{Nayyeri2016}, which we model with \texttt{ISOFIT}. The background lensed source is located at $z_{\rm s}^{\rm spec}=1.50$ \citep{Cox2023} and gives rise to two bright knots along the galaxy minor axis, opposite to each other with respect to the galaxy nucleus.

\textbf{HERSJ011722.3$+$005624$_{1,2}$~(HERS10; S\_6)}: This type 3 foreground lensing system is formed by a main galaxy (labelled as `1' in Fig.~\ref{fig:models_full}) at $z_{\rm l}^{\rm phot}=0.87$ \citep{Nayyeri2016} and possibly a secondary fainter galaxy (labelled as `2'), for which the redshift is not yet available. This prevents us from being more conclusive about the lensing nature of the secondary galaxy. We model the two galaxies with \texttt{GASP2D} using one Sérsic and one Gaussian component for the brighter galaxy and a single Sérsic component for the fainter one. The background lensed source is located at $z_{\rm s}^{\rm spec}=2.47$ \citep{Cox2023} and forms an arc on the E side of the lens and a diffuse secondary image on the NW side.

\textbf{HERSJ012620.5$+$012950~(HERS5; S\_7)}: We model this type 1 foreground lens at $z_{\rm l}^{\rm phot}=0.43$ \citep{Nayyeri2016} with one Sérsic and one Gaussian component using \texttt{GASP2D}. The background lensed source is located at $z_{\rm s}^{\rm spec}=1.45$ \citep{Cox2023} and forms two arcs on the NE and SW sides of the lens. The SW arc shows an additional diffuse radial component, which extends westward.

\textbf{HERSJ020941.2$+$001558$_{1,2}$~(HERS1; S\_8)}: This type 3 foreground lensing system is formed by a main galaxy (labelled as `1' in Fig.~\ref{fig:models}) and a second fainter galaxy (labelled as `2'), which are both located at $z_{\rm l}^{\rm spec}=0.20$ \citep{Nayyeri2016, Liu2022}. We use \texttt{GASP2D} to model the brighter galaxy with three Sérsic components and the fainter one with a single Sérsic component. The background lensed source at $z_{\rm s}^{\rm spec}=2.55$ \citep{Geach2015} forms a bright extended arc on the SW side of the lens and two knots on the E and SE sides. This system was studied in detail by \citet{Geach2015} and \citet{Liu2022}, who measured an Einstein radius of $\theta_{\rm E}=2.48^{+0.02}_{-0.01}$ arcsec. It has also been observed with SMA \citep[340 GHz,][]{Liu2022}, ALMA \citep[Band 7,][]{Liu2022}, VLA \citep[1.4 GHz and 5 GHz,][]{Geach2015}, eMERLIN \citep[1.52 GHz,][]{Geach2015} in the sub-mm/mm wavelength range and with \textit{HST} \citep[1.6 $\mu$m,][]{Liu2022} and Keck/NIRC2 AO \citep[\textit{H} and \textit{K}$_{s}$ band,][]{Liu2022} in the near-IR. The sub-mm/mm follow-up observations show no significant difference in morphology with those performed in the near-IR.

\textbf{HERMESJ032637$-$270044~(HECDFS05; S\_9)}: We model this type 1 foreground lens with a single Sérsic component using \texttt{GASP2D}. The background lensed source forms two compact images on the E and W sides of the lens, which is circled by a fainter incomplete ring. This system was studied in detail by \citet{Calanog2014}, who measured an Einstein radius of $\theta_{{\rm E}}=0.96^{+0.02}_{-0.03}$ arcsec.

\textbf{HERMESJ033732$-$295353~(HECDFS02; S\_10)}: We model this type 1 foreground lens at $z_{\rm l}^{\rm phot}=0.19$ \citep{Shirley2021} with three Sérsic components using \texttt{GASP2D}. The background lensed source forms a bright extended arc (divided into two knots on the NE side of the lens) and an inner secondary image on the SW side. This system was studied in detail by \citet{Calanog2014}, who measured an Einstein radius of $\theta_{\rm E}=1.65^{+0.03}_{-0.05}$ arcsec.

\textbf{HATLASJ083051$+$013225$_{1,2}$~(HerBS4; S\_11)}: This type 2 foreground lens is formed by two galaxies, one (labelled as `1' in Fig.~\ref{fig:models}) at $z_{\rm l}^{\rm spec}=0.63$ and the other (labelled as `2') at $z_{\rm l}^{\rm spec}=1.00$ \citep{Bussmann2013}. We deblend them with \texttt{GALFIT} by simultaneously modelling their surface brightness with a single Sérsic component each. The background lensed source at $z_{\rm s}^{\rm spec}=3.11$ \citep{Yang2017} shows a very complex structure with two intersecting arcs on the NW side of the lens. This system was studied in detail by \citet{Enia2018}, \citet{Yang2019}, and \citet{Maresca2022} who modelled high-resolution SMA \citep[340 GHz,][]{Bussmann2013}, ALMA (Band 7, \citealt{Amvrosiadis2018}; and Band 4 \citealt{Yang2019}) observations. This system was also observed with Keck/NIRC2 AO \citep[\textit{K}$_{s}$ band,][]{Calanog2014}. The sub-mm/mm follow-up observations show a ring around both lenses that splits into two arcs on the SE and S sides of the system, a third compact image on the N side of the second galaxy, and a further compact image between the two lenses.

\textbf{HERMESJ100144$+$025709~(HCOSMOS01; S\_12)}: We model this type 1 foreground lens at $z_{\rm l}^{\rm spec}=0.61$ \citep[][]{Calanog2014} with a single Sérsic component using \texttt{GASP2D}. The background lensed source forms three compact images on the NE, S, and NW sides of the lens, which is surrounded by a fainter ring. This system was studied in detail by \citet{Calanog2014}, who measured an Einstein radius of $\theta_{{\rm E}}=0.91^{+0.01}_{-0.01}$ arcsec. It was also observed with ALMA \citep[Band 7,][]{Bussmann2015} in the sub-mm/mm wavelength range and with Keck/NIRC2 AO \citep[\textit{K}$_{s}$ band,][]{Calanog2014} in the near-IR. The sub-mm/mm follow-up observations show only two lensing features on the NW and SE sides of the lens, with the first one having an arc-like shape. \citet{Bussmann2015} used the ALMA dataset to measure an Einstein radius of $\theta_{{\rm E}}=0.956\pm 0.005$ arcsec.

\textbf{HERMESJ103827$+$581544~(HLock04; S\_13)}: We model this type 1 foreground lens at $z_{\rm l}^{\rm spec}=0.59$ \citep[SDSS DR13,][]{Albareti2017} with two Sérsic components using \texttt{GASP2D}. The background lensed source forms two knots on the SE and NW sides of the lens, which is also surrounded by other fainter and more extended arcs. This system was studied in detail by \citet{Calanog2014}, who measured an Einstein radius of $\theta_{{\rm E}}=2.40^{+0.01}_{-0.05}$ arcsec and it was also observed with SMA \citep[340 GHz,][]{Bussmann2013}. The sub-mm/mm follow-up observations show no significant difference in morphology with respect to the near-IR observations, with the exception of a positional offset. \citet{Bussmann2013} used the SMA dataset to measure an Einstein radius of $\theta_{{\rm E}}=2.0\pm0.2$ arcsec.

\textbf{HERMESJ110016$+$571736~(HLock12; S\_14)}: We model this type 1 foreground lens at $z_{\rm l}^{\rm spec}=0.78$ \citep[SDSS DR13,][]{Albareti2017} with a single Sérsic component using \texttt{GASP2D}. The background lensed source gives rise to an extended arc on the NW side of the lens and a very faint compact secondary image on the SE side. This system was studied in detail by \citet{Calanog2014}, who measured an Einstein radius of $\theta_{{\rm E}}=1.14^{+0.04}_{-0.07}$ arcsec. 

\textbf{HATLASJ114638$-$001132$_{1,2}$~(HerBS2; S\_15)}: This type 2 foreground lens is formed by two galaxies at $z_{\rm s}^{\rm spec}=1.22$ \citep{Bussmann2013}, which we deblend by simultaneously modelling them with \texttt{GALFIT} using two Sérsic components each. The background lensed source at $z_{\rm s}^{\rm spec}=3.26$ \citep{Yang2017} shows a very complex structure with multiple arcs and compact images. This system was studied in detail by \citet{Fu2012}. It was also observed with ALMA \citep[Band 6,][]{Amvrosiadis2018}, SMA \citep[340 GHz,][]{Fu2012}, PdBI \citep[232 GHz,][]{Omont2013}, and the JVLA \citep[\textit{Ka}-band][]{Fu2012} in the sub-mm/mm wavelength range and Keck/NIRC2 AO \citep[\textit{J} and \textit{K}$_{s}$ band,][]{Fu2012} in the near-IR. The sub-mm/mm follow-up observations show three images: two arcs that both split into two knots on the S and N sides of the lenses and a compact image between the two lenses. There is only a partial overlap between the near-IR and sub-mm/mm observations.

\textbf{HATLASJ125126$+$254928~(HerBS52; S\_16)}: We model this type 1 foreground lens at $z_{\rm l}^{\rm phot}=0.62$ \citep{Negrello2017}, with a single Sérsic component using \texttt{GASP2D}. The background lensed source at $z_{\rm s}^{\rm spec}=3.44$ \citep{Bakx2020} forms four knots located N, NE, S, and NW with respect to the lens, three of which are connected by a fainter arc extending on the N side. This object was also observed with ALMA (Band 6, Prop. ID 2018.1.00526.S, PI I. Oteo). The sub-mm/mm follow-up observations show the counterparts of the N, NE, and S images.

\textbf{HATLASJ125760$+$224558~(S\_17)}: We model this type 1 foreground lens at $z_{\rm l}^{\rm spec}=0.55$ \citep[SDSS DR13,][]{Albareti2017} with a single Sérsic component using \texttt{GASP2D}. The background lensed source at $z_{\rm s}^{\rm phot}=1.53$ \citep{Negrello2017} forms a compact ring and a brighter knot on the NW side of the lens, which is slightly offset with respect to the ring.

\textbf{HATLASJ133008$+$245860~(HerBS12; S\_18)}: We model this type 1 foreground lens at $z_{\rm l}^{\rm spec}=0.43$ \citep{Bussmann2013} with one Sérsic and one Gaussian component using \texttt{GASP2D}. The background lensed source at $z_{\rm s}^{\rm spec}=3.11$ \citep{Calanog2014} forms three compact images on the N, SE, and SW sides of the lens, which is also surrounded by a fainter incomplete ring. This system was studied in detail by \citet{Calanog2014}, who measured an Einstein radius of $\theta_{{\rm E}}=0.944^{+0.002}_{-0.001}$ arcsec. It was also observed with ALMA (Band 7, Prop. ID 2018.1.00966.S, N. Indriolo), SMA \citep[340 GHz,][]{Bussmann2013}, PdBI \citep[240 GHz,][]{Omont2013}, and NOEMA \citep[283 GHz,][]{Yang2016} in the sub-mm/mm wavelength range and with Keck/NIRC2 AO \citep[\textit{K}$_{s}$ band,][]{Calanog2014} in the near-IR. The sub-mm/mm follow-up observations show no significant difference in morphology with respect to the near-IR observations. \citet{Bussmann2013} used the ALMA dataset to measure an Einstein radius of $\theta_{{\rm E}}=0.88\pm0.02$ arcsec.

\textbf{HATLASJ133846$+$255057~(HerBS29; S\_19)}: We model this type 1 foreground lens at $z_{\rm l}^{\rm phot}=0.42$ \citep{Negrello2017} with a single Sérsic component using \texttt{GASP2D}. The background lensed source at $z_{\rm s}^{\rm phot}=2.34$ \citep{Bakx2018} forms two extended arcs with a sharp break on the S and N sides of the lens. This system was observed with SMA \citep[340 GHz,][]{Bussmann2013} too. The sub-mm/mm follow-up observations show no significant difference in morphology with respect to the near-IR ones.

\textbf{HERMESJ171451$+$592634~(HFLS02; S\_21)}: We model this type 1 foreground lens at $z_{\rm l}^{\rm spec}=1.24$ \citep{Shirley2021} with a single de Vaucouleurs component using \texttt{GASP2D}. The background lensed source at $z_{\rm s}^{\rm spec}=3.18$ \citep{Hernan-Caballero2016} forms two arcs on the NE and SW sides of the lens, which is surrounded by a fainter ring. This system was studied in detail by \citet{Calanog2014}, who measured an Einstein radius of $\theta_{\rm E}=0.87^{+0.02}_{-0.05}$ arcsec.

\textbf{HERMESJ171545$+$601238$_{1,2}$~(HFLS08; S\_22)}: This type 3 foreground lensing system is formed by a main galaxy (labelled as `1' in Fig.~\ref{fig:models_full}) at $z_{\rm l}^{\rm phot}=0.40$ \citep{Shirley2021} and possibly a secondary unresolved galaxy (labelled as `2'), for which the redshift is not yet available. This prevents us from being more conclusive about the lensing nature of the secondary galaxy. We used \texttt{GASP2D} to model the brighter galaxy with two Sérsic components and one PSF and the fainter one with a single Sérsic component. The background lensed source at $z_{\rm s}^{\rm spec}=2.26$ \citep{Calanog2014} forms an extended and irregular arc E of the lens and a compact secondary image on its W side. This system was studied in detail by \citet{Calanog2014}, who measured an Einstein radius of $\theta_{{\rm E}}=1.95^{+0.05}_{-0.04}$ arcsec.

\textbf{HATLASJ225844.8$-$295124~(HerBS26; S\_23)}: This type 2 configuration is consistent with two blended galaxies at $z_{\rm l}^{\rm phot}=0.69$ \citep{Shirley2021},
which we deblend by simultaneously fitting them with \texttt{GALFIT} using one and two Sérsic components, respectively. The background lensed source is located at $z_{\rm s}^{\rm phot}=2.48$ \citep{Bakx2018}. It gives rise to an extended arc encircling the lens on the SW side but without showing any clear secondary image. This is likely due to the shear effect caused by the multiple-lens system.

\textbf{HELMSJ232210.3$-$033559$_{1,2}$~(HELMS19; S\_24)}: This type 3 foreground lensing system is formed by a main galaxy (labelled as `1' in Fig.~\ref{fig:models_full}) at $z_{\rm l}^{\rm phot}=0.14$ \citep{Nayyeri2016} and possibly a secondary fainter galaxy (labelled as `2'), for which the redshift is not yet available. This prevents us from being more conclusive about the lensing nature of the secondary galaxy. We model the two galaxies with \texttt{GASP2D} using two and one Sérsic components, respectively. The background lensed source is located at $z_{\rm s}^{\rm spec}=4.69$ \citep{Cox2023} and forms an extended irregular arc on the E side of the lens, a compact secondary image on the SW side, and possibly a third fainter image on the NW side. 

\textbf{HATLASJ233037.2$-$331217$_{1,2}$~(HerBS123; S\_25)}: This type 3 foreground lensing system is formed by a main galaxy (labelled as `1' in Fig.~\ref{fig:models_full}) at $z_{\rm l}^{\rm phot}=0.66$ \citep{Shirley2021} and possibly a secondary unresolved galaxy (labelled as `2'), for which the redshift is not yet available. We deblend and model the two galaxies with \texttt{GALFIT} by using two Sérsic components and a single unresolved component, respectively. The background lensed source is located at $z_{\rm s}^{\rm spec}=2.17$, as measured by the BEARS survey \citep{Urquhart2022}. It shows a very complex structure with multiple knots embedded into a main arc and a second inner elongated image on the SE side. This kind of morphology is indicative of strong shear, supporting the idea of a multiple lensing system. 

\subsubsection{Lenses confirmed after the lens subtraction}

\textbf{HELMSJ001626.0$+$042613~(HELMS22; S\_26)}: We model this type 1 foreground lens at $z_{\rm s}^{\rm spec}=0.22$ \citep{Amvrosiadis2018} with two Sérsic components using \texttt{GASP2D}. The background lensed source at $z_{\rm s}^{\rm spec}=2.51$ \citep{Nayyeri2016} forms an arc on the S side of the lens. This system was observed with ALMA (Band 6, \citealt{Amvrosiadis2018}; Band 7, Prop. ID 2016.1.01188.S, S. Eales) in the sub-mm/mm wavelength range. The sub-mm morphology shows two compact images on top of the \textit{HST} arc and a secondary image closer to the lens on its NE side. This system was studied in detail by \citet{Dye2018}, who measured an Einstein radius of $\theta_{\rm E}=0.98\pm0.07$ arcsec. The sub-mm/mm follow-up observations show both the S arc that splits into two images and a fainter secondary image on the NE side. 

\textbf{HATLASJ002624.8$-$341737~(HerBS22; S\_27)}: We model this type 1 foreground lens at $z_{\rm l}^{\rm phot}=0.93$ \citep{Ward2022} with two Sérsic components using \texttt{GASP2D}. The background lensed source at $z_{\rm s}^{\rm phot}=2.70$ \citep{Zhang2018} forms an arc on the SW of the lens and a fainter secondary image on the opposite side. The SW arc is split into two components: a main one and a fainter one, which are W and S of the lens, respectively.

\textbf{HELMSJ004723.6$+$015751$_{1,2}$~(HELMS9; S\_28)}: This type 2 foreground lens is formed by two galaxies at $z_{\rm s}^{\rm spec}=0.37$ \citep{Amvrosiadis2018}, which we deblend by simultaneously modelling them with \texttt{GALFIT} using two Sérsic components for the main galaxy (labelled as '1` in Table~\ref{tab:models}) and one de Vaucouleurs component for the second galaxy (labelled as '2`). The background lensed source at $z_{\rm s}^{\rm spec}=1.44$ \citep{Nayyeri2016} shows a very complex structure with one confirmed arc on the N side of the lens and possibly multiple compact images. The background lensed source morphology is likely to be disturbed by additional features possibly related to the interaction with the two lenses. It was observed with ALMA \citep[Band 6,][]{Amvrosiadis2018} in the sub-mm/mm wavelength range. This system was studied in detail by \citet{Dye2018}, who measured an Einstein radius of $\theta_{\rm E}=2.16\pm0.10$ arcsec. The sub-mm/mm follow-up observations show clearly the N arc we detected in the \textit{HST} image and some hints for further lensed images on the S side of the lens galaxies. 

\textbf{HERSJ012041.6$-$002705~(HERS2; S\_29)}: We model this type 1 foreground lens at $z_{\rm l}^{\rm phot}=0.73$ \citep{Nayyeri2016} with two Sérsic components using \texttt{GASP2D}. The background lensed source forms an arc on the NE of the lens, a clump eastward of the arc and a fainter secondary image on the SW side of the lens.

\textbf{HATLASJ085112$+$004934~(S\_30)}: We model this type 1 foreground lens at $z_{\rm l}^{\rm phot}=0.66$ \citep{Shirley2021} with three Sérsic components using \texttt{GASP2D}. The background lensed source at $z_{\rm s}^{\rm phot}=1.77$ \citep{Manjon-garcia2019} forms three arcs on the SW, SE, and NE sides of the lens. The SW arc is divided into two parts by a bright clump. This is likely due to the shear effect caused by nearby galaxies. 

\textbf{HATLASJ085359$+$015537~(G09v1.40; S\_31)}: We model this type 1 foreground lens with two Sérsic components using \texttt{GASP2D}. The background lensed source at $z_{\rm s}^{\rm spec}=2.09$ \citep{Yang2016} forms a ring that splits into two arcs on the E and W sides of the lens. This system was studied in detail by \citet{Bussmann2013, Calanog2014}, and \citet{Butler2021} with ALMA (Band 6, \citealt{Butler2021} and Prop. ID 2017.1.00027.S, S. Eales; Band 7, \citealt{Amvrosiadis2018} and \citealt{Butler2021}), and SMA \citep[340 GHz][]{Bussmann2013} in the sub-mm/mm wavelength range and with Keck/NIRC2 AO \citep[\textit{K}$_{s}$ band,][]{Calanog2014} in the near-IR, respectively. These works measured an Einstein radius of $\theta_{{\rm E}}=0.56^{+0.01}_{-0.02}$ arcsec from the Keck observations and $\theta_{{\rm E}}=0.553\pm0.004$ arcsec from SMA observations. The sub-mm/mm follow-up observations show no significant difference in morphology with respect to the near-IR ones.

\textbf{HERMESJ104549$+$574512~(HLock06; S\_32)}: We model this type 1 foreground lens at $z_{\rm l}^{\rm phot}=0.20$ \citep{Wardlow2013} with two Sérsic components using \texttt{GASP2D}. The background lensed source at $z_{\rm s}^{\rm spec}=2.91$ \citep{Calanog2014} forms two images on the NW and SE sides of the lens. \citet{Bussmann2013} used the SMA dataset to measure an Einstein radius of $\theta_{{\rm E}}=0.10\pm0.03$ arcsec.

\textbf{HERMESJ105551$+$592845~(HLock08; S\_)33}: This type 2 foreground lens is formed by two galaxies (labelled as `1' and `2' in Fig.~\ref{fig:models_full}) at $z_{\rm l}^{\rm phot}=0.38$ \citep{Wardlow2013}, which we deblend by simultaneously modelling them with \texttt{GALFIT} using one Sérsic component each. The background lensed source at $z_{\rm s}^{\rm spec}=1.70$ \citep{Calanog2014} shows a complex morphology with a bright arc and a faint secondary image SW and NE of the galaxy 1 respectively, and a possible third image E of the galaxy 2.

\textbf{HERMESJ105751$+$573026$_{1,2,3,4,5,6}$~(HLock01; S\_34)}: This type 2 foreground lens is formed by a group of six galaxies (labelled from `1' to `6' in Fig.~\ref{fig:models_full}) at $z_{\rm l}^{\rm phot}=0.60$ \citep{Gavazzi2011}. We simultaneously model them with \texttt{GALFIT} using respectively two Sérsic components, one de Vaucouleurs and one Sérsic component, one Sérsic component, one de Vaucouleurs component, one Sérsic component, and a PSF. With respect to \citet{Gavazzi2011} we divide galaxy 5 into two separate components. The background lensed source at $z_{\rm s}^{\rm spec}=2.96$ \citep{Gavazzi2011} forms a bright arc on the NE side of galaxy 1 that splits into two images in the proximity of galaxy 6, two additional bright compact images on the S and NW of galaxy 1, and, a faint counter image near the NE side galaxy 1 (which was predicted but not observed by \citet{Gavazzi2011}. This system was also observed with SMA \citep[340 GHz,][]{Bussmann2013} and with Keck/NIRC2 AO \citep[\textit{K}$_{s}$ band,][]{Gavazzi2011}. 

\textbf{HATLASJ132630$+$334410~(NAv1.195; S\_35)}: We model this type 1 foreground lens at $z_{\rm l}^{\rm spec}=0.79$ \citep[][]{Bussmann2013} with one Sérsic component using \texttt{GASP2D}. The background lensed source at $z_{\rm s}^{\rm spec}=2.95$ \citep[][]{Bussmann2013} forms two images on the NW and SE sides of the lens. This system was observed with ALMA (Band 3, \citealt{Berman2022} and Band 6, Prop. ID 2017.1.01214.S, Y. Min), and SMA \citep[340 GHz][]{Bussmann2013}. The sub-mm/mm follow-up observations show no significant difference in morphology with respect to the near-IR ones. \citet{Bussmann2013} and \citet{Kamieneski2023} used the SMA and ALMA band 3 datasets to measure an Einstein radius of $\theta_{{\rm E}}= 1.80\pm0.02$ arcsec and $\theta_{{\rm E}}= 1.78^{+0.21}_{-0.14}$ arcsec, respectively.

\textbf{HATLASJ133543$+$300404$_{1,2,3,4}$}~(HerBS35; S\_36): This type 2 foreground lens is formed by a cluster of galaxies (labelled as `1', `2', `3', and `4' in Fig.~\ref{fig:models_full}) at $z_{\rm l}^{\rm spec}=0.98$ \citep{Stanford2014, Nayyeri2017}, which we deblend by simultaneously modelling them with \texttt{GALFIT} using one Sérsic and one PSF component for galaxy 1 and one Sérsic component each for the remaining ones. The background lensed source at $z_{\rm s}^{\rm spec}=2.69$ \citep{Nayyeri2017} shows a complex morphology with a compact bright arc and a faint secondary image NE and SW of the galaxy 1, one arc S of the galaxies 1 and 2, and another arc N of the galaxies 3 and 4. This system was observed with SMA \citep[228 GHz,][]{Nayyeri2017}, the JVLA \citep[4 GHz,][]{Nayyeri2017}, and Keck/NIRC2 AO \citep[\textit{K}$_{s}$ and \textit{H} band,][]{Nayyeri2017}. The sub-mm/mm follow-up observations show a complex morphology with various arcs and images that are not detected in the near-IR observations.

\textbf{HATLASJ142140$+$000448~(HerBS140; S\_37)}: We model this type 1 foreground lens at $z_{\rm s}^{\rm PHOT}=1.11$ \citep{Shirley2021} with a single Sérsic component using \texttt{GASP2D}. The background lensed source is located at $z_{\rm s}^{\rm spec}=2.78$ \citep{Cox2023} and forms a faint ring that splits on the N and S sides of the lens and a clump slightly offset of the ring on the NE side of the lens. 

\textbf{HERMESJ142824$+$352620~(HBootes03; S\_38)}: We model this type 1 foreground lens at $z_{\rm s}^{\rm spec}=1.03$ \citep{Borys2006} with two Sérsic components using \texttt{GASP2D}. The background lensed source at $z_{\rm s}^{\rm spec}=1.33$ \citep{Borys2006} forms a closer arc on the NE of the lens and a diffuse secondary image on the SW side of the lens. It was observed with SMA \citep[340 GHz,][]{Bussmann2013}, and ALMA \citep[Band 6,][]{Amvrosiadis2018} in the sub-mm/mm wavelength range. The sub-mm/mm follow-up observations show no significant difference in morphology with respect to the near-IR observations. This system was studied in detail by \citet{Bussmann2013}, who measured an Einstein radius of $\theta_{{\rm E}}=2.46^{+0.01}_{-0.01}$ arcsec.

\textbf{HATLASJ223753.8$-$305828~(HerBS68; S\_39)}: We model this type 1 foreground lens with one Sérsic component using \texttt{GASP2D}. The background lensed source at $z_{\rm s}^{\rm phot}=2.13$ \citep{Ward2022} or $z_{\rm s}^{\rm phot}=2.26$ \citep{Manjon-garcia2019} forms an arc on the SE side of the lens and a faint secondary image on the E side. The SE arc shows a secondary faint farther arc that extends from the N tip of the main arc. 

\textbf{HATLASJ225250.7$-$313657~(HerBS47; S\_40)}: We model this type 1 foreground lens with one Sérsic component using \texttt{GASP2D}. The background lensed source at $z_{\rm s}^{\rm phot}=2.70$ \citep{Manjon-garcia2019} forms a bright arc on the NW side of the lens and a faint secondary image on the SE side. Both the arc and secondary image are split into two knots. 

\textbf{HELMSJ233441.0$-$065220~(HELMS1; S\_41)}: We model this type 1 foreground lens with two Sérsic components using \texttt{GASP2D}. For this case, we first model and subtract the main images of the background lensed sources. Then, we mask the residual signal of the source and model the lens. The background lensed source is located at $z_{\rm s}^{\rm spec}=1.90$ \citep{Cox2023} and forms an arc split into three bright unresolved images on the S side of the lens and a faint secondary image on the N side of the lens. 

\textbf{HELMSJ233633.5$-$032119$_{1,2}$~(HELMS41; S\_42)}: This type 2 foreground lens is formed by two galaxies, which we deblend by simultaneously modelling them with \texttt{GALFIT} using two Sérsic components for the first galaxy (labelled as '1` in Fig.~\ref{fig:models_full}) and one Sérsic component for the second galaxy (labelled as '2`). The background lensed source is located at $z_{\rm s}^{\rm spec}=2.34$ \citep{Cox2023} and forms a ring (split into three arcs on the N, E, and W of the lens), as well as a secondary faint, more distant arc and a more diffuse component, which are located close to the N arc.

\subsubsection{sub-mm/mm confirmed lenses}

\textbf{HELMSJ001615.7$+$032435~(HELMS13; S\_43)}: We model this type 1 foreground lens at $z_{\rm l}^{\rm spec}=0.66$ \citep{Nayyeri2016} with one Sérsic component and one Gaussian using \texttt{GASP2D}. The background lensed source at $z_{\rm s}^{\rm spec}=2.77$ \citep{Nayyeri2016} is detected by ALMA \citep[Band 7,][]{Amvrosiadis2018}, and forms a wide arc on the NE side of the lens. No secondary image is found. This system was studied in detail by \citet{Dye2018}, who measured an Einstein radius of $\theta_{\rm E}=2.79\pm0.10$ arcsec.

\textbf{HELMSJ002220.9$-$015524~(HELMS29; S\_44)}: We model this type 1 foreground lens with one Sérsic component using \texttt{GASP2D}. The background lensed source at $z_{\rm s}^{\rm spec}=5.16$ \citep{Asboth2016} was detected by ALMA (Band 6, Prop. ID 2015.1.01486.S, PI D. Riechers; Band 7, Prop. ID 2015.1.01486.S, PI D. Riechers; Band 8, Prop. ID 2015.1.01486.S and 2017.1.00043.S, PI D. Riechers). It forms a ring split into two brighter knots on the N and S sides of the lens. \textit{HST} shows an arc-like structure at $\sim1.5-2$ times the Einstein radius that is not detected in the sub-mm. We argue that it is part of the foreground galaxy. This system has also available deeper \textit{HST F105W} (Prop. ID 14083, PI I. P\'{e}rez-Fournon), \textit{F125W} (Prop. ID 15464, PI A. Long), and \textit{F160W} observations (Prop. ID 14083, PI I. P\'{e}rez-Fournon). No counterpart is found in the \textit{HST} after subtracting the lens.

\textbf{HELMSJ003814.1$-$002252~(HELMS24; S\_45)}: We model this type 1 foreground lens at $z_{\rm l}^{\rm phot}=0.17$ \citep{Nayyeri2016} with one PSF using \texttt{GASP2D}. The background lensed source at $z_{\rm s}^{\rm spec}=4.98$ \citep{Su2017} is detected by ALMA \citep[Band 7,][]{Ma2019}. It forms an arc and a closer faint secondary image on the NE and SW sides of the lens, respectively. 

\textbf{HELMSJ003929.6$+$002426~(HELMS11; S\_46)}: The foreground lens is a type 1 edge-on galaxy, which we model with \texttt{ISOFIT}. The background lensed source is located at $z_{\rm s}^{\rm spec}=2.85$ \citep{Cox2023} and was detected by ALMA \citep[Band 7,][]{Ma2019}. It forms a ring with two brighter knots on the N and S sides of the lens. No \textit{HST} counterpart is found after subtracting the lens. 

\textbf{HELMSJ004714.2$+$032454~(HELMS8; S\_47)}: We model this type 1 foreground lens at $z_{\rm l}^{\rm phot}=0.48$ \citep{Nayyeri2016} with one Sérsic component and one exponential component using \texttt{GALFIT} including also the spiral structure. Additional $m=2,\, 3$ Fourier components are added to the spiral arms in order to model their N-S asymmetry. The background lensed source at $z_{\rm s}^{\rm spec}=1.19$ \citep{Nayyeri2016} was detected by ALMA \citep[Band 7,][]{Amvrosiadis2018}. It forms one arc on the S side of the lens and a compact secondary image on the N side. Near-IR counterparts of the background lensed source are visible only for the S arc. This system was studied in detail by \citet{Dye2018}, who measured an Einstein radius of $\theta_{\rm E}= 0.59\pm0.03$ arcsec.

\textbf{HELMSJ005159.4$+$062240$_{1,2,3}$~(HELMS18; S\_48)}: This type 2 foreground lens is formed by a group three galaxies at $z_{\rm l}^{\rm spec}=0.60$ \citep{Okido2021}, which we deblend by simultaneously modelling them with \texttt{GALFIT} using two Sérsic components for the first galaxy (labelled as `1' in Fig.~\ref{fig:models_full}), one de Vaucouleurs component for the second galaxy (labelled as `2'), and one exponential profile, a PSF, and spiral arms for the third galaxy (labelled as `3'). The background lensed source at $z_{\rm s}^{\rm spec}=2.39$ \citep{Nayyeri2016} was detected by ALMA(Band 6, Prop. ID 2017.1.00027.S, PI S. Eales; Band 7, \citealt{Amvrosiadis2018}). It forms an arc on the SW side of galaxy 3, and a secondary image on the E side of galaxies 1 and 2. This system was studied in detail by \citet{Maresca2022}, who modelled the band 7 observations and measured the Einstein radii to be $\theta_{\rm E}=3.80\pm0.02$ arcsec for the galaxies 1 and 2, and $\theta_{\rm E}=1.46\pm0.02$ arcsec for the galaxy 3. No \textit{HST} counterpart is found after subtracting the lens. 

\textbf{HATLASJ005724.2$-$273122~(HerBS60; S\_49)}: We model this type 1 foreground lens at $z_{\rm l}^{\rm phot}=0.89$ \citep{Ward2022} with one Sérsic component using \texttt{GASP2D}. The background lensed source at $z_{\rm s}^{\rm spec}=3.26$ \citep{Urquhart2022} is detected by ALMA (Band 6, Prop. ID 2018.1.00526.S, PI I. Oteo). It forms a ring that splits into two knots on the E and the W side of the lens, respectively. No \textit{HST} counterpart is found after subtracting the lens. 

\textbf{HERMESJ021831$-$053131~(HXMM02; S\_50)}: We model this type 1 foreground lens at $z_{\rm l}^{\rm spec}=1.35$ \citep{Wardlow2013} with one Sérsic component using \texttt{GASP2D}. The background lensed source at $z_{\rm s}^{\rm spec}=3.40$ \citep{Wardlow2013} is detected by ALMA (Band 6, Prop. ID 2013.1.00781.S, PI B. Hatsukade; Band 7, Prop. ID 2015.1.01528.S, PI I. Smail; Band 8, Prop. ID 2013.1.00749.S, PI D. Riechers), and SMA \citep[340 GHz,][]{Bussmann2013}. It forms an arc that splits into two knots on the N side of the lens, an elongated second image on the S side of the lens, and a third elongated image on the W side of the lens. This system was studied in detail by \citet{Bussmann2013} and \citet{Bussmann2015}, who modelled the ALMA band 6 and SMA observations, respectively. They measured an Einstein radius of $\theta_{\rm E} = 0.44\pm0.02$ arcsec and $\theta_{\rm E} = 0.507\pm0.004$ arcsec, respectively. No \textit{HST} counterpart is found after subtracting the lens.

\textbf{HERMESJ033211$-$270536~(HECDFS04; S\_51)}: We model this type 1 foreground lens with one PSF using \texttt{GASP2D}. The background lensed source is detected by ALMA \citep[Band 7,][]{Bussmann2015}. It forms an arc and a faint secondary image on the S and the NW sides of the lens, respectively. This system was studied in detail by \citet{Bussmann2015}, who measured an Einstein radius of $\theta_{{\rm E}} = 0.5$ arcsec.

\textbf{HERMESJ044154$-$540352~(HADFS01; S\_52)}: We model this type 1 foreground lens with one Sérsic component using \texttt{GASP2D}. The background lensed source was detected by ALMA \citep[Band 7,][]{Bussmann2015}. It forms two arcs: one is more extended, splits into two knots, and is located on the E side of the lens, while the other is on the W side. This system was studied in detail by \citet{Bussmann2015}, who measured an Einstein radius of $\theta_{{\rm E}}= 1.006\pm0.004$ arcsec. No \textit{HST} counterpart is found after subtracting the lens. 

\textbf{HATLASJ083932$-$011760~(HerBS105; S\_53)}: The foreground lens is a type 1 edge-on galaxy, which we model with \texttt{ISOFIT}. The background lensed source is located at $z_{\rm s}^{\rm spec}=2.67$ \citep{Cox2023} and was detected by ALMA (Band 6, Prop. ID 2018.1.00526.S, PI I. Oteo). It forms two blended images separated by the disk of the lens. No \textit{HST} counterpart is found after subtracting the lens. 

\textbf{HATLASJ091841$+$023048$_{1,2}$~(HerBS32; S\_54)}: This type 2 foreground lens is formed by two galaxies which we deblend by simultaneously modelling them with \texttt{GALFIT} using two Sérsic components for first galaxy (labelled as `1' in Fig.~\ref{fig:models_full}) and one de Vaucouleurs component for the second one (labelled as `2'). The background lensed source at $z_{\rm s}^{\rm spec}=2.58$ \citep{Harris2012} is detected by ALMA (Band 3, Prop. ID 2017.1.01694.S, PI I. Oteo; and Band 7, \citealt{Giulietti2022}). It forms an arc on the E side of galaxy 1 and a secondary image on the W side of galaxy 2. An additional third ALMA detection is found E of the two galaxies near a \textit{HST} clump and shows a faint near-IR counterpart. 

\textbf{HATLASJ113526$-$014606~(HerBS10; S\_55)}: The background lensed source at $z_{\rm s}^{\rm spec}=3.13$ was detected by ALMA \citep[Band 3, 6, 7, and 8,][]{Giulietti2023}. It forms a wide arc on the SE and a secondary image on the NW. No foreground lens is detected in the near-IR, whereas there is a tentative detection for the background source. This system was studied in detail by \citet{Giulietti2023}, who measured an Einstein radius of $\theta_{\rm E}= 0.4241^{+0.0005}_{-0.0005}$ arcsec.

\textbf{HATLASJ115433.6$+$005042~(HerBS177; S\_56)}: We model this type 1 foreground lens at $z_{\rm l}^{\rm phot}=0.52$ \citep{Shirley2021} with one Sérsic component using \texttt{GASP2D}. The background lensed source at $z_{\rm s}^{\rm phot}=3.90$ \citep{Bakx2020} is detected by ALMA (Band 7, Prop. ID 2019.1.01784.S, PI T. Bakx). It forms an arc on the E side of the lens and a secondary image on the W side. Potential near-IR counterparts of the background lensed source are visible on the S of the lens.

\textbf{HATLASJ120127.6$-$014043~(HerBS61; S\_57)}: We model this type 1 (or 2) foreground lens at $z_{\rm l}^{\rm phot}=0.88$ \citep{Shirley2021} with one Sérsic component and one exponential component using \texttt{GASP2D}. The background lensed source at $z_{\rm s}^{\rm phot}=4.06$ \citep{Manjon-garcia2019} is detected by SMA \citep[340 GHz,][]{Enia2018}. It forms an arc on the NW side of the lens and a bright secondary image on the SE side. This system was studied in detail by \citet{Enia2018}, who measured an Einstein radius of $\theta_{\rm E}=0.82\pm0.04$ arcsec.

\textbf{HATLASJ131611$+$281220~(HerBS89; S\_58)}: We model this type 1 foreground lens at $z_{\rm l}^{\rm phot}=0.90$ \citep[][]{Berta2021} with one Sérsic component using \texttt{GASP2D}. The background lensed source at $z_{\rm s}^{\rm spec}=2.9497$ \citep[][]{Neri2020} was resolved by NOEMA \citep[255 GHz,][]{Berta2021}. It forms an arc on the S side of the lens and a compact image on the N side. This system was studied and modelled in detail by \citet{Berta2021}, who measured an Einstein radius of $\theta_{\rm E} = 0.4832\pm0.0006$ arcsec. There is a tentative detection of the background lensed source in \textit{HST} on the E and W sides of the lens.

\textbf{HATLASJ134429$+$303036~(HerBS1; S\_59)}: We model this type 1 foreground lens at $z_{\rm l}^{\rm spec}=0.67$ \citep{Bussmann2013} with one Sérsic component and one exponential component using \texttt{GASP2D}. The background lensed source at $z_{\rm s}^{\rm spec}=2.30$ \citep{Harris2012} was detected by ALMA \citep[Band 7,][]{Falgarone2017} and SMA \citep[340 GHz,][]{Bussmann2013}. It forms an arc on the W side of the lens and a more distant secondary image on the E side. This system was studied in detail by \citet{Bussmann2013} with the SMA data. They measured an Einstein radius of $\theta_{{\rm E}}= 0.92\pm0.02$ arcsec.

\textbf{HATLASJ141352$-$000027$_{1,2,3}$~(HerBS15; S\_60)}: This type 3 foreground lens is formed by a cluster of galaxies at $z_{\rm l}^{\rm spec}=0.55$ \citep{Bussmann2013}. We model three of the spectroscopically confirmed members, which are likely contributing to lensing, with GASP2D using one Sérsic and one exponential component for the main lens (labelled as `1' in Fig.~\ref{fig:models_full}), and one Sérsic component for the second (labelled as `2') and the third (labelled as `3'). The background lensed source at $z_{\rm s}^{\rm spec}=2.48$ \citep{Harris2012} is detected by ALMA (Band 3, 6, and 7, Prop. ID, 2018.1.00861.S PI C. Yang) and forms two arcs, one on the N of the galaxy 3 and the other on the E side of the galaxy 1. The background lensed source is only partly detected in \textit{HST} images. 

\textbf{HATLASJ142414$+$022304$_{1,2}$~(HerBS13, ID 141; S\_61)}: This type 2 foreground lens is formed by two galaxies at $z_{\rm l}^{\rm spec}=0.60$ \citep{Bussmann2012}, which we deblend by simultaneously modelling them with \texttt{GALFIT} using one Sérsic component and a PSF for the first galaxy (labelled as `1' in Fig.~\ref{fig:models_full}) and one Sérsic component for the second one (labelled as `2'). The background lensed source at $z_{\rm s}^{\rm spec}=4.24$ \citep{Cox2011} is detected by ALMA (Band 3 and 4, \citealt{Dye2022}; Band 5, 6, and 7, \citealt{Dye2018}; and Band 8 Prop. ID 2016.1.00284.S, PI J. Bernard-Salas), and SMA (340 GHz, \citealt{Bussmann2012}; and 880 Ghz, \citealt{Cox2011}). No \textit{HST} counterpart is found after subtracting the lens. 

\textbf{HERMESJ142826$+$345547~(HBootes02; S\_62)}: The foreground lens at $z_{\rm l}^{\rm spec}=0.41$ \citep{Wardlow2013} is a type 1 edge-on galaxy which we model with \texttt{ISOFIT}. The background lensed source at $z_{\rm s}^{\rm spec}=2.41$ \citep{Wardlow2013} was detected by SMA \citep[340 GHz,][]{Bussmann2013} and JVLA \citep[7 GHz,][]{Wardlow2013}. It forms an arc that splits into two knots and is located on the SE side of the lens, and a fainter secondary image on the NW side of the lens. No \textit{HST} counterpart is found after subtracting the lens. 

\textbf{HATLASJ230815.5$-$343801~(HerBS28; S\_63)}: We model this type 1 foreground lens at $z_{\rm l}^{\rm phot}=0.72$ \citep{Ward2022} with two Sérsic components using \texttt{GASP2D}. The background lensed source at $z_{\rm s}^{\rm phot}=4.03$ \citep{Manjon-garcia2019} is detected by ALMA (Band 6, Prop. ID 2018.1.00526.S, PI I. Oteo). It forms an arc that splits into two knots and is located on the N side of the lens and a secondary image on the W side of the lens. No \textit{HST} counterpart is found after subtracting the lens. 

\textbf{HELMSJ232439.5$-$043936~(HELMS7; S\_64)}: We model this type 1 foreground lens with two Sérsic components using \texttt{GASP2D}. The background lensed source at $z_{\rm s}^{\rm spec}=2.47$ \citep{Nayyeri2016} is detected by ALMA (Band 6, \citealt{Amvrosiadis2018}; and 7, Prop. ID 2017.1.00027.S, PI S. Eales). It forms three arcs, two of them are located close to the lens on its W and E sides, whereas the third arc is near an \textit{HST} clump on the E side of the lens. This kind of morphology is indicative of the presence of multiple lenses, which possibly are the \textit{HST} clump and a further clump one on the NW side of the main lens. This system was studied in detail by \citet{Maresca2022}, who adopted two lenses and measured their Einstein radii to be $\theta_{{\rm E}}= 0.54\pm0.01$  arcsec and $\theta_{{\rm E}}= 0.40\pm0.01$ arcsec, respectively. At the position of the second lens, there is no \textit{HST} detection.

\textbf{HELMSJ233620.8$-$060828~(HELMS6; S\_65)}: We model this type 1 foreground lens at $z_{\rm l}^{\rm spec}=0.40$ \citep{Nayyeri2016} with two Sérsic components using \texttt{GASP2D}. The lens is also part of the cluster that hosts other strong lensing events (see \citealt{Carrasco2017} for details). The background lensed source at $z_{\rm s}^{\rm spec}=3.43$ \citep{Nayyeri2016} was detected by ALMA (Band 6, Prop. ID 2021.1.01116.S, PI D. Riechers and \citealt{Amvrosiadis2018}; and Band 8, Prop. ID 2013.1.00749.S, PI D. Riechers). It forms four images: two of them are on the S side of the lens and are connected by an arc, a third one is on the W side, and a fourth one is on the NE side. Near-IR counterparts of the background lensed source are visible only for the S images.

\subsubsection{Uncertain lenses}

\textbf{HATLASJ013840.5$-$281855~(HerBS14; S\_81}): It is unclear whether this source at $z^{\rm spec}=3.78$ \citep{Urquhart2022} is strongly lensed. It was detected by ALMA (Band 6, Prop. ID 2018.1.00526.S, PI I. Oteo) and shows one elongated component. Due to the $\sim0.6$ arcsec resolution, it is not possible to exclude the presence of background lensed sources with very low angular separation. The candidate lens is an edge-on spiral galaxy at $z^{\rm phot}=0.61\pm0.28$ \citep{Ward2022}.

\textbf{HERMESJ022017$-$060143$_{1,2}$~(HXMM01; S\_85)}: This system is confirmed to be a DSFG pair that is weakly lensed by a pair of foreground galaxies. This type 3 foreground system is formed by an edge-on disk galaxy (labelled as `1' in Fig.~\ref{fig:models_full}) at $z_{\rm l}^{\rm phot}=0.87$ \citep{Nayyeri2016} and a second rounder and ringed galaxy (labelled as `2'). The sources at $z_{\rm s}^{\rm phot}=0.87$ \citep{Fu2012} were detected by ALMA (Band 6, \citealt{Bussmann2015} and Prop. ID 2015.1.00723.S, PI I. Oteo; Band 7, Prop. ID 2011.0.00539.S, PI D. Riechers), SMA \citep[340 GHz][]{Fu2012}, and JVLA \citep[30 GHz][]{Fu2012} in the sub-mm/mm and by Keck/NIRC2 AO \citep[\textit{K}$_{s}$][]{Fu2012} in the near-IR. The background sources are located between the two lenses, one on the S side of them and the other on the N side.

\textbf{HERMESJ022135$-$062617~(HXMM03; S\_88)}: It is unclear whether this background source at $z_{\rm s}^{\rm spec}=2.72$ \citep{Bussmann2015} is strongly lensed. We model the brightest cluster galaxy (BCG) of a candidate lensing cluster at $z_{\rm l}^{\rm spec}=0.31$ \citep{Albareti2017} with two Sérsic components using \texttt{GASP2D}. The source was detected by ALMA \citep[Band 7,][]{Bussmann2015}. It forms a bright arc and is located on the E side of the lens. No secondary image was picked up by ALMA. Unfortunately, the W side of the candidate lens, where the counter image is expected to be found, is outside the ALMA Band 7 FOV. A near-IR counterpart of the ALMA arc is clearly visible and shows a complex morphology with multiple knots, while the secondary image is not found even after subtracting the potential lens. One possibility is that the source is only weakly lensed by the foreground structure.

\textbf{HERMESJ045058$-$531654~(HADFS03; S\_91)}: This system was proposed to be a group of three weakly lensed DSFGs by a foreground edge-on disk galaxy \citep{Bussmann2015}. It was detected by ALMA \citep[Band 7,][]{Bussmann2015} and forms three compact clumps near the candidate lens. All the ALMA clumps are on the S on the candidate lens and, as such, are not consistent with the usual strong lensing morphology. 

\textbf{HERMESJ143331$+$345440~(HBootes01; S\_116)}: It is unclear whether this background source at $z^{\rm spec}=3.27$ \citep{Wardlow2013} is strongly lensed. It was detected by SMA \citep[340 GHz,][]{Bussmann2013}, and shows one elongated component. Due to the $\sim0.6$ arcsec resolution, it is not possible to exclude the presence of background lensed sources with very low angular separation. We model the candidate lensing system at $z^{\rm phot}=0.59$ with one Sérsic component.

\textbf{HELMSJ235331.7$+$031717~(HELMS40; S\_139)}: It is unclear whether this source located at $z_{\rm s}^{\rm spec}=$ \citep{Cox2023} is strongly lensed. The source was detected by ALMA \citep[Band 7,][]{Amvrosiadis2018}. It forms two pairs of compact sources. The first pair (on the SE side of the cutout) shows a possible near-IR detection but no foreground lens candidate. The members of the second pair are located on the N and S sides of a possible lensing galaxy, respectively, and they do not show any \textit{HST} counterpart. We model the candidate lensing system with two Sérsic components and a Gaussian using \texttt{GALFIT} to account for the different locations of their centres. This system was studied in detail by \citet{Maresca2022}, who modelled the SE pair and measured an Einstein radius of $\theta_{\rm E}=0.21\pm0.01$ arcsec. We note that the lens redshift adopted by \citet{Maresca2022} is likely associated with a close-by spiral galaxy on these sides of the SE pair.

\textbf{HATLASJ011014.5$-$314813~(HerBS160; S\_153)}: It is unclear whether this background source at $z^{\rm spec}=3.96$ \citep{Urquhart2022} is strongly lensed. It is detected by ALMA (Band 6, Prop. ID 2018.1.00526.S, PI I. Oteo) and forms an elongated, arc-like structure near a faint \textit{HST} counterpart. We find no secondary image. 

\textbf{HERMESJ023006$-$034153~(HXMM12; S\_170)}: This system was proposed to be a weakly lensed DSFG by a group of foreground galaxies \citep{Bussmann2015}. It was detected by ALMA \citep[Band 7,][]{Bussmann2015} and forms an arc-like structure near a faint \textit{HST} counterpart. We find no secondary image. 

\textbf{HERMESJ043341$-$540338~(HADFS04; S\_175)}: This system was proposed to be a system of three DSFGs that is weakly lensed by a foreground group of galaxies \citep{Bussmann2015}. It was detected by ALMA in \citep[Band 7,][]{Bussmann2015}, the two DSFGs are located on the SE and NE sides of the foreground galaxy. This system was studied in detail by \citet{Bussmann2015}, who assumed an Einstein radius of 0.5 arcsec. A near-IR counterpart of the ALMA southernmost source is visible. 

\textbf{HATLASJ144556.1$-$004853~(HERBS46; S\_249)}: It is unclear whether this background source is strongly lensed. It was detected by SMA \citep[340 GHz,][]{Bussmann2013}, and shows one elongated component near a faint \textit{HST} counterpart. Due to the $\sim0.6$ arcsec resolution, it is not possible to exclude the presence of background lensed sources with very low angular separation. 

\textbf{HATLASJ224207.2$-$324159~(HerBS67; S\_262)}: It is unclear whether this background source at $z^{\rm phot}=3.57$ \citep{Bakx2018} is strongly lensed. It was detected by ALMA (Band 6, Prop. ID 2018.1.00526.S, PI I. Oteo) and shows one elongated component. Due to the $\sim0.6$ arcsec resolution, it is not possible to exclude the presence of background lensed sources with very low angular separation.

\begin{table*}
\centering
\caption{Structural parameters of four representative confirmed lensing systems obtained from a parametric fit of their surface brightness distributions.}
\label{tab:models}
{\scriptsize
\begin{tabular}{c l c c c c c c c c c c c}
\hline
\hline
	 No. & IAU Name & Type & Components & $\mu_{\rm e}$ & $R_{\rm e}$ & $n$ & $q$ & $PA$ & $m_{\rm PSF}$ & $C/T$ & $\chi^2$ & $N_{\rm dof}$  \\ 
	   &   &   &   & [mag arcsec$^{-2}$] & [arcsec] &   &   & [deg] & [mag] &   &   &    \\ 
	 (1) & (2) & (3) & (4) & (5) & (6) & (7) & (8) & (9) & (10) & (11) & (12) & (13)  \\ 
\hline
	 S\_12 & HERMESJ100144$+$025709 & 1 & Sérsic & $21.74\pm0.05$ & $0.82\pm0.02$ & $5.35^{+0.11}_{-0.10}$ & $0.606\pm0.005$ & $71.89^{+0.73}_{-0.76}$ & -- & $[1]$ & 1.05 & 43906  \\ 
 \hline 
	 \multirow{3}{*}{S\_11} & HATLASJ083051$+$013225$_{1}$ & \multirow{3}{*}{2} & Sérsic & $22.60^{+0.13}_{-0.12}$ & $0.36\pm0.02$ & $4.86\pm0.36$ & $0.84\pm0.01$ & $33.02^{+1.44}_{-1.45}$ & -- & $[1]$ & \multirow{3}{*}{1.04} & \multirow{3}{*}{32063} \\ 
 \\ 
	   & HATLASJ083051$+$013225$_{2}$ &   & Sérsic & $23.28^{+0.15}_{-0.13}$ & $0.69\pm0.04$ & $1.34^{+0.13}_{-0.12}$ & $0.48\pm0.02$ & $22.24^{+1.87}_{-1.83}$ & -- & $[1]$ &   &    \\ 
 \hline 
	 \multirow{5}{*}{S\_8} & \multirow{3}{*}{HERSJ020941.2$+$001558$_{1}$} & \multirow{5}{*}{3} & Sérsic & $21.13^{+0.19}_{-0.16}$ & $1.54\pm0.18$ & $5.11\pm0.23$ & $0.916\pm0.004$ & $82.42^{+0.93}_{-0.99}$ & -- & $0.56^{+0.07}_{-0.08}$ & \multirow{3}{*}{0.97} & \multirow{3}{*}{121439} \\ 
	   &   &   & Sérsic & $23.16^{+0.13}_{-0.12}$ & $5.03\pm0.13$ & $0.75^{+0.07}_{-0.06}$ & $0.75\pm0.02$ & $81.91^{+1.29}_{-1.26}$ & -- & $0.31^{+0.06}_{-0.05}$ &   &    \\ 
	   &   &   & Sérsic & $22.28^{+0.15}_{-0.13}$ & $2.10^{+0.06}_{-0.07}$ & $0.47\pm0.04$ & $0.92\pm0.02$ & $66.03^{+3.11}_{-3.18}$ & -- & $0.13^{+0.03}_{-0.02}$ &   &    \\ 
 \\ 
	   & HERSJ020941.2$+$001558$_{2}$ &   & Sérsic & $20.49\pm0.06$ & $0.315^{+0.010}_{-0.009}$ & $2.86\pm0.07$ & $0.969\pm0.007$ & $156.46^{+0.93}_{-0.92}$ & -- & $[1]$ & 0.97 & 124612  \\ 
 \hline 
	 S\_50 & HERMESJ021831$-$053131 & 1 & Sérsic & $23.60^{+0.13}_{-0.12}$ & $0.83\pm0.05$ & $5.12\pm0.38$ & $0.38\pm0.01$ & $148.55^{+1.47}_{-1.43}$ & -- & $[1]$ & 1.07 & 49835  \\  
\hline
\hline
\end{tabular}
}
\begin{flushleft}
\textit{Notes}: Col.~(1): Source reference number. Col.~(2): IAU name of the {\it Herschel} detection. Indices 1 and 2 refer to the two components of the lens candidate. Col.~(3): Type of the system. Col.~(4): Adopted model for the lens component. Col.~(5): Effective surface brightness (i.e., the surface brightness at the effective radius). Col.~(6): Effective radius (i.e., the semi-major axis of the isophote containing half of the light of the component). Col.~(7): Sérsic index.  Col.~(8): Axis ratio. Col.~(9): Position angle. Col.~(10): Total magnitude of the unresolved component. Col.~(11): Luminosity ratio of the component with respect to the galaxy. Col.~(12): Reduced $\chi^2$ of the fit. Col.~(13): Number of degrees of freedom of the fit. The values in brackets are left fixed in the fit.
\end{flushleft}
\end{table*}

\begin{figure*}
    \centering
    \includegraphics[width=\textwidth]{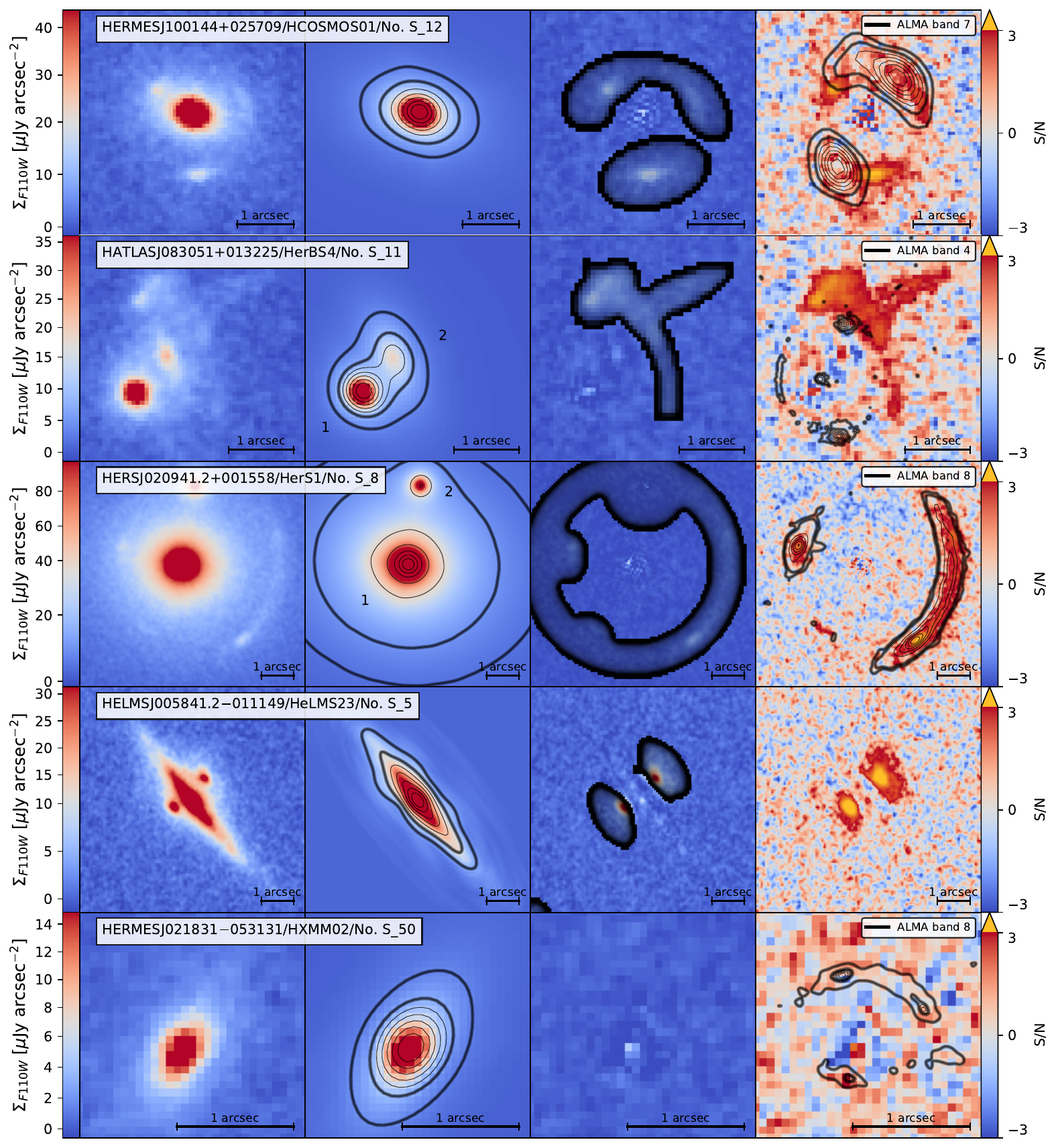}
    \caption{{\em From left to right panels:\/} Observed \textit{F110W} \textit{HST} image, best-fitting lens model, residual, and SNR residual map of five representative lensing systems: one for each type plus a confirmed one by multiwavelength data only. The remaining sample systems are shown in Fig. \ref{fig:models_full}). The best-fitting lens model and residual images result from the surface brightness modelling, while the SNR map is the ratio between the residual image and noise map. The contours in the model images are taken at two levels corresponding to $\textit{SNR} = 5$ and 10 (thick curves), and five uniformly spaced levels between the $\textit{SNR} = 10$ and the maximum SNR in the model image (thin curves). The residual map shows the pixel mask (corresponding to the black-shaded regions) adopted for the surface brightness modelling. The residual maps show the contours of available high-resolution multiwavelength data taken at two levels corresponding to $\textit{SNR} = 5$ and 10 (thick curves), and five uniformly spaced levels between the $\textit{SNR} = 10$ and the maximum SNR in the multiwavelength image (thin black curves). The images are oriented such that N is up and E is to the left.}
    \label{fig:models}
\end{figure*}

\subsection{Properties of the lenses}
\label{sec4.5} 

Since $\sim47\%$ of the modelled lensing galaxies are composed of multiple components, it is necessary to use a set of parameters that describes the global properties of the surface brightness distribution and does not depend on the number and type of components used to model the galaxy. We therefore use the total effective radius $R_{50}$ (i.e., the radius and surface brightness of the circularised isophote that contains half of the light of the galaxy), total effective surface brightness $\mu_{50}$, and total magnitude $m_{F110W}$. For each galaxy, we sampled $10^4$ different combinations of its components to take into account the uncertainties of the best-fitting model of the surface brightness distribution. This is done through a set of truncated normal distributions (one for each component parameter) with central value being the best-fitting value from the surface brightness model, standard deviation the uncertainty of the best-fitting value, and ranges $[0,\infty)$ ADU pixel$^{-1}$, $[0,\infty)$ pixel, $[0.3,10]$, $[0,1]$, and $[0,360]$ deg for $\mu_{\rm e}$, $R_{\rm e}$, $n$, $q$, and PA, respectively. For all the combinations, we estimate $R_{50}$ by computing the curve of growth (COG) and retrieving the radius within which half of the total flux is encircled. The $\mu_{50}$ value is derived from the surface brightness profile at $R_{50}$, while the $m_{F110W}$ value is obtained by summing the luminosities of all the components. For each parameter, we calculate the median and the 16\% and 84\% percentiles from the cumulative distribution function. Additionally, we estimate the $C_{31}=R_{75}/R_{25}$ concentration index \citep[e.g.][]{Watkins2022} in order to compare our results with the literature. To derive the distributions of $R_{25}$ and $R_{75}$, which correspond to the radii containing 25\% and 75\% of the total luminosity, respectively, we apply the same procedure as done for $R_{50}$.

For the edge-on lenses of S\_5, S\_46, S\_53, and S\_62, we run \texttt{ISOFIT} on the model image (Fig.~\ref{fig:models}) after having deconvolved it with the PSF, and then we use the result to produce the COG. We also measure the surface brightness radial profile. We use the COG to estimate the values of $R_{25}$, $R_{50}$, and $R_{75}$ and to assess their uncertainties, we build a set of $10^4$ COGs for each galaxy. We sample the flux within each circularised radius with a Gaussian distribution centred on the measured flux value and with standard deviation being the flux Poisson uncertainty. The circularised radius is sampled with a Gaussian distribution centred on the measured radius value and with standard deviation being the uncertainty on the ellipticity as measured by \texttt{ISOFIT}. We use the surface radial profile to estimate the values of $\mu_{50}$, and to assess its uncertainty, we build a set of 10$^4$ surface brightness radial profiles. At each radius, we adopt a Gaussian distribution centred on the measured value of surface brightness, with standard deviation being the uncertainty measured by \texttt{ISOFIT}.

\begin{figure*}
    \centering
    \includegraphics[width=0.9\textwidth]{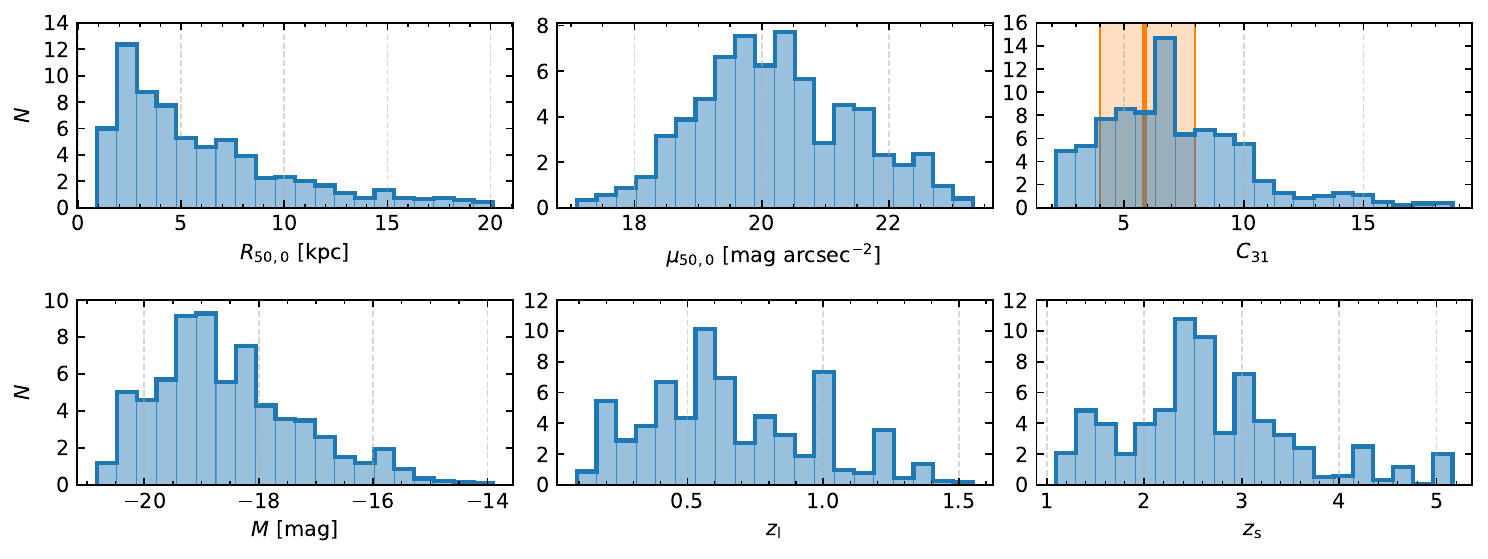}
    \caption{Distribution of the total effective radii $R_{50,0}$ (top left panel), total effective surface brightnesses $\mu_{50,0}$ (top central panel), concentrations $C_{31}$ (top right panel), total absolute magnitudes $M_{F110W}$ (bottom left panel), redshifts $z_{\rm l}$ of the foreground lenses (bottom central panel), and redshifts $z_{\rm s}$ of the background sources (bottom right panel). All the foreground lenses and background sources with a measured redshift are included. The orange line corresponds to the median of the values of $C_{31}$ of the ETGs sample of \citet{Watkins2022}, while the orange shaded region marks the corresponding $\pm 1\sigma$ confidence interval.}
    \label{fig:hists}
\end{figure*}

In Fig.~\ref{fig:hists}, we show the $C_{31}$ distribution of our lenses. The peak at $C_{31} \simeq 7$ is due to the systems that we model with a de Vaucouleurs profile and have all the same concentration. 
We measure a median $\langle C_{31} \rangle=6.8\pm0.4$, which is slightly higher than the median value $\langle C_{31} \rangle=5.2\pm0.1$ measured by \citet{Watkins2022} for a sample of nearby ETGs observed with \textit{Spitzer} at 3.6\,$\mu$m. This finding and the observed morphology of smooth and featureless elliptical systems provide further evidence that our lenses are mostly ETGs. Lastly, we measure the flux densities at 1.1 $\mu$m $S_{1.1}$ of the background sources by taking the aperture photometry in the regions dominated by the lensing features on the lens subtracted residuals. The uncertainties $\sigma_{S_{1.1}}$ are computed with $\sigma_{S_{1.1}}^2 = \sum_{i, j} \sigma_{i, j}^2$, where $\sigma_{i, j}$ is the noise map. The values of $S_{1.1}$ and their uncertainties are given in Tables~\ref{tab:params} and \ref{tab:params_full}.

For the 68 lensing foreground galaxies with known photometric or spectroscopic redshift, we convert the size of the total effective radius from arcsec to kpc through the angular diameter distance and the total magnitude to absolute magnitude $M_{F110W}$ through the luminosity distance. We also correct the total effective surface brightness for cosmological dimming by multiplying the \textit{F110W} fluxes by a factor of $(1+z)^4$. This is done for the parameter distributions as derived before. To take into account the uncertainties on the photometric redshifts, we sample a truncated normal distribution centred on the redshift measurement, with standard deviation being the uncertainty and range $[0,\infty)$. When more than one potential lens is present (i.e. for type 2 and 3 systems), we correct for the redshift only if the secondary potential lens has a spectroscopic redshift (S\_8, S\_11, S\_15, and S\_36) or none of the lenses dominates the available photometric redshift (S\_23, S\_28, and S\_33). To denote the parameters corrected for the lens redshifts, we add a subscript `0'. In Table~\ref{tab:params}, we list the measured values of $m_{F110W}$, $M_{F110W}$, $R_{50}$, $R_{50,0}$, $\mu_{50}$, $\mu_{50,0}$, and $C_{31}$ for the five representative systems shown in Fig.~\ref{fig:models}, while in Table~\ref{tab:params_full} we show the parameters for the rest of the sample.

\begin{table}
\centering
\caption{Derived properties of five representative lensing systems.}
\label{tab:params}
{\scriptsize
\begingroup
\setlength{\tabcolsep}{1pt}
\begin{tabular}{c l c c c c c}
\hline
\hline
	 No. & IAU Name & $m_{F110W}$ & $\mu_{\rm 50}$ & $R_{\rm 50}$ & $C_{31}$ & $\mu S_{1.1}$  \\ 
	   &   & [mag] & [mag arcsec$^{-2}$] & [arcsec] & &  [$\mu$Jy]  \\ 
	   &   & $M_{F110W}$ & $\mu_{\rm 50,\, 0}$ & $R_{\rm 50,\, 0}$ & &  \\ 
	   &   & [mag] & [mag kpc$^{-2}$] & [kcp] &  &  \\ 
	 (1) & (2) & (3) & (4) & (5) & (6) & (7)  \\ 
\hline
	 \multirow{2}{*}{S\_12} & \multirow{2}{*}{HERMESJ100144$+$025709} & $19.17^{+0.08}_{-0.08}$ & $21.74^{+0.05}_{-0.05}$ & $0.64^{+0.02}_{-0.02}$ & \multirow{2}{*}{$9.37^{+0.20}_{-0.20}$} & \multirow{2}{*}{$10.14^{+0.17}_{-0.17}$} \\ 
	   & 	   & $-20.17^{+0.08}_{-0.08}$ & $18.96^{+0.05}_{-0.05}$ & $5.10^{+0.13}_{-0.13}$ &  &  \\ 
\hline
	 \multirow{4}{*}{S\_11} & \multirow{2}{*}{HATLASJ083051$+$013225$_{1}$} & $21.54^{+0.18}_{-0.18}$ & $22.60^{+0.13}_{-0.12}$ & $0.33^{+0.02}_{-0.02}$ & \multirow{2}{*}{$8.47^{+0.66}_{-0.64}$} & \multirow{4}{*}{$6.79^{+0.11}_{-0.11}$} \\ 
	   & 	   & $-16.86^{+0.18}_{-0.18}$ & $20.48^{+0.13}_{-0.12}$ & $2.30^{+0.13}_{-0.13}$ &  &  \\ 
	   & \multirow{2}{*}{HATLASJ083051$+$013225$_{2}$} & $22.05^{+0.21}_{-0.19}$ & $23.28^{+0.15}_{-0.13}$ & $0.48^{+0.03}_{-0.03}$ & \multirow{2}{*}{$3.21^{+0.16}_{-0.15}$}  &   \\ 
	   & 	   & $-17.59^{+0.21}_{-0.20}$ & $20.27^{+0.15}_{-0.13}$ & $3.94^{+0.26}_{-0.25}$ &  &  \\ 
\hline
	 \multirow{4}{*}{S\_8} & \multirow{2}{*}{HERSJ020941.2$+$001558$_{1}$} & $16.15^{+0.18}_{-0.17}$ & $21.27^{+0.19}_{-0.17}$ & $2.47^{+0.13}_{-0.12}$ & \multirow{2}{*}{$5.35^{+0.45}_{-0.47}$} & \multirow{4}{*}{$18.72^{+0.34}_{-0.34}$} \\ 
	   & 	   & $-19.38^{+0.18}_{-0.17}$ & $20.47^{+0.19}_{-0.17}$ & $8.48^{+0.46}_{-0.40}$ &  &  \\ 
	   & \multirow{2}{*}{HERSJ020941.2$+$001558$_{2}$} & $19.82^{+0.09}_{-0.09}$ & $20.49^{+0.06}_{-0.06}$ & $0.310^{+0.010}_{-0.009}$ & \multirow{2}{*}{$5.24^{+0.10}_{-0.10}$}  &  \\ 
	   & 	   & $-15.71^{+0.09}_{-0.09}$ & $19.69^{+0.06}_{-0.06}$ & $1.06^{+0.03}_{-0.03}$ &  &  \\ 
\hline
	 \multirow{2}{*}{S\_5} & \multirow{2}{*}{HELMSJ005841.2$-$011149} & $18.910^{+0.001}_{-0.001}$ & $20.26^{+0.02}_{-0.02}$ & $0.478^{+0.005}_{-0.006}$ & \multirow{2}{*}{$3.05^{+0.04}_{-0.04}$} & \multirow{2}{*}{$14.55^{+0.10}_{-0.10}$} \\ 
	   & 	   & $-18.15^{+0.57}_{-0.48}$ & $18.88^{+0.25}_{-0.24}$ & $2.54^{+0.31}_{-0.35}$ &  &  \\ 
\hline
	 \multirow{2}{*}{S\_50} & \multirow{2}{*}{HERMESJ021831$-$053131} & $21.55^{+0.19}_{-0.18}$ & $23.60^{+0.13}_{-0.12}$ & $0.51^{+0.03}_{-0.03}$ & \multirow{2}{*}{$8.95^{+0.71}_{-0.69}$} & \multirow{2}{*}{$<0.04$} \\ 
	   & 	   & $-18.89^{+0.19}_{-0.18}$ & $19.89^{+0.13}_{-0.12}$ & $4.40^{+0.27}_{-0.26}$ &  &  \\ 
\hline
\hline
\end{tabular}
\endgroup
}
\begin{flushleft}
\textit{Notes}: Col.~(1): Source reference number. Col.~(2): IAU name of the {\it Herschel} detection. Indices 1 and 2 refer to the two components of the lens candidate. Col.~(3): Apparent magnitude of the model (first row), absolute magnitude of the model (second row). Col.~(4): Total effective radius in arcsec (first row) and in kpc (second row). Col.~(5): Total effective surface brightness before (first row) and after correcting for cosmological dimming (second row). Col.~(6): Concentration index of the model.
\end{flushleft}
\end{table}
Figure~\ref{fig:hists} shows the distribution of $R_{50,0}$ in kpc, $\mu_{50,0}$ in mag arcsec$^{-2}$, $C_{31}$, $M_{F110W}$ in mag, $z_{\rm l}$ and $z_{\rm s}$. For each parameter, we stack all the single-system parameter distributions to take into account the varying uncertainties between different systems. We bin them together and normalised for the total number of realisations (i.e., $10^4$ for each system). 

\section{Lens modelling}
\label{sec5} 

For the 34 systems for which we can confidently identify a galaxy acting as the main lens and the background source in the \textit{HST} images, we proceed to estimate the Einstein radius $\theta_{E}$ from the residual images. Depending on the morphology of the lensing features, we define $\theta_{\rm E}$ as either the radius of the circle centred on the lens that best fits the Einstein ring (e.g., S\_4), or as half of the separation between the multiple images (e.g., S\_2) following the approach of \citet[][]{Amvrosiadis2018}. For the uncertainty on $\theta_{\rm E}$, we adopt the value corresponding to one-third of the observed width of the lensing feature. We use these rough estimates as priors for proper lens modelling.

The lens modelling is performed with \texttt{PyAutoLens}\footnote{\texttt{PyAutoLens} is available from \url{https://github.com/Jammy2211/PyAutoLens}} \citep{Nightingale2018, Nightingale2021}, by using a non-linear search trough \texttt{dynesty} \citep{Speagle2020}\footnote{\texttt{dynesty} is available from \url{https://dynesty.readthedocs.io/en/stable/}}. In order to reduce the complexity of the lens model, we adopt a singular isothermal ellipsoid (SIE) for the mass model (see \citealt{Nightingale2018} for details). It has five free parameters, the differences $\delta x_{\rm SIE}$ and $\delta y_{\rm SIE}$ of the centre coordinates of the mass distribution relative to the centre coordinates of the lens surface brightness, Einstein radius $\theta_{\rm E}$, and elliptical components $\epsilon_{\rm SIE,1} = f_{\rm SIE} \sin(2PA_{\rm SIE})$, $\epsilon_{\rm SIE,2} = f_{\rm SIE} \cos(2PA_{\rm SIE})$, where $f_{\rm SIE} = (1-q_{\rm SIE})/(1+q_{\rm SIE})$, $PA_{\rm SIE}$ is the position angle of the mass profile, and $q_{\rm SIE}$ is its flattening.

From the lens-subtracted images and noise maps as obtained in Sec.~\ref{sec4}, we set up the inputs for the lens modelling. We crop the images and we mask the regions that do not include lensing features in order to reduce the computation time. To this aim we also crop the PSF to $\sim5\,{\rm FWHM} \times\,5\,{\rm FWHM}$ cutouts. We then mask any residual from the lens subtraction and non-modelled nearby contaminants. 
We estimate the position of the lensing features by locating the brightest pixel in $5\,{\rm pixel} \times\,5\, {\rm pixel}$ boxes encompassing the lensing features. These estimates are used by \texttt{PyAutoLens} to constrain the location of the background source in the source plane (see \citealt{Nightingale2018} for details). 

We proceed with the lens modelling with the following approach. First, we perform a parametric fit by fitting both the mass model and background source with parametric functions. We use a single Sérsic component for the parametric background source. This fit has a total of twelve free parameters: five for the mass model and seven for the background source. The priors for the mass model are a Gaussian distribution centred on the cutout centre with $\sigma = 0.1$ arcsec for $\delta x_{\rm SIE}$ and $\delta y_{\rm SIE}$; a Gaussian distribution centred on the estimated $\theta_{\rm E}$ with $\sigma$ three times its error; a Gaussian distribution centred on 0 with $\sigma = 0.3$ for $\epsilon_{\rm SIE,1}$ and $\epsilon_{\rm SIE,2}$, respectively. The priors for the background source are Gaussian distribution centred on the cutout centre with $\sigma = 0.3$ arcsec for $x_{\rm s}$ and $y_{\rm s}$; a log-uniform distribution between $10^{-6}$ ADU and $10^6$ ADU for $I_{\rm{s,eff}}$; a uniform distribution between 0 arcsec and 30 arcsec for $R_{\rm{s,eff}}$; a uniform distribution between 0.8 and 5.0 for $n_{\rm s}$; a Gaussian distribution centred on 0 with $\sigma = 0.3$ for $\epsilon_{\rm s,1}$ and $\epsilon_{\rm s,2}$ (defined in a similar way as $\epsilon_{\rm SIE,1}$ and $\epsilon_{\rm SIE,2}$), respectively. As position threshold, we use a Gaussian distribution centred on the previously identified position with $\sigma = 0.5$ arcsec for all the systems, except for S\_41. For the latter, we adopt $\sigma = 0.1$ due to the small angular separation of the lensed images. We then use the posteriors of the mass model obtained in this parametric fit and the priors on the position of the background sources on the image plane to initialize a Voronoi pixelization with a constant regularization coefficient $\lambda$ (see \citealt{Nightingale2018} for details). For the pixelization, we use as priors a uniform distribution between 20 and 45 pixels for the shape of the pixelization and a log-uniform distribution between $10^{-6}$ and $10^6$ for $\lambda$.
We do this pixelized initialization fit by marginalizing the lens model over the mass model and only fitting the pixelization. In a similar fashion, we marginalize over the resulting pixelization to fit the mass model with priors defined by the parametric fit posteriors. We consider this fit as the best guess for the lens model and source reconstruction.

In the cases of S\_6, S\_11, S\_22, S\_24, S\_27, and S\_31 the SNR of the background lensed sources is not high enough to produce reliable results. S\_2, S\_5,  S\_25, and S\_41 show very complex and/or clumpy lensed background source morphologies that needed additional information on what is part of the background source or additional multiple images. Whereas, S\_26 shows residuals related to the lens subtraction that prevent us from building the model.
For S\_3, S\_19, S\_40, the SIE mass model is not sufficient to correctly reproduce the observations, so we add a shear component, defined, as for the mass model, through the elliptical components $\epsilon_{\rm shear,\, 1}$ and $\epsilon_{\rm shear,\, 2}$. This is done to improve the mass model and not to model an actual external shear (see \citealt{Etherington2023} for details). The priors for the shear elliptical components are uniform distributions between $-0.2$ and $0.2$.
%
%

\begin{figure}
    \centering
    \includegraphics[width=0.45\textwidth]{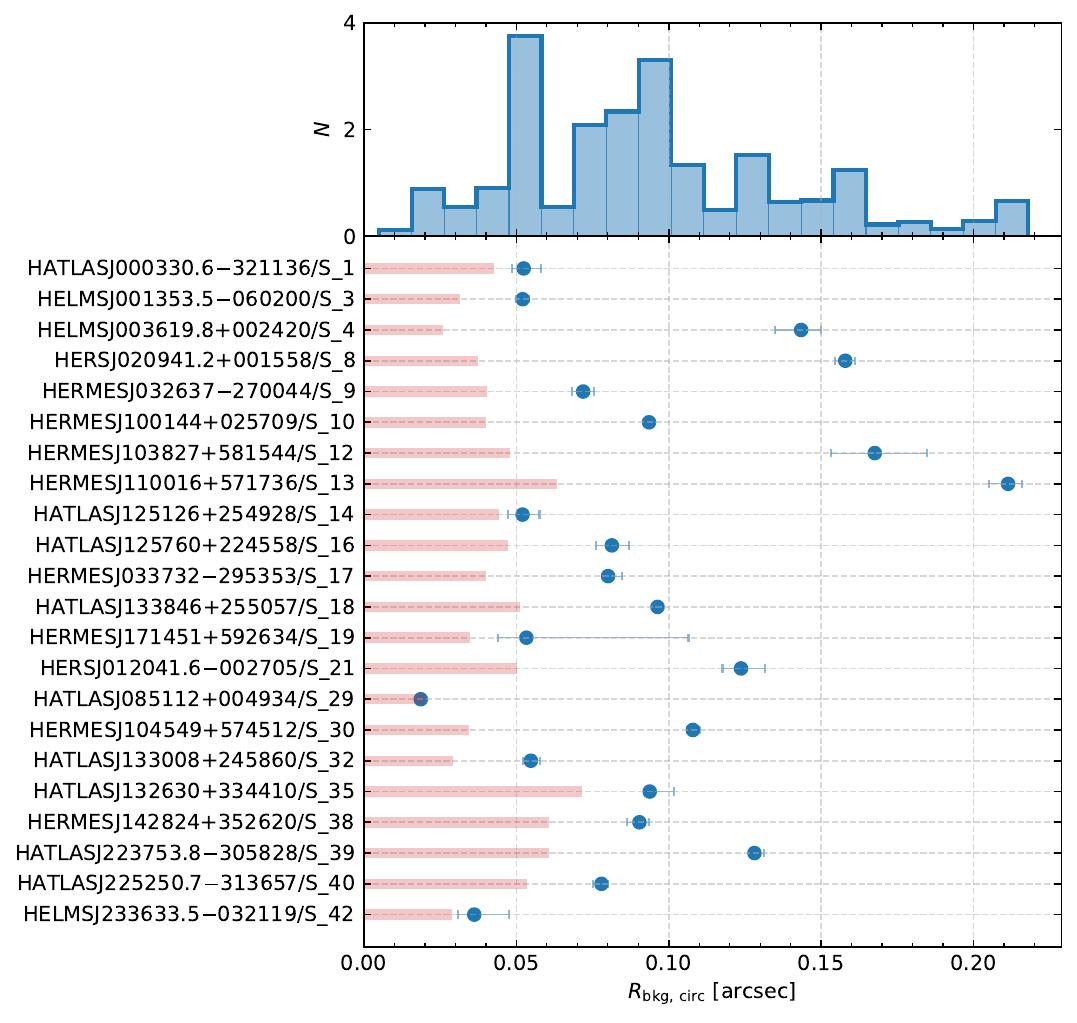}
    \caption{\textit{Top panel}: Distribution of the circularised radius of the background sources from the lens modelling and taking into account the uncertainties on the radii. \textit{Bottom panel}: Circularised radius of the background sources (blues circles) and HWHM of the PSF of the image corrected for the lensing magnification (pink line).}
    \label{fig:sizes}
\end{figure}

\begin{figure*}
    \centering
    \includegraphics[width=\textwidth]{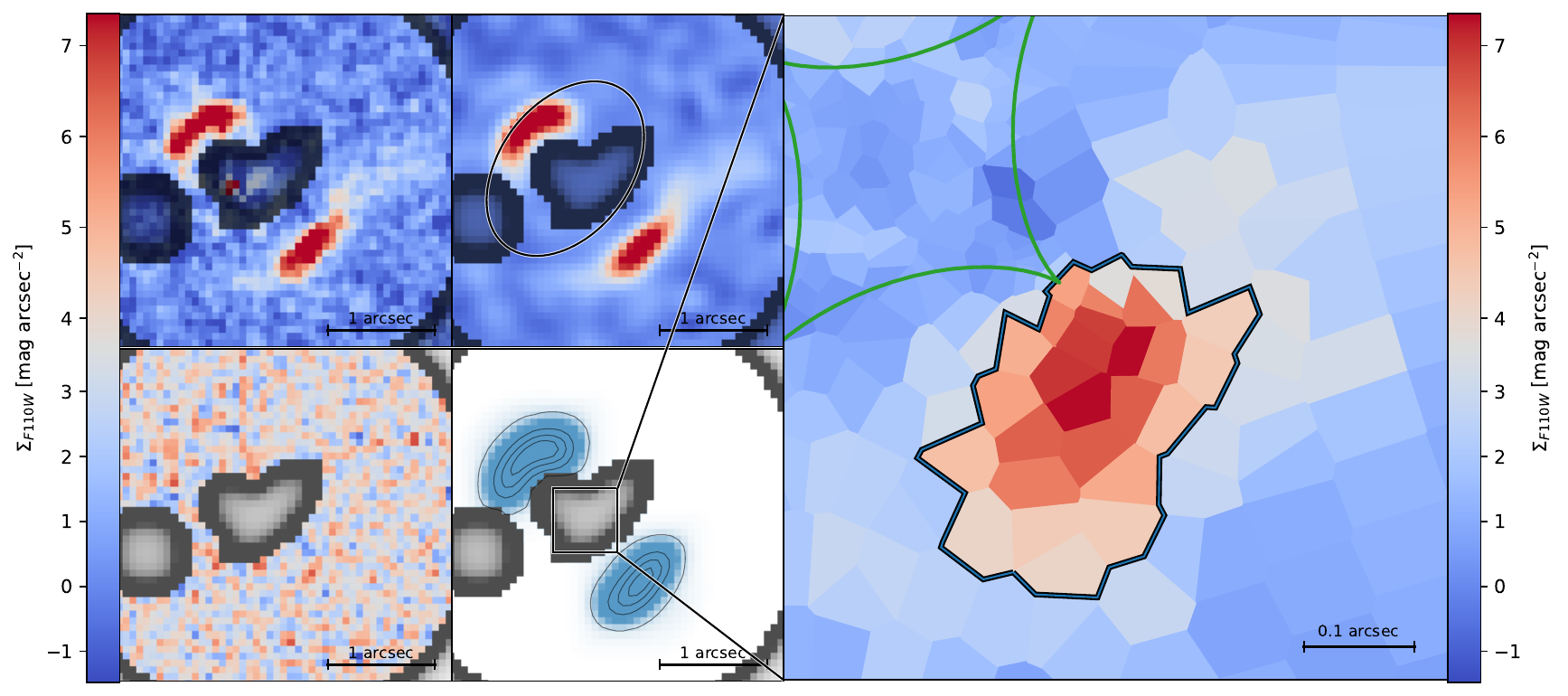}
    \caption{\textit{Left panels:\/} Lens subtracted image (top left panel), best-fitting background lensed source (top right panel), SNR residual map (bottom left panel), and ${\rm SNR}>3$ regions of the source plane lensed back to the image plane of S\_7. Each panel shows the pixel mask (corresponding to the black-shaded regions) adopted for the lens modelling. The top right panel shows the critical curves (black line). The bottom right panel show the region of the source plane displayed in the right panel. \textit{Right panel:\/} Source plane reconstruction of the background source and caustics curves (green line). We highlight with black and blue contour the $\textit{SNR}>3$ region adopted for computing the lensing magnification and source size.}
    \label{fig:lm}
\end{figure*}

For each modelled lens, we extract 500 fits from the lens modelling parameters posteriors, and, for each fit, we compute the regions of the source plane with $\textit{SNR}>3$. We compute the magnification by taking the ratio between the fluxes of the regions on the source plane and the image plane after mapping them back. We measure the sizes of the source plane by computing the radii of the circularised regions that we used to compute the magnifications. The distribution of the physical sizes of our systems is shown in Fig.~\ref{fig:sizes}.
As before, the results of the lens modelling for S\_7 are shown in Fig.~\ref{fig:lm}, while the results for the remaining lenses are reported in Table~\ref{tab:LM} and Fig.~\ref{fig:LM1}. 

\begin{landscape}
\begin{table}
\centering
\caption{Results of the lens modelling.}
\label{tab:LM}
{\scriptsize
\begingroup
\setlength{\tabcolsep}{1pt}
\begin{tabular}{c l c c c c c c c c c c c c c c}
\hline
\hline
	 No. & IAU name & $\delta x$ & $\delta y$ & $\theta_{E}$ & $q$ & $PA$ & $\epsilon_{shear,\, 1}$ & $\epsilon_{shear,\, 2}$ & $x_{\rm pix}$ & $y_{\rm pix}$ & $\lambda$ & $\mu$ & $R_{\rm bkg,\,circ}$ & $\kappa_0$ &    \\ 
	   &   & [arcsec] & [arcsec] & [arcsec] &   &   &   &   &   &   &   &   & [arcsec] & [] &    \\ 
	   &   & $\delta x_0$ & $\delta y_0$ & $\theta_{E,\, 0}$ &   &   &   &   &   &   &   &   & $R_{\rm bkg,\,circ,\,0}$ & $M$ & $M/L$  \\ 
	   &   & [kpc] & [kpc] & [kpc] &   &   &   &   &   &   &   &   & [kpc] & [$10^{11}$ M$_{\odot}$] &    \\ 
	 (1) & (2) & (3) & (4) & (5) & (6) & (7) & (8) & (9) & (10) & (11) & (12) & (13) & (14) & (15) & (16)  \\ 
\hline
	 \multirow{2}{*}{S\_1} & \multirow{2}{*}{HATLASJ000330.6$-$321136} & $-0.13^{+0.03}_{-0.02}$ & $-0.05^{+0.01}_{-0.02}$ & $0.733^{+0.006}_{-0.009}$ & $0.85\pm0.03$ & $80.11^{+3.78}_{-10.85}$ & -- & -- & $39^{+3}_{-9}$ & $33^{+8}_{-7}$ & $0.0101^{+0.0011}_{-0.0009}$ & $5.48^{+0.62}_{-0.48}$ & $0.052^{+0.006}_{-0.004}$ & $1.68^{+0.03}_{-0.04}$ & --  \\ 
	   & 	   & $-0.65^{+0.19}_{-0.16}$ & $-0.24^{+0.08}_{-0.12}$ & $3.85^{+0.61}_{-0.85}$ &   &   &   &   &   &   &   &   & $0.41^{+0.04}_{-0.03}$ & $0.92^{+0.23}_{-0.28}$ & $2.93^{+1.65}_{-0.98}$  \\ 
\hline
	 \multirow{2}{*}{S\_3} & \multirow{2}{*}{HELMSJ001353.5$-$060200} & $-0.038^{+0.004}_{-0.003}$ & $0.066^{+0.004}_{-0.006}$ & $0.658^{+0.004}_{-0.003}$ & $0.85^{+0.03}_{-0.04}$ & $-49.56^{+8.51}_{-8.32}$ & $0.017^{+0.008}_{-0.010}$ & $0.191\pm0.009$ & $40\pm4$ & $31\pm9$ & $0.0084\pm0.0006$ & $10.37^{+0.47}_{-0.56}$ & $0.052\pm0.002$ & $1.49^{+0.02}_{-0.01}$ & --  \\ 
	   & 	   & $-0.26^{+0.05}_{-0.04}$ & $0.45^{+0.07}_{-0.08}$ & $4.56^{+0.50}_{-0.74}$ &   &   &   &   &   &   &   &   & $0.45\pm0.02$ & $1.48^{+0.52}_{-0.44}$ & $4.19^{+3.00}_{-1.48}$  \\ 
\hline
	 \multirow{2}{*}{S\_4} & \multirow{2}{*}{HELMSJ003619.8$+$002420} & $-0.015^{+0.002}_{-0.001}$ & $0.122^{+0.001}_{-0.002}$ & $1.9349^{+0.0017}_{-0.0008}$ & $0.813^{+0.001}_{-0.003}$ & $-79.38^{+0.23}_{-0.21}$ & -- & -- & $36$ & $40$ & $0.0164^{+0.0009}_{-0.0008}$ & $15.82^{+1.23}_{-0.72}$ & $0.143^{+0.006}_{-0.008}$ & $12.16\pm0.3$ & --  \\ 
	   & 	   & $-0.063^{+0.007}_{-0.005}$ & $0.504^{+0.006}_{-0.008}$ & $7.977^{+0.007}_{-0.004}$ &   &   &   &   &   &   & $1.25^{+0.06}_{-0.07}$ & $5.234\pm0.001$ & $3.40^{+0.68}_{-0.54}$  \\ 
\hline
	 \multirow{2}{*}{S\_7} & \multirow{2}{*}{HERSJ012620.5$+$012950} & $-0.006^{+0.015}_{-0.007}$ & $0.14^{+0.01}_{-0.02}$ & $0.766\pm0.004$ & $0.60^{+0.07}_{-0.03}$ & $52.13^{+0.85}_{-0.58}$ & -- & -- & $31^{+4}_{-0}$ & $40^{+4}_{-7}$ & $0.0114\pm0.0006$ & $5.19^{+0.56}_{-0.20}$ & $0.142^{+0.003}_{-0.010}$ & $1.73^{+0.02}_{-0.01}$ & --  \\ 
	   & 	   & $-0.04^{+0.08}_{-0.04}$ & $0.79^{+0.09}_{-0.13}$ & $4.44^{+0.28}_{-0.32}$ &   &   &   &   &   &   &   &   & $1.23^{+0.03}_{-0.08}$ & $1.32^{+0.18}_{-0.17}$ & $2.36^{+0.73}_{-0.54}$  \\ 
\hline
	 \multirow{2}{*}{S\_8} & \multirow{2}{*}{HERSJ020941.2$+$001558} & $-0.259^{+0.005}_{-0.010}$ & $0.046^{+0.008}_{-0.002}$ & $2.579^{+0.002}_{-0.003}$ & $0.704^{+0.003}_{-0.004}$ & $-10.81^{+0.55}_{-0.21}$ & -- & -- & $43$ & $43$ & $0.0093\pm0.0004$ & $7.29^{+0.30}_{-0.10}$ & $0.158\pm0.003$ & $18.61^{+0.08}_{-0.19}$ & --  \\ 
	   & 	   & $-0.89^{+0.02}_{-0.03}$ & $0.159^{+0.028}_{-0.006}$ & $8.844^{+0.007}_{-0.011}$ &   &   &   &   &   &   &   &   & $1.30\pm0.03$ & $5.99^{+0.03}_{-0.06}$ & $4.96^{+0.95}_{-0.77}$  \\ 
\hline
	 \multirow{2}{*}{S\_9} & \multirow{2}{*}{HERMESJ032637$-$270044} & $0.024^{+0.009}_{-0.013}$ & $-0.12\pm0.01$ & $0.981^{+0.005}_{-0.006}$ & $0.87\pm0.01$ & $-40.54^{+1.64}_{-2.25}$ & -- & -- & $42^{+0}_{-9}$ & $28^{+11}_{-5}$ & $0.0110^{+0.0010}_{-0.0009}$ & $6.15^{+0.32}_{-0.38}$ & $0.072\pm0.004$ & $3.01^{+0.03}_{-0.04}$ & --  \\ 
	   & 	   & -- & -- & -- &   &   &   &   &   &   &   &   & -- & -- & --  \\ 
\hline
	 \multirow{2}{*}{S\_10} & \multirow{2}{*}{HERMESJ033732$-$295353} & $0.1754^{+0.0007}_{-0.0005}$ & $0.0576^{+0.0012}_{-0.0008}$ & $1.9718^{+0.0003}_{-0.0006}$ & $0.6495^{+0.0003}_{-0.0006}$ & $-80.49^{+0.04}_{-0.15}$ & -- & -- & $41$ & $38$ & $0.0057^{+0.0002}_{-0.0003}$ & $6.27^{+0.50}_{-0.05}$ & $0.0801^{+0.0045}_{-0.0002}$ & $11.662^{+0.004}_{-0.007}$ & --  \\ 
	   & 	   & $0.57^{+0.11}_{-0.12}$ & $0.18^{+0.04}_{-0.05}$ & $6.40^{+1.18}_{-1.30}$ &   &   &   &   &   &   &   &   & -- & -- & --  \\ 
\hline
	 \multirow{2}{*}{S\_12} & \multirow{2}{*}{HERMESJ100144$+$025709} & $-0.033^{+0.003}_{-0.004}$ & $0.042^{+0.009}_{-0.003}$ & $0.930\pm0.002$ & $0.686^{+0.006}_{-0.008}$ & $-16.66^{+0.39}_{-0.56}$ & -- & -- & $33^{+11}_{-10}$ & $30^{+9}_{-0}$ & $0.0076^{+0.0004}_{-0.0003}$ & $6.33^{+0.15}_{-0.17}$ & $0.094^{+0.002}_{-0.001}$ & $2.63\pm0.02$ & --  \\ 
	   & 	   & $-0.26^{+0.02}_{-0.03}$ & $0.34^{+0.08}_{-0.02}$ & $7.44^{+0.02}_{-0.01}$ &   &   &   &   &   &   &   & -- & -- & --  \\ 
\hline
	 \multirow{2}{*}{S\_13} & \multirow{2}{*}{HERMESJ103827$+$581544} & $0.228^{+0.016}_{-0.006}$ & $0.27\pm0.01$ & $2.408\pm0.010$ & $0.582\pm0.007$ & $-70.47^{+0.76}_{-0.47}$ & -- & -- & $44$ & $44$ & $0.025\pm0.002$ & $4.29^{+0.15}_{-0.25}$ & $0.17^{+0.02}_{-0.01}$ & $15.68\pm0.4$ & --  \\ 
	   & 	   & $1.56^{+0.11}_{-0.04}$ & $1.84^{+0.09}_{-0.10}$ & $16.47\pm0.07$ &   &   &   &   &   &   &   & -- & -- & --  \\ 
\hline
	 \multirow{2}{*}{S\_14} & \multirow{2}{*}{HERMESJ110016$+$571736} & $-0.324^{+0.007}_{-0.009}$ & $-0.251^{+0.007}_{-0.011}$ & $1.081\pm0.003$ & $0.71\pm0.01$ & $11.63^{+1.13}_{-0.88}$ & -- & -- & $32^{+5}_{-7}$ & $42^{+1}_{-0}$ & $0.0151^{+0.0008}_{-0.0009}$ & $2.43^{+0.03}_{-0.04}$ & $0.211^{+0.004}_{-0.006}$ & $3.56\pm0.02$ & --  \\ 
	   & 	   & $-2.48^{+0.05}_{-0.07}$ & $-1.92^{+0.05}_{-0.08}$ & $8.28^{+0.02}_{-0.03}$ &   &   &   &   &   &   &   &   & $1.84^{+0.04}_{-0.06}$ & $5.39^{+0.03}_{-0.04}$ & $1.95^{+0.10}_{-0.09}$  \\ 
\hline
	 \multirow{2}{*}{S\_16} & \multirow{2}{*}{HATLASJ125126$+$254928} & $-0.40\pm0.02$ & $0.056^{+0.009}_{-0.014}$ & $1.133^{+0.016}_{-0.007}$ & $0.58^{+0.01}_{-0.02}$ & $20.05^{+2.41}_{-3.88}$ & -- & -- & $39\pm3$ & $37^{+5}_{-9}$ & $0.015\pm0.002$ & $5.10^{+0.37}_{-0.67}$ & $0.052^{+0.006}_{-0.005}$ & $3.73^{+0.13}_{-0.04}$ & --  \\ 
	   & 	   & $-2.75^{+0.24}_{-0.23}$ & $0.38^{+0.08}_{-0.09}$ & $7.94^{+0.50}_{-0.66}$ &   &   &   &   &   &   &   &   & $0.39\pm0.04$ & $3.20^{+0.45}_{-0.46}$ & $3.29^{+1.07}_{-0.72}$  \\ 
\hline
	 \multirow{2}{*}{S\_17} & \multirow{2}{*}{HATLASJ125760$+$224558} & $0.104^{+0.012}_{-0.007}$ & $-0.06^{+0.01}_{-0.04}$ & $0.681^{+0.005}_{-0.006}$ & $0.70^{+0.07}_{-0.02}$ & $60.10^{+1.21}_{-2.50}$ & -- & -- & $38$ & $43$ & $0.0059\pm0.0004$ & $4.46^{+0.56}_{-0.52}$ & $0.081^{+0.006}_{-0.005}$ & $1.41^{+0.04}_{-0.03}$ & --  \\ 
	   & 	   & $0.69^{+0.08}_{-0.04}$ & $-0.40^{+0.08}_{-0.30}$ & $4.52^{+0.03}_{-0.04}$ &   &   &   &   &   &   &   &   & $0.70^{+0.05}_{-0.04}$ & $1.43^{+0.22}_{-0.13}$ & $1.74^{+0.28}_{-0.17}$  \\ 
\hline
	 \multirow{2}{*}{S\_18} & \multirow{2}{*}{HATLASJ133008$+$245860} & $0.037^{+0.003}_{-0.010}$ & $-0.287^{+0.010}_{-0.005}$ & $1.037^{+0.001}_{-0.002}$ & $0.616^{+0.009}_{-0.007}$ & $-14.02^{+0.61}_{-0.43}$ & -- & -- & $39^{+2}_{-5}$ & $36^{+7}_{-5}$ & $0.0084\pm0.0006$ & $12.31^{+1.10}_{-0.72}$ & $0.055\pm0.003$ & $3.184^{+0.009}_{-0.008}$ & --  \\ 
	   & 	   & $0.21^{+0.02}_{-0.06}$ & $-1.65^{+0.06}_{-0.03}$ & $5.974^{+0.009}_{-0.011}$ &   &   &   &   &   &   &   &   & $0.43\pm0.02$ & $1.990^{+0.006}_{-0.005}$ & $3.47^{+1.14}_{-0.80}$  \\ 
\hline
	 \multirow{2}{*}{S\_19} & \multirow{2}{*}{HATLASJ133846$+$255057} & $0.0905^{+0.0005}_{-0.0028}$ & $-0.1092^{+0.0001}_{-0.0008}$ & $0.6697^{+0.0004}_{-0.0002}$ & $0.895^{+0.002}_{-0.011}$ & $79.56^{+0.59}_{-0.58}$ & $0.281^{+0.008}_{-0.001}$ & $-0.0560^{+0.0024}_{-0.0002}$ & $42$ & $41$ & $0.0046^{+0.0003}_{-0.0002}$ & $3.78^{+0.04}_{-0.02}$ & $0.0963^{+0.0006}_{-0.0008}$ & $1.706^{+0.012}_{-0.002}$ & --  \\ 
	   & 	   & $0.51^{+0.07}_{-0.08}$ & $-0.63^{+0.10}_{-0.08}$ & $3.83^{+0.48}_{-0.62}$ &   &   &   &   &   &   &   &   & $0.80\pm0.03$ & $1.12\pm0.26$ & $7.95^{+3.82}_{-2.14}$  \\ 
\hline
	 \multirow{2}{*}{S\_21} & \multirow{2}{*}{HERMESJ171451$+$592634} & $-0.23^{+0.05}_{-0.01}$ & $0.21^{+0.03}_{-0.01}$ & $0.976^{+0.006}_{-0.007}$ & $0.72\pm0.02$ & $-76.63^{+1.34}_{-0.90}$ & -- & -- & $42^{+1}_{-6}$ & $35^{+5}_{-2}$ & $0.012\pm0.001$ & $8.40^{+5.93}_{-6.00}$ & $0.053^{+0.053}_{-0.009}$ & $2.46^{+0.02}_{-0.01}$ & --  \\ 
	   & 	   & $-2.01^{+0.39}_{-0.09}$ & $1.80^{+0.27}_{-0.09}$ & $8.35^{+0.05}_{-0.06}$ &   &   &   &   &   &   &   &   & $0.41^{+0.41}_{-0.07}$ & $4.1435\pm0.0008$ & $2.83^{+0.18}_{-0.17}$  \\ 
\hline
\hline

\end{tabular}
\endgroup
}
\begin{flushleft}
\textit{Notes}: Col.~(1): Source reference number. Col.~(2): IAU name of the {\it Herschel} detection. Col.~(3) and (4): RA and Dec. offsets in arcsec (first row) and kpc (second row). Col.~(5): Einstein radius in arcsec (first row) and kpc (second row). Col.~(6): Axis ratio of the mass profile. Col.~(7): Position angle of the mass profile. Col.~(8) and (9): Shear elliptical components of the mass profile. Col.~(10) and (11): Size of the source plane pixelization. Col.~(12): Regularization coefficient. Col.~(13): Magnification computed on the regions of the source plane with $\textit{SNR}>3$. Col.~(14): Circularised radius equivalent to the regions of the source plane with $\textit{SNR}>3$. Col.~(15): Mass within the Einstein radius. Col.~(16): Mass-to-light ratio in $F110W$ within the Einstein radius. 
\end{flushleft}
\end{table}
\end{landscape}

\begin{landscape}
\begin{table}
\centering
\contcaption{}
{\scriptsize
\begingroup
\setlength{\tabcolsep}{0.5pt}
\begin{tabular}{c l c c c c c c c c c c c c c c}
\hline
\hline
	 No. & IAU name & $\delta x$ & $\delta y$ & $\theta_{E}$ & $q$ & $PA$ & $\epsilon_{shear,\, 1}$ & $\epsilon_{shear,\, 2}$ & $x_{\rm pix}$ & $y_{\rm pix}$ & $\lambda$ & $\mu$ & $R_{\rm bkg,\,circ}$ & $k_0$ &    \\ 
	   &   & [arcsec] & [arcsec] & [arcsec] &   &   &   &   &   &   &   &   & [arcsec] & [] &    \\ 
	   &   & $dx_0$ & $dy_0$ & $\theta_{E,\, 0}$ &   &   &   &   &   &   &   &   & $R_{\rm bkg,\,circ,\,0}$ & $M$ & $M/L$  \\ 
	   &   & [kpc] & [kpc] & [kpc] &   &   &   &   &   &   &   &   & [kpc] & [$10^{11}$ M$_{\odot}$] &    \\ 
	 (1) & (2) & (3) & (4) & (5) & (6) & (7) & (8) & (9) & (10) & (11) & (12) & (13) & (14) & (15) & (16)  \\ 

\hline
	 \multirow{2}{*}{S\_29} & \multirow{2}{*}{HERSJ012041.6$-$002705} & $-0.07\pm0.01$ & $-0.095^{+0.010}_{-0.011}$ & $0.928^{+0.005}_{-0.006}$ & $0.76^{+0.02}_{-0.01}$ & $-41.77^{+2.16}_{-2.17}$ & -- & -- & $38^{+5}_{-6}$ & $35\pm6$ & $0.022\pm0.002$ & $3.91^{+0.43}_{-0.34}$ & $0.124^{+0.008}_{-0.006}$ & $2.65\pm0.03$ & --  \\ 
	   & 	   & $-0.55^{+0.08}_{-0.09}$ & $-0.71\pm0.08$ & $6.94\pm0.15$ &   &   &   &   &   &   &   &   & -- & -- & --  \\ 
\hline
	 \multirow{2}{*}{S\_30} & \multirow{2}{*}{HATLASJ085112$+$004934} & $0.03406^{+0.00009}_{-0.00014}$ & $0.1967^{+0.0010}_{-0.0002}$ & $1.24359^{+0.00004}_{-0.00003}$ & $0.34221^{+0.00034}_{-0.00007}$ & $-41.033^{+0.008}_{-0.033}$ & $-0.29461^{+0.00011}_{-0.00006}$ & $0.28098^{+0.00005}_{-0.00043}$ & $44$ & $44$ & $0.00112^{+0.00004}_{-0.00005}$ & $27.80^{+0.50}_{-0.59}$ & $0.01863^{+0.00004}_{-0.00009}$ & $4.0578^{+0.0014}_{-0.0003}$ & --  \\ 
	   & 	   & $0.24^{+0.03}_{-0.07}$ & $1.41^{+0.20}_{-0.39}$ & $8.90^{+1.26}_{-2.45}$ &   &   &   &   &   &   &   &   & $0.161^{+0.001}_{-0.004}$ & $4.60^{+3.73}_{-2.22}$ & $1.86^{+2.07}_{-0.72}$  \\ 
\hline
	 \multirow{2}{*}{S\_32} & \multirow{2}{*}{HERMESJ104549$+$574512} & $-0.47^{+0.03}_{-0.04}$ & $-0.129^{+0.006}_{-0.009}$ & $2.631^{+0.007}_{-0.012}$ & $0.677^{+0.005}_{-0.003}$ & $18.85^{+0.69}_{-0.23}$ & -- & -- & $41$ & $40\pm4$ & $0.014\pm0.001$ & $8.63^{+0.37}_{-0.21}$ & $0.108\pm0.002$ & $17.23^{+0.21}_{-0.12}$ & --  \\ 
	   & 	   & $-1.59^{+0.15}_{-0.17}$ & $-0.44\pm0.04$ & $8.97^{+0.67}_{-0.73}$ &   &   &   &   &   &   &   &   & $0.85\pm0.02$ & $5.45^{+0.50}_{-0.52}$ & $63.80^{+12.97}_{-10.76}$  \\ 
\hline
	 \multirow{2}{*}{S\_35} & \multirow{2}{*}{HATLASJ132630$+$334410} & $0.004^{+0.004}_{-0.006}$ & $-0.057^{+0.009}_{-0.013}$ & $1.684^{+0.005}_{-0.004}$ & $0.577^{+0.008}_{-0.006}$ & $-47.54^{+0.42}_{-0.40}$ & -- & -- & $42^{+2}_{-4}$ & $34^{+8}_{-2}$ & $0.0095\pm0.0006$ & $1.89\pm0.02$ & $0.0938^{+0.0079}_{-0.0004}$ & $8.26^{+0.03}_{-0.02}$ & --  \\ 
	   & 	   & $0.03^{+0.03}_{-0.05}$ & $-0.44^{+0.07}_{-0.10}$ & $12.92^{+0.04}_{-0.03}$ &   &   &   &   &   &   &   &   & $0.743^{+0.063}_{-0.003}$ & $9.09\pm0.03$ & $3.83^{+0.20}_{-0.19}$  \\ 
\hline

  \multirow{2}{*}{S\_38} & \multirow{2}{*}{HERMESJ142824$+$352620} & $-0.0008^{+0.0065}_{-0.0074}$ & $0.016^{+0.005}_{-0.006}$ & $0.215^{+0.006}_{-0.003}$ & $0.66^{+0.06}_{-0.05}$ & $-5.44^{+5.95}_{-8.75}$ & -- & -- & $42^{+1}_{-8}$ & $25^{+5}_{-4}$ & $0.0064^{+0.0006}_{-0.0004}$ & $2.64^{+0.15}_{-0.14}$ & $0.090^{+0.003}_{-0.004}$ & $0.137^{+0.010}_{-0.005}$ & --  \\ 
	   & 	   & $-0.007^{+0.054}_{-0.061}$ & $0.13^{+0.04}_{-0.05}$ & $1.78^{+0.05}_{-0.03}$ &   &   &   &   &   &   &   &   & $0.78\pm0.03$ & $0.58^{+0.04}_{-0.02}$ & $0.98^{+0.11}_{-0.09}$  \\ 
\hline
	 \multirow{2}{*}{S\_39} & \multirow{2}{*}{HATLASJ223753.8$-$305828} & $0.19^{+0.02}_{-0.01}$ & $-0.091^{+0.010}_{-0.017}$ & $1.26^{+0.02}_{-0.01}$ & $0.54^{+0.01}_{-0.03}$ & $65.49^{+0.77}_{-0.62}$ & -- & -- & $44$ & $20$ & $0.0098^{+0.0006}_{-0.0005}$ & $2.65^{+0.04}_{-0.10}$ & $0.1281^{+0.0032}_{-0.0010}$ & $4.53^{+0.07}_{-0.05}$ & --  \\ 
	   & 	   & $1.24^{+0.18}_{-0.20}$ & $-0.60^{+0.11}_{-0.13}$ & $8.28^{+0.90}_{-1.31}$ &   &   &   &   &   &   &   &   & $1.10\pm0.03$ & $3.88^{+1.15}_{-1.02}$ & $4.00^{+2.08}_{-1.12}$  \\ 
\hline
	 \multirow{2}{*}{S\_40} & \multirow{2}{*}{HATLASJ225250.7$-$313657} & $0.182^{+0.012}_{-0.005}$ & $0.08^{+0.01}_{-0.02}$ & $0.605^{+0.008}_{-0.004}$ & $0.633^{+0.020}_{-0.009}$ & $-73.84^{+2.35}_{-0.69}$ & -- & -- & $41^{+3}_{-9}$ & $40^{+4}_{-14}$ & $0.0031^{+0.0002}_{-0.0003}$ & $3.44^{+0.18}_{-0.11}$ & $0.078^{+0.002}_{-0.003}$ & $1.09^{+0.03}_{-0.01}$ & --  \\ 
	   & 	   & $1.33^{+0.17}_{-0.27}$ & $0.54^{+0.14}_{-0.18}$ & $4.41^{+0.51}_{-0.89}$ &   &   &   &   &   &   &   &   & $0.65\pm0.02$ & $1.12^{+0.47}_{-0.40}$ & $4.89^{+4.71}_{-1.81}$  \\ 
\hline
	 \multirow{2}{*}{S\_41} & \multirow{2}{*}{HELMSJ233633.5$-$032119} & $0.051^{+0.001}_{-0.005}$ & $-0.0851^{+0.0106}_{-0.0004}$ & $1.7281^{+0.0130}_{-0.0007}$ & $0.7347^{+0.0008}_{-0.0196}$ & $-48.44^{+1.47}_{-0.11}$ & -- & -- & $39$ & $43$ & $0.0084^{+0.0005}_{-0.0004}$ & $12.39^{+2.16}_{-2.48}$ & $0.036^{+0.011}_{-0.005}$ & $9.161^{+0.105}_{-0.006}$ & --  \\ 
	   & 	   & -- & -- & -- &   &   &   &   &   &   &   &   & $0.30^{+0.10}_{-0.05}$ & -- & --  \\ 
\hline
\hline
\end{tabular}
\endgroup
}

\end{table}
\end{landscape}

\begin{figure}
    \centering
    \includegraphics[width=0.45\textwidth]{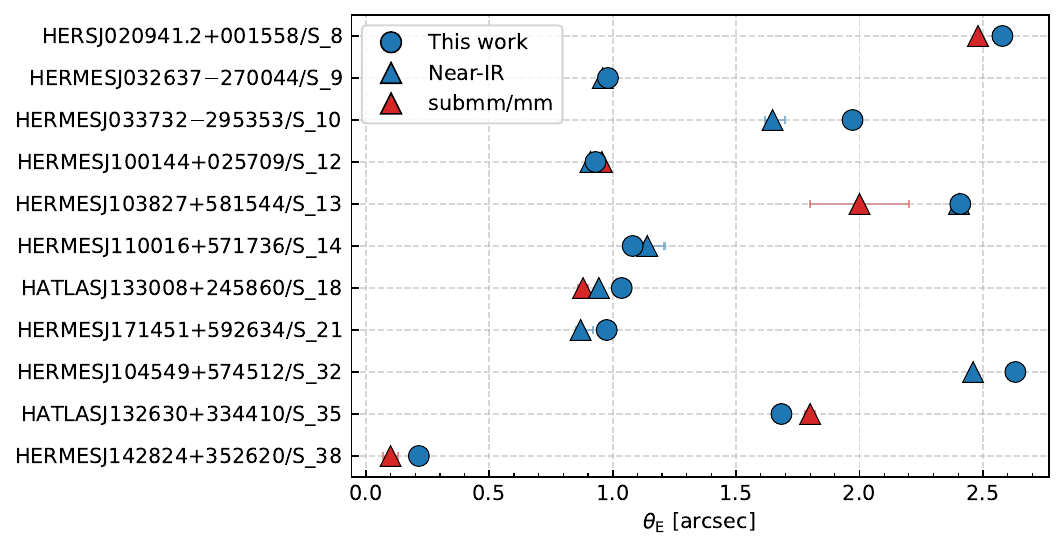}
    \caption{Comparison between our estimated Einstein radii (blue circles) and the ones from literature measured either in near-IR (blue triangles) or sub-mm/mm (red triangles).}
    \label{fig:lit_my_rein}
\end{figure}

A comparison between our Einstein radii and those available in the literature \citep{Bussmann2013, Bussmann2015, Calanog2014, Geach2015, Enia2018, Dye2022, Maresca2022, Kamieneski2023} is shown in Table~\ref{tab:lm_lit} and in Fig.~\ref{fig:lit_my_rein}. Our values are comparable with what is found in the literature, with only two exceptions: S\_38 which \citet{Bussmann2013} did not resolve with SMA, due to the very low angular separation of the multiple images and assumed to have Einstein radius of $0.1\pm0.3$ arcsec; and S\_10 for which \citet{Calanog2014} fixed the flattening and position angle of the mass model to be circular and aligned with the lens.

\begin{table*}
    \centering
    \caption{Comparison between our Einstein radii and those available in the literature in the near-IR and in the sub-mm/mm if available.}
    \label{tab:lm_lit}
    \begin{tabular}{c l c c c c c c c}
    \hline
    \hline
No. &  IAU Name & $\theta_{\rm{E}}$ & $\theta_{\rm{E, near-IR lit.}}$ & Ref. & $\theta_{\rm{E, sub-mm/mm. lit.}}$ & Ref. & Multiw. & Ref  \\
& & [arcsec] & [arcsec] &  & [arcsec] & & & \\

(1) & (2) & (3) & (4) & (5) & (6) & (7) & (8) & (9) \\
\hline
S\_8 & HERSJ020941.2$+$001558 & $2.579^{+0.002}_{-0.003}$ & $2.48^{+0.02}_{-0.01}$ & Ge15 & $2.55^{+0.14}_{-0.20}$ & Ka23 & ALMA band 6 & Li22 \\  

S\_9 & HERMESJ032637$-$270044 & $0.981^{+0.005}_{-0.006}$ & $0.96^{+0.02}_{-0.03}$ & Ca14 & -- & -- & -- & -- \\

S\_10 & HERMESJ033732$-$295353 & $1.9718^{+0.0004}_{-0.0005}$ & $1.65^{+0.03}_{-0.054}$ & Ca14 & -- & -- & -- & -- \\

S\_12 & HERMESJ100144$-$025709 & $0.930\pm0.002$ & $0.91\pm0.01$ & Ca14 & $0.956\pm 0.005$ & Bu15 & ALMA band 7 & Bu15 \\

S\_13 & HERMESJ103827$+$581544 & $2.408\pm0.010$ & $2.40^{+0.01}_{-0.05}$ & Ca14 & $2.0\pm0.2$ & Bu13 & SMA 340 GHz & Bu13 \\ 

S\_14 & HERMESJ110016$+$571736 & $1.081\pm0.003$ & $1.14^{+0.04}_{-0.07}$ & Ca14 & -- & -- & -- & -- \\

S\_18 & HATLASJ133008$+$24586 & $1.037\pm0.002$ & $0.944^{+0.002}_{-0.001}$ & Ca14 & $0.88\pm0.02$ & Bu13 & SMA 340 GHz & Bu13 \\

S\_21 & HERMESJ171451$+$592634 & $0.976^{+0.006}_{-0.007}$ & $0.87^{+0.02}_{-0.05}$ & Ca14 & -- & -- & -- & -- \\

S\_32 & HERMESJ104549$+$574512 & $2.631^{+0.008}_{-0.012}$ & $2.46\pm0.01$ & Ca14 & -- & -- & -- & -- \\

S\_35 & HATLASJ132630$+$334410 & $1.684^{+0.005}_{-0.004}$ & -- & -- & $1.80\pm0.02$ & Bu13 & SMA 340 GHz & Bu13 \\

S\_38 & HERMESJ142824$+$35262 & $0.37\pm0.02$ & -- & -- & $0.10\pm0.03$ & Bu13 & SMA 340 GHz & Bu13 \\

\hline
\hline
\end{tabular}
\begin{flushleft}
\textit{Notes}: Col.~(1): Source reference number. Col.~(2): IAU name of the {\it Herschel} detection. Col.~(3): Value of the Einstein radii measured in this work. Col.~(4) and (5): Value of the Einstein radii measured in the literature from the near-IR observations and reference. Col.~(6) and (7): Value of the Einstein radii measured in the literature from the sub-mm/mm observations and reference. Col.~(8) and (9): Multiwavelength observations and reference.  
\end{flushleft}
\end{table*}
 
\section{Discussion}
\label{sec6}

Consistent with expectations \citep{Perrotta2002, Lapi2012, Negrello2017} our 65 confirmed lenses mostly comprise systems where the lens is a single galaxy (51). Only 4 are made of spectroscopically confirmed groups or clusters (S\_36, S\_48, S\_60, and S\_61), and 9 systems have a morphology consistent with a group (S\_6, S\_15, S\_23, S\_25, S\_28, S\_33, S\_34, S\_54, and S\_63). The lensing galaxies are nearly all ETGs, with the only exceptions of S\_47 and one of the galaxies in the lensing cluster of S\_48. Both of them show a clear spiral structure. The ETG nature of the lensing galaxies is further confirmed by their concentrations, with $92\%$ of our sample lenses being more concentrated than expected for an exponential disk. 

The mass within the Einstein radii of the foreground lenses ranges between $9.9\times10^{10}$ M$_{\odot}$ (1-st percentile) and $6.4\times10^{11}$ M$_{\odot}$ (99-th percentile). The mass-to-light ratios in the \textit{F110W} filter within the Einstein radius range between 1.25 and 8.86 M$_\odot$ L$_\odot^{-1}$. 
The background sources host a variety of morphologies, including single compact sources (e.g. S\_12), more diffuse sources (e.g. S\_18), and multiple clumps (S\_17). We find that 14 out of 22 reconstructed background sources are single sources, whereas the remaining 8 are multiple sources. The present analysis alone is not enough to determine whether these morphologies indicate different populations of DSFGs or variations in the stellar and/or dust distributions. 
The physical sizes of the background sources, measured as the circularised radii of the circles equivalent to the regions of the source plane with ${\rm SNR}>3$, range between 0.34 kpc (1-st percentile) and 1.30 kpc (99-th percentile). We compare our stellar sizes with the dust ones by \citet{Enia2018}, who modelled the SMA sub-mm/mm emission of a sample of 12 lenses similarly selected and partially included in our sample. They found dust sizes between 1.52 kpc (1-st percentile) and 4.07 kpc (99-th percentile).
We can understand this result by considering that the \textit{HST} observations, by sampling the optical/near-UV rest-frame emission, are expected to pick up only the intrinsically brightest and more compact regions of the background sources whose light can pass through the dust. This is consistent with an inside-out growth and later inside-out quenching, a scenario where the central regions of these DSFGs begin to clear out the dust first, causing the more compact stellar emission to shine through.
Interestingly, the \textit{HST} sizes are comparable to bright nuclei and off-nuclear knots in local ULIRGs \citep[][]{Surace1998, Farrah2001}.
This is somewhat in contrast with previous literature \citep[e.g.,][]{Pantoni2021} that argued for the dust emission to be more compact than the stellar one. Although, recently, \citet{Kamieneski2023b} studied a $z\sim2.3$ lensed galaxies with properties similar to those of the lensed DSFGs in our sample and found rest frame near-UV/optical/near-IR emitting regions comparable or even slightly smaller in size than the ones measured in the sub-mm, thus providing evidence for an inside-out quenching scenario. It is worth noticing that \citet{Kamieneski2023b} has also considered another possible, but less likely, explanation related to inclination effects and uneven dust distribution.
We also want to point out that the difference in sizes we observe between the stellar emitting region, as sampled by \textit{HST}, and the dust emitting region, as probed by SMA, may be affected by the lower angular resolution of the SMA observations compared to \textit{HST} data. 
In fact, by projecting the half width at half maximum (HWHM) of the SMA beam to the redshift of the source and then dividing it by the square root of the total magnification within the $>3\sigma$ regions, we get a physical resolution (once averaged over the whole sample) of $0.85$ kpc, which is larger than what we get for near-IR size for about 50 per cent of our sample. While \citet{Enia2018} sources are resolved ($R_{\rm bkg,\,circ}/{\rm PSF_{HWHM}} \simgt 1$), any emission coming from regions smaller than the projected HWHM would inevitably be broadened. 
Moreover, the depth of our observations does not allow us to exclude that other, more obscured and undetected, star-forming knots are present throughout the disk, implying that mass buildup could happen disk-wide.

Finally, it is worth mentioning that the optical/near-IR sizes quoted in the literature are often measured through the effective radius obtained by fitting a Sérsic profile to the surface brightness distribution of the background sources. For lensed galaxies, this is usually done directly during the lens modelling by including a parametric background source \citep[e.g.][]{Calanog2014}. Such an estimate of size takes into account the concentration of the source and averages out the surface brightness distribution in the case of clumpy emission. In comparison, our measure is SNR-dependent and is not affected by the smoothness of the surface brightness distribution. 
We remark that the relation between the observed-frame near-IR sizes and sub-mm ones is still up to debate and could depend on additional factors we do not discuss here, like the stellar mass of the galaxy \citep[see,][for details]{Zhaoran2023} or its evolutionary stage.
At the same time, we find magnifications ranging between 2.4 (1-st percentile) and 15.8 (99-th percentile), with a median of 5.5 for our sample. The magnifications by \citet{Enia2018} range between 3.20 (1-st percentile) and 8.35 (99-th percentile) with a median of 5.7. This comparison suggests a minor, or even negligible, differential magnification. 

\begin{figure}
    \centering
    \includegraphics[width=0.45\textwidth]{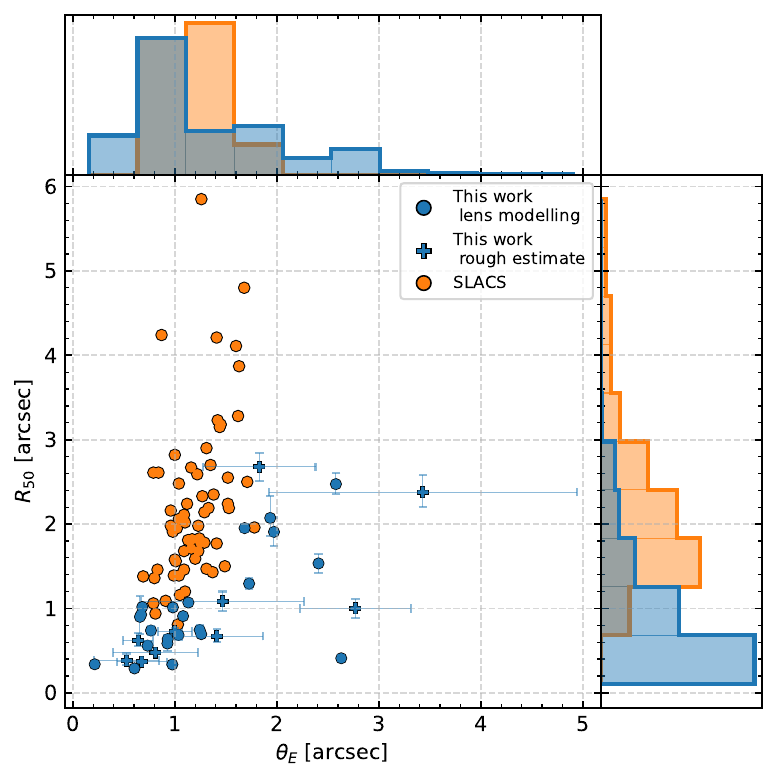}
    \caption{\textit{Top panel}: Normalised distribution of the values of the Einstein radius of the lenses in our (blue histogram) and in the SLACS samples (orange histogram). \textit{Bottom left panel:\/} Total effective radius as a function of Einstein radius of the lenses in our (blue symbols) and in the SLACS sample (orange circles). The blue circles and crosses correspond to systems for which we derived the Einstein radius from the lens modelling or the separation between multiple images, respectively. \textit{Bottom right panel\/} Normalised distribution of the values of the total effective radius of the lenses.}
    \label{fig:SLACS_2}
\end{figure}

\begin{figure}
    \centering
    \includegraphics[width=0.45\textwidth]{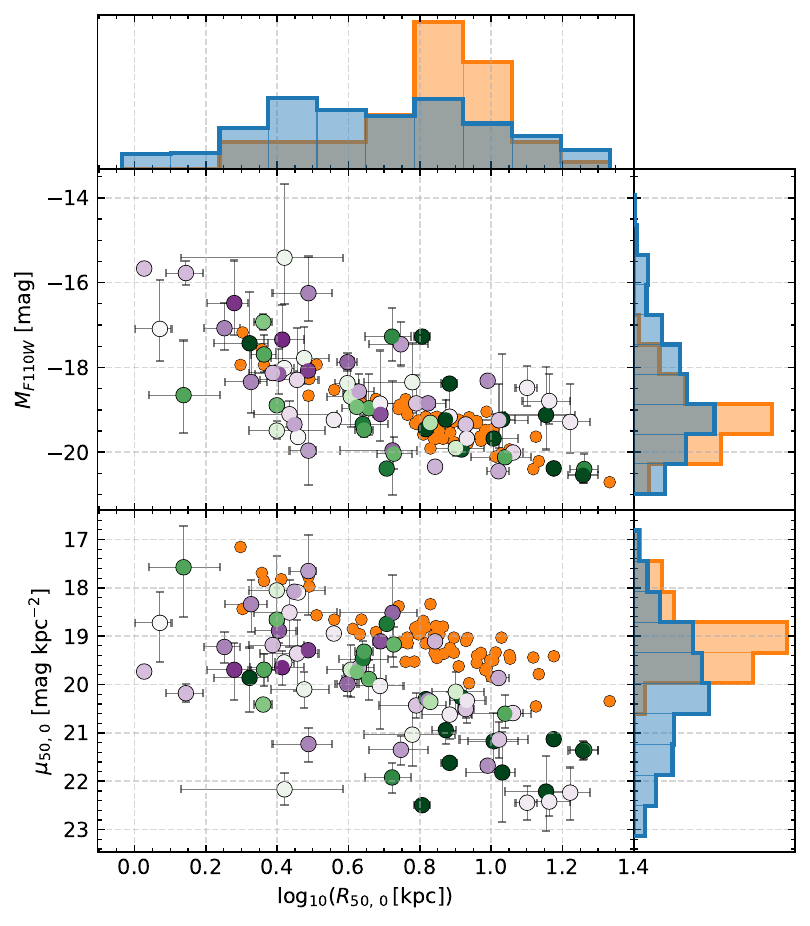}
    \caption{\textit{Top panel:\/} Normalized distribution of the total effective radii in kpc of our (blue histogram) and SLACS samples (orange histogram). \textit{Central left panel:} Absolute magnitude corrected for $K$-correction against the total effective radii in kpc of the two samples. \textit{Central right panel:\/} Normalized distribution of the absolute magnitude of the two samples with the $K$-correction applied. Our lensing galaxies are color-coded according to their concentration, with purple corresponding to the less-concentrated systems and green to the most-concentrated ones. White marks the concentration of a de Vaucouleurs profile. \textit{Bottom left panel:\/} Kormendy relation of the two samples. Our lensing galaxies are color-coded according to their concentration. \textit{Bottom right panel:\/} Normalized distribution of the total effective surface brightness of the two samples with the $K$-correction applied.}
    \label{fig:SLACS_1}
\end{figure}

We compare the properties of our lenses (i.e., total effective radii, Einstein radii, and total effective surface brightnesses) to those of the SLACS sample of confirmed lenses \citep[][]{Bolton2008}. This is the largest available sample of strong lenses followed up by \textit{HST} and with lens modelling. The SLACS lensing candidates were identified from archival SDSS spectroscopic data as all the systems that show the presence of two separate redshift estimates in the same spectrum: a lower absorption-line redshift for the foreground lens and a higher emission-line redshift for the background source. These candidates were then followed up and confirmed by means of \textit{HST ACS/F814W} imaging. We consider the 63 confirmed lenses out of 131 candidates, for which \citet{Bolton2008} were able to subtract the surface brightness distribution of the foreground lens and perform the lens modelling. They modelled the surface brightness distribution of the lens, adopting the de Vaucouleurs profile. 
We perform a K-correction of the total effective surface brightness of all the galaxies of our and SLACS samples with available redshifts. 
As a template for the ETG SED, we use the 17 local elliptical galaxies available in the Brown Atlas \citep{Brown2014}. To estimate the \textit{F110W} magnitude of the SLACS lenses, we redshift and normalize the template to match the \textit{F814W} magnitude using \texttt{PySynphot} package\footnote{\url{https://pysynphot.readthedocs.io/en/latest/}}. Then, we compute the \textit{F110W} magnitude of each template at $z=0$ and at the redshift of the foreground lenses of both samples. Finally, we adopt the \textit{F110W} magnitude difference as the K-correction term. This correction does not take into account possible differences between the SEDs of the templates and the ones of the lenses. The median K-correction terms for our sample and SLACS are 0.04 mag and 0.03 mag, respectively.
From Fig.~\ref{fig:SLACS_2}, it is possible to see that our sample shows different $R_{50}$ and $\theta_{E}$ distributions than SLACS. Our selection is able to pick up more systems with both smaller or larger values of $\theta_{E}$. Moreover, at similar Einstein radii, the effective radii of our lenses are significantly smaller than the SLACS, making our background sources less contaminated by the emission of the lenses.
Figure~\ref{fig:SLACS_1} shows that the two samples have similar $R_{50,0}$ and absolute magnitude distributions, although our lenses show a tail for smaller and fainter galaxies. At the same time, our lenses have lower $\mu_{50,0,{\rm K_{corr}}}$ for the same radii, which implies that our lenses either are fainter or more concentrated than the de Vaucouleurs profiles. The increased scatter we find in both the $R_{50,0}$ and $M_{F110W}$, and $R_{50,0}$ and $\mu_{50,0,{\rm K_{corr}}}$ distributions can be accounted for by considering the variation in concentration and the wide range of foreground lens redshifts we observed. We can try to further explain these differences by considering the different selections of the two samples. The SLACS sample, being constructed from SDSS spectroscopic data and needing a robust detection of both the foreground lens and background source, is limited in magnitude to brighter lenses at lower redshift. In particular, the SLACS spectroscopic selection was done through 3 arcsec diameter fibres and comprises either local galaxies or massive ETGs at $z\lesssim0.6$.

\section{Conclusions}
\label{sec7}
We have carried out a snapshot \textit{HST} F110W observing campaign to follow-up 281 {\it Herschel}-selected candidate strongly lensed galaxies. We visually inspect them and classify them according to the presence of the background sources, confirming 25 candidates as strong lenses. We model all the systems where we identify a suitable lens candidate exploiting three different surface-brightness modelling algorithms, namely \texttt{GASP2D}, \texttt{GALFIT}, and \texttt{ISOFIT+CMODEL}, based on the morphology of the foreground lensing galaxies. By combining the visual inspection, available multi-wavelength follow-ups, and results of the lens subtraction, we are able to confirm as lensed a total of 65 systems. Lastly, we perform lens modelling and source reconstruction of all the systems for which we can confidently identify both a galaxy acting as the main lens and the background source in the \textit{HST} images (34 systems). We obtain successful results for 23 galaxies. The main results of our analysis are the following:
\begin{itemize}
    \item the overall surface brightness distribution of the lenses and their morphology follows the relations expected for ETGs; however, the foreground lensing galaxies often needed multiple surface brightness components, with properties significantly varying from galaxy to galaxy;
    \item most of the lensing systems are consistent with single galaxy lenses, with only $\sim7\%$ of them being confirmed as a group or cluster lenses;
    \item the estimated magnifications and Einstein radii are consistent with previous analysis conducted on data at different wavelengths; however, the inferred size of the background sources is about 3 times smaller than the ones measured in the sub-mm/mm with SMA by \citet{Enia2018} for a similarly selected, and partially overlapping sample. This difference in size between the stellar and the dust-emitting regions is suggestive of an inside-out quenching scenario;
    \item the sub-mm selection used to build our sample revealed lensing systems with fainter lenses and larger Einstein radii than the SLACS survey.
\end{itemize}

To conclude, we point out that the \textit{HST} snapshots alone, even after carefully identifying and subtracting the candidate lenses, might miss some of the fainter lensed background sources. We argue that this is due to their dust obscuration and high redshift.
This underlines the importance of dedicated, deeper near-IR observations (e.g., with \textit{HST}, Keck, \textit{Euclid} or \textit{JWST}) and high-resolution sub-mm observations (e.g., with ALMA and NOEMA) to effectively determine the nature of the majority of {\it Herschel}-selected strong lensing candidates. This work provides the basis for further detailed analysis of the background sources by allowing not only more complex lens modelling but also providing constraints on the stellar content through SED-fitting techniques.

\section*{Acknowledgements}
This work benefited from the support of the project Z-GAL ANR-AAPG2019 of the French National Research Agency (ANR). EB acknowledges the School of Physics and Astronomy of Cardiff University for hospitality while this paper was in progress. EB, EMC, and GR are supported by Padua University grants Dotazione Ordinaria Ricerca (DOR) 2019-2021 and by the Italian Ministry for Education, University, and Research (MIUR) grant Progetto di Ricerca di Interesse Nazionale (PRIN) 2017 20173ML3WW-001. EMC is also funded by the Istituto Nazionale di Astrofisica (INAF) through grant PRIN 2022 C53D23000850006. LM and MV acknowledge financial support from the Inter-University Institute for Data Intensive Astronomy (IDIA), a partnership of the University of Cape Town, the University of Pretoria and the University of the Western Cape, and from the South African Department of Science and Innovation's National Research Foundation under the ISARP RADIOSKY2020 and RADIOMAP+ Joint Research Schemes (DSI-NRF Grant Numbers 113121 and 150551) and the SRUG Projects (DSI-NRF Grant Numbers 121291, SRUG22031677 and  SRUG2204254729). DW acknowledges support from program number HST-GO-15242, provided by NASA through a grant from the Space Telescope Science Institute, which is operated by the Association of Universities for Research in Astronomy, Incorporated, under NASA contract NAS5-26555. AA is supported by ERC Advanced Investigator grant, DMIDAS [GA 786910], to C. S. Frenk. PC acknowledges the support of the project Z-GAL ANR-AAPG2019 of the French National Research Agency (ANR). HD acknowledges financial support from the Agencia Estatal de Investigación del Ministerio de Ciencia e Innovación (AEI-MCINN) under grant (La evolución de los cíumulos de galaxias desde el amanecer hasta el mediodía cósmico) with reference (PID2019-105776GB-I00/DOI:10.13039/501100011033) and acknowledge support from the ACIISI, Consejería de Economía, Conocimiento y Empleo del Gobierno de Canarias and the European Regional Development Fund (ERDF) under grant with reference PROID2020010107. JGN acknowledges financial support from the PGC projects PGC2018-101948-B-I00 and PID2021-125630NB-I00 (MICINN, FEDER). SJ is supported by the European Union's Horizon Europe research and innovation program under the Marie Sk\l{}odowska-Curie grant agreement No. 101060888. AL is partly supported by the PRIN MIUR 2017 prot. 20173ML3WW 002 ‘Opening the ALMA window on the cosmic evolution of gas, stars, and massive black holes’. IPF acknowledges support from the Spanish State Research Agency (AEI) under grant number PID2019-105552RB-C43. SS was partly supported by ESCAPE – The European Science Cluster of Astronomy \& Particle Physics ESFRI Research Infrastructures, which in turn received funding from the European Union’s Horizon 2020 research and innovation programme under Grant Agreement no. 824064. SS thanks the Science and Technology Facilities Council for support under grants ST/P000584/1. CY acknowledges the support from the ERC Advanced Grant 789410.

\section*{Data Availability}

The reduced images, models, residuals, and PSF cutouts are available on Zenodo at the following link \url{https://doi.org/10.5281/zenodo.10007041}. Future versions will be available at \url{https://zenodo.org/doi/10.5281/zenodo.10007040}.



\bibliographystyle{mnras}
\bibliography{bib} 




\appendix

\section{The Full `A' sample}

\begin{landscape}
\begin{figure}
    
    \caption{{\em From left to right panels:\/} Observed \textit{HST} F110W image, best-fitting surface-brightness model of the lens, residual obtained after the subtraction of the lens model from the image, and SNR map of the residuals for the A class candidates. The contours in the model images are taken at two levels corresponding to $\textit{SNR} = 5$ and 10 (thick curves), and five uniformly spaced levels between the $\textit{SNR} = 10$ and the maximum SNR in the model image (thin curves). The residual map shows the pixel mask (corresponding to the black-shaded regions) adopted for the surface brightness modelling. The residual maps show the contours of available high-resolution multiwavelength data taken at two levels corresponding to $\textit{SNR} = 5$ and 10 (thick curves), and five uniformly spaced levels between the $\textit{SNR} = 10$ and the maximum SNR in the multiwavelength image (thin black curves). The images are oriented such that N is up and E is to the left. For S\_55, we show the image and SNR map panels alone since we do not model the lensing galaxy that is undetected in the \textit{HST} snapshot.}
    \includegraphics[height=0.85\textwidth]{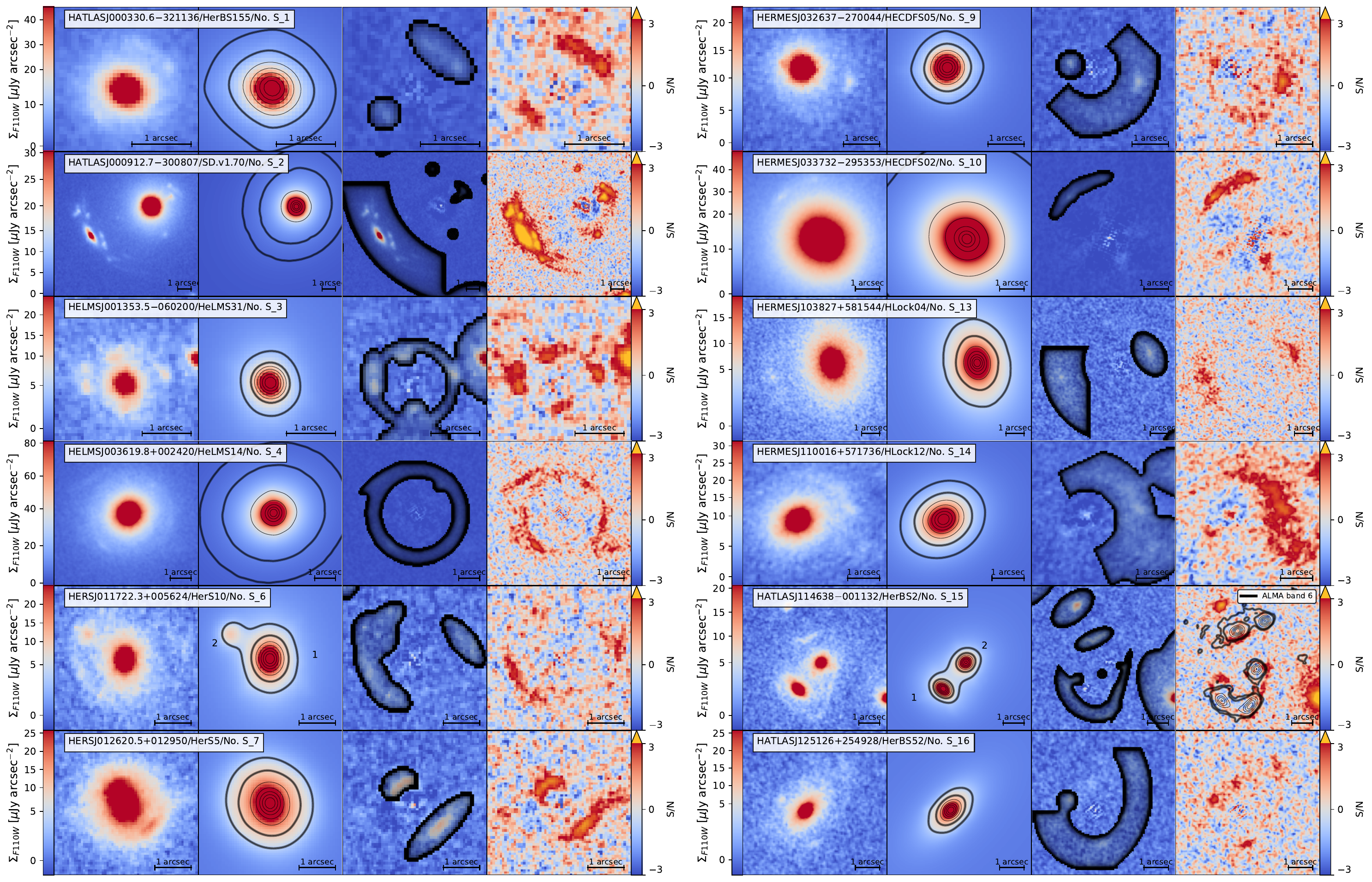}
    \label{fig:models_full}
    \end{figure}
\end{landscape}
\begin{landscape}
\begin{figure}
    \includegraphics[height=0.85\textwidth]{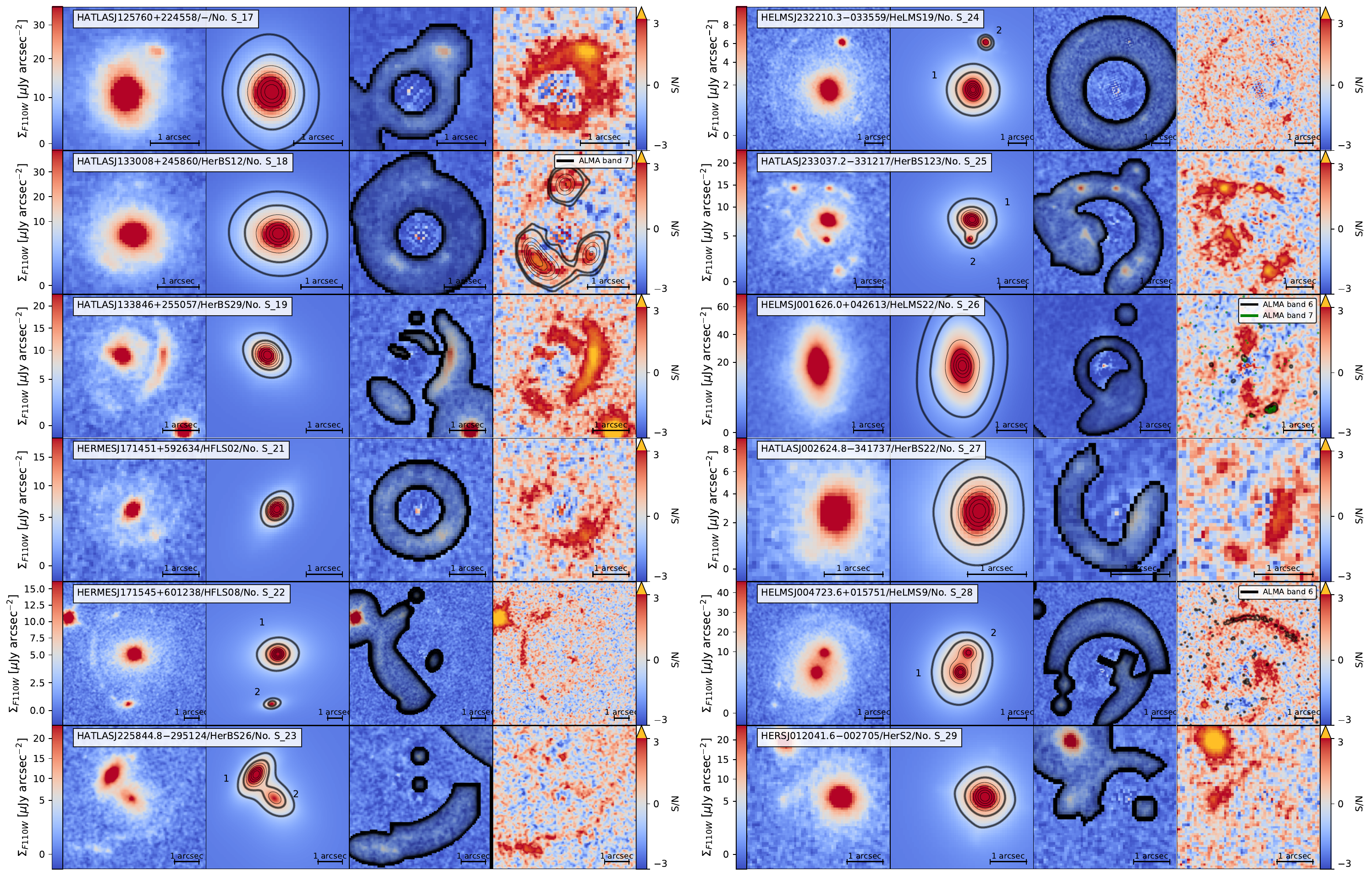}
    \contcaption{}
\end{figure}
\end{landscape}
\begin{landscape}
\begin{figure}
    \includegraphics[height=0.85\textwidth]{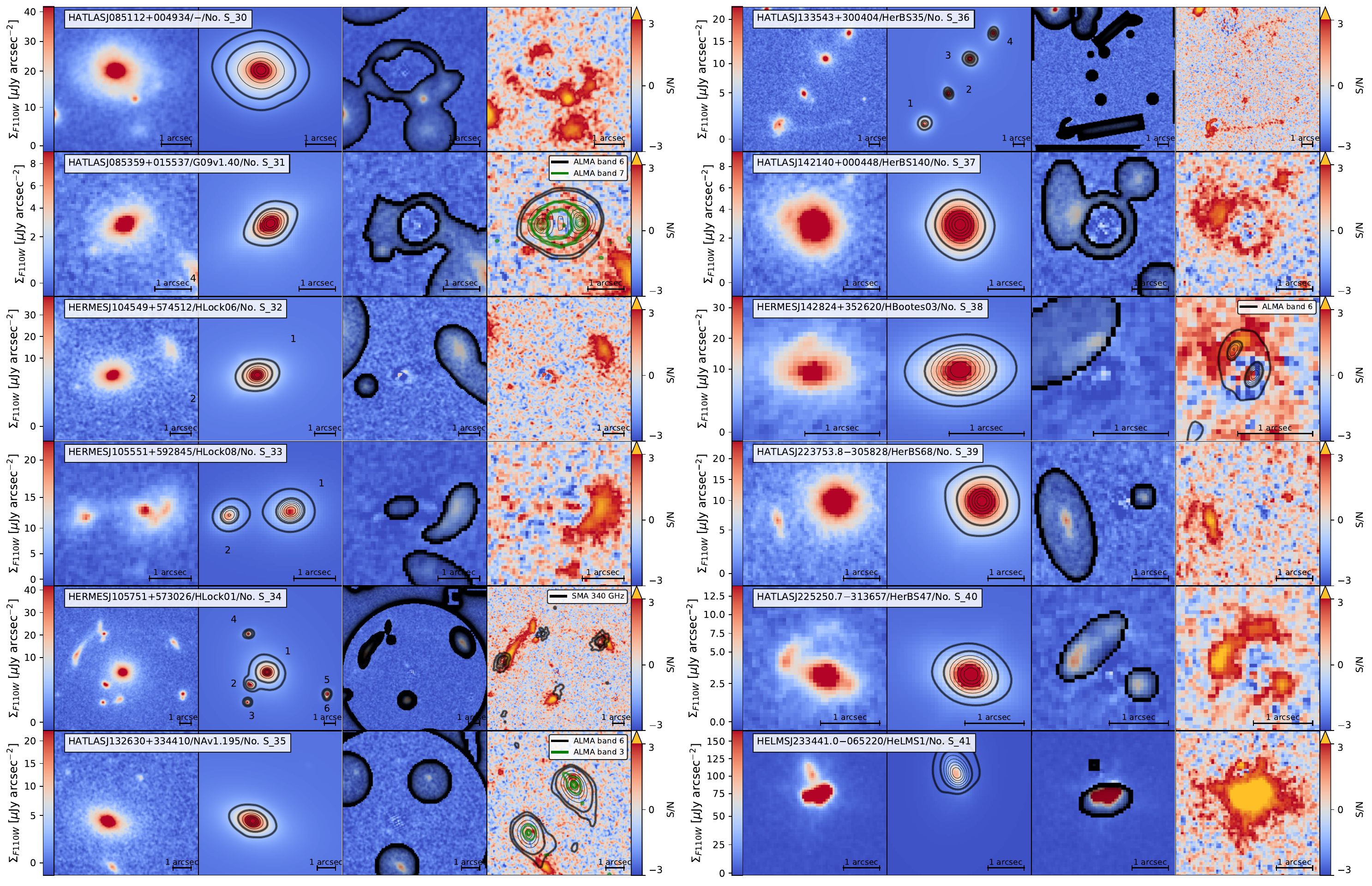}
    \contcaption{}
\end{figure}
\end{landscape}
\begin{landscape}
\begin{figure}
    \includegraphics[height=0.85\textwidth]{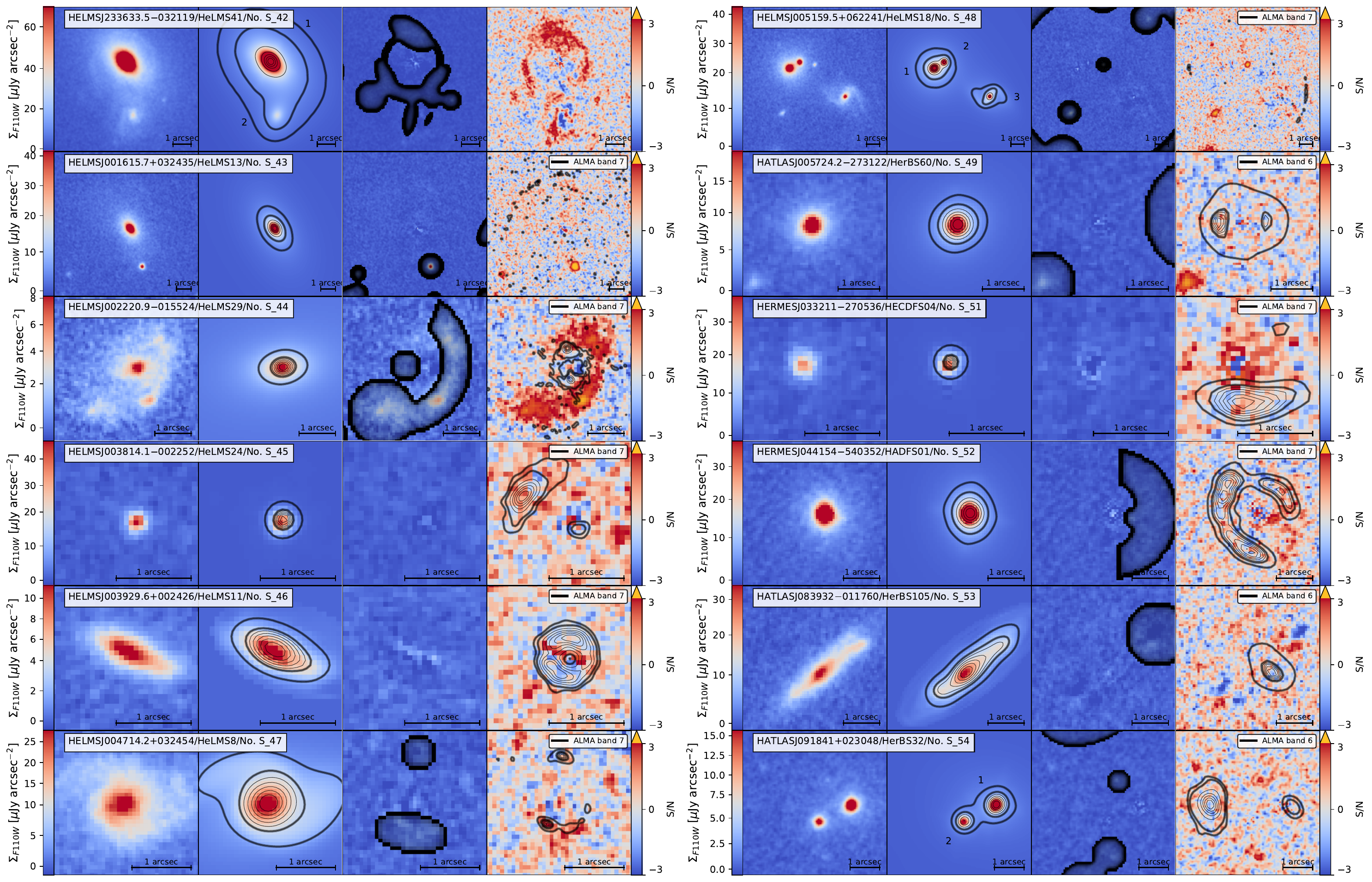}
    \contcaption{}
\end{figure}
\end{landscape}
\begin{landscape}
\begin{figure}
    \includegraphics[height=0.85\textwidth]{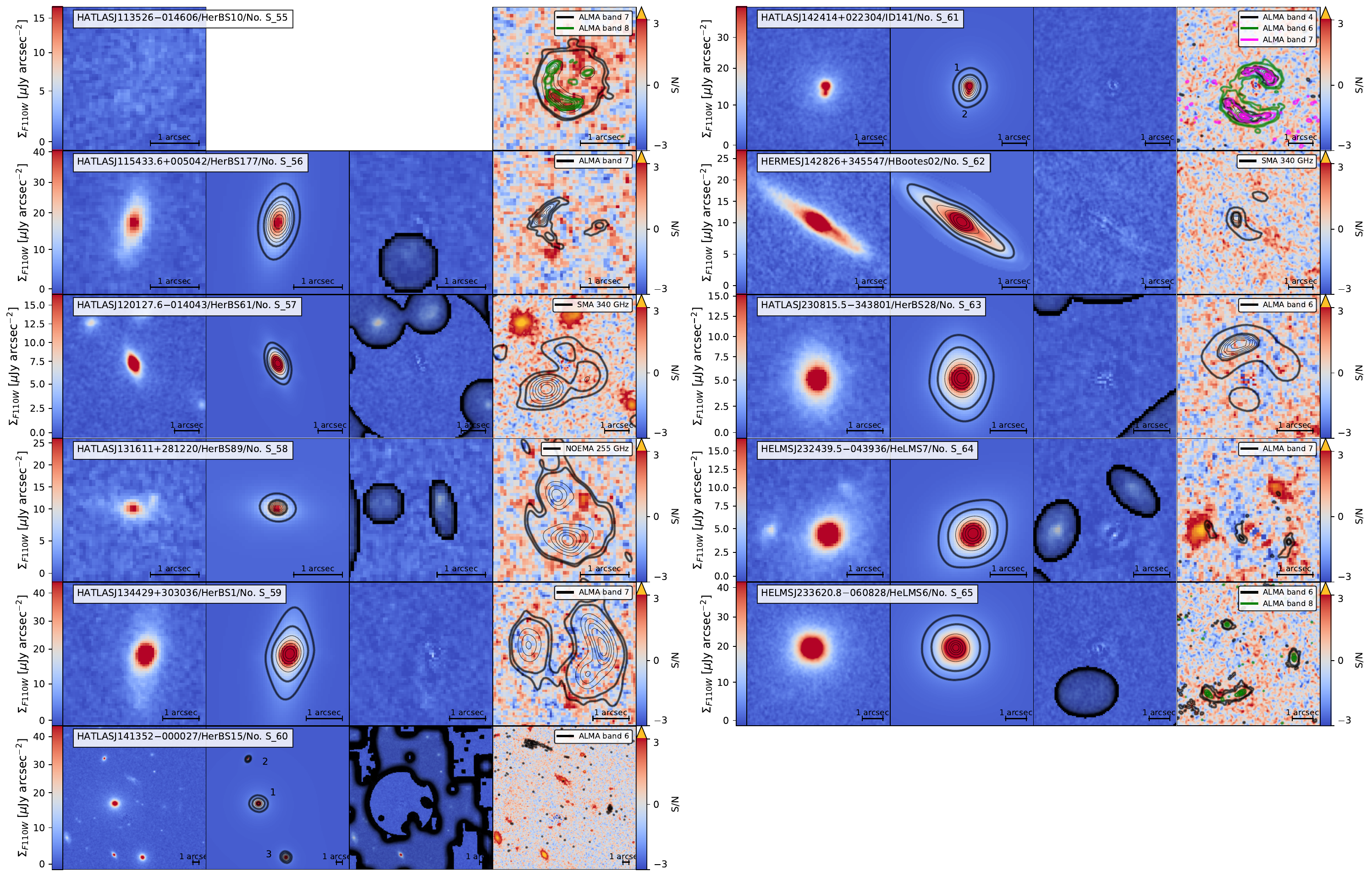}
    \contcaption{}
\end{figure}
\end{landscape}

\begin{table*}
\centering
\caption{Structural parameters of the systems classified as A obtained from a parametric fit of their surface brightness distributions.}
\label{tab:SBmodels_full}
{\scriptsize
\begingroup
\setlength{\tabcolsep}{3pt}
\begin{tabular}{c l c c c c c c c c c c c}
\hline
\hline
	 No. & IAU Name & Type & Components & $\mu_{\rm e}$ & $R_{\rm e}$ & $n$ & $q$ & $PA$ & $m_{\rm PSF}$ & $C/T$ & $\chi^2$ & $N_{\rm dof}$  \\ 
	   &   &   &   & [mag arcsec$^{-2}$] & [arcsec] &   &   & [deg] & [mag] &   &   &    \\ 
	 (1) & (2) & (3) & (4) & (5) & (6) & (7) & (8) & (9) & (10) & (11) & (12) & (13)  \\ 
\hline
	 \multirow{2}{*}{S\_1} & \multirow{2}{*}{HATLASJ000330.6$-$321136} & \multirow{2}{*}{1} & Sérsic & $20.69^{+0.12}_{-0.11}$ & $0.36\pm0.02$ & $4.52^{+0.28}_{-0.27}$ & $0.803\pm0.009$ & $73.05^{+0.96}_{-0.95}$ & -- & $0.71\pm0.05$ & \multirow{2}{*}{1.03} & \multirow{2}{*}{35501} \\ 
	   &   &   & Sérsic & $23.39^{+0.15}_{-0.13}$ & $1.17\pm0.06$ & $0.57\pm0.05$ & $0.96\pm0.02$ & $0.003^{+1.933}_{-1.884}$ & -- & $0.29\pm0.05$ &   &    \\ 
 \hline 
	 \multirow{3}{*}{S\_2} & \multirow{3}{*}{HATLASJ000912.7$-$300807} & \multirow{3}{*}{1} & Sérsic & $21.63^{+0.23}_{-0.20}$ & $1.16^{+0.17}_{-0.18}$ & $3.96\pm0.27$ & $0.929\pm0.006$ & $34.92^{+1.39}_{-1.42}$ & -- & $0.42^{+0.09}_{-0.10}$ & \multirow{3}{*}{1.03} & \multirow{3}{*}{135697} \\ 
	   &   &   & Sérsic & $26.24^{+0.14}_{-0.12}$ & $5.24^{+0.19}_{-0.18}$ & $0.80\pm0.07$ & $0.27\pm0.03$ & $121.76^{+6.00}_{-6.17}$ & -- & $0.018^{+0.005}_{-0.004}$ &   &    \\ 
	   &   &   & Sérsic & $23.31^{+0.14}_{-0.12}$ & $3.93\pm0.11$ & $1.54\pm0.13$ & $0.78\pm0.02$ & $139.43^{+3.96}_{-3.97}$ & -- & $0.56^{+0.10}_{-0.09}$ &   &    \\ 
 \hline 
	 \multirow{2}{*}{S\_3} & \multirow{2}{*}{HELMSJ001353.5$-$060200} & \multirow{2}{*}{1} & Sérsic & $21.21^{+0.28}_{-0.22}$ & $0.26^{+0.04}_{-0.03}$ & $4.03\pm0.60$ & $0.90\pm0.02$ & $20.44^{+3.85}_{-3.82}$ & -- & $0.45\pm0.11$ & \multirow{2}{*}{1.03} & \multirow{2}{*}{61093} \\ 
	   &   &   & Exp. disk & $24.41^{+0.22}_{-0.18}$ & $2.02\pm0.19$ & $[1]$ & $0.66\pm0.05$ & $2.58^{+4.59}_{-4.42}$ & -- & $0.55\pm0.11$ &   &    \\ 
 \hline 
	 \multirow{3}{*}{S\_4} & \multirow{3}{*}{HELMSJ003619.8$+$002420} & \multirow{3}{*}{1} & Sérsic & $20.59^{+0.23}_{-0.19}$ & $1.35\pm0.17$ & $1.26\pm0.07$ & $0.905\pm0.005$ & $109.96^{+1.05}_{-1.06}$ & -- & $0.46^{+0.08}_{-0.09}$ & \multirow{3}{*}{1.01} & \multirow{3}{*}{171928} \\ 
	   &   &   & Sérsic & $17.78^{+0.16}_{-0.14}$ & $0.177^{+0.006}_{-0.005}$ & $1.54\pm0.15$ & $0.90\pm0.02$ & $159.86^{+2.55}_{-2.57}$ & -- & $0.11^{+0.03}_{-0.02}$ &   &    \\ 
	   &   &   & Sérsic & $23.65^{+0.14}_{-0.13}$ & $5.68\pm0.17$ & $0.88^{+0.08}_{-0.07}$ & $0.92\pm0.02$ & $128.85^{+4.37}_{-4.42}$ & -- & $0.42^{+0.08}_{-0.07}$ &   &    \\ 
 \hline 
	 \multirow{4}{*}{S\_6} & \multirow{2}{*}{HERSJ011722.3$+$005624$_{1}$} & \multirow{4}{*}{3} & Sérsic & $21.64^{+0.29}_{-0.23}$ & $0.60^{+0.08}_{-0.07}$ & $5.19^{+0.42}_{-0.41}$ & $0.70\pm0.02$ & $2.90^{+1.96}_{-1.92}$ & -- & $0.85^{+0.05}_{-0.07}$ & \multirow{2}{*}{1.01} & \multirow{2}{*}{55924} \\ 
	   &   &   & Gauss & $24.97^{+0.21}_{-0.17}$ & $1.79\pm0.16$ & $[0.5]$ & $0.83^{+0.10}_{-0.11}$ & $180.53^{+36.51}_{-36.53}$ & -- & $0.15^{+0.07}_{-0.05}$ &   &    \\ 
 \\ 
	   & HERSJ011722.3$+$005624$_{2}$ &   & Sérsic & $22.52^{+0.17}_{-0.15}$ & $0.25\pm0.02$ & $1.12\pm0.15$ & $0.93\pm0.03$ & $-0.001^{+3.108}_{-3.000}$ & -- & $[1]$ & 1.01 & 56315  \\ 
 \hline 
	 \multirow{2}{*}{S\_7} & \multirow{2}{*}{HERSJ012620.5$+$012950} & \multirow{2}{*}{1} & Sérsic & $22.23^{+0.31}_{-0.24}$ & $0.89\pm0.12$ & $5.86^{+0.48}_{-0.47}$ & $0.70\pm0.02$ & $13.18^{+1.91}_{-1.92}$ & -- & $0.66^{+0.08}_{-0.11}$ & \multirow{2}{*}{1.04} & \multirow{2}{*}{66653} \\ 
	   &   &   & Gauss & $21.72^{+0.09}_{-0.08}$ & $0.76\pm0.05$ & $[0.5]$ & $0.88\pm0.05$ & $74.75\pm3.74$ & -- & $0.34^{+0.11}_{-0.08}$ &   &    \\ 
 \hline 
	 S\_9 & HERMESJ032637$-$270044 & 1 & Sérsic & $22.70\pm0.05$ & $1.01^{+0.02}_{-0.03}$ & $5.73\pm0.11$ & $0.997^{+0.002}_{-0.003}$ & $46.45^{+0.69}_{-0.70}$ & -- & $[1]$ & 1.04 & 58583  \\ 
 \hline 
	 \multirow{3}{*}{S\_10} & \multirow{3}{*}{HERMESJ033732$-$295353} & \multirow{3}{*}{1} & Sérsic & $20.23^{+0.25}_{-0.21}$ & $0.50\pm0.09$ & $4.51^{+0.37}_{-0.38}$ & $0.804\pm0.009$ & $53.64^{+1.65}_{-1.68}$ & -- & $0.15^{+0.06}_{-0.05}$ & \multirow{3}{*}{1.05} & \multirow{3}{*}{177156} \\ 
	   &   &   & Sérsic & $22.12^{+0.13}_{-0.11}$ & $2.96^{+0.08}_{-0.07}$ & $1.73\pm0.15$ & $0.96\pm0.02$ & $149.69^{+1.32}_{-1.33}$ & -- & $0.71^{+0.05}_{-0.06}$ &   &    \\ 
	   &   &   & Sérsic & $20.61^{+0.16}_{-0.14}$ & $0.83\pm0.03$ & $0.69\pm0.06$ & $0.90\pm0.02$ & $54.94^{+2.84}_{-2.94}$ & -- & $0.14^{+0.03}_{-0.02}$ &   &    \\ 
 \hline 
	 \multirow{2}{*}{S\_13} & \multirow{2}{*}{HERMESJ103827$+$581544} & \multirow{2}{*}{1} & Sérsic & $20.25^{+0.10}_{-0.09}$ & $0.47\pm0.03$ & $1.56\pm0.07$ & $0.762\pm0.008$ & $6.07\pm0.82$ & -- & $0.27\pm0.03$ & \multirow{2}{*}{1.02} & \multirow{2}{*}{107445} \\ 
	   &   &   & Sérsic & $22.65^{+0.07}_{-0.06}$ & $2.56^{+0.07}_{-0.08}$ & $1.07\pm0.05$ & $0.74\pm0.01$ & $12.31^{+1.06}_{-1.07}$ & -- & $0.73\pm0.03$ &   &    \\ 
 \hline 
	 S\_14 & HERMESJ110016$+$571736 & 1 & Sérsic & $21.76\pm0.04$ & $1.01\pm0.02$ & $2.68\pm0.04$ & $0.807\pm0.004$ & $120.78^{+0.60}_{-0.59}$ & -- & $[1]$ & 1.06 & 46997  \\ 
 \hline 
	 \multirow{5}{*}{S\_15} & \multirow{2}{*}{HATLASJ114638$-$001132$_{1}$} & \multirow{5}{*}{2} & Sérsic & $20.30^{+0.24}_{-0.20}$ & $0.19\pm0.02$ & $1.36\pm0.17$ & $0.53\pm0.03$ & $50.27^{+1.72}_{-1.69}$ & -- & $0.58^{+0.08}_{-0.09}$ & \multirow{5}{*}{1.03} & \multirow{5}{*}{44201} \\ 
	   &   &   & Sérsic & $24.52^{+0.16}_{-0.13}$ & $1.14\pm0.07$ & $0.45^{+0.07}_{-0.08}$ & $0.81^{+0.05}_{-0.04}$ & $53.88^{+2.83}_{-2.87}$ & -- & $0.42^{+0.09}_{-0.08}$ &   &    \\ 
 \\ 
	   & \multirow{2}{*}{HATLASJ114638$-$001132$_{2}$} &   & Sérsic & $21.17^{+0.25}_{-0.21}$ & $0.18\pm0.02$ & $3.26\pm0.43$ & $0.81\pm0.03$ & $128.86^{+1.77}_{-1.75}$ & -- & $0.29^{+0.08}_{-0.07}$ &   &    \\ 
	   &   &   & Sérsic & $25.49^{+0.14}_{-0.12}$ & $2.29\pm0.10$ & $1.73^{+0.15}_{-0.14}$ & $0.86\pm0.03$ & $132.15^{+1.64}_{-1.66}$ & -- & $0.71^{+0.07}_{-0.08}$ &   &    \\ 
 \hline 
	 S\_16 & HATLASJ125126$+$254928 & 1 & Sérsic & $23.11^{+0.06}_{-0.05}$ & $1.50\pm0.04$ & $7.96\pm0.16$ & $0.504\pm0.005$ & $135.36\pm0.77$ & -- & $[1]$ & 1.07 & 50036  \\ 
 \hline 
	 S\_17 & HATLASJ125760$+$224558 & 1 & Sérsic & $22.33\pm0.04$ & $1.13\pm0.02$ & $4.63\pm0.08$ & $0.814\pm0.004$ & $6.39^{+0.58}_{-0.61}$ & -- & $[1]$ & 1.07 & 50512  \\ 
 \hline 
	 \multirow{2}{*}{S\_18} & \multirow{2}{*}{HATLASJ133008$+$245860} & \multirow{2}{*}{1} & Sérsic & $21.81^{+0.30}_{-0.24}$ & $0.69\pm0.09$ & $6.34^{+0.49}_{-0.50}$ & $0.80\pm0.01$ & $81.56^{+1.97}_{-1.95}$ & -- & $0.83^{+0.05}_{-0.07}$ & \multirow{2}{*}{1.04} & \multirow{2}{*}{44862} \\ 
	   &   &   & Gauss & $22.85^{+0.11}_{-0.10}$ & $0.83\pm0.05$ & $[0.5]$ & $0.91^{+0.05}_{-0.06}$ & $96.76^{+10.85}_{-11.16}$ & -- & $0.17^{+0.07}_{-0.05}$ &   &    \\ 
 \hline 
	 S\_19 & HATLASJ133846$+$255057 & 1 & Sérsic & $23.46^{+0.07}_{-0.06}$ & $1.11\pm0.04$ & $5.25^{+0.15}_{-0.16}$ & $0.696\pm0.007$ & $50.61^{+0.87}_{-0.90}$ & -- & $[1]$ & 1.04 & 57037  \\ 
 \hline 
	 S\_21 & HERMESJ171451$+$592634 & 1 & de Vauc. & $22.05^{+0.05}_{-0.04}$ & $0.45\pm0.01$ & $[4]$ & $0.55\pm0.01$ & $144.39^{+1.69}_{-1.64}$ & -- & $[1]$ & 1.10 & 31356  \\ 
 \hline 
	 \multirow{5}{*}{S\_22} & \multirow{3}{*}{HERMESJ171545$+$601238$_{1}$} & \multirow{5}{*}{3} & Sérsic & $22.47^{+0.17}_{-0.15}$ & $0.56\pm0.04$ & $4.43^{+0.33}_{-0.34}$ & $0.992^{+0.006}_{-0.009}$ & $180.00^{+1.08}_{-1.07}$ & -- & $0.17^{+0.04}_{-0.03}$ & \multirow{3}{*}{1.05} & \multirow{3}{*}{55119} \\ 
	   &   &   & PSF & -- & -- & -- & -- & -- & $[24.97]$ & $0.0023\pm0.0002$ &   &    \\ 
	   &   &   & Sérsic & $24.43\pm0.08$ & $4.10^{+0.14}_{-0.13}$ & $2.72\pm0.16$ & $0.69\pm0.01$ & $96.24^{+1.29}_{-1.27}$ & -- & $0.83^{+0.03}_{-0.04}$ &   &    \\ 
 \\ 
	   & HERMESJ171545$+$601238$_{2}$ &   & Sérsic & $23.87^{+0.10}_{-0.09}$ & $1.23\pm0.06$ & $3.76\pm0.20$ & $0.39\pm0.01$ & $99.53^{+1.02}_{-0.99}$ & -- & $[1]$ & 1.05 & 54970  \\ 
 \hline 
	 \multirow{4}{*}{S\_23} & HATLASJ225844.8$-$295124$_{1}$ & \multirow{4}{*}{2} & Sérsic & $20.69^{+0.07}_{-0.06}$ & $0.44^{+0.01}_{-0.02}$ & $1.84\pm0.06$ & $0.448\pm0.008$ & $145.47^{+0.87}_{-0.83}$ & -- & $[1]$ & \multirow{4}{*}{1.05} & \multirow{4}{*}{15053} \\ 
 \\ 
	   & \multirow{2}{*}{HATLASJ225844.8$-$295124$_{2}$} &   & Sérsic & $22.59^{+0.24}_{-0.20}$ & $0.61\pm0.07$ & $1.12\pm0.15$ & $0.39\pm0.03$ & $58.30\pm1.77$ & -- & $0.10\pm0.03$ &   &    \\ 
	   &   &   & Sérsic & $24.93^{+0.10}_{-0.09}$ & $3.29^{+0.15}_{-0.14}$ & $2.83^{+0.27}_{-0.28}$ & $0.69^{+0.03}_{-0.02}$ & $26.28^{+3.28}_{-3.25}$ & -- & $0.90\pm0.03$ &   &    \\ 
 \hline 
	 \multirow{4}{*}{S\_24} & \multirow{2}{*}{HELMSJ232210.3$-$033559$_{1}$} & \multirow{4}{*}{3} & Sérsic & $23.38^{+0.10}_{-0.09}$ & $1.42\pm0.08$ & $2.25\pm0.14$ & $0.963^{+0.008}_{-0.009}$ & $80.29^{+0.86}_{-0.89}$ & -- & $0.81\pm0.03$ & \multirow{2}{*}{1.08} & \multirow{2}{*}{54469} \\ 
	   &   &   & Sérsic & $19.99^{+0.13}_{-0.12}$ & $0.214\pm0.008$ & $1.29\pm0.12$ & $0.57\pm0.03$ & $7.88^{+1.32}_{-1.33}$ & -- & $0.19\pm0.03$ &   &    \\ 
 \\ 
	   & HELMSJ232210.3$-$033559$_{2}$ &   & de Vauc. & $21.44\pm0.06$ & $0.134\pm0.006$ & $[4]$ & $0.74\pm0.03$ & $64.64^{+4.21}_{-4.33}$ & -- & $[1]$ & 1.07 & 50334  \\ 
\hline
\hline
\end{tabular}
\endgroup
}
\begin{flushleft}
\textit{Notes}: Col.~(1): Source reference number. Col.~(2): IAU name of the {\it Herschel} detection. Indices 1 and 2 refer to the two components of the lens candidate. Col.~(3): Type of the system. Col.~(4): Adopted model for the lens components. Col.~(5): Effective surface brightness, i.e., the surface brightness at the effective radius. Col.~(6): Effective radius, i.e., the semi-major axis of the isophote containing half of the light of the component. Col.~(7): Sérsic index. Values fixed in the fit are bracketed. Col.~(8): Axial ratio. Col.~(9): Position angle. Col.~(10): Magnitude of the unresolved component. Col.~(11): Relative flux of the component $C/T$. Col.~(12): Reduced $\chi^2$ of the fit. Col.~(13): Number of degrees of freedom of the fit.
\end{flushleft}
\end{table*}

\begin{table*}
\centering
\contcaption{}
 
{\scriptsize
\begingroup
\setlength{\tabcolsep}{3pt}
\begin{tabular}{c l c c c c c c c c c c c}
\hline
\hline
	 No. & IAU Name & Type & Components & $\mu_{\rm e}$ & $R_{\rm e}$ & $n$ & $q$ & $PA$ & $m_{\rm PSF}$ & $C/T$ & $\chi^2$ & $N_{\rm dof}$  \\ 
	   &   &   &   & [mag arcsec$^{-2}$] & [arcsec] &   &   & [deg] & [mag] &   &   &    \\ 
	 (1) & (2) & (3) & (4) & (5) & (6) & (7) & (8) & (9) & (10) & (11) & (12) & (13)  \\ 
\hline
	 \multirow{4}{*}{S\_25} & \multirow{2}{*}{HATLASJ233037.2$-$331217$_{1}$} & \multirow{4}{*}{3} & Sérsic & $20.79^{+0.23}_{-0.19}$ & $0.25\pm0.02$ & $7.05^{+0.58}_{-0.57}$ & $0.58\pm0.02$ & $75.25^{+1.03}_{-1.05}$ & -- & $0.32^{+0.07}_{-0.06}$ & \multirow{4}{*}{1.05} & \multirow{4}{*}{47114} \\ 
	   &   &   & Sérsic & $23.71^{+0.12}_{-0.10}$ & $1.66\pm0.06$ & $1.44\pm0.14$ & $0.89\pm0.03$ & $139.93^{+3.41}_{-3.29}$ & -- & $0.68^{+0.06}_{-0.07}$ &   &    \\ 
 \\ 
	   & HATLASJ233037.2$-$331217$_{2}$ &   & PSF & -- & -- & -- & -- & -- & $[23.14]$ & $[1]$ &   &    \\ 

\\
\multicolumn{13}{l}{\textit{Confirmed after the lens subtraction}} \\
\\

 	 \multirow{2}{*}{S\_26} & \multirow{2}{*}{HELMSJ001626.0$+$042613} & \multirow{2}{*}{1} & Sérsic & $19.39\pm0.05$ & $0.52\pm0.01$ & $4.23\pm0.09$ & $0.467\pm0.004$ & $7.10^{+0.26}_{-0.25}$ & -- & $0.59\pm0.03$ & \multirow{2}{*}{1.05} & \multirow{2}{*}{55249} \\ 
	   &   &   & Sérsic & $22.22^{+0.09}_{-0.08}$ & $1.83\pm0.06$ & $1.27\pm0.08$ & $0.63\pm0.01$ & $162.98^{+1.24}_{-1.26}$ & -- & $0.41\pm0.03$ &   &    \\ 
 \hline 
	 \multirow{2}{*}{S\_27} & \multirow{2}{*}{HATLASJ002624.8$-$341737} & \multirow{2}{*}{1} & Sérsic & $21.92^{+0.25}_{-0.20}$ & $0.31\pm0.03$ & $4.94^{+0.51}_{-0.52}$ & $0.63\pm0.02$ & $172.06^{+1.73}_{-1.74}$ & -- & $0.54^{+0.08}_{-0.09}$ & \multirow{2}{*}{1.05} & \multirow{2}{*}{52636} \\ 
	   &   &   & Sérsic & $24.33^{+0.13}_{-0.12}$ & $1.08\pm0.04$ & $0.69^{+0.09}_{-0.08}$ & $0.97^{+0.02}_{-0.03}$ & $180.01\pm1.36$ & -- & $0.46^{+0.09}_{-0.08}$ &   &    \\ 
 \hline 
	 \multirow{4}{*}{S\_28} & \multirow{2}{*}{HELMSJ004723.6$+$015751$_{1}$} & \multirow{4}{*}{2} & de Vauc. & $18.60^{+0.30}_{-0.23}$ & $0.09\pm0.01$ & $[4]$ & $0.84^{+0.04}_{-0.03}$ & $43.47^{+4.28}_{-4.38}$ & -- & $0.11^{+0.04}_{-0.03}$ & \multirow{4}{*}{1.03} & \multirow{4}{*}{58679} \\ 
	   &   &   & Sérsic & $21.92\pm0.02$ & $1.57\pm0.02$ & $1.37^{+0.08}_{-0.07}$ & $0.851\pm0.004$ & $165.60^{+1.09}_{-1.13}$ & -- & $0.89^{+0.03}_{-0.04}$ &   &    \\ 
 \\ 
	   & HELMSJ004723.6$+$015751$_{2}$ &   & Sérsic & $22.17\pm0.06$ & $0.80\pm0.02$ & $3.14\pm0.07$ & $0.856^{+0.006}_{-0.007}$ & $79.44^{+0.83}_{-0.82}$ & -- & $[1]$ &   &    \\ 
 \hline
	 \multirow{2}{*}{S\_29} & \multirow{2}{*}{HERSJ012041.6$-$002705} & \multirow{2}{*}{1} & Sérsic & $19.91^{+0.20}_{-0.17}$ & $0.19\pm0.02$ & $4.39\pm0.34$ & $0.71\pm0.01$ & $86.79\pm1.04$ & -- & $0.44\pm0.07$ & \multirow{2}{*}{1.03} & \multirow{2}{*}{50744} \\ 
	   &   &   & Sérsic & $23.67^{+0.13}_{-0.12}$ & $1.37^{+0.05}_{-0.06}$ & $1.98\pm0.18$ & $0.79\pm0.02$ & $3.03^{+2.85}_{-2.88}$ & -- & $0.56\pm0.07$ &   &    \\ 
 \hline 
	 \multirow{3}{*}{S\_30} & \multirow{3}{*}{HATLASJ085112$+$004934} & \multirow{3}{*}{1} & Sérsic & $20.24^{+0.41}_{-0.30}$ & $0.27\pm0.06$ & $4.45^{+0.72}_{-0.71}$ & $0.61\pm0.02$ & $81.52^{+1.84}_{-1.83}$ & -- & $0.25^{+0.12}_{-0.10}$ & \multirow{3}{*}{1.06} & \multirow{3}{*}{40914} \\ 
	   &   &   & Sérsic & $21.51^{+0.17}_{-0.14}$ & $0.78\pm0.02$ & $0.44\pm0.04$ & $0.94\pm0.03$ & $158.68^{+6.02}_{-6.11}$ & -- & $0.37\pm0.07$ &   &    \\ 
	   &   &   & Sérsic & $22.35^{+0.14}_{-0.12}$ & $1.39\pm0.04$ & $0.72\pm0.07$ & $0.55\pm0.02$ & $81.26^{+1.61}_{-1.71}$ & -- & $0.38\pm0.07$ &   &    \\ 
 \hline 
	 \multirow{2}{*}{S\_31} & \multirow{2}{*}{HATLASJ085359$+$015537} & \multirow{2}{*}{1} & Sérsic & $22.69^{+0.26}_{-0.20}$ & $0.45\pm0.05$ & $6.78\pm0.96$ & $0.30\pm0.03$ & $117.52^{+1.82}_{-1.81}$ & -- & $0.30^{+0.08}_{-0.07}$ & \multirow{2}{*}{1.13} & \multirow{2}{*}{22254} \\ 
	   &   &   & Sérsic & $26.34^{+0.14}_{-0.12}$ & $2.34\pm0.09$ & $5.14\pm0.48$ & $0.85\pm0.03$ & $156.50^{+1.28}_{-1.29}$ & -- & $0.70^{+0.07}_{-0.08}$ &   &    \\ 
 \hline 
	 \multirow{2}{*}{S\_32} & \multirow{2}{*}{HERMESJ104549$+$574512} & \multirow{2}{*}{1} & Sérsic & $19.59^{+0.23}_{-0.19}$ & $0.16\pm0.02$ & $3.86^{+0.40}_{-0.39}$ & $0.37\pm0.02$ & $121.59^{+1.47}_{-1.38}$ & -- & $0.22^{+0.06}_{-0.05}$ & \multirow{2}{*}{1.04} & \multirow{2}{*}{60465} \\ 
	   &   &   & Sérsic & $21.54^{+0.12}_{-0.11}$ & $0.64\pm0.03$ & $1.74\pm0.17$ & $0.68\pm0.03$ & $89.59^{+3.16}_{-3.13}$ & -- & $0.78^{+0.05}_{-0.06}$ &   &    \\ 
 \hline 
	 \multirow{3}{*}{S\_33} & HERMESJ105551$+$592845$_{1}$ & \multirow{3}{*}{2} & Sérsic & $22.62\pm0.08$ & $0.66\pm0.03$ & $1.82\pm0.08$ & $0.762\pm0.009$ & $92.59\pm0.85$ & -- & $[1]$ & \multirow{3}{*}{1.15} & \multirow{3}{*}{6328} \\ 
 \\ 
	   & HERMESJ105551$+$592845$_{2}$ &   & Sérsic & $21.85^{+0.15}_{-0.13}$ & $0.28\pm0.02$ & $1.25\pm0.12$ & $0.79\pm0.02$ & $123.08^{+1.84}_{-1.82}$ & -- & $[1]$ &   &    \\ 
 \hline 
	 \multirow{13}{*}{S\_34} & \multirow{2}{*}{HERMESJ105751$+$573026$_{1}$} & \multirow{13}{*}{2} & Sérsic & $20.36\pm0.09$ & $0.41\pm0.02$ & $3.90\pm0.18$ & $0.752\pm0.008$ & $91.89^{+0.83}_{-0.84}$ & -- & $0.27\pm0.03$ & \multirow{13}{*}{1.11} & \multirow{13}{*}{31547} \\ 
	   &   &   & Sérsic & $24.58\pm0.07$ & $5.16\pm0.15$ & $2.63^{+0.13}_{-0.12}$ & $0.74\pm0.01$ & $30.55^{+1.03}_{-1.05}$ & -- & $0.73\pm0.03$ &   &    \\ 
 \\ 
	   & \multirow{2}{*}{HERMESJ105751$+$573026$_{2}$} &   & de Vauc. & $19.30^{+0.27}_{-0.22}$ & $0.10\pm0.02$ & $[4]$ & $0.37\pm0.04$ & $32.23^{+4.35}_{-4.53}$ & -- & $0.42^{+0.36}_{-0.24}$ &   &    \\ 
	   &   &   & Sérsic & $22.80^{+0.67}_{-0.42}$ & $0.43^{+0.23}_{-0.21}$ & $5.95^{+2.42}_{-2.72}$ & $0.76^{+0.12}_{-0.13}$ & $94.68^{+7.37}_{-7.46}$ & -- & $0.58^{+0.24}_{-0.36}$ &   &    \\ 
 \\ 
	   & HERMESJ105751$+$573026$_{3}$ &   & Sérsic & $20.72^{+0.15}_{-0.13}$ & $0.137\pm0.009$ & $3.63^{+0.31}_{-0.32}$ & $0.82\pm0.02$ & $74.54^{+1.82}_{-1.84}$ & -- & $[1]$ &   &    \\ 
 \\ 
	   & HERMESJ105751$+$573026$_{4}$ &   & de Vauc. & $21.16\pm0.05$ & $0.220\pm0.007$ & $[4]$ & $0.61\pm0.02$ & $100.13^{+2.05}_{-2.07}$ & -- & $[1]$ &   &    \\ 
 \\ 
	   & HERMESJ105751$+$573026$_{5}$ &   & Sérsic & $20.30^{+0.13}_{-0.12}$ & $0.131^{+0.008}_{-0.007}$ & $3.11\pm0.23$ & $0.89\pm0.01$ & $175.89^{+1.42}_{-1.45}$ & -- & $[1]$ &   &    \\ 
 \\ 
	   & HERMESJ105751$+$573026$_{6}$ &   & PSF & -- & -- & -- & -- & -- & $[23.60]$ & $[1]$ &   &    \\ 
 \hline 
	 S\_35 & HATLASJ132630$+$334410 & 1 & Sérsic & $23.80\pm0.04$ & $2.53\pm0.05$ & $5.73\pm0.09$ & $0.595\pm0.004$ & $72.04^{+0.57}_{-0.59}$ & -- & $[1]$ & 1.05 & 39273  \\ 
 \hline 
	 \multirow{8}{*}{S\_36} & \multirow{2}{*}{HATLASJ133543$+$300404$_{1}$} & \multirow{8}{*}{2} & Sérsic & $22.42\pm0.08$ & $0.65\pm0.03$ & $0.88\pm0.04$ & $0.902\pm0.009$ & $127.84\pm0.89$ & -- & $0.954\pm0.005$ & \multirow{8}{*}{1.03} & \multirow{8}{*}{41523} \\ 
	   &   &   & PSF & -- & -- & -- & -- & -- & $[24.11]$ & $0.046\pm0.005$ &   &    \\ 
 \\ 
	   & HATLASJ133543$+$300404$_{2}$ &   & Sérsic & $21.90^{+0.09}_{-0.08}$ & $0.41\pm0.02$ & $4.97^{+0.23}_{-0.24}$ & $0.567\pm0.010$ & $24.77^{+0.94}_{-0.95}$ & -- & $[1]$ &   &    \\ 
 \\ 
	   & HATLASJ133543$+$300404$_{3}$ &   & Sérsic & $23.50\pm0.06$ & $1.10\pm0.03$ & $5.58\pm0.15$ & $0.844\pm0.007$ & $64.17^{+0.92}_{-0.94}$ & -- & $[1]$ &   &    \\ 
 \\ 
	   & HATLASJ133543$+$300404$_{4}$ &   & Sérsic & $23.58^{+0.07}_{-0.06}$ & $1.02\pm0.04$ & $6.21\pm0.21$ & $0.609\pm0.008$ & $165.10^{+0.80}_{-0.84}$ & -- & $[1]$ &   &    \\ 
 \hline 
	 S\_37 & HATLASJ142140$+$000448 & 1 & Sérsic & $21.27^{+0.07}_{-0.06}$ & $0.38\pm0.01$ & $2.28\pm0.05$ & $0.925\pm0.005$ & $0.67\pm0.77$ & -- & $[1]$ & 1.11 & 44658  \\ 
 \hline 
	 S\_38 & HERMESJ142824$+$352620 & 1 & Sérsic & $21.47^{+0.08}_{-0.07}$ & $0.49\pm0.02$ & $2.43^{+0.10}_{-0.09}$ & $0.487\pm0.008$ & $92.76^{+0.89}_{-0.91}$ & -- & $[1]$ & 1.17 & 7204  \\ 
 \hline 
	 S\_39 & HATLASJ223753.8$-$305828 & 1 & Sérsic & $21.81\pm0.05$ & $0.72\pm0.02$ & $5.03^{+0.10}_{-0.09}$ & $0.940\pm0.005$ & $87.28^{+0.70}_{-0.71}$ & -- & $[1]$ & 1.03 & 63365  \\ 
 \hline 
	 S\_40 & HATLASJ225250.7$-$313657 & 1 & Sérsic & $22.23^{+0.12}_{-0.10}$ & $0.40\pm0.02$ & $6.21\pm0.40$ & $0.53\pm0.01$ & $75.70^{+1.21}_{-1.25}$ & -- & $[1]$ & 1.04 & 34943  \\ 
 \hline 
	 \multirow{2}{*}{S\_41} & \multirow{2}{*}{HELMSJ233441.0$-$065220} & \multirow{2}{*}{1} & Sérsic & $19.47^{+0.20}_{-0.17}$ & $0.28\pm0.02$ & $1.14\pm0.09$ & $0.38\pm0.01$ & $22.16^{+1.00}_{-1.01}$ & -- & $0.49^{+0.07}_{-0.08}$ & \multirow{2}{*}{1.57} & \multirow{2}{*}{35357} \\ 
	   &   &   & Sérsic & $22.59^{+0.15}_{-0.13}$ & $1.00\pm0.06$ & $0.62\pm0.06$ & $0.68\pm0.03$ & $167.72^{+2.29}_{-2.37}$ & -- & $0.51^{+0.08}_{-0.07}$ &   &    \\ 	
\hline
\hline
\end{tabular}
\endgroup
}
\end{table*}

\begin{table*}
\centering
\contcaption{}
 
{\scriptsize
\begingroup
\setlength{\tabcolsep}{3pt}
\begin{tabular}{c l c c c c c c c c c c c}
\hline
\hline
	 No. & IAU Name & Type & Components & $\mu_{\rm e}$ & $R_{\rm e}$ & $n$ & $q$ & $PA$ & $m_{\rm PSF}$ & $C/T$ & $\chi^2$ & $N_{\rm dof}$  \\ 
	   &   &   &   & [mag arcsec$^{-2}$] & [arcsec] &   &   & [deg] & [mag] &   &   &    \\ 
	 (1) & (2) & (3) & (4) & (5) & (6) & (7) & (8) & (9) & (10) & (11) & (12) & (13)  \\ 
  \hline
   \multirow{4}{*}{S\_42} & \multirow{2}{*}{HELMSJ233633.5$-$032119$_{1}$} & \multirow{4}{*}{2} & Sérsic & $18.38\pm0.05$ & $0.332\pm0.009$ & $2.82\pm0.06$ & $0.577^{+0.004}_{-0.003}$ & $42.72^{+0.24}_{-0.25}$ & -- & $0.34\pm0.02$ & \multirow{4}{*}{1.06} & \multirow{4}{*}{58206} \\ 
	   &   &   & Sérsic & $21.94\pm0.04$ & $2.55^{+0.03}_{-0.04}$ & $1.25\pm0.04$ & $0.738\pm0.006$ & $46.04\pm0.84$ & -- & $0.66\pm0.02$ &   &    \\ 
 \\ 
	   & HELMSJ233633.5$-$032119$_{2}$ &   & Sérsic & $23.16^{+0.06}_{-0.05}$ & $1.79\pm0.04$ & $2.21\pm0.04$ & $0.723\pm0.005$ & $160.63^{+0.74}_{-0.75}$ & -- & $[1]$ &   &    \\ 
 
 \\
\multicolumn{13}{l}{\textit{Confirmed through sub-mm/mm follow-up}} \\
\\

	 \multirow{2}{*}{S\_43} & \multirow{2}{*}{HELMSJ001615.7$+$032435} & \multirow{2}{*}{1} & Sérsic & $24.18^{+0.31}_{-0.25}$ & $3.46^{+0.45}_{-0.44}$ & $7.32^{+0.38}_{-0.39}$ & $0.579^{+0.007}_{-0.006}$ & $31.91^{+0.67}_{-0.65}$ & -- & $0.86^{+0.04}_{-0.06}$ & \multirow{2}{*}{1.09} & \multirow{2}{*}{57290} \\ 
	   &   &   & Gauss & $24.61\pm0.09$ & $3.25\pm0.21$ & $[0.5]$ & $0.50\pm0.07$ & $17.05^{+3.68}_{-3.72}$ & -- & $0.14^{+0.06}_{-0.04}$ &   &    \\ 
 \hline 
	 S\_44 & HELMSJ002220.9$-$015524 & 1 & Sérsic & $25.43^{+0.10}_{-0.09}$ & $2.09\pm0.10$ & $3.77^{+0.19}_{-0.20}$ & $0.58\pm0.01$ & $98.62\pm0.99$ & -- & $[1]$ & 1.13 & 37251  \\ 
 \hline 
	 S\_45 & HELMSJ003814.1$-$002252 & 1 & PSF & -- & -- & -- & -- & -- & $[22.45]$ & $[1]$ & 1.04 & 18416  \\ 
\hline
	 \multirow{2}{*}{S\_47} & \multirow{2}{*}{HELMSJ004714.2$+$032454} & \multirow{2}{*}{1} & Sérsic & $20.79^{+0.18}_{-0.15}$ & $0.20\pm0.02$ & $0.60\pm0.10$ & $0.75\pm0.03$ & $103.11^{+4.11}_{-4.17}$ & -- & $0.17^{+0.06}_{-0.05}$ & \multirow{2}{*}{1.01} & \multirow{2}{*}{26442} \\ 
	   &   &   & Exp. disk & $23.42^{+0.19}_{-0.17}$ & $1.47^{+0.14}_{-0.13}$ & $[1]$ & $0.60^{+0.04}_{-0.05}$ & $111.21^{+4.52}_{-4.42}$ & -- & $0.83^{+0.05}_{-0.06}$ &   &    \\ 
 \hline 
	 \multirow{7}{*}{S\_48} & \multirow{2}{*}{HELMSJ005159.5$+$062241$_{1}$} & \multirow{7}{*}{2} & Sérsic & $18.59^{+0.21}_{-0.17}$ & $0.113^{+0.009}_{-0.010}$ & $2.28\pm0.18$ & $0.77\pm0.01$ & $119.44\pm1.05$ & -- & $0.14\pm0.03$ & \multirow{7}{*}{1.03} & \multirow{7}{*}{48792} \\ 
	   &   &   & Sérsic & $23.10^{+0.08}_{-0.07}$ & $2.10\pm0.06$ & $1.95^{+0.10}_{-0.09}$ & $0.96\pm0.01$ & $9.11^{+1.24}_{-1.22}$ & -- & $0.86\pm0.03$ &   &    \\ 
 \\ 
	   & HELMSJ005159.5$+$062241$_{2}$ &   & de Vauc. & $20.48\pm0.03$ & $0.201\pm0.004$ & $[4]$ & $0.76\pm0.01$ & $173.86^{+1.76}_{-1.70}$ & -- & $[1]$ &   &    \\ 
 \\ 
	   & \multirow{2}{*}{HELMSJ005159.5$+$062241$_{3}$} &   & Sérsic & $24.00^{+0.07}_{-0.06}$ & $1.92\pm0.06$ & $1.33\pm0.04$ & $0.703^{+0.007}_{-0.008}$ & $37.42^{+0.90}_{-0.87}$ & -- & $0.899^{+0.008}_{-0.009}$ &   &    \\ 
	   &   &   & PSF & -- & -- & -- & -- & -- & $[22.52]$ & $0.101^{+0.009}_{-0.008}$ &   &    \\ 
 \hline 
	 S\_49 & HATLASJ005724.2$-$273122 & 1 & Sérsic & $24.83^{+0.07}_{-0.06}$ & $1.46\pm0.05$ & $9.11\pm0.30$ & $0.858\pm0.008$ & $126.93^{+0.84}_{-0.83}$ & -- & $[1]$ & 1.05 & 37643  \\ 
 \hline 
	 S\_51 & HERMESJ033211$-$270536 & 1 & PSF & -- & -- & -- & -- & -- & $[23.01]$ & $[1]$ & 1.07 & 9735  \\ 
 \hline 
	 S\_52 & HERMESJ044154$-$540352 & 1 & Sérsic & $21.75\pm0.06$ & $0.54\pm0.02$ & $7.49\pm0.17$ & $0.783\pm0.007$ & $9.27\pm0.86$ & -- & $[1]$ & 1.03 & 44932  \\ 
 \hline 
	 \multirow{4}{*}{S\_54} & \multirow{2}{*}{HATLASJ091841$+$023048$_{1}$} & \multirow{4}{*}{2} & Sérsic & $20.11^{+0.34}_{-0.27}$ & $0.11\pm0.02$ & $1.83^{+0.30}_{-0.31}$ & $0.69\pm0.03$ & $152.52^{+1.94}_{-1.93}$ & -- & $0.19^{+0.07}_{-0.06}$ & \multirow{4}{*}{1.07} & \multirow{4}{*}{43760} \\ 
	   &   &   & Sérsic & $25.92^{+0.14}_{-0.12}$ & $2.99\pm0.14$ & $2.44\pm0.20$ & $0.76\pm0.02$ & $70.78^{+1.87}_{-1.88}$ & -- & $0.81^{+0.06}_{-0.07}$ &   &    \\ 
 \\ 
	   & HATLASJ091841$+$023048$_{2}$ &   & de Vauc. & $21.80\pm0.06$ & $0.155\pm0.006$ & $[4]$ & $0.91\pm0.03$ & $115.92^{+4.12}_{-4.11}$ & -- & $[1]$ &   &    \\ 
 \hline 
	 S\_56 & HATLASJ115433.6$+$005042 & 1 & Sérsic & $21.10^{+0.08}_{-0.07}$ & $0.48\pm0.02$ & $1.83\pm0.08$ & $0.337\pm0.009$ & $169.27^{+0.85}_{-0.87}$ & -- & $[1]$ & 1.01 & 34603  \\ 
 \hline 
	 \multirow{2}{*}{S\_57} & \multirow{2}{*}{HATLASJ120127.6$-$014043} & \multirow{2}{*}{1} & Sérsic & $25.61^{+0.35}_{-0.26}$ & $1.51\pm0.23$ & $1.00\pm0.18$ & $0.74\pm0.03$ & $42.98^{+1.91}_{-1.83}$ & -- & $0.25^{+0.09}_{-0.08}$ & \multirow{2}{*}{1.06} & \multirow{2}{*}{28398} \\ 
	   &   &   & Sérsic & $19.56^{+0.14}_{-0.12}$ & $0.225\pm0.009$ & $2.12\pm0.20$ & $0.27\pm0.03$ & $20.79^{+1.31}_{-1.29}$ & -- & $0.75^{+0.08}_{-0.09}$ &   &    \\ 
 \hline 
 S\_58 & HATLASJ131611$+$281220 & 1 & Sérsic & $22.68^{+0.15}_{-0.13}$ & $0.45\pm0.03$ & $5.08^{+0.44}_{-0.43}$ & $0.40\pm0.02$ & $1.47^{+1.77}_{-1.73}$ & -- & [1] & 1.02 & 28595 \\
 \hline 
	 \multirow{2}{*}{S\_59} & \multirow{2}{*}{HATLASJ134429$+$303036} & \multirow{2}{*}{1} & Sérsic & $19.44^{+0.37}_{-0.28}$ & $0.20\pm0.03$ & $2.50\pm0.29$ & $0.65\pm0.01$ & $161.16^{+2.23}_{-2.21}$ & -- & $0.54^{+0.13}_{-0.14}$ & \multirow{2}{*}{1.04} & \multirow{2}{*}{48806} \\ 
	   &   &   & Exp. disk & $22.71^{+0.20}_{-0.16}$ & $1.64\pm0.15$ & $[1]$ & $0.26^{+0.04}_{-0.05}$ & $173.10^{+4.55}_{-4.49}$ & -- & $0.46^{+0.14}_{-0.13}$ &   &    \\ 
 \hline 
	 \multirow{6}{*}{S\_60} & \multirow{2}{*}{HATLASJ141352$-$000027$_{1}$} & \multirow{6}{*}{3} & Sérsic & $20.47^{+0.18}_{-0.15}$ & $0.51\pm0.05$ & $3.36^{+0.21}_{-0.22}$ & $0.825\pm0.006$ & $83.00\pm1.08$ & -- & $0.55^{+0.07}_{-0.08}$ & \multirow{2}{*}{1.01} & \multirow{2}{*}{73852} \\ 
	   &   &   & Exp. disk & $23.99\pm0.11$ & $2.87\pm0.17$ & $[1]$ & $0.96\pm0.03$ & $-0.01^{+4.33}_{-4.12}$ & -- & $0.45^{+0.08}_{-0.07}$ &   &    \\ 
 \\ 
	   & HATLASJ141352$-$000027$_{2}$ &   & Sérsic & $21.05\pm0.08$ & $0.31\pm0.01$ & $3.64^{+0.17}_{-0.16}$ & $0.536\pm0.010$ & $148.77^{+0.88}_{-0.89}$ & -- & $[1]$ & 1.01 & 63073  \\ 
 \\ 
	   & HATLASJ141352$-$000027$_{3}$ &   & Sérsic & $21.69\pm0.05$ & $0.68\pm0.02$ & $4.91\pm0.10$ & $0.866\pm0.005$ & $41.57^{+0.79}_{-0.75}$ & -- & $[1]$ & 1.01 & 64764  \\ 
 \hline 
	 \multirow{4}{*}{S\_61} & \multirow{2}{*}{HATLASJ142414$+$022304$_{1}$} & \multirow{4}{*}{2} & Sérsic & $24.20^{+0.07}_{-0.06}$ & $1.49\pm0.05$ & $3.76\pm0.11$ & $0.910\pm0.007$ & $15.97^{+0.94}_{-0.89}$ & -- & $0.880^{+0.009}_{-0.010}$ & \multirow{4}{*}{1.04} & \multirow{4}{*}{22914} \\ 
	   &   &   & PSF & -- & -- & -- & -- & -- & $[22.25]$ & $0.120^{+0.010}_{-0.009}$ &   &    \\ 
 \\ 
	   & HATLASJ142414$+$022304$_{2}$ &   & Sérsic & $24.59^{+0.10}_{-0.09}$ & $1.13\pm0.06$ & $7.02\pm0.36$ & $0.68\pm0.01$ & $149.22^{+1.02}_{-0.98}$ & -- & $[1]$ &   &    \\ 
 \hline 
	 \multirow{2}{*}{S\_62} & \multirow{2}{*}{HATLASJ230815.5$-$343801} & \multirow{2}{*}{1} & Sérsic & $22.86^{+0.21}_{-0.17}$ & $0.56\pm0.05$ & $8.73^{+0.65}_{-0.69}$ & $0.89\pm0.01$ & $157.26\pm1.05$ & -- & $0.76^{+0.05}_{-0.06}$ & \multirow{2}{*}{1.13} & \multirow{2}{*}{21244} \\ 
	   &   &   & Sérsic & $22.89^{+0.14}_{-0.13}$ & $0.75\pm0.03$ & $0.52\pm0.07$ & $0.56^{+0.03}_{-0.04}$ & $7.75^{+1.49}_{-1.48}$ & -- & $0.24^{+0.06}_{-0.05}$ &   &    \\ 
 \hline 
	 \multirow{2}{*}{S\_63} & \multirow{2}{*}{HELMSJ232439.5$-$043936} & \multirow{2}{*}{1} & Sérsic & $21.14^{+0.22}_{-0.18}$ & $0.27^{+0.02}_{-0.03}$ & $3.26\pm0.24$ & $0.98\pm0.01$ & $89.07^{+1.07}_{-1.03}$ & -- & $0.67^{+0.06}_{-0.08}$ & \multirow{2}{*}{1.06} & \multirow{2}{*}{53698} \\ 
	   &   &   & Sérsic & $23.95^{+0.13}_{-0.12}$ & $1.10\pm0.04$ & $1.61\pm0.18$ & $0.52\pm0.03$ & $129.07^{+1.13}_{-1.16}$ & -- & $0.33^{+0.08}_{-0.06}$ &   &    \\ 
 \hline 
	 \multirow{2}{*}{S\_64} & \multirow{2}{*}{HELMSJ233620.8$-$060828} & \multirow{2}{*}{1} & Sérsic & $20.53\pm0.05$ & $0.64\pm0.02$ & $4.76^{+0.10}_{-0.09}$ & $0.878\pm0.004$ & $52.62\pm0.24$ & -- & $0.85\pm0.02$ & \multirow{2}{*}{1.04} & \multirow{2}{*}{27656} \\ 
	   &   &   & Sérsic & $22.22^{+0.12}_{-0.10}$ & $0.96\pm0.04$ & $0.87\pm0.08$ & $0.73\pm0.03$ & $127.25^{+3.16}_{-3.17}$ & -- & $0.15\pm0.02$ &   &    \\ 

\hline
\hline
\end{tabular}
\endgroup
}
\end{table*}

\begin{table}
\centering
\caption{Derived properties of the lensing systems.}
\label{tab:params_full}
{\tiny
\begingroup
\setlength{\tabcolsep}{0pt}
\begin{tabular}{c l c c c c c}
\hline
\hline
	 No. & IAU Name & $m_{F110W}$ & $\mu_{\rm 50}$ & $R_{\rm 50}$ & $C_{31}$ & $\mu S_{1.1}$  \\ 
	   &   & [mag] & [mag arcsec$^{-2}$] & [arcsec] & &  [$\mu$Jy]  \\ 
	   &   & $M_{F110W}$ & $\mu_{\rm 50,\, 0}$ & $R_{\rm 50,\, 0}$ & &  \\ 
	   &   & [mag] & [mag kpc$^{-2}$] & [kcp] &  &  \\ 
	 (1) & (2) & (3) & (4) & (5) & (6) & (7)  \\ 
\hline
	 \multirow{2}{*}{S\_1} & \multirow{2}{*}{HATLASJ000330.6$-$321136} & $19.30\pm0.14$ & $21.49^{+0.21}_{-0.20}$ & $0.56\pm0.05$ & \multirow{2}{*}{$7.24^{+0.43}_{-0.40}$} & \multirow{2}{*}{$1.81\pm0.06$} \\ 
	   & 	   & $-17.78^{+0.73}_{-0.59}$ & $20.10^{+0.37}_{-0.35}$ & $2.99^{+0.52}_{-0.55}$ & &   \\ 
\hline
	 \multirow{2}{*}{S\_2} & \multirow{2}{*}{HATLASJ000912.7$-$300807} & $17.09\pm0.19$ & $22.18\pm0.26$ & $2.38^{+0.20}_{-0.17}$ & \multirow{2}{*}{$5.34^{+0.38}_{-0.40}$} & \multirow{2}{*}{$45.35\pm0.12$} \\ 
	   & 	   & $-19.26^{+0.83}_{-0.68}$ & $21.10^{+0.38}_{-0.37}$ & $10.49^{+2.19}_{-2.36}$ & &    \\ 
\hline
	 \multirow{2}{*}{S\_3} & \multirow{2}{*}{HELMSJ001353.5$-$060200} & $19.99\pm0.24$ & $23.16^{+0.41}_{-0.50}$ & $0.90^{+0.25}_{-0.21}$ & \multirow{2}{*}{$7.26^{+1.15}_{-1.34}$} & \multirow{2}{*}{$3.00\pm0.10$} \\ 
	   & 	   & $-18.29^{+0.92}_{-0.72}$ & $21.09^{+0.67}_{-0.64}$ & $5.98^{+1.98}_{-1.57}$ & &   \\ 
\hline
	 \multirow{2}{*}{S\_4} & \multirow{2}{*}{HELMSJ003619.8$+$002420} & $16.41\pm0.18$ & $21.29^{+0.46}_{-0.36}$ & $2.07^{+0.26}_{-0.20}$ & \multirow{2}{*}{$6.38^{+1.30}_{-0.97}$} & \multirow{2}{*}{$13.58\pm0.25$} \\ 
	   & 	   & $-19.72^{+0.18}_{-0.17}$ & $20.29^{+0.46}_{-0.36}$ & $8.55^{+1.06}_{-0.84}$ &  &  \\ 
\hline
	 \multirow{4}{*}{S\_6} & \multirow{2}{*}{HERSJ011722.3$+$005624$_{1}$} & $19.45^{+0.34}_{-0.31}$ & $22.10^{+0.47}_{-0.34}$ & $0.67^{+0.08}_{-0.07}$ & \multirow{2}{*}{$8.68^{+0.74}_{-0.68}$} & \multirow{4}{*}{$2.23\pm0.10$} \\ 
	   & 	   & $-19.82^{+0.37}_{-0.36}$ & $19.38^{+0.47}_{-0.37}$ & $5.33^{+0.66}_{-0.55}$ &  &  \\ 
	   & \multirow{2}{*}{HERSJ011722.3$+$005624$_{2}$} & $22.87^{+0.23}_{-0.22}$ & $22.52^{+0.17}_{-0.15}$ & $0.24\pm0.02$ & \multirow{2}{*}{$2.95\pm0.18$} \\ 
	   & 	   & -- & -- & -- &  &  \\ 
\hline
	 \multirow{2}{*}{S\_7} & \multirow{2}{*}{HERSJ012620.5$+$012950} & $18.86^{+0.29}_{-0.28}$ & $21.22^{+0.13}_{-0.11}$ & $0.74\pm0.05$ & \multirow{2}{*}{$4.64^{+1.03}_{-0.76}$} & \multirow{2}{*}{$5.72\pm0.09$} \\ 
	   & 	   & $-18.55^{+0.42}_{-0.40}$ & $19.67^{+0.20}_{-0.19}$ & $4.27^{+0.43}_{-0.42}$ &   & \\ 
\hline
	 \multirow{2}{*}{S\_9} & \multirow{2}{*}{HERMESJ032637$-$270044} & $19.10\pm0.07$ & $22.70\pm0.05$ & $1.01^{+0.02}_{-0.03}$ & \multirow{2}{*}{$10.12\pm0.22$}  & \multirow{2}{*}{$3.81\pm0.13$} \\ 
	   & 	   & -- & -- & -- &  &  \\ 
\hline
	 \multirow{2}{*}{S\_10} & \multirow{2}{*}{HERMESJ033732$-$295353} & $16.47^{+0.13}_{-0.12}$ & $21.15^{+0.19}_{-0.20}$ & $1.91\pm0.16$ & \multirow{2}{*}{$5.36^{+0.33}_{-0.32}$} & \multirow{2}{*}{$2.93\pm0.07$} \\ 
	   & 	   & $-18.89^{+0.68}_{-0.55}$ & $20.40^{+0.25}_{-0.27}$ & $6.15\pm1.32$ &  &  \\ 
\hline
	 \multirow{2}{*}{S\_13} & \multirow{2}{*}{HERMESJ103827$+$581544} & $17.86^{+0.09}_{-0.08}$ & $21.95^{+0.12}_{-0.13}$ & $1.53\pm0.11$ & \multirow{2}{*}{$4.75^{+0.19}_{-0.20}$} & \multirow{2}{*}{$10.12\pm0.23$} \\ 
	   & 	   & $-20.38^{+0.09}_{-0.08}$ & $19.93^{+0.12}_{-0.13}$ & $10.48^{+0.76}_{-0.72}$ &  &  \\ 
\hline
	 \multirow{2}{*}{S\_14} & \multirow{2}{*}{HERMESJ110016$+$571736} & $18.78\pm0.06$ & $21.76\pm0.04$ & $0.91\pm0.02$ & \multirow{2}{*}{$4.99\pm0.06$} & \multirow{2}{*}{$12.71\pm0.19$} \\ 
	   & 	   & $-20.19\pm0.06$ & $19.25\pm0.04$ & $6.97\pm0.15$ &  &  \\ 
\hline
	 \multirow{4}{*}{S\_15} & \multirow{2}{*}{HATLASJ114638$-$001132$_{1}$} & $21.15^{+0.21}_{-0.20}$ & $21.97^{+1.01}_{-0.70}$ & $0.29^{+0.10}_{-0.05}$ & \multirow{2}{*}{$7.50^{+0.98}_{-1.17}$} & \multirow{4}{*}{$6.33\pm0.14$} \\ 
	   & 	   & $-19.03^{+0.21}_{-0.20}$ & $18.50^{+1.01}_{-0.70}$ & $2.51^{+0.86}_{-0.42}$ &  &  \\ 
	   & \multirow{2}{*}{HATLASJ114638$-$001132$_{2}$} & $20.52^{+0.16}_{-0.15}$ & $24.52^{+0.29}_{-0.37}$ & $1.28\pm0.24$ & \multirow{2}{*}{$8.90^{+1.82}_{-1.84}$} & \\ 
	   & 	   & $-19.66^{+0.16}_{-0.15}$ & $21.05^{+0.29}_{-0.37}$ & $10.90^{+2.06}_{-2.09}$ &  &  \\ 
\hline
	 \multirow{2}{*}{S\_16} & \multirow{2}{*}{HATLASJ125126$+$254928} & $19.21\pm0.08$ & $23.11^{+0.06}_{-0.05}$ & $1.07\pm0.03$ & \multirow{2}{*}{$15.14^{+0.42}_{-0.41}$} &  \multirow{2}{*}{$4.18\pm0.15$} \\ 
	   & 	   & $-19.16^{+0.46}_{-0.40}$ & $21.02^{+0.27}_{-0.26}$ & $7.46^{+0.52}_{-0.61}$ &  &  \\ 
\hline
	 \multirow{2}{*}{S\_17} & \multirow{2}{*}{HATLASJ125760$+$224558} & $18.83\pm0.06$ & $22.33\pm0.04$ & $1.02\pm0.02$ & \multirow{2}{*}{$8.05\pm0.13$} & \multirow{2}{*}{$10.13\pm0.12$} \\ 
	   & 	   & $-19.25\pm0.06$ & $20.42\pm0.04$ & $6.76\pm0.15$ &  &  \\ 
\hline
	 \multirow{2}{*}{S\_18} & \multirow{2}{*}{HATLASJ133008$+$245860} & $19.05^{+0.35}_{-0.32}$ & $21.55^{+0.23}_{-0.20}$ & $0.68\pm0.05$ & \multirow{2}{*}{$7.33^{+1.09}_{-1.03}$} & \multirow{2}{*}{$3.54\pm0.11$} \\ 
	   & 	   & $-18.36^{+0.35}_{-0.32}$ & $20.01^{+0.23}_{-0.20}$ & $3.95\pm0.31$ &    \\ 
\hline
	 \multirow{2}{*}{S\_19} & \multirow{2}{*}{HATLASJ133846$+$255057} & $20.09^{+0.10}_{-0.09}$ & $23.46^{+0.07}_{-0.06}$ & $0.93\pm0.03$ & \multirow{2}{*}{$9.19\pm0.29$} & \multirow{2}{*}{$12.23\pm0.12$} \\ 
	   & 	   & $-17.25^{+0.70}_{-0.55}$ & $21.94^{+0.32}_{-0.30}$ & $5.27^{+0.68}_{-0.84}$ &  &  \\ 
\hline
	 \multirow{2}{*}{S\_21} & \multirow{2}{*}{HERMESJ171451$+$592634} & $21.03\pm0.08$ & $22.05^{+0.05}_{-0.04}$ & $0.336^{+0.010}_{-0.009}$ & \multirow{2}{*}{$6.9887^{+0.0005}_{-0.0003}$} & \multirow{2}{*}{$7.00\pm0.13$} \\ 
	   & 	   & $-19.17\pm0.08$ & $18.56^{+0.05}_{-0.04}$ & $2.87\pm0.08$ &  &  \\ 
\hline
	 \multirow{4}{*}{S\_22} & \multirow{2}{*}{HERMESJ171545$+$601238$_{1}$} & $18.38^{+0.11}_{-0.10}$ & $23.86\pm0.12$ & $2.68\pm0.17$ & \multirow{2}{*}{$6.51^{+0.37}_{-0.35}$} & \multirow{4}{*}{$6.65\pm0.14$} \\ 
	   & 	   & $-18.78^{+0.64}_{-0.53}$ & $22.43^{+0.30}_{-0.29}$ & $14.54^{+2.11}_{-2.40}$ &   & \\ 
	   & \multirow{2}{*}{HERMESJ171545$+$601238$_{2}$} & $21.09^{+0.15}_{-0.14}$ & $23.87^{+0.10}_{-0.09}$ & $0.77\pm0.04$ & \multirow{2}{*}{$6.60\pm0.32$} & \\ 
	   & 	   & -- & -- & -- &  &  \\ 
\hline
\hline
\end{tabular}
\endgroup
}
\begin{flushleft}
\textit{Notes}: Col.~(1): Source reference number. Col.~(2): IAU name of the {\it Herschel} detection. Indices 1 and 2 refer to the two components of the lens candidate. Col.~(3): Apparent magnitude of the model (first row), absolute magnitude of the model (second row). Col.~(4): Total effective radius in arcsec (first row) and in kpc (second row). Col.~(5): Total effective surface brightness before (first row) and after correcting for cosmological dimming (second row). Col.~(6): $C_{31}$ of the model. Col.~(7): Flux density of the background source uncorrected for the magnification.
\end{flushleft}
\end{table}

\begin{table}
\centering
\contcaption{}
 
{\tiny
\begingroup
\setlength{\tabcolsep}{0pt}
\begin{tabular}{c l c c c c c}
\hline
\hline
	 No. & IAU Name & $m_{F110W}$ & $\mu_{\rm 50}$ & $R_{\rm 50}$ & $C_{31}$ & $\mu S_{1.1}$  \\ 
	   &   & [mag] & [mag arcsec$^{-2}$] & [arcsec] & &  [$\mu$Jy]  \\ 
	   &   & $M_{F110W}$ & $\mu_{\rm 50,\, 0}$ & $R_{\rm 50,\, 0}$ & &  \\ 
	   &   & [mag] & [mag kpc$^{-2}$] & [kcp] &  &  \\ 
	 (1) & (2) & (3) & (4) & (5) & (6) & (7)  \\ 
\hline
	 \multirow{4}{*}{S\_23} & \multirow{2}{*}{HATLASJ225844.8$-$295124$_{1}$} & $20.37\pm0.10$ & $20.69^{+0.07}_{-0.06}$ & $0.29\pm0.01$ & \multirow{2}{*}{$3.86\pm0.08$} & \multirow{4}{*}{$1.41\pm0.08$} \\ 
	   & 	   & $-18.28^{+0.93}_{-0.69}$ & $18.41^{+0.57}_{-0.49}$ & $2.13^{+0.22}_{-0.33}$ &  &  \\ 
	   & \multirow{2}{*}{HATLASJ225844.8$-$295124$_{2}$} & $19.43\pm0.14$ & $24.60^{+0.12}_{-0.14}$ & $2.32^{+0.19}_{-0.20}$ & \multirow{2}{*}{$6.47^{+0.56}_{-0.53}$} & \\ 
	   & 	   & $-19.18^{+0.92}_{-0.73}$ & $22.33^{+0.56}_{-0.52}$ & $16.60^{+2.34}_{-2.76}$ &  &  \\ 
\hline
	 \multirow{4}{*}{S\_24} & \multirow{2}{*}{HELMSJ232210.3$-$033559$_{1}$} & $19.33^{+0.14}_{-0.13}$ & $22.74^{+0.13}_{-0.14}$ & $1.00^{+0.11}_{-0.12}$ & \multirow{2}{*}{$7.18^{+0.50}_{-0.53}$} & \multirow{4}{*}{$3.47\pm0.08$} \\ 
	   & 	   & $-15.46^{+1.76}_{-1.08}$ & $22.13\pm0.33$ & $2.64^{+1.20}_{-1.30}$ &   & \\ 
	   & \multirow{2}{*}{HELMSJ232210.3$-$033559$_{2}$} & $22.74\pm0.12$ & $21.44\pm0.06$ & $0.116^{+0.006}_{-0.005}$ & \multirow{2}{*}{$6.982^{+0.004}_{-0.001}$} & \\ 
	   & 	   & -- & -- & -- &  &  \\ 
\hline
	 \multirow{4}{*}{S\_25} & \multirow{2}{*}{HATLASJ233037.2$-$331217$_{1}$} & $19.45^{+0.14}_{-0.13}$ & $22.91^{+0.19}_{-0.21}$ & $1.08\pm0.12$ & \multirow{2}{*}{$6.58^{+1.32}_{-1.03}$} & \multirow{4}{*}{$19.93\pm0.21$} \\ 
	   & 	   & $-19.07^{+0.82}_{-0.63}$ & $20.71^{+0.52}_{-0.47}$ & $7.66^{+1.16}_{-1.32}$ &  &  \\ 
	   & \multirow{2}{*}{HATLASJ233037.2$-$331217$_{2}$} & $[23.14]$ & -- & -- & \multirow{2}{*}{--} & \\ 
	   & 	   &$-15.40^{+0.80}_{-0.60}$ & -- & -- &  &  \\ 

\\
\multicolumn{6}{l}{\textit{Confirmed after the lens subtraction}} \\
\\

	 \multirow{2}{*}{S\_26} & \multirow{2}{*}{HELMSJ001626.0$+$042613} & $17.63\pm0.07$ & $20.35\pm0.11$ & $0.73\pm0.04$ & \multirow{2}{*}{$7.11\pm0.17$} & \multirow{2}{*}{$1.46\pm0.08$} \\ 
	   & 	   & $-18.06\pm0.07$ & $19.50\pm0.11$ & $2.63^{+0.16}_{-0.15}$ &  &  \\ 
\hline
	 \multirow{2}{*}{S\_27} & \multirow{2}{*}{HATLASJ002624.8$-$341737} & $20.82^{+0.20}_{-0.19}$ & $23.13^{+0.30}_{-0.31}$ & $0.62^{+0.08}_{-0.07}$ & \multirow{2}{*}{$6.69^{+0.86}_{-0.94}$} & \multirow{2}{*}{$0.90\pm0.03$} \\ 
	   & 	   & $-18.61^{+1.20}_{-0.88}$ & $20.28^{+0.88}_{-0.78}$ & $4.88^{+0.81}_{-0.85}$ &  &  \\ 
\hline
	 \multirow{4}{*}{S\_28} & \multirow{2}{*}{HELMSJ004723.6$+$015751$_{1}$} & $18.14\pm0.06$ & $21.67^{+0.09}_{-0.11}$ & $1.27^{+0.06}_{-0.07}$ & \multirow{2}{*}{$4.10^{+0.49}_{-0.35}$} & \multirow{4}{*}{$10.99\pm0.18$} \\ 
	   & 	   & $-18.86\pm0.06$ & $20.32^{+0.09}_{-0.11}$ & $6.65^{+0.33}_{-0.38}$ &  &  \\ 
	   & \multirow{2}{*}{HELMSJ004723.6$+$015751$_{2}$} & $19.58^{+0.09}_{-0.08}$ & $22.17\pm0.06$ & $0.74\pm0.02$ & \multirow{2}{*}{$5.65\pm0.10$} & \\ 
	   & 	   & -- & -- & -- &  &  \\ 
\hline
	 \multirow{2}{*}{S\_29} & \multirow{2}{*}{HERSJ012041.6$-$002705} & $19.59\pm0.15$ & $21.97^{+0.34}_{-0.37}$ & $0.58^{+0.10}_{-0.09}$ & \multirow{2}{*}{$9.37^{+0.72}_{-0.85}$} & \multirow{2}{*}{$2.57\pm0.10$} \\ 
	   & 	   & $-19.21^{+0.21}_{-0.20}$ & $19.59^{+0.35}_{-0.38}$ & $4.36^{+0.75}_{-0.66}$ &  &  \\ 
\hline
	 \multirow{2}{*}{S\_30} & \multirow{2}{*}{HATLASJ085112$+$004934} & $18.68^{+0.17}_{-0.19}$ & $20.81^{+0.18}_{-0.20}$ & $0.75\pm0.06$ & \multirow{2}{*}{$3.25^{+0.66}_{-0.47}$} & \multirow{2}{*}{$9.37\pm0.15$} \\ 
	   & 	   & $-19.86^{+1.58}_{-1.07}$ & $18.60^{+0.88}_{-0.79}$ & $5.27^{+0.91}_{-1.47}$ &  &  \\ 
\hline
	 \multirow{2}{*}{S\_31} & \multirow{2}{*}{HATLASJ085359$+$015537} & $20.77^{+0.16}_{-0.15}$ & $24.92^{+0.36}_{-0.38}$ & $1.22^{+0.20}_{-0.18}$ & \multirow{2}{*}{$14.19^{+1.47}_{-1.44}$} & \multirow{2}{*}{$0.87\pm0.03$} \\ 
	   & 	   & $-19.26^{+0.58}_{-0.49}$ & $21.58\pm0.57$ & $10.17^{+1.76}_{-1.54}$ &  &  \\ 
\hline
	 \multirow{2}{*}{S\_32} & \multirow{2}{*}{HERMESJ104549$+$574512} & $19.69\pm0.14$ & $20.94^{+0.17}_{-0.18}$ & $0.41\pm0.04$ & \multirow{2}{*}{$5.16^{+0.46}_{-0.43}$} & \multirow{2}{*}{$7.20\pm0.10$} \\ 
	   & 	   & $-15.81^{+0.29}_{-0.28}$ & $20.15^{+0.18}_{-0.19}$ & $1.39^{+0.17}_{-0.16}$ &  &  \\ 
\hline
	 \multirow{4}{*}{S\_33} & \multirow{2}{*}{HERMESJ105551$+$592845$_{1}$} & $20.84\pm0.12$ & $22.62\pm0.08$ & $0.58\pm0.02$ & \multirow{2}{*}{$3.83\pm0.10$} & \multirow{4}{*}{$5.60\pm0.08$} \\ 
	   & 	   & $-16.26^{+0.85}_{-0.66}$ & $21.22^{+0.37}_{-0.33}$ & $3.09^{+0.51}_{-0.63}$ &  &  \\ 
	   & \multirow{2}{*}{HERMESJ105551$+$592845$_{2}$} & $22.06^{+0.21}_{-0.20}$ & $21.86^{+0.15}_{-0.13}$ & $0.25\pm0.02$ & \multirow{2}{*}{$3.10\pm0.14$} & \\ 
	   & 	   & -- & -- & -- &  &  \\ 
\hline
	 \multirow{12}{*}{S\_34} & \multirow{2}{*}{HERMESJ105751$+$573026$_{1}$} & $17.82\pm0.08$ & $23.48^{+0.15}_{-0.17}$ & $2.64\pm0.24$ & \multirow{2}{*}{$9.89^{+0.72}_{-0.71}$} & \multirow{12}{*}{$43.86\pm0.21$} \\ 
	   & 	   & $-20.47^{+0.20}_{-0.19}$ & $21.43^{+0.19}_{-0.20}$ & $18.16^{+1.76}_{-1.72}$ &  &  \\ 
	   & \multirow{2}{*}{HERMESJ105751$+$573026$_{2}$} & $20.98^{+0.70}_{-0.81}$ & $20.67^{+0.99}_{-1.12}$ & $0.16^{+0.19}_{-0.08}$ & \multirow{2}{*}{$11.33^{+4.02}_{-2.77}$} & \\ 
	   & 	   & -- & -- & -- &  &  \\ 
	   & \multirow{2}{*}{HERMESJ105751$+$573026$_{3}$} & $21.93^{+0.22}_{-0.20}$ & $20.72^{+0.15}_{-0.13}$ & $0.124\pm0.008$ & \multirow{2}{*}{$6.39\pm0.49$} & \\ 
	   & 	   & -- & -- & -- &  &  \\ 
	   & \multirow{2}{*}{HERMESJ105751$+$573026$_{4}$} & $21.59^{+0.10}_{-0.09}$ & $21.16\pm0.05$ & $0.172\pm0.006$ & \multirow{2}{*}{$6.987^{+0.002}_{-0.001}$} & \\ 
	   & 	   & -- & -- & -- &  &  \\ 
	   & \multirow{2}{*}{HERMESJ105751$+$573026$_{5}$} & $21.58^{+0.19}_{-0.18}$ & $20.30^{+0.13}_{-0.12}$ & $0.124\pm0.007$ & \multirow{2}{*}{$5.60\pm0.34$} & \\ 
	   & 	   & -- & -- & -- &  &  \\ 
	   & \multirow{2}{*}{HERMESJ105751$+$573026$_{6}$} & $[23.60]$ & -- & -- & \multirow{2}{*}{--} & \\ 
	   & 	   & -- & -- & -- &  &  \\ 
\hline
	 \multirow{2}{*}{S\_35} & \multirow{2}{*}{HATLASJ132630$+$334410} & $18.76\pm0.06$ & $23.80\pm0.04$ & $1.95\pm0.04$ & \multirow{2}{*}{$10.11^{+0.19}_{-0.18}$} & \multirow{2}{*}{$1.34\pm0.06$} \\ 
	   & 	   & $-20.23\pm0.06$ & $21.28\pm0.04$ & $15.00^{+0.32}_{-0.31}$ &    \\ 
\hline
	 \multirow{8}{*}{S\_36} & \multirow{2}{*}{HATLASJ133543$+$300404$_{1}$} & $20.76^{+0.12}_{-0.11}$ & $22.35^{+0.08}_{-0.07}$ & $0.60\pm0.03$ & \multirow{2}{*}{$2.88^{+0.06}_{-0.05}$} & \multirow{8}{*}{$15.09\pm0.23$} \\ 
	   & 	   & $-18.83^{+0.12}_{-0.11}$ & $19.38^{+0.08}_{-0.07}$ & $4.90\pm0.22$ &   &    \\ 
	   & \multirow{2}{*}{HATLASJ133543$+$300404$_{2}$} & $20.98\pm0.13$ & $21.90^{+0.09}_{-0.08}$ & $0.31\pm0.01$ & \multirow{2}{*}{$8.66^{+0.43}_{-0.42}$} \\ 
	   & 	   & $-18.62\pm0.13$ & $18.92^{+0.09}_{-0.08}$ & $2.50^{+0.12}_{-0.11}$ &  &  \\ 
	   & \multirow{2}{*}{HATLASJ133543$+$300404$_{3}$} & $19.92\pm0.09$ & $23.50\pm0.06$ & $1.01\pm0.03$ & \multirow{2}{*}{$9.81\pm0.30$} & \\ 
	   & 	   & $-19.67\pm0.09$ & $20.53\pm0.06$ & $8.27^{+0.25}_{-0.26}$ &   &   \\ 
	   & \multirow{2}{*}{HATLASJ133543$+$300404$_{4}$} & $20.45^{+0.11}_{-0.10}$ & $23.58^{+0.07}_{-0.06}$ & $0.80\pm0.03$ & \multirow{2}{*}{$11.10^{+0.45}_{-0.44}$} & \\ 
	   & 	   & $-19.18^{+0.11}_{-0.10}$ & $20.58^{+0.07}_{-0.06}$ & $6.56\pm0.23$ &  &  \\

\hline
\hline
\end{tabular}
\endgroup
}
\end{table}

\begin{table}
\centering
\contcaption{}
 
{\tiny
\begingroup
\setlength{\tabcolsep}{0pt}
\begin{tabular}{c l c c c c c}
\hline
\hline
	 No. & IAU Name & $m_{F110W}$ & $\mu_{\rm 50}$ & $R_{\rm 50}$ & $C_{31}$ & $\mu S_{1.1}$  \\ 
	   &   & [mag] & [mag arcsec$^{-2}$] & [arcsec] & &  [$\mu$Jy]  \\ 
	   &   & $M_{F110W}$ & $\mu_{\rm 50,\, 0}$ & $R_{\rm 50,\, 0}$ & &  \\ 
	   &   & [mag] & [mag kpc$^{-2}$] & [kcp] &  &  \\ 
	 (1) & (2) & (3) & (4) & (5) & (6) & (7)  \\ 
  \hline
	 \multirow{2}{*}{S\_37} & \multirow{2}{*}{HATLASJ142140$+$000448} & $20.34\pm0.10$ & $21.27^{+0.07}_{-0.06}$ & $0.36\pm0.01$ & \multirow{2}{*}{$4.44\pm0.10$} & \multirow{2}{*}{$3.24\pm0.05$} \\ 
	   & 	   & $-19.62^{+1.19}_{-0.80}$ & $18.00^{+0.91}_{-0.73}$ & $3.08^{+0.17}_{-0.34}$ &  &  \\ 
\hline
	 \multirow{2}{*}{S\_38} & \multirow{2}{*}{HERMESJ142824$+$352620} & $20.69^{+0.12}_{-0.11}$ & $21.47^{+0.08}_{-0.07}$ & $0.34\pm0.01$ & \multirow{2}{*}{$4.64\pm0.13$} & \multirow{2}{*}{$1.93\pm0.05$} \\ 
	   & 	   & $-19.04^{+0.12}_{-0.11}$ & $18.38^{+0.08}_{-0.07}$ & $2.81\pm0.11$ &   & \\ 
\hline
	 \multirow{2}{*}{S\_39} & \multirow{2}{*}{HATLASJ223753.8$-$305828} & $19.09^{+0.08}_{-0.07}$ & $21.81\pm0.05$ & $0.70\pm0.02$ & \multirow{2}{*}{$8.77^{+0.18}_{-0.17}$}  &  \multirow{2}{*}{$3.56\pm0.08$} \\ 
	   & 	   & $-18.92^{+0.79}_{-0.63}$ & $19.93^{+0.43}_{-0.39}$ & $4.55^{+0.53}_{-0.72}$ &  &  \\ 
\hline
	 \multirow{2}{*}{S\_40} & \multirow{2}{*}{HATLASJ225250.7$-$313657} & $21.31^{+0.17}_{-0.16}$ & $22.23^{+0.12}_{-0.10}$ & $0.29\pm0.02$ & \multirow{2}{*}{$11.10^{+0.86}_{-0.81}$}  &  \multirow{2}{*}{$2.04\pm0.03$} \\ 
	   & 	   & $-17.32^{+1.21}_{-0.86}$ & $19.97^{+0.72}_{-0.63}$ & $2.09^{+0.27}_{-0.43}$ &    \\ 
\hline
	 \multirow{2}{*}{S\_41} & \multirow{2}{*}{HELMSJ233441.0$-$065220} & $19.79^{+0.17}_{-0.16}$ & $21.04\pm0.51$ & $0.38^{+0.08}_{-0.05}$ & \multirow{2}{*}{$5.13^{+0.39}_{-0.40}$} & \multirow{2}{*}{$172.09\pm0.22$} \\ 
	   & 	   & $-18.09^{+0.17}_{-0.16}$ & $19.23\pm0.51$ & $2.44^{+0.53}_{-0.35}$ &  &  \\ 
\hline
	 \multirow{4}{*}{S\_42} & \multirow{2}{*}{HELMSJ233633.5$-$032119$_{1}$} & $16.99^{+0.05}_{-0.04}$ & $20.94\pm0.08$ & $1.29\pm0.06$ & \multirow{2}{*}{$7.24^{+0.28}_{-0.29}$} & \multirow{4}{*}{$14.41\pm0.21$} \\ 
	   & 	   & -- & -- & -- &  &  \\ 
	   & \multirow{2}{*}{HELMSJ233633.5$-$032119$_{2}$} & $19.16\pm0.08$ & $23.16^{+0.06}_{-0.05}$ & $1.53\pm0.04$ & \multirow{2}{*}{$4.34\pm0.06$} &  \\ 
	   & 	   & -- & -- & -- &  &  \\ 

 \\
\multicolumn{6}{l}{\textit{Confirmed through sub-mm/mm follow-up}} \\
\\    

	 \multirow{2}{*}{S\_43} & \multirow{2}{*}{HELMSJ001615.7$+$032435} & $18.24^{+0.36}_{-0.34}$ & $23.64^{+0.18}_{-0.16}$ & $2.54^{+0.24}_{-0.22}$ & \multirow{2}{*}{$9.16^{+1.47}_{-1.63}$} & \multirow{2}{*}{$<0.04$} \\ 
	   & 	   & $-20.31^{+0.36}_{-0.34}$ & $21.43^{+0.18}_{-0.16}$ & $18.25^{+1.76}_{-1.62}$ &  &  \\ 
\hline
	 \multirow{2}{*}{S\_44} & \multirow{2}{*}{HELMSJ002220.9$-$015524} & $21.07^{+0.15}_{-0.14}$ & $25.43^{+0.10}_{-0.09}$ & $1.59\pm0.08$ & \multirow{2}{*}{$6.62\pm0.31$} & \multirow{2}{*}{$<0.04$} \\ 
	   & 	   & $-18.27^{+0.50}_{-0.44}$ & $22.66^{+0.36}_{-0.34}$ & $12.62^{+0.82}_{-0.90}$ &  &  \\ 
\hline
	 \multirow{2}{*}{S\_45} & \multirow{2}{*}{HELMSJ003814.1$-$002252} & $[22.45]$ & -- & -- & \multirow{2}{*}{--} & \multirow{2}{*}{$<0.04$} \\ 
	   & 	   & $-12.66^{+1.43}_{-0.95}$ & -- & -- &  &  \\ 
\hline
	 \multirow{2}{*}{S\_46} & \multirow{2}{*}{HELMSJ003929.6$+$002426} & $22.398\pm0.003$ & $22.17^{+0.02}_{-0.01}$ & $0.256\pm0.003$ & \multirow{2}{*}{$2.38^{+0.08}_{-0.09}$} & \multirow{2}{*}{$<0.04$} \\ 
	   & 	   & $-16.34^{+0.99}_{-0.73}$ & $19.83^{+0.61}_{-0.53}$ & $1.90^{+0.18}_{-0.31}$ &  &  \\ 
\hline
	 \multirow{2}{*}{S\_47} & \multirow{2}{*}{HELMSJ004714.2$+$032454} & $20.25^{+0.24}_{-0.23}$ & $23.07^{+0.17}_{-0.16}$ & $0.92\pm0.14$ & \multirow{2}{*}{$4.40^{+0.61}_{-0.60}$} & \multirow{2}{*}{$<0.04$} \\ 
	   & 	   & $-17.43^{+0.54}_{-0.48}$ & $21.38^{+0.31}_{-0.30}$ & $5.57^{+1.05}_{-1.04}$ &   & \\ 
\hline
	 \multirow{6}{*}{S\_48} & \multirow{2}{*}{HELMSJ005159.5$+$062241$_{1}$} & $18.36\pm0.09$ & $22.71^{+0.10}_{-0.12}$ & $1.67^{+0.11}_{-0.12}$ & \multirow{2}{*}{$5.95^{+0.91}_{-0.64}$} & \multirow{6}{*}{$<0.04$} \\ 
	   & 	   & $-19.94\pm0.09$ & $20.66^{+0.10}_{-0.12}$ & $11.54^{+0.74}_{-0.80}$ &    \\ 
	   & \multirow{2}{*}{HELMSJ005159.5$+$062241$_{2}$} & $20.87\pm0.06$ & $20.48\pm0.03$ & $0.176\pm0.004$ & \multirow{2}{*}{$6.987^{+0.001}_{-0.002}$} & \\ 
	   & 	   & -- & -- & -- &   & \\ 
	   & \multirow{2}{*}{HELMSJ005159.5$+$062241$_{3}$} & $20.03\pm0.09$ & $23.78\pm0.05$ & $1.42\pm0.06$ & \multirow{2}{*}{$4.08^{+0.13}_{-0.11}$} & \\ 
	   & 	   & $-18.24\pm0.09$ & $21.75\pm0.05$ & $9.78^{+0.41}_{-0.42}$ &  &  \\ 
\hline
	 \multirow{2}{*}{S\_49} & \multirow{2}{*}{HATLASJ005724.2$-$273122} & $20.36\pm0.10$ & $24.83^{+0.07}_{-0.06}$ & $1.35\pm0.05$ & \multirow{2}{*}{$18.23^{+0.86}_{-0.84}$} & \multirow{2}{*}{$<0.05$} \\ 
	   & 	   & $-19.00^{+1.52}_{-0.99}$ & $22.05^{+1.01}_{-0.83}$ & $10.76^{+0.88}_{-2.22}$ &  &  \\ 
\hline
	 \multirow{2}{*}{S\_51} & \multirow{2}{*}{HERMESJ033211$-$270536} & $[23.01]$ & -- & -- & \multirow{2}{*}{--} & \multirow{2}{*}{$<0.03$} \\ 
	   & 	   & -- & -- & -- &  &  \\ 
\hline
	 \multirow{2}{*}{S\_52} & \multirow{2}{*}{HERMESJ044154$-$540352} & $19.64^{+0.09}_{-0.08}$ & $21.75\pm0.06$ & $0.48\pm0.01$ & \multirow{2}{*}{$13.98^{+0.40}_{-0.41}$} &  \multirow{2}{*}{$<0.03$} \\ 
	   & 	   & -- & -- & -- &   & \\ 
\hline
	 \multirow{2}{*}{S\_53} & \multirow{2}{*}{HATLASJ083932$-$011760} & $19.992\pm0.002$ & $21.16\pm0.01$ & $0.458\pm0.005$ & \multirow{2}{*}{$2.18\pm0.03$}  &  \multirow{2}{*}{$<0.03$} \\ 
	   & 	   & $-17.34^{+0.84}_{-0.65}$ & $19.65^{+0.38}_{-0.35}$ & $2.60^{+0.38}_{-0.50}$ &   & \\ 
\hline
	 \multirow{4}{*}{S\_54} & \multirow{2}{*}{HATLASJ091841$+$023048$_{1}$} & $20.48^{+0.17}_{-0.16}$ & $25.27^{+0.24}_{-0.30}$ & $1.85^{+0.30}_{-0.34}$ & \multirow{2}{*}{$9.93^{+5.55}_{-2.95}$}  &  \multirow{4}{*}{$<0.04$} \\ 
	   & 	   & $-18.92^{+1.12}_{-0.80}$ & $22.43^{+0.82}_{-0.71}$ & $14.30^{+2.89}_{-3.32}$ &  &  \\ 
	   & \multirow{2}{*}{HATLASJ091841$+$023048$_{2}$} & $22.56\pm0.11$ & $21.80\pm0.06$ & $0.148\pm0.006$ & \multirow{2}{*}{$6.9849^{+0.0029}_{-0.0010}$}  &   \\ 
	   & 	   & $-16.84^{+1.12}_{-0.80}$ & $18.97^{+0.78}_{-0.67}$ & $1.18^{+0.09}_{-0.17}$ &  &   \\ 
\hline
	 \multirow{2}{*}{S\_56} & \multirow{2}{*}{HATLASJ115433.6$+$005042} & $20.89\pm0.13$ & $21.10^{+0.08}_{-0.07}$ & $0.28\pm0.01$ & \multirow{2}{*}{$3.83^{+0.10}_{-0.11}$}  &  \multirow{2}{*}{$<0.03$} \\ 
	   & 	   & $-17.03^{+0.62}_{-0.51}$ & $19.28^{+0.33}_{-0.31}$ & $1.79^{+0.19}_{-0.23}$ &  &  \\ 
\hline
	 \multirow{2}{*}{S\_57} & \multirow{2}{*}{HATLASJ120127.6$-$014043} & $20.85\pm0.19$ & $20.44^{+0.60}_{-0.42}$ & $0.18^{+0.04}_{-0.03}$ & \multirow{2}{*}{$8.27^{+3.41}_{-1.67}$} & \multirow{2}{*}{$<0.04$} \\ 
	   & 	   & $-18.43^{+1.34}_{-0.91}$ & $17.81^{+1.06}_{-0.91}$ & $1.37^{+0.36}_{-0.29}$ &    \\ 
\hline
	 \multirow{2}{*}{S\_58} & \multirow{2}{*}{HATLASJ131611$+$281220} & $21.87^{+0.22}_{-0.20}$ & $22.68^{+0.15}_{-0.13}$ & $0.29\pm0.02$ & \multirow{2}{*}{$8.86^{+0.84}_{-0.78}$} & \multirow{2}{*}{$<0.03$} \\ 
	   & 	   & $-17.48^{+0.46}_{-0.42}$ & $19.91^{+0.34}_{-0.33}$ & $2.31^{+0.18}_{-0.19}$ &  &  \\ 
\hline
	 \multirow{2}{*}{S\_59} & \multirow{2}{*}{HATLASJ134429$+$303036} & $19.57^{+0.32}_{-0.30}$ & $20.85^{+0.73}_{-0.69}$ & $0.38^{+0.14}_{-0.08}$ & \multirow{2}{*}{$6.09^{+0.69}_{-0.68}$} & \multirow{2}{*}{$<0.04$} \\ 
	   & 	   & $-19.01^{+0.32}_{-0.30}$ & $18.61^{+0.73}_{-0.69}$ & $2.74^{+1.00}_{-0.58}$ &  &  \\ 

\hline
\hline
\end{tabular}
\endgroup
}
\end{table}

\begin{table}
\centering
\contcaption{}
 
{\scriptsize
\begingroup
\setlength{\tabcolsep}{0pt}
\begin{tabular}{c l c c c c c}
\hline
\hline
	 No. & IAU Name & $m_{F110W}$ & $\mu_{\rm 50}$ & $R_{\rm 50}$ & $C_{31}$  \\ 
	   &   & [mag] & [mag arcsec$^{-2}$] & [arcsec] &    \\ 
	   &   & $M_{F110W}$ & $\mu_{\rm 50,\, 0}$ & $R_{\rm 50,\, 0}$ &    \\ 
	   &   & [mag] & [mag kpc$^{-2}$] & [kcp] &    \\ 
	 (1) & (2) & (3) & (4) & (5) & (6)  \\ 
\hline
	 \multirow{6}{*}{S\_60} & \multirow{2}{*}{HATLASJ141352$-$000027$_{1}$} & $18.19^{+0.18}_{-0.17}$ & $22.11^{+0.43}_{-0.42}$ & $1.22^{+0.23}_{-0.18}$ & \multirow{2}{*}{$7.67\pm0.48$} & \multirow{6}{*}{$<0.04$} \\ 
	   & 	   & $-19.85^{+0.18}_{-0.17}$ & $20.21^{+0.43}_{-0.42}$ & $8.01^{+1.52}_{-1.17}$ &  &  \\ 
	   & \multirow{2}{*}{HATLASJ141352-000027$_{2}$} & $20.91^{+0.13}_{-0.12}$ & $21.05\pm0.08$ & $0.229\pm0.010$ & \multirow{2}{*}{$6.40^{+0.27}_{-0.26}$}  &  \\ 
	   & 	   & -- & -- & -- &  &  \\ 
	   & \multirow{2}{*}{HATLASJ141352$-$000027$_{3}$} & $19.18\pm0.08$ & $21.69\pm0.05$ & $0.64\pm0.02$ & \multirow{2}{*}{$8.56\pm0.18$}  &   \\ 
	   & 	   & $-18.87\pm0.08$ & $19.79\pm0.05$ & $4.20^{+0.10}_{-0.11}$ &  &  \\ 
\hline
	 \multirow{4}{*}{S\_61} & \multirow{2}{*}{HATLASJ142414$+$022304$_{1}$} & $19.95^{+0.09}_{-0.08}$ & $23.04\pm0.09$ & $1.11\pm0.06$ & \multirow{2}{*}{$10.30^{+0.64}_{-0.53}$}  &  \multirow{4}{*}{$<0.03$} \\ 
	   & 	   & $-18.31\pm0.09$ & $21.68\pm0.05$ & $7.65^{+0.39}_{-0.40}$ &  &  \\ 
	   & \multirow{2}{*}{HATLASJ142414$+$022304$_{2}$} & $21.05^{+0.15}_{-0.14}$ & $24.59^{+0.10}_{-0.09}$ & $0.93\pm0.05$ & \multirow{2}{*}{$12.87^{+0.84}_{-0.80}$}  &  \\ 
	   & 	   & $-17.21^{+0.15}_{-0.14}$ & $22.56^{+0.10}_{-0.09}$ & $6.41^{+0.31}_{-0.30}$ &  &  \\ 
\hline

	 \multirow{2}{*}{S\_62} & \multirow{2}{*}{HERMESJ142826+345547} & $19.254\pm0.001$ & $20.801\pm0.007$ & $0.542\pm0.003$ & \multirow{2}{*}{$2.56\pm0.03$} &  \multirow{2}{*}{$<0.03$} \\ 
	   & 	   & $-18.065\pm0.001$ & $19.297\pm0.007$ & $3.07\pm0.02$ &  &  \\ 
\hline
	 \multirow{2}{*}{S\_63} & \multirow{2}{*}{HATLASJ230815.5$-$343801} & $20.18^{+0.21}_{-0.20}$ & $22.15^{+0.13}_{-0.11}$ & $0.55^{+0.02}_{-0.03}$ & \multirow{2}{*}{$7.71^{+1.59}_{-1.35}$}  & \multirow{2}{*}{$<0.04$} \\ 
	   & 	   & $-18.58^{+0.93}_{-0.72}$ & $19.80^{+0.59}_{-0.52}$ & $4.05^{+0.42}_{-0.60}$ &  &  \\ 
\hline
	 \multirow{2}{*}{S\_64} & \multirow{2}{*}{HELMSJ232439.5$-$043936} & $20.34^{+0.21}_{-0.20}$ & $21.64^{+0.31}_{-0.26}$ & $0.40\pm0.03$ & \multirow{2}{*}{$5.96^{+0.32}_{-0.29}$} & \multirow{2}{*}{$<0.04$} \\ 
	   & 	   & $-18.21^{+1.27}_{-0.89}$ & $19.44^{+0.77}_{-0.67}$ & $2.86^{+0.43}_{-0.65}$ &  &  \\ 
\hline
	 \multirow{2}{*}{S\_65} & \multirow{2}{*}{HELMSJ233620.8$-$060828} & $17.97\pm0.07$ & $20.39\pm0.05$ & $0.66\pm0.02$ & \multirow{2}{*}{$6.58^{+0.20}_{-0.19}$}  & \multirow{2}{*}{$<0.04$} \\ 
	   & 	   & $-19.24\pm0.07$ & $18.94\pm0.05$ & $3.62\pm0.09$ &  &  \\ 

\hline
\hline
\end{tabular}
\endgroup
}
\end{table}

\section{Full candidates table}

Below, we present the figures and tables that did not enter the full body of the paper for the sake of readability.

In Table~\ref{tab:full_candidate_table}, we summarise the properties for all the systems classified as B, C and D after the inclusion of the multiwavelength follow-ups and the lens subtraction. Figure~\ref{fig:CDs} shows the candidates classified as C or D.

\begin{table*}
\centering
\caption{Properties of the 217 systems classified as B, C and D by visual inspection}.
\label{tab:full_candidate_table}
{\scriptsize
\begingroup
\setlength{\tabcolsep}{4pt}
\begin{tabular}{c l l l c c c c c c c l}
\hline
\hline
	 No. & IAU Name & Alt. Name & Ref. & RA & Dec & Vis. Class. & Multiw. Class. & $S_{250}$ & $S_{350}$ & $S_{500}$ & Ref.  \\ 
	   &   &   &   & [h m s] & [d m s] &   &   & [mJy] & [mJy] & [mJy] &    \\ 
	 (1) & (2) & (3) & (4) & (5) & (6) & (7) & (8) & (9) & (10) & (11) & (12)  \\ 
\hline
	 S\_66 & HELMSJ000215.9$-$012829 & HELMS3 & Na16 & 00:02:16 & $-$01:28:29.00 & B & -- & $643.0\pm7.0$ & $510.0\pm6.0$ & $258.0\pm7.0$ & Na16  \\ 
	 S\_67 & HELMSJ001800.1$-$060235 & HELMS21 & Na16 & 00:18:00 & $-$06:02:35.00 & B & -- & $206.0\pm6.0$ & $186.0\pm7.0$ & $130.0\pm7.0$ & Na16  \\ 
	 S\_68 & HELMSJ002208.1$+$034044 & HELMS38 & Na16 & 00:22:08 & $+$03:40:44.00 & B & -- & $190.0\pm6.0$ & $157.0\pm6.0$ & $113.0\pm7.0$ & Na16  \\ 
	 S\_69 & HATLASJ002533.5$-$333825 & HerBS87 & Ba18 & 00:25:34 & $-$33:38:26.00 & B & -- & $114.7\pm5.2$ & $127.8\pm6.1$ & $96.0\pm7.3$ & Ba18  \\ 
	 S\_70 & HERMESJ002906$-$421419 & HELAISS01 & Wa13 & 00:29:06 & $-$42:14:19.00 & B & -- & $129.0$ & $116.0$ & $81.0$ & Ca14  \\ 
	 S\_71 & HATLASJ003207.7$-$303724 & HerBS56 & Ba18 & 00:32:08 & $-$30:37:24.00 & B & -- & $80.3\pm5.0$ & $106.2\pm5.2$ & $105.8\pm6.3$ & Ne17  \\ 
	 S\_72 & HERMESJ003824$-$433705 & HELAISS02 & Ca14 & 00:38:24 & $-$43:37:05.00 & B & D & $115.0\pm6.0$ & $124.0\pm6.0$ & $108.0\pm6.0$ & Bu15  \\ 
	 S\_73 & HELMSJ004747.1$+$061444 & HELMS26 & Na16 & 00:47:47 & $+$06:14:44.00 & B & -- & $85.0\pm7.0$ & $119.0\pm6.0$ & $125.0\pm8.0$ & Na16  \\ 
	 S\_74 & HERSJ010301.2$-$003300 & HERS6 & Na16 & 01:03:01 & $-$00:33:01.00 & B & -- & $121.0\pm7.0$ & $147.0\pm6.0$ & $130.0\pm8.0$ & Na16  \\ 
	 S\_75 & HERSJ011640.1$-$000454 & HERS4 & Na16 & 01:16:40 & $-$00:04:54.00 & B & -- & $137.5\pm6.75$ & $196.11\pm6.56$ & $189.91\pm7.81$ & Su17  \\ 
	 S\_76 & HATLASJ012415.9$-$310500 & HerBS69 & Ba18 & 01:24:16 & $-$31:05:00.00 & B & -- & $140.4\pm5.4$ & $154.5\pm5.7$ & $100.3\pm7.0$ & Ne17  \\ 
	 S\_77 & HERSJ012521.0$+$011724 & HERS13 & Na16 & 01:25:21 & $+$01:17:24.00 & B & -- & $165.0\pm8.0$ & $153.0\pm7.0$ & $114.0\pm10.0$ & Na16  \\ 
	 S\_78 & HERSJ012546.3$-$001143 & HERS12 & Na16 & 01:25:46 & $-$00:11:43.00 & B & -- & $152.0\pm8.0$ & $135.0\pm7.0$ & $114.0\pm9.0$ & Na16  \\ 
	 S\_79 & HERSJ012754.1$+$004940 & HERS3 & Na16 & 01:27:54 & $+$00:49:40.00 & B & -- & $253.0\pm6.0$ & $250.0\pm6.0$ & $191.0\pm7.0$ & Na16  \\ 
  S\_80 & HATLASJ012853.0$-$332719 & HerBS73 & Ba18 & 01:28:53 & $-$33:27:19.00 & B & -- & $117.1\pm6.0$ & $129.0\pm6.2$ & $99.6\pm7.4$ & Ba18  \\ 
	 S\_81 & HATLASJ013840.5$-$281855 & HerBS14 & Ba18 & 01:38:41 & $-$28:18:56.00 & B & Unc. & $116.3\pm5.7$ & $177.0\pm6.0$ & $179.4\pm7.1$ & Ne17  \\ 
	 S\_82 & HATLASJ013951.9$-$321446 & HerBS55 & Ba18 & 01:39:52 & $-$32:14:46.00 & B & -- & $109.0\pm4.9$ & $116.5\pm5.3$ & $107.1\pm6.2$ & Ne17  \\ 
	 S\_83 & HERSJ021402.6$-$004612 & HERS17 & Na16 & 02:14:03 & $-$00:46:12.00 & B & -- & $110.0\pm8.0$ & $130.0\pm8.0$ & $105.0\pm9.0$ & Na16  \\ 
	 S\_84 & HERSJ021434.4$+$005926 & HERS16 & Na16 & 02:14:34 & $+$00:59:26.00 & B & -- & $110.0\pm9.0$ & $134.0\pm8.0$ & $109.0\pm10.0$ & Na16  \\ 
	 S\_85 & HERMESJ022017$-$060143 & HXMM01 & Wa13 & 02:20:17 & $-$06:01:44.00 & B & Unc. & $179.0\pm7.0$ & $188.0\pm8.0$ & $134.0\pm7.0$ & Bu15  \\ 
	 S\_86 & HERMESJ022022$-$015329 & HXMM04 & Wa13 & 02:20:22 & $-$01:53:29.00 & B & D & $162.0\pm7.0$ & $157.0\pm8.0$ & $125.0\pm11.0$ & Bu15  \\ 
	 S\_87 & HERMESJ022029$-$064846 & HXMM09 & Wa13 & 02:20:29 & $-$06:48:46.00 & B & D & $129.0\pm7.0$ & $118.0\pm8.0$ & $85.0\pm7.0$ & Bu15  \\ 
	 S\_88 & HERMESJ022135$-$062617 & HXMM03 & Wa13 & 02:21:35 & $-$06:26:17.00 & B & Unc. & $114.0\pm7.0$ & $134.0\pm8.0$ & $116.0\pm7.0$ & Bu15  \\ 
	 S\_89 & HERMESJ022548$-$041750 & HXMM05 & Wa13 & 02:25:48 & $-$04:17:50.00 & B & D & $103.0\pm7.0$ & $118.0\pm8.0$ & $97.0\pm7.0$ & Bu15  \\ 
	 S\_90 & HERMESJ045027$-$524126 & HADFS08 & Ca14 & 04:50:27 & $-$52:41:26.00 & B & D & $142.0\pm6.0$ & $133.0\pm6.0$ & $90.0\pm6.0$ & Bu15  \\ 
	 S\_91 & HERMESJ045058$-$531654 & HADFS03 & Ca14 & 04:50:58 & $-$53:16:54.00 & B & Unc. & $119.0\pm6.0$ & $102.0\pm6.0$ & $63.0\pm6.0$ & Bu15  \\ 
	 S\_92 & HATLASJ084933$+$021443 & G09-v1.124 &   & 08:49:33 & $+$02:14:43.14 & B & D & $216.7\pm7.5$ & $248.5\pm8.2$ & $208.6\pm8.6$ & Ne17  \\ 
	   &   &  HerBS8 & Ba18 &   &   &   &   &   &   &   &   \\ 
	 S\_93 & HATLASJ084958$+$010713 & G09v1.1259 &   & 08:49:58 & $+$01:07:12.73 & B & -- & $81.2\pm7.3$ & $98.9\pm8.2$ & $85.2\pm8.7$ & Ba18  \\ 
	   &   &  HerBS157 & Ba18 &   &   &   &   &   &   &   &     \\ 
	 S\_94 & HATLASJ090453.2$+$022017 & HerBS183 & Ba18 & 09:04:53 & $+$02:20:17.87 & B & -- & $87.0\pm7.2$ & $98.2\pm8.0$ & $82.3\pm8.8$ & Ba18  \\ 
	   &   &   SDP.392 &    &   &   &   &   &   &   &   &    \\ 
	 S\_95 & HATLASJ091331$-$003644 & SDP.44 &   & 09:13:31 & $-$00:36:44.00 & B & -- & $175.5\pm6.5$ & $142.2\pm7.4$ & $85.8\pm7.8$ & Va16  \\ 
	 S\_96 & HATLASJ092136$+$000132 & HerBS91 & Ba18 & 09:21:36 & $+$00:01:31.92 & B & -- & $139.2\pm7.3$ & $128.8\pm8.1$ & $95.1\pm8.6$ & Ba18  \\ 
	 S\_97 & HATLASJ092409$-$005018 & HerBS185 &   & 09:24:09 & $-$00:50:18.00 & B & -- & $71.8\pm7.4$ & $87.7\pm8.2$ & $82.2\pm8.5$ & Ba18  \\ 
	 S\_98 & HERMESJ103618$+$585454 & HLock05 & Wa13 & 10:36:18 & $+$58:54:54.36 & B & -- & $71.0$ & $102.0$ & $98.0$ & Ca14  \\ 
	 S\_99 & HATLASJ113804$-$011736 & HerBS96 & Ba18 & 11:38:04 & $-$01:17:36.00 & B & -- & $85.1\pm7.3$ & $98.4\pm8.2$ & $94.8\pm8.8$ & Ba18  \\ 
	 S\_100 & HATLASJ113833.3$+$004909 & HerBS100 & Ba18 & 11:38:33 & $+$00:49:09.68 & B & -- & $96.8\pm7.3$ & $106.4\pm8.1$ & $93.4\pm8.7$ & Ba18  \\ 
	 S\_101 & HATLASJ113841$-$020237 & -- & -- & 11:38:41 & $-$02:02:37.00 & B & -- & $45.2\pm7.0$ & $72.7\pm7.7$ & $71.7\pm8.1$ & Ma18  \\ 
	 S\_102 & HATLASJ115112$-$012638 & HerBS53 & Ba18 & 11:51:12 & $-$01:26:38.00 & B & -- & $141.2\pm7.4$ & $137.7\pm8.2$ & $108.4\pm8.8$ & Ne17  \\ 
	 S\_103 & HATLASJ115120$-$003322 & -- & -- & 11:51:20 & $-$00:33:22.00 & B & -- & $146.9\pm7.8$ & $150.3\pm8.5$ & $88.0\pm8.8$ & Ma18  \\ 
	 S\_104 & HATLASJ120127.8$-$021648 & -- & -- & 12:01:28 & $-$02:16:48.00 & B & -- & $207.9\pm7.3$ & $160.9\pm8.2$ & $103.6\pm8.7$ & Ne17  \\ 
	 S\_105 & HATLASJ120319.1$-$011253 & HerBS50 & Ba18 & 12:03:19 & $-$01:12:54.00 & B & -- & $114.3\pm7.4$ & $142.8\pm8.2$ & $110.2\pm8.6$ & Ne17  \\ 	  
	 S\_106 & HATLASJ121301.5$-$004922 & HerBS48 & Ba18 & 12:13:02 & $-$00:49:23.00 & B & -- & $136.6\pm6.6$ & $142.6\pm7.4$ & $110.9\pm7.7$ & Ne17  \\ 
	 S\_107 & HATLASJ121334.9$-$020323 & -- & -- & 12:13:35 & $-$02:03:23.00 & B & -- & $211.0$ & $197.9$ & $129.9$ & Ne17  \\ 
	 S\_108 & HATLASJ121542.7$-$005220 & HerBS62 & Ba18 & 12:15:43 & $-$00:52:20.00 & B & -- & $119.7\pm7.4$ & $135.5\pm8.2$ & $103.4\pm8.6$ & Ne17  \\ 
	 S\_109 & HATLASJ130054$+$260303 & HerBS129 & Ba18 & 13:00:54 & $+$26:03:02.68 & B & -- & $59.4\pm5.9$ & $85.4\pm5.9$ & $89.0\pm7.0$ & Ba18  \\ 
	 S\_110 & HATLASJ131322$+$285836 & -- & -- & 13:13:22 & $+$28:58:35.61 & B & -- & $160.8\pm7.4$ & $116.0\pm8.4$ & $68.0\pm8.8$ & Ma18  \\ 
	 S\_111 & HATLASJ134159$+$292833 & -- & -- & 13:41:59 & $+$29:28:32.54 & B & -- & $174.4\pm6.7$ & $172.3\pm7.7$ & $109.2\pm8.1$ & Ne17  \\ 
	 S\_112 & HATLASJ141833$+$010212 & HerBS110 & Ba18 & 14:18:33 & $+$01:02:12.40 & B & -- & $66.0$ & $106.5$ & $92.8$ & Ba18  \\ 
	 S\_113 & HERMESJ142201$+$533214 & HEGS01 &   & 14:22:01 & $+$53:32:13.70 & B & -- & $74.0$ & $98.0$ & $89.0$ & Ca14  \\ 
	 S\_114 & HATLASJ142318.3$+$013913 & -- & -- & 14:23:18 & $+$01:39:13.80 & B & -- & $61.8\pm7.5$ & $83.0\pm8.2$ & $65.1\pm8.8$ & Ma18  \\ 
	 S\_115 & HATLASJ143203$-$005219 & -- & -- & 14:32:03 & $-$00:52:19.00 & B & -- & $160.6\pm7.6$ & $125.0\pm8.3$ & $76.1\pm8.9$ & Ma18  \\ 
	 S\_116 & HERMESJ143331$+$345440 & HBootes01 &   & 14:33:31 & $+$34:54:39.60 & B & Unc. & $158.0$ & $191.0$ & $160.0$ & Ca14  \\ 
	 S\_117 & HATLASJ144243$+$015506 & HerBS153 & Ba18 & 14:42:43 & $+$01:55:05.86 & B & -- & $123.2\pm7.2$ & $133.4\pm8.1$ & $85.7\pm8.8$ & Ba18  \\ 
	 S\_118 & HATLASJ144715$-$012114 & -- & -- & 14:47:15 & $-$01:21:14.00 & B & -- & $132.7\pm7.3$ & $118.9\pm8.2$ & $74.7\pm8.6$ & Ma18  \\ 
	 S\_119 & HATLASJ222503.7$-$304847 & HerBS166 & Ba18 & 22:25:04 & $-$30:48:48.00 & B & D & $32.4\pm7.2$ & $50.1\pm8.5$ & $84.3\pm10.3$ & Ba18  \\ 
	   &   &  UR085S &    &   &   &   &   &   &   &   &      \\ 
	 S\_120 & HATLASJ223942.4$-$333304 & HerBS111 & Ba18 & 22:39:42 & $-$33:33:04.00 & B & -- & $105.9\pm6.5$ & $115.6\pm6.2$ & $92.7\pm7.4$ & Ba18  \\ 
	 S\_121 & HATLASJ224026.5$-$315154 & HerBS148 & Ba18 & 22:40:27 & $-$31:51:55.00 & B & -- & $120.6\pm5.0$ & $121.2\pm5.5$ & $86.3\pm6.8$ & Ba18  \\ 
	 S\_122 & HATLASJ224759.6$-$310134 & HerBS141 & Ba18 & 22:48:00 & $-$31:01:35.00 & B & -- & $122.1\pm6.1$ & $124.4\pm6.5$ & $87.3\pm7.5$ & Ba18  \\ 
	 
\hline
\hline
\end{tabular}
\endgroup
}
\begin{flushleft}
\textit{Notes}: Col.~(1): Source reference number. Col.~(2): IAU name of the {\it Herschel} detection. Cols.~(3) and (4): Alternative name and reference. Cols.~(5) and (6): ICRS RA and Dec coordinate (J2000.0) of the {\it Herschel} detection. Cols.~(7) and (8): Visual and multiwavelength lens classifications. Col.~(9), (10), (11), and (11): SPIRE flux density at 250 $\mu {\rm m}$, 350 $\mu {\rm m}$, 500 $\mu {\rm m}$ and reference. 
\end{flushleft}
\end{table*}

\begin{table*}
\centering
\contcaption{}
{\scriptsize
\begingroup
\setlength{\tabcolsep}{4pt}
\begin{tabular}{c l l l c c c c c c c l}
\hline
\hline
	 No. & IAU Name & Alt. Name & Ref. & RA & Dec & Vis. Class. & Multiw. Class. & $S_{250}$ & $S_{350}$ & $S_{500}$ & Ref.  \\ 
	   &   &   &   & [h m s] & [d m s] &   &   & [mJy] & [mJy] & [mJy] &    \\ 
	 (1) & (2) & (3) & (4) & (5) & (6) & (7) & (8) & (9) & (10) & (11) & (12)  \\ 
\hline

S\_123 & HATLASJ224805.3$-$335820 & HerBS33 & Ba18 & 22:48:05 & $-$33:58:20.00 & B & -- & $122.3\pm5.7$ & $135.5\pm6.3$ & $126.9\pm7.2$ & Ne17  \\ 
	 S\_124 & HATLASJ225045.5$-$304719 & HerBS168 & Ba18 & 22:50:46 & $-$30:47:19.00 & B & -- & $65.5\pm6.1$ & $88.1\pm6.1$ & $84.0\pm7.5$ & Na16  \\ 
	 S\_125 & HATLASJ230546.2$-$331038 & HerBS49 & Ba18 & 23:05:46 & $-$33:10:39.00 & B & -- & $76.8\pm5.6$ & $110.9\pm5.9$ & $110.4\pm7.0$ & Ne17  \\ 
	 S\_126 & HELMSJ231447.5$-$045658 & HELMS44 & Na16 & 23:14:48 & $-$04:56:58.00 & B & -- & $220.0\pm8.0$ & $141.0\pm7.0$ & $106.0\pm8.0$ & Na16  \\ 
	 S\_127 & HATLASJ232210.9$-$333749 & HerBS146 & Ba18 & 23:22:11 & $-$33:37:49.00 & B & -- & $122.4\pm5.2$ & $134.6\pm5.4$ & $86.6\pm6.8$ & Ba18  \\ 
	 S\_128 & HATLASJ232531.3$-$302235 & HerBS17 & Ba18 & 23:25:31 & $-$30:22:36.00 & B & -- & $175.5\pm4.3$ & $227.0\pm4.7$ & $175.7\pm5.7$ & Ne17  \\ 
	 S\_129 & HELMSJ232617.5$-$025319 & HELMS51 & Na16 & 23:26:18 & $-$02:53:19.00 & B & -- & $86.0\pm6.0$ & $109.0\pm6.0$ & $104.0\pm7.0$ & Na16  \\ 
	 S\_130 & HELMSJ232831.8$-$004035 & HELMS55 & Na16 & 23:28:32 & $-$00:40:35.00 & B & -- & $95.0\pm7.0$ & $120.0\pm6.0$ & $102.0\pm7.0$ & Na16  \\ 
	S\_131 & HELMSJ232833.6$-$031416 & HELMS48 & Na16 & 23:28:34 & $-$03:14:16.00 & B & -- & $49.0\pm6.0$ & $104.0\pm6.0$ & $105.0\pm8.0$ & Na16  \\ 
	 S\_132 & HELMSJ233420.4$-$003458 & HELMS43 & Na16 & 23:34:20 & $-$00:34:58.00 & B & -- & $156.0\pm7.0$ & $141.0\pm5.0$ & $109.0\pm8.0$ & Na16  \\ 
	 S\_133 & HELMSJ233721.9$-$064740 & HELMS49 & Na16 & 23:37:22 & $-$06:47:40.00 & B & -- & $173.0\pm6.0$ & $161.0\pm7.0$ & $105.0\pm8.0$ & Na16  \\ 
	 S\_134 & HELMS233728.8$-$045106 & HELMS20 & Na16 & 23:37:29 & $-$04:51:06.00 & B & -- & $162.0\pm6.0$ & $178.0\pm7.0$ & $132.0\pm8.0$ & Na16  \\ 
	 S\_135 & HELMSJ234314.0$+$012152 & HELMS36 & Na16 & 23:43:14 & $+$01:21:52.00 & B & -- & $115.0\pm6.0$ & $115.0\pm6.0$ & $113.0\pm8.0$ & Na16  \\ 
	 S\_136 & HATLASJ234955.7$-$330833 & HerBS184 & Ba18 & 23:49:56 & $-$33:08:33.00 & B & -- & $91.9\pm5.9$ & $107.6\pm6.0$ & $82.3\pm7.1$ & Ba18  \\ 
	 S\_137 & HELMSJ235101.7$-$024425 & HELMS50 & Na16 & 23:51:02 & $-$02:44:26.00 & B & -- & $112.0\pm6.0$ & $124.0\pm6.0$ & $105.0\pm7.0$ & Na16  \\ 
	 S\_138 & HATLASJ235121.9$-$332902 & HerBS159 & Ba18 & 23:51:22 & $-$33:29:02.00 & B & Unc. & $92.1\pm5.9$ & $98.3\pm5.9$ & $85.0\pm7.1$ & Ba18  \\ 
	 S\_139 & HELMSJ235331.7$+$031717 & HELMS40 & Na16 & 23:53:32 & $+$03:17:18.00 & B & -- & $102.0\pm6.0$ & $123.0\pm7.0$ & $111.0\pm7.0$ & Na16  \\   
	 S\_140 & HATLASJ000455.3$-$330811 & HerBS170 & Ba18 & 00:04:55 & $-$33:08:12.00 & C & D & $61.9\pm5.4$ & $78.8\pm6.0$ & $83.8\pm7.0$ & Ba18  \\ 
	 S\_141 & HATLASJ001030.1$-$330621 & HerBS98 & Ba18 & 00:10:30 & $-$33:06:22.00 & C & -- & $56.3\pm4.9$ & $51.7\pm5.0$ & $94.4\pm6.5$ & Ba18  \\ 
	 S\_142 & HELMSJ001226.9$+$020810 & HELMS45 & Na16 & 00:12:27 & $+$02:08:10.00 & C & -- & $107.0\pm6.0$ & $142.0\pm6.0$ & $106.0\pm7.0$ & Na16  \\ 
	 S\_143 & HELMSJ001325.7$+$042509 & HELMS56 & Na16 & 00:13:26 & $+$04:25:09.00 & C & -- & $89.0\pm6.0$ & $98.0\pm6.0$ & $102.0\pm7.0$ & Na16  \\ 
	 S\_144 & HATLASJ002144.8$-$295217 & HerBS156 & Ba18 & 00:21:45 & $-$29:52:18.00 & C & -- & $103.7\pm5.7$ & $91.3\pm6.1$ & $85.4\pm6.9$ & Ba18  \\ 
	 S\_145 & HELMSJ002719.5$+$001204 & HELMS34 & Na16 & 00:27:20 & $+$00:12:04.00 & C & -- & $248.0\pm6.0$ & $206.0\pm7.0$ & $116.0\pm8.0$ & Na16  \\ 
	 S\_146 & HERMESJ002854$-$420457 & HELAISS04 & Ca14 & 00:28:54 & $-$42:04:57.00 & C & -- & $131.0$ & $102.0$ & $58.0$ & Ca14  \\ 
	 S\_147 & HELMSJ003519.7$+$072806 & HELMS57 & Na16 & 00:35:20 & $+$07:28:06.00 & C & -- & $134.0\pm7.0$ & $135.0\pm7.0$ & $101.0\pm8.0$ & Na16  \\ 
	 S\_148 & HATLASJ003728.7$-$284124 & HerBS174 & Ba18 & 00:37:29 & $-$28:41:25.00 & C & -- & $95.6\pm5.7$ & $84.8\pm5.9$ & $83.2\pm7.4$ & Ba18  \\ 
	 S\_149 & HELMSJ004622.3$+$073509 & HELMS46 & Na16 & 00:46:22 & $+$07:35:09.00 & C & -- & $82.0\pm9.0$ & $113.0\pm9.0$ & $105.0\pm10.0$ & Na16  \\ 
	 S\_150 & HATLASJ005132.8$-$301848 & HerBS45 & Ba18 & 00:51:33 & $-$30:18:48.00 & C & -- & $164.6\pm5.4$ & $160.2\pm5.8$ & $113.0\pm7.2$ & Ne17  \\ 
	 S\_151 & HELMSJ005258.6$+$061319 & HELMS10 & Na16 & 00:52:59 & $+$06:13:19.00 & C & D & $88.0\pm6.0$ & $129.0\pm6.0$ & $155.0\pm7.0$ & Na16  \\ 
	 S\_152 & HATLASJ005849.9$-$290122 & HerBS181 & Ba18 & 00:58:50 & $-$29:01:22.00 & C & -- & $92.5\pm5.7$ & $116.6\pm6.0$ & $82.6\pm7.2$ & Ba18  \\ 
	 S\_153 & HATLASJ011014.5$-$314813 & HerBS160 & Ba18 & 01:10:15 & $-$31:48:14.00 & C & Unc. & $48.6\pm5.6$ & $84.2\pm6.0$ & $84.8\pm7.1$ & Ba18  \\ 
	S\_154 & HATLASJ011730.3$-$320719 & HerBS138 & Ba18 & 01:17:30 & $-$32:07:19.00 & C & -- & $120.4\pm5.8$ & $111.2\pm6.4$ & $87.4\pm7.8$ & Ba18  \\ 
	 S\_155 & HATLASJ012209.4$-$273824 & HerBS114 & Ba18 & 01:22:09 & $-$27:38:24.00 & C & -- & $81.7\pm5.9$ & $93.8\pm6.0$ & $91.8\pm7.7$ & Ba18  \\ 
	 S\_156 & HATLASJ012335.1$-$314618 & HerBS145 & Ba18 & 01:23:35 & $-$31:46:19.00 & C & D & $54.7\pm6.0$ & $67.4\pm6.2$ & $86.8\pm7.7$ & Ba18  \\ 
	 S\_157 & HERSJ013212.2$+$001754 & HERS18 & Na16 & 01:32:12 & $+$00:17:54.00 & C & -- & $176.0\pm7.0$ & $175.0\pm6.0$ & $104.0\pm8.0$ & Na16  \\ 
	 S\_158 & HATLASJ013239.9$-$330906 & HerBS40 & Ba18 & 01:32:40 & $-$33:09:07.00 & C & -- & $112.0\pm5.5$ & $148.8\pm6.2$ & $117.7\pm7.0$ & Ne17  \\ 
	 S\_159 & HERSJ014057.3$-$010547 & HERS14 & Na16 & 01:40:57 & $-$01:05:47.00 & C & -- & $136.0\pm8.0$ & $143.0\pm8.0$ & $112.0\pm9.0$ & Na16  \\ 
	 S\_160 & HATLASJ014520.0$-$313834 & HerBS107 & Ba18 & 01:45:20 & $-$31:38:35.00 & C & -- & $97.3\pm6.1$ & $99.1\pm6.4$ & $93.1\pm7.8$ & Ba18  \\ 
	 S\_161 & HERSJ020529.1$+$000501 & HERS19 & Na16 & 02:05:29 & $+$00:05:01.00 & C & -- & $89.0\pm6.0$ & $112.0\pm6.0$ & $102.0\pm8.0$ & Na16  \\ 
	 S\_162 & HERMESJ021632$-$053421 & HXMM14 & Ca14 & 02:16:32 & $-$05:34:21.00 & C & -- & $98.0$ & $98.0$ & $78.0$ & Ca14  \\ 
	 S\_163 & HERMESJ021837$-$035315 & HXMM13 & Ca14 & 02:18:37 & $-$03:53:15.00 & C & -- & $55.0$ & $88.0$ & $94.0$ & Ca14  \\ 
	 S\_164 & HERMESJ021943$-$052433 & HXMM20 & Ca14 & 02:19:43 & $-$05:24:33.00 & C & D & $85.0$ & $79.0$ & $67.0$ & Ca14  \\ 
	 S\_165 & HERMESJ022206$-$070727 & HXMM23 & Ca14 & 02:22:06 & $-$07:07:27.00 & C & D & $137.0$ & $108.0$ & $57.0$ & Ca14  \\ 
	 S\_166 & HERMESJ022213$-$070222 & HXMM28 & Ca14 & 02:22:13 & $-$07:02:22.00 & C & -- & $27.0$ & $47.0$ & $87.0$ & Ca14  \\ 
	 S\_167 & HERMESJ022251$-$032414 & HXMM22 & Ca14 & 02:22:51 & $-$03:24:14.00 & C & D & $101.0\pm6.0$ & $85.0\pm6.0$ & $61.0\pm6.0$ & Bu15  \\ 
	 S\_168 & HERMESJ022515$-$024707 & HXMM19 & Ca14 & 02:25:15 & $-$02:47:07.00 & C & -- & $43.0$ & $67.0$ & $70.0$ & Ca14  \\ 
	 S\_169 & HERMESJ022518$-$044610 & HXMM27 & Ca14 & 02:25:17 & $-$04:46:10.00 & C & -- & $0.0$ & $48.0$ & $43.0$ & Ca14  \\ 
	 S\_170 & HERMESJ023006$-$034153 & HXMM12 & Ca14 & 02:30:06 & $-$03:41:53.00 & C & Unc. & $98.0\pm7.0$ & $106.0\pm8.0$ & $82.0\pm7.0$ & Bu15  \\ 
	 S\_171 & HERMESJ032434$-$292646 & HECDFS08 & Ca14 & 03:24:34 & $-$29:26:46.00 & C & -- & $104.0$ & $67.0$ & $54.0$ & Ca14  \\ 
	 S\_172 & HERMESJ032443$-$282134 & HECDFS03 & Ca14 & 03:24:43 & $-$28:21:34.00 & C & -- & $83.0$ & $118.0$ & $113.0$ & Ca14  \\ 
	 S\_173 & HERMESJ032713$-$285106 & HECDFS09 & Ca14 & 03:27:13 & $-$28:51:06.00 & C & -- & $77.0$ & $66.0$ & $51.0$ & Ca14  \\ 
	 S\_174 & HERMESJ033118$-$272015 & HECDFS11 & Ca14 & 03:31:18 & $-$27:20:15.00 & C & -- & $45.0$ & $52.0$ & $42.0$ & Ca14  \\ 
	 S\_175 & HERMESJ043341$-$540338 & HADFS04 & Ca14 & 04:33:41 & $-$54:03:38.00 & C & Unc. & $74.0\pm6.0$ & $93.0\pm6.0$ & $84.0\pm6.0$ & Bu15  \\ 
	 S\_176 & HERMESJ043830$-$541832 & HADFS02 & Ca14 & 04:38:30 & $-$54:18:32.00 & C & D & $19.0\pm6.0$ & $39.0\pm5.0$ & $52.0\pm6.0$ & Bu15  \\ 
	 S\_177 & HERMESJ044947$-$525427 & HADFS09 & Ca14 & 04:49:47 & $-$52:54:27.00 & C & D & $98.0\pm6.0$ & $102.0\pm6.0$ & $72.0\pm6.0$ & Bu15  \\ 
	 S\_178 & HATLASJ083153$+$014014 & -- & -- & 08:31:53 & $+$01:40:14.43 & C & D & $69.8\pm7.3$ & $93.0\pm8.1$ & $82.0\pm8.6$ & Va16  \\ 
	 S\_179 & HATLASJ083546$+$002804 & -- & -- & 08:35:46 & $+$00:28:03.75 & C & -- & $86.6\pm6.6$ & $94.5\pm7.5$ & $79.5\pm8.0$ & Va16  \\ 
	 S\_180 & HATLASJ083817$-$004134 & HerBS108 & Ba18 & 08:38:17 & $-$00:41:34.00 & C & -- & $84.5\pm7.4$ & $106.1\pm8.2$ & $93.0\pm8.8$ & Ba18  \\ 
	 S\_181 & HATLASJ083859.3$+$021325 & HerBS169 & Ba18 & 08:38:59 & $+$02:13:25.87 & C & -- & $95.2\pm7.5$ & $105.2\pm8.2$ & $84.0\pm8.7$ & Ba18  \\ 
	 S\_182 & HATLASJ083945$+$021023 & HerBS171 & Ba18 & 08:39:45 & $+$02:10:22.83 & C & -- & $71.3\pm7.3$ & $97.4\pm8.1$ & $83.4\pm8.6$ & Ba18  \\ 
	 S\_183 & HATLASJ084259.9$+$024959 & HerBS188 & Ba18 & 08:43:00 & $+$02:49:59.02 & C & -- & $84.2\pm7.4$ & $101.5\pm8.1$ & $81.8\pm8.6$ & Ba18  \\ 
	 S\_184 & HATLASJ085033$+$012914 & -- & -- & 08:50:33 & $+$01:29:14.10 & C & -- & $182.5\pm6.9$ & $131.6\pm7.7$ & $74.7\pm8.0$ & Ma18  \\ 
	 S\_185 & HATLASJ085126$+$014638 & -- & -- & 08:51:26 & $+$01:46:38.37 & C & -- & $66.4\pm7.4$ & $92.3\pm8.1$ & $79.1\pm8.5$ & Va16  \\ 
	 S\_186 & HATLASJ085309$-$005727 & HerBS136 & Ba18 & 08:53:09 & $-$00:57:27.00 & C & -- & $68.3\pm7.5$ & $97.5\pm8.2$ & $87.7\pm8.6$ & Ne17  \\ 
	 S\_187 & HATLASJ090953$-$010811 & SDP.60 &   & 09:09:53 & $-$01:08:11.00 & C & -- & $152.1\pm7.2$ & $109.3\pm8.1$ & $83.5\pm8.4$ & Va16  \\ 
	 S\_188 & HATLASJ091454$-$010357 & HerBS142 & Ba18 & 09:14:54 & $-$01:03:57.00 & C & D & $69.0\pm7.3$ & $72.2\pm8.1$ & $87.2\pm8.5$ & Ba18  \\ 
	 S\_189 & HATLASJ091809$+$001927 & HerBS99 & Ba18 & 09:18:09 & $+$00:19:27.27 & C & -- & $93.2\pm7.4$ & $116.6\pm8.2$ & $94.3\pm8.7$ & Ba18  \\ 
	 S\_190 & HATLASJ091857$-$000047 & -- & -- & 09:18:57 & $-$00:00:47.00 & C & -- & $171.6\pm6.4$ & $145.7\pm7.4$ & $85.1\pm7.7$ & Va16  \\ 
	 S\_191 & HATLASJ091949$-$005037 & -- & -- & 09:19:49 & $-$00:50:37.00 & C & -- & $150.7\pm7.5$ & $141.4\pm8.4$ & $90.4\pm8.7$ & Va16  \\ 
	 S\_192 & HATLASJ092141$+$005356 & -- & -- & 09:21:41 & $+$00:53:55.58 & C & -- & $71.2\pm7.3$ & $85.4\pm8.2$ & $79.6\pm8.8$ & Va16  \\ 
	 S\_193 & HERMESJ100057$+$022014 & HCOSMOS02 & Ca14 & 10:00:57 & $+$02:20:13.70 & C & D & $70.0\pm6.0$ & $85.0\pm6.0$ & $71.0\pm6.0$ & Bu15  \\ 
	    &   &  COSBO3 &    &   &   &   &   &   &   &   &      \\

\hline
\hline
\end{tabular}
\endgroup
}
\end{table*}

\begin{table*}
\centering
\contcaption{}
{\scriptsize
\begingroup
\setlength{\tabcolsep}{3pt}
\begin{tabular}{c l l l c c c c c c c l}

\hline
\hline
	 No. & IAU Name & Alt. Name & Ref. & RA & Dec & Vis. Class. & Multiw. Class. & $S_{250}$ & $S_{350}$ & $S_{500}$ & Ref.  \\ 
	   &   &   &   & [h m s] & [d m s] &   &   & [mJy] & [mJy] & [mJy] &    \\ 
	 (1) & (2) & (3) & (4) & (5) & (6) & (7) & (8) & (9) & (10) & (11) & (12)  \\ 
\hline
	 S\_194 & HERMESJ103958$+$563120 & HLock17 & Wa13 & 10:39:58 & $+$56:31:19.92 & C & -- & $62.0$ & $82.0$ & $67.0$ & Ca14  \\ 
	 S\_195 & HERMESJ104051$+$560654 & HLock02 & Wa13 & 10:40:51 & $+$56:06:54.14 & C & -- & $53.0$ & $115.0$ & $140.0$ & Ca14  \\ 
	   &   &  LSW102 &    &   &   &   &   &   &   &   &      \\ 
	 S\_196 & HERMESJ104140$+$570859 & HLock11 & Wa13 & 10:41:40 & $+$57:08:59.42 & C & -- & $97.0$ & $112.0$ & $80.0$ & Ca14  \\ 
	 S\_197 & HERMESJ105712$+$565458 & HLock03 & Wa13 & 10:57:12 & $+$56:54:58.25 & C & -- & $114.0\pm7.0$ & $147.0\pm10.0$ & $114.0\pm8.0$ & Ca14  \\ 
	 S\_198 & HATLASJ113243$-$005109 & HerBS71 & Ba18 & 11:32:43 & $-$00:51:09.00 & C & -- & $68.0\pm7.0$ & $106.0\pm8.0$ & $100.0\pm9.0$ & Ma19  \\ 
	 S\_199 & HATLASJ113834$-$014657 & HerBS119 & Ba18 & 11:38:34 & $-$01:46:57.00 & C & -- & $68.5\pm7.2$ & $85.6\pm8.1$ & $91.2\pm8.6$ & Ba18  \\ 
  S\_200 & HATLASJ114753$-$005832 & HerBS85 & Ba18 & 11:47:53 & $-$00:58:32.00 & C & -- & $92.1\pm6.6$ & $104.2\pm7.4$ & $96.0\pm7.7$ & Ba18  \\ 
	 S\_201 & HATLASJ115521$-$021332 & HERBS179 &   & 11:55:21 & $-$02:13:32.00 & C & D & $67.3\pm7.2$ & $84.9\pm8.1$ & $75.7\pm8.9$ & Va16  \\ 
	 S\_202 & HATLASJ115820$-$013754 & HerBS66 & Ba18 & 11:58:20 & $-$01:37:54.00 & C & -- & $120.0\pm7.0$ & $124.0\pm8.0$ & $101.0\pm8.0$ & Ma19  \\ 
	 S\_203 & HATLASJ120600.7$+$003459 & HerBS74 & Ba18 & 12:06:01 & $+$00:34:59.51 & C & -- & $88.7\pm7.4$ & $104.1\pm8.1$ & $98.8\pm8.7$ & Ba18  \\ 
  S\_204 & HATLASJ120709.2$-$014702 & HerBS51 & Ba18 & 12:07:09 & $-$01:47:03.00 & C & -- & $143.2\pm7.4$ & $149.2\pm8.1$ & $110.3\pm8.7$ & Ne17  \\
	 S\_205 & HATLASJ121348.0$+$010812 & HerBS116 & Ba18 & 12:13:48 & $+$01:08:12.59 & C & -- & $65.1\pm7.4$ & $96.6\pm8.2$ & $93.6\pm8.5$ & Ba18  \\ 
	 S\_206 & HATLASJ121416.3$-$013704 & HerBS164 & Ba18 & 12:14:16 & $-$01:37:04.00 & C & -- & $88.0\pm6.4$ & $99.3\pm7.4$ & $84.3\pm7.7$ & Ba18  \\ 
	 S\_207 & HATLASJ121812.8$+$011841 & HerBS83 & Ba18 & 12:18:13 & $+$01:18:41.67 & C & D & $49.5\pm7.2$ & $79.7\pm8.1$ & $94.1\pm8.8$ & Ba18  \\ 
	 S\_208 & HATLASJ122034.2$-$003805 & HerBS197 & Ba18 & 12:20:34 & $-$00:38:06.00 & C & -- & $81.9\pm7.5$ & $93.8\pm8.2$ & $84.8\pm8.7$ & Ba18  \\ 
	 S\_209 & HATLASJ122117.0$-$014924 & -- & -- & 12:21:17 & $-$01:49:24.00 & C & -- & $146.6\pm7.4$ & $128.9\pm8.2$ & $87.2\pm8.3$ & Va16  \\ 
	 S\_210 & HATLASJ122158.5$+$003326 & HerBS124 & Ba18 & 12:21:59 & $+$00:33:26.15 & C & -- & $135.7\pm7.3$ & $116.1\pm8.2$ & $89.8\pm8.6$ & Ba18  \\ 
	 S\_211 & HATLASJ122407.4$-$003247 & HerBS161 & Ba18 & 12:24:07 & $-$00:32:47.00 & C & D & $56.5\pm7.3$ & $75.7\pm8.1$ & $82.4\pm8.8$ & Ba18  \\ 
	 S\_212 & HATLASJ122459.1$-$005647 & HerBS150 & Ba18 & 12:24:59 & $-$00:56:47.00 & C & D & $53.6\pm7.2$ & $81.3\pm8.3$ & $92.0\pm8.9$ & Ba18  \\ 
	S\_213 & HATLASJ125105$+$261653 & NGP.2070 &   & 12:51:05 & $+$26:16:52.62 & C & -- & $116.5\pm5.5$ & $125.3\pm6.5$ & $81.3\pm7.7$ & Va16  \\ 
	 S\_214 & HATLASJ125653$+$275903 & HerBS31 & Ba18 & 12:56:53 & $+$27:59:02.98 & C & -- & $133.9\pm7.5$ & $164.1\pm8.2$ & $131.8\pm8.9$ & Ne17  \\ 
	 S\_215 & HATLASJ125810$+$263710 & -- & -- & 12:58:10 & $+$26:37:09.99 & C & -- & $131.6\pm7.3$ & $122.9\pm8.3$ & $78.8\pm8.5$ & Ma18  \\ 
	 S\_216 & HATLASJ130601$+$231322 & -- & -- & 13:06:01 & $+$23:13:21.65 & C & -- & $86.0\pm7.3$ & $88.0\pm8.2$ & $76.5\pm8.6$ & Va16  \\ 
	 S\_217 & HATLASJ131001$+$264759 & -- & -- & 13:10:01 & $+$26:47:59.05 & C & -- & $142.1\pm7.3$ & $124.1\pm8.2$ & $81.2\pm8.9$ & Va16  \\ 
	 S\_218 & HATLASJ131020$+$253731 & -- & -- & 13:10:20 & $+$25:37:31.47 & C & -- & $186.4\pm7.3$ & $158.7\pm8.1$ & $93.7\pm8.5$ & Va16  \\ 
	 S\_219 & HATLASJ131609$+$254931 & -- & -- & 13:16:09 & $+$25:49:31.15 & C & -- & $121.4\pm7.3$ & $108.2\pm8.1$ & $76.8\pm8.7$ & Va16  \\ 
	 S\_220 & HATLASJ131642$+$251158 & -- & -- & 13:16:42 & $+$25:11:58.38 & C & -- & $62.7\pm6.9$ & $71.5\pm7.7$ & $63.9\pm8.1$ & Ma18  \\ 
	 S\_221 & HATLASJ131805$+$325018 & HerBS173 & Ba18 & 13:18:05 & $+$32:50:18.05 & C & -- & $73.3\pm5.6$ & $92.7\pm6.0$ & $83.3\pm7.2$ & Ba18  \\ 
	S\_222 & HATLASJ132128$+$282023 & HerBS127 & Ba18 & 13:21:28 & $+$28:20:22.74 & C & -- & $110.0\pm5.5$ & $122.7\pm6.1$ & $89.5\pm6.9$ & Ba18  \\ 
	 S\_223 & HATLASJ132227$+$300723 & -- & -- & 13:22:27 & $+$30:07:22.54 & C & -- & $48.0\pm7.3$ & $78.6\pm8.2$ & $25.0\pm8.8$ & Va16  \\ 
	 S\_224 & HATLASJ132302$+$341650 & HerBS30 & Ba18 & 13:23:02 & $+$34:16:49.66 & C & -- & $124.2\pm7.3$ & $144.6\pm8.2$ & $137.0\pm8.7$ & Ne17  \\ 
	 S\_225 & HATLASJ132419$+$320754 & HerBS43 & Ba18 & 13:24:19 & $+$32:07:54.43 & C & -- & $84.5\pm6.8$ & $116.0\pm7.6$ & $115.4\pm8.0$ & Ne17  \\ 
	 S\_226 & HATLASJ132909$+$300958 & HerBS204 & Ba18 & 13:29:09 & $+$30:09:58.23 & C & -- & $57.9\pm5.5$ & $95.3\pm6.1$ & $80.1\pm7.1$ & Ba18  \\ 
	 S\_227 & HATLASJ133255$+$265529 & -- & -- & 13:32:55 & $+$26:55:29.23 & C & -- & $192.5\pm7.4$ & $167.4\pm8.1$ & $116.6\pm8.6$ & Ne17  \\ 
	 S\_228 & HATLASJ133256$+$342210 & HerBS44 & Ba18 & 13:32:56 & $+$34:22:09.50 & C & -- & $164.3\pm7.5$ & $186.8\pm8.1$ & $114.9\pm8.7$ & Ne17  \\ 
	   &   &  NA.v1.267 &    &   &   &   &   &   &   &   &      \\ 
	 S\_229 & HATLASJ133440$+$353140 & HerBS134 & Ba18 & 13:34:40 & $+$35:31:40.17 & C & -- & $69.9\pm5.9$ & $97.3\pm6.2$ & $87.9\pm7.3$ & Ba18  \\ 
	 S\_230 & HATLASJ133534$+$341837 & HerBS76 & Ba18 & 13:35:34 & $+$34:18:36.97 & C & -- & $108.5\pm5.9$ & $124.3\pm6.0$ & $98.5\pm7.0$ & Ba18  \\ 
	 S\_231 & HATLASJ133623$+$343806 & -- & -- & 13:36:23 & $+$34:38:05.88 & C & -- & $78.8\pm7.4$ & $83.2\pm8.3$ & $79.1\pm8.6$ & Ma18  \\ 
	 S\_232 & HATLASJ133715$+$352058 & -- & -- & 13:37:15 & $+$35:20:58.04 & C & -- & $69.9\pm7.5$ & $88.1\pm8.3$ & $69.4\pm8.8$ & Ma18  \\ 
	 S\_233 & HATLASJ133905$+$340820 & -- & -- & 13:39:05 & $+$34:08:20.20 & C & -- & $59.3\pm7.3$ & $82.7\pm8.2$ & $70.1\pm8.7$ & Va16  \\ 
	 S\_234 & HATLASJ134124$+$354007 & -- & -- & 13:41:24 & $+$35:40:06.58 & C & -- & $62.9\pm7.3$ & $79.9\pm8.2$ & $82.2\pm8.5$ & Va16  \\ 
	 S\_235 & HATLASJ134139$+$322837 & -- & -- & 13:41:39 & $+$32:28:37.34 & C & -- & $39.8\pm7.4$ & $42.7\pm8.1$ & $80.1\pm8.8$ & Va16  \\ 
	 S\_236 & HATLASJ134403$+$242627 & HerBS196 & Ba18 & 13:44:03 & $+$24:26:26.95 & C & -- & $86.9\pm5.7$ & $92.3\pm6.3$ & $81.0\pm7.1$ & Ba18  \\ 
	 S\_237 & HATLASJ134442$+$240346 & HerBS133 & Ba18 & 13:44:42 & $+$24:03:46.12 & C & -- & $85.4\pm5.5$ & $98.5\pm6.1$ & $88.1\pm7.3$ & Ba18  \\ 
	 S\_238 & HATLASJ134654$+$295659 & -- & -- & 13:46:54 & $+$29:56:59.06 & C & -- & $85.0\pm7.6$ & $82.0\pm8.3$ & $77.8\pm8.6$ & Va16  \\ 
	 S\_239 & HATLASJ140422$-$001218 & HerBS206 & Ba18 & 14:04:22 & $-$00:12:18.00 & C & -- & $79.3\pm7.4$ & $102.6\pm8.4$ & $80.2\pm8.8$ & Ba18  \\ 
	 S\_240 & HATLASJ141118$-$010655 & HerBS201 & Ba18 & 14:11:18 & $-$01:06:55.00 & C & D & $52.2\pm7.2$ & $78.6\pm8.2$ & $80.5\pm8.7$ & Ba18  \\ 
	 S\_241 & HATLASJ141810.0$-$003747 & HerBS143 & Ba18 & 14:18:10 & $-$00:37:47.00 & C & -- & $77.7\pm6.5$ & $97.3\pm7.4$ & $87.1\pm7.9$ & Ba18  \\ 
	 S\_242 & HATLASJ141955.5$-$003449 & -- & -- & 14:19:56 & $-$00:34:49.00 & C & -- & $76.4\pm7.5$ & $94.9\pm8.0$ & $80.2\pm8.9$ & Va16  \\ 
	 S\_243 & HERMESJ142549$+$345024 & -- & -- & 14:25:49 & $+$34:50:23.57 & C & -- & $48.0\pm7.4$ & $64.4\pm8.3$ & $52.9\pm9.0$ & Ma18  \\ 
	 S\_244 & HERMESJ142558$+$332549 & HBootes09 &   & 14:25:58 & $+$33:25:48.68 & C & -- & $69.0$ & $81.0$ & $60.0$ & Ca14  \\ 
	 S\_245 & HATLASJ142707$+$002258 & HerBS130 & Ba18 & 14:27:07 & $+$00:22:57.60 & C & -- & $119.4\pm7.3$ & $118.7\pm8.1$ & $88.8\pm8.6$ & Ba18  \\ 
	 S\_246 & HATLASJ143403.5$+$000234 & HerBS147 & Ba18 & 14:34:04 & $+$00:02:34.34 & C & -- & $103.3\pm7.4$ & $103.3\pm8.1$ & $86.6\pm8.5$ & Ba18  \\ 
	 S\_247 & HERMESJ143544$+$344743 & HBootes12 &   & 14:35:44 & $+$34:47:43.37 & C & -- & $11.0$ & $52.0$ & $21.0$ & Ca14  \\ 
	 S\_248 & HERMESJ144030$+$333843 & HBootes07 &   & 14:40:30 & $+$33:38:42.68 & C & -- & $86.0$ & $88.0$ & $72.0$ & Ca14  \\ 
	 S\_249 & HATLASJ144556.1$-$004853 & G15v2.481 &   & 14:45:56 & $-$00:48:53.00 & C & Unc. & $126.7\pm7.3$ & $132.6\pm8.4$ & $111.8\pm8.7$ & Ne17  \\ 
	   &   &  HerBS46 & Ba18 &    &   &   &   &   &   &   &     \\ 
	 S\_250 & HATLASJ144608.6$+$021927 & HerBS38 & Ba18 & 14:46:09 & $+$02:19:27.01 & C & -- & $73.4\pm7.1$ & $111.7\pm8.1$ & $122.1\pm8.7$ & Ne17  \\ 
	 S\_251 & HATLASJ145135.2$-$011418 & HerBS126 & Ba18 & 14:51:35 & $-$01:14:18.00 & C & -- & $81.9\pm7.2$ & $95.9\pm8.2$ & $89.8\pm8.8$ & Ba18  \\ 
	 S\_252 & HATLASJ145337.2$+$000407 & HerBS137 & Ba18 & 14:53:37 & $+$00:04:07.94 & C & -- & $86.0\pm7.2$ & $103.6\pm8.0$ & $87.7\pm8.6$ & Ba18  \\ 
	 S\_253 & HATLASJ145754$+$000017 & HerBS195 & Ba18 & 14:57:54 & $+$00:00:17.07 & C & -- & $70.3\pm7.3$ & $92.7\pm8.1$ & $81.0\pm8.8$ & Ba18  \\ 
	 S\_254 & HERMESJ161332$+$544358 & HELAISN01 &   & 16:13:32 & $+$54:43:57.76 & C & -- & $123.0$ & $129.0$ & $88.0$ & Ca14  \\ 
	 S\_255 & HERMESJ170508$+$594056 & HFLS07 & Ca14 & 17:05:08 & $+$59:40:56.32 & C & -- & $115.0$ & $92.0$ & $69.0$ & Ca14  \\ 
	 S\_256 & HERMESJ170608$+$590921 & HFLS03 & Ca14 & 17:06:08 & $+$59:09:21.28 & C & -- & $98.0$ & $105.0$ & $81.0$ & Ca14  \\ 
	 S\_257 & HERMESJ170818$+$582845 & HFLS05 & Ca14 & 17:08:18 & $+$58:28:45.41 & C & -- & $40.0$ & $75.0$ & $74.0$ & Ca14  \\ 
	 S\_258 & HERMESJ172222$+$582611 & HFLS10 & Ca14 & 17:22:22 & $+$58:26:10.82 & C & -- & $52.0$ & $50.0$ & $32.0$ & Ca14  \\ 
	 S\_259 & HERMESJ172612$+$583742 & HFLS01 & Ca14 & 17:26:12 & $+$58:37:42.24 & C & -- & $107.0$ & $123.0$ & $98.0$ & Ca14  \\ 
	 S\_260 & HATLASJ222629.4$-$321111 & HerBS144 & Ba18 & 22:26:29 & $-$32:11:12.00 & C & -- & $98.9\pm8.4$ & $116.5\pm8.2$ & $87.0\pm11.5$ & Ba18  \\ 
	 S\_261 & HATLASJ224027.7$-$343134 & HerBS97 & Ba18 & 22:40:28 & $-$34:31:35.00 & C & -- & $96.1\pm6.0$ & $98.5\pm6.3$ & $94.4\pm7.7$ & Ba18  \\ 
	 S\_262 & HATLASJ224207.2$-$324159 & HerBS67 & Ba18 & 22:42:07 & $-$32:41:59.00 & C & Unc. & $73.0\pm5.5$ & $88.1\pm6.2$ & $100.8\pm7.7$ & Ne17  \\ 
	 S\_263 & HATLASJ224400.8$-$340030 & HerBS84 & Ba18 & 22:44:01 & $-$34:00:31.00 & C & -- & $105.1\pm5.9$ & $123.0\pm6.4$ & $97.0\pm7.6$ & Ba18  \\ 
	 S\_264 & HATLASJ225324.2$-$323504 & HerBS103 & Ba18 & 22:53:24 & $-$32:35:04.00 & C & -- & $126.1\pm5.3$ & $131.2\pm5.7$ & $93.5\pm7.0$ & Ba18  \\ 

\hline
\hline
\end{tabular}
\endgroup
}
\end{table*}

\begin{table*}
\centering
\contcaption{}
{\scriptsize
\begingroup
\setlength{\tabcolsep}{4pt}
\begin{tabular}{c l l l c c c c c c c l}
\hline
\hline
	 No. & IAU Name & Alt. Name & Ref. & RA & Dec & Vis. Class. & Multiw. Class. & $S_{250}$ & $S_{350}$ & $S_{500}$ & Ref.  \\ 
	   &   &   &   & [h m s] & [d m s] &   &   & [mJy] & [mJy] & [mJy] &    \\ 
	 (1) & (2) & (3) & (4) & (5) & (6) & (7) & (8) & (9) & (10) & (11) & (12)  \\ 
\hline
	 S\_265 & HATLASJ225339.1$-$325549 & HerBS131 & Ba18 & 22:53:39 & $-$32:55:50.00 & C & -- & $85.5\pm5.2$ & $99.7\pm5.5$ & $88.0\pm6.9$ & Ba18  \\ 
	S\_266 & HATLASJ225611.6$-$325652 & HerBS135 & Ba18 & 22:56:12 & $-$32:56:53.00 & C & -- & $85.4\pm5.5$ & $96.7\pm6.2$ & $87.8\pm7.5$ & Ba18  \\ 
	 S\_267 & HATLASJ230002.6$-$315005 & HerBS80 & Ba18 & 23:00:03 & $-$31:50:05.00 & C & -- & $122.7\pm5.7$ & $122.1\pm6.3$ & $97.7\pm7.6$ & Ba18  \\ 
	 S\_268 & HATLASJ230538.5$-$312204 & HerBS182 & Ba18 & 23:05:39 & $-$31:22:04.00 & C & -- & $89.0\pm5.7$ & $109.1\pm6.2$ & $82.3\pm7.9$ & Ba18  \\ 
	 S\_269 & HATLASJ231205.1$-$295026 & HerBS132 & Ba18 & 23:12:05 & $-$29:50:27.00 & C & -- & $86.7\pm5.8$ & $102.6\pm6.0$ & $90.6\pm7.8$ & Ba18  \\ 
	 S\_270 & HELMSJ231857.2$-$053035 & HELMS16 & Na16 & 23:18:57 & $-$05:30:35.00 & C & -- & $143.0\pm7.0$ & $183.0\pm7.0$ & $146.0\pm8.0$ & Na16  \\ 
	 S\_271 & HATLASJ232200.0$-$355622 & HerBS118 & Ba18 & 23:22:00 & $-$35:56:22.00 & C & D & $60.0\pm6.3$ & $84.3\pm6.6$ & $90.9\pm7.7$ & Ba18  \\ 
	 S\_272 & HATLASJ232419.8$-$323926 & HerBS18 & Ba18 & 23:24:20 & $-$32:39:27.00 & C & -- & $213.0\pm4.4$ & $244.2\pm4.8$ & $169.4\pm5.8$ & Ne17  \\ 
	    &   &   SC.v1.128 &    &   &   &   &   &   &   &   &     \\ 
	 S\_273 & HELMSJ232558.3$-$044525 & HELMS17 & Na16 & 23:25:58 & $-$04:45:25.00 & C & -- & $190.0\pm6.0$ & $189.0\pm6.0$ & $142.0\pm8.0$ & Na16  \\ 
	 S\_274 & HATLASJ232623.0$-$342642 & HerBS37 & Ba18 & 23:26:23 & $-$34:26:42.00 & C & -- & $153.7\pm4.4$ & $178.3\pm5.0$ & $123.5\pm6.2$ & Ne17  \\ 
	    &   &   SB.v1.202 &    &   &   &   &   &   &   &   &  \\ 
	 S\_275 & HELMSJ234014.6$-$070738 & HELMS42 & Na16 & 23:40:15 & $-$07:07:38.00 & C & -- & $158.0\pm6.0$ & $154.0\pm6.0$ & $110.0\pm8.0$ & Na16  \\ 
	 S\_276 & HELMSJ234951.6$-$030019 & HELMS47 & Na16 & 23:49:52 & $-$03:00:19.00 & C & -- & $186.0\pm7.0$ & $167.0\pm6.0$ & $105.0\pm8.0$ & Na16  \\ 
	 S\_277 & HERSJ010911.7$-$011733 & HERS9 & Na16 & 01:09:12 & $-$01:17:33.00 & D & -- & $393.0\pm8.0$ & $220.0\pm8.0$ & $118.0\pm9.0$ & Na16  \\ 
	 S\_278 & HATLASJ083345$+$000109 & HerBS88 & Ba18 & 08:33:45 & $+$00:01:09.41 & D & D & $71.0\pm7.6$ & $96.0\pm8.1$ & $95.9\pm8.8$ & Ba18  \\ 
	 S\_279 & HATLASJ090613.8$-$010042 & HerBS165 & Ba18 & 09:06:14 & $-$01:00:43.00 & D & D & $73.4\pm7.4$ & $80.2\pm8.0$ & $84.3\pm8.7$ & Ba18  \\ 
	 S\_280 & HATLASJ091238$+$020050 & -- & -- & 09:12:38 & $+$02:00:49.71 & D & -- & $173.2\pm7.7$ & $140.2\pm8.2$ & $97.6\pm9.1$ & Ma18  \\ 
	 S\_281 & HATLASJ142004$+$014045 & -- & -- & 14:20:04 & $+$01:40:44.73 & D & -- & $191.7\pm7.5$ & $150.1\pm8.3$ & $65.2\pm8.7$ & Ma18  \\ 
\hline
\hline
\end{tabular}
\endgroup
}
\end{table*}

\begin{figure*}
    \centering
    \includegraphics[width=0.9\textwidth]{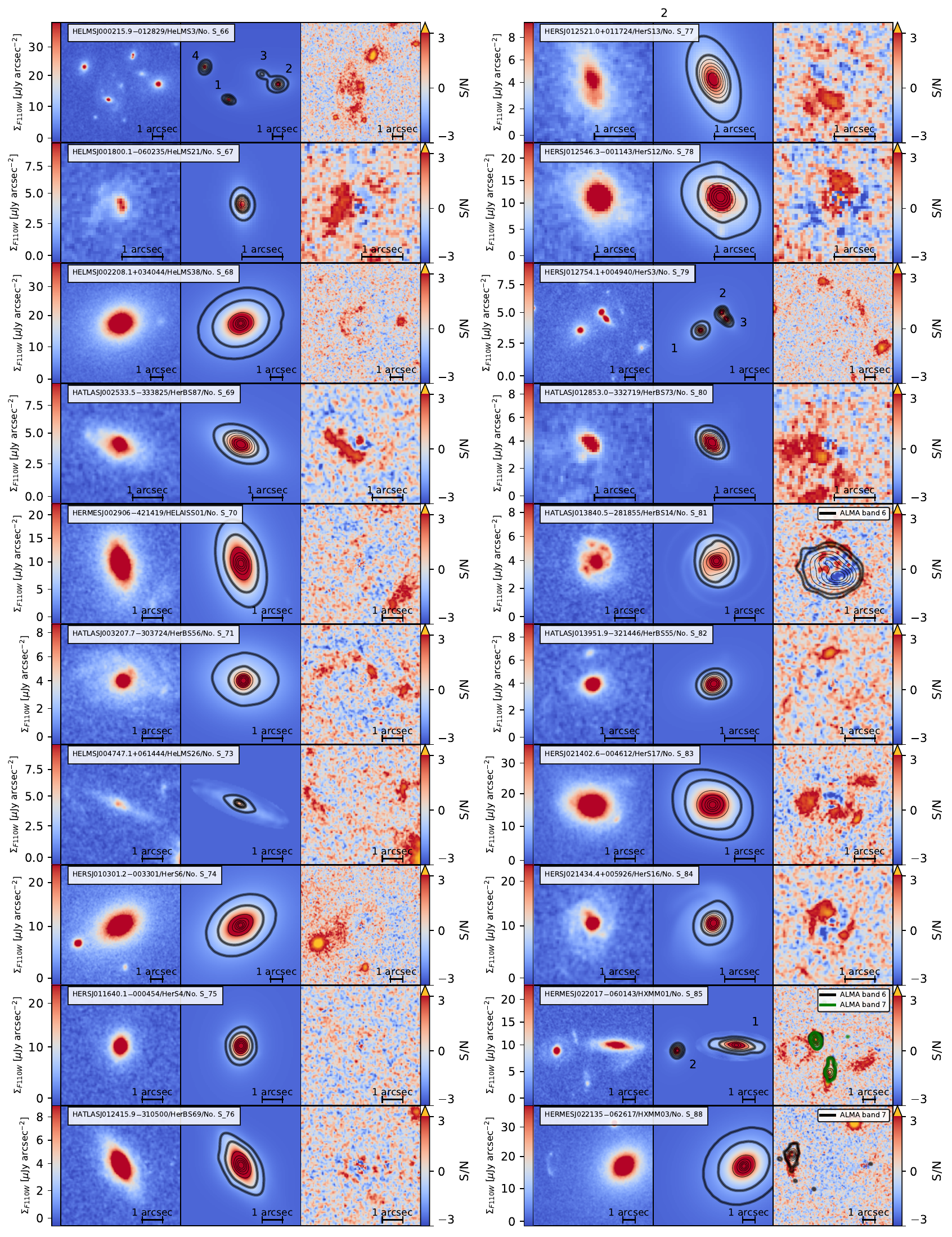}
    \caption{{\em From left to right panels:\/} Observed \textit{HST} F110W image, best-fitting surface-brightness model of the lens and SNR map of the residuals for the B class candidates. The contours in the model images are taken at two levels corresponding to $\textit{SNR} = 5$ and 10 (thick curves), and five uniformly spaced levels between the $\textit{SNR} = 10$ and the maximum SNR in the model image (thin curves). The residual maps show the contours of available high-resolution multiwavelength data taken at two levels corresponding to $\textit{SNR} = 5$ and 10 (thick curves), and five uniformly spaced levels between the $\textit{SNR} = 10$ and the maximum SNR in the multiwavelength image (thin black curves). The images are oriented such that N is up and E is to the left.}
    \label{fig:Bs}
\end{figure*}

\begin{figure*}
    \centering
    \includegraphics[width=0.9\textwidth]{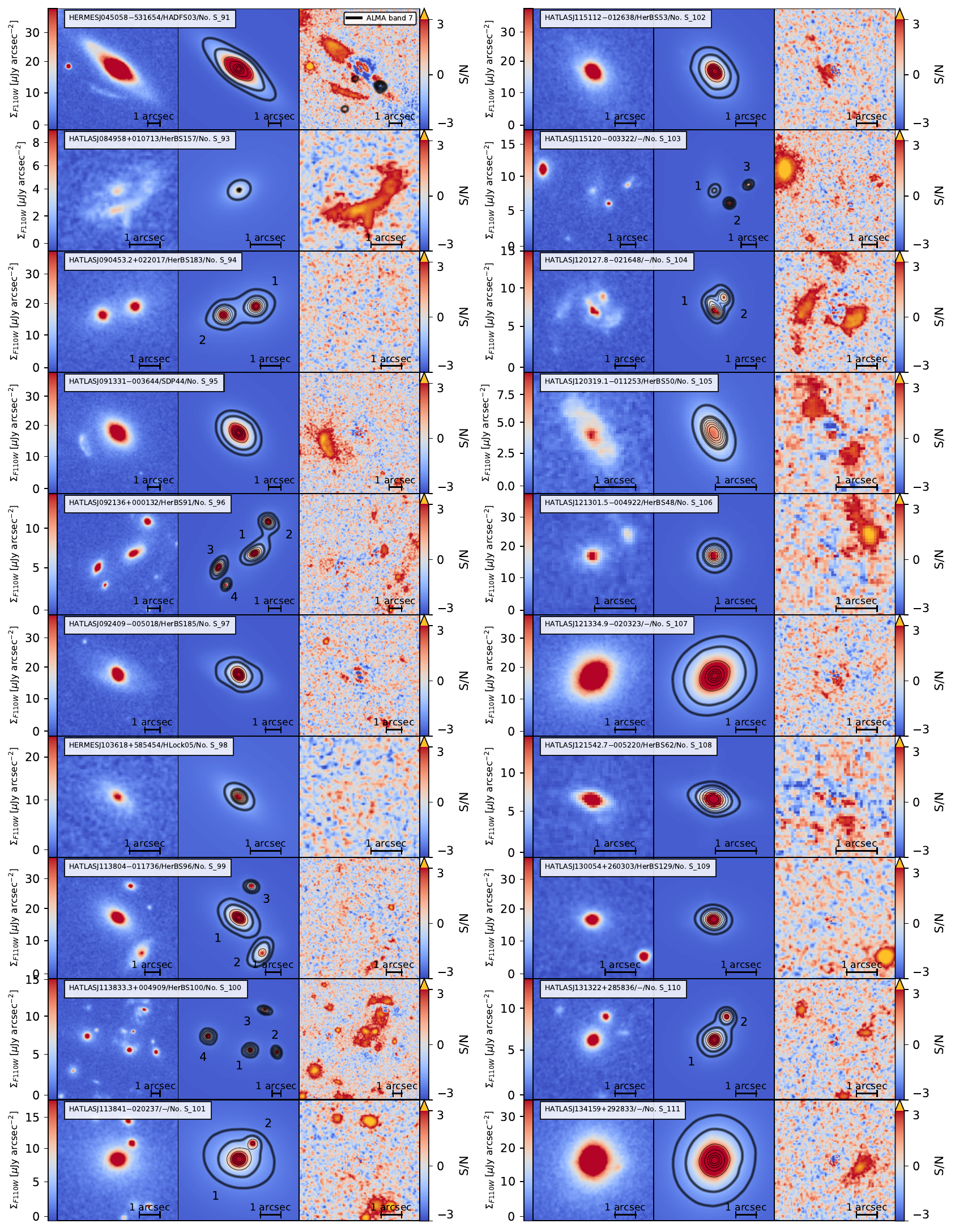}
    \contcaption{}
\end{figure*}

\begin{figure*}
    \centering
    \includegraphics[width=0.9\textwidth]{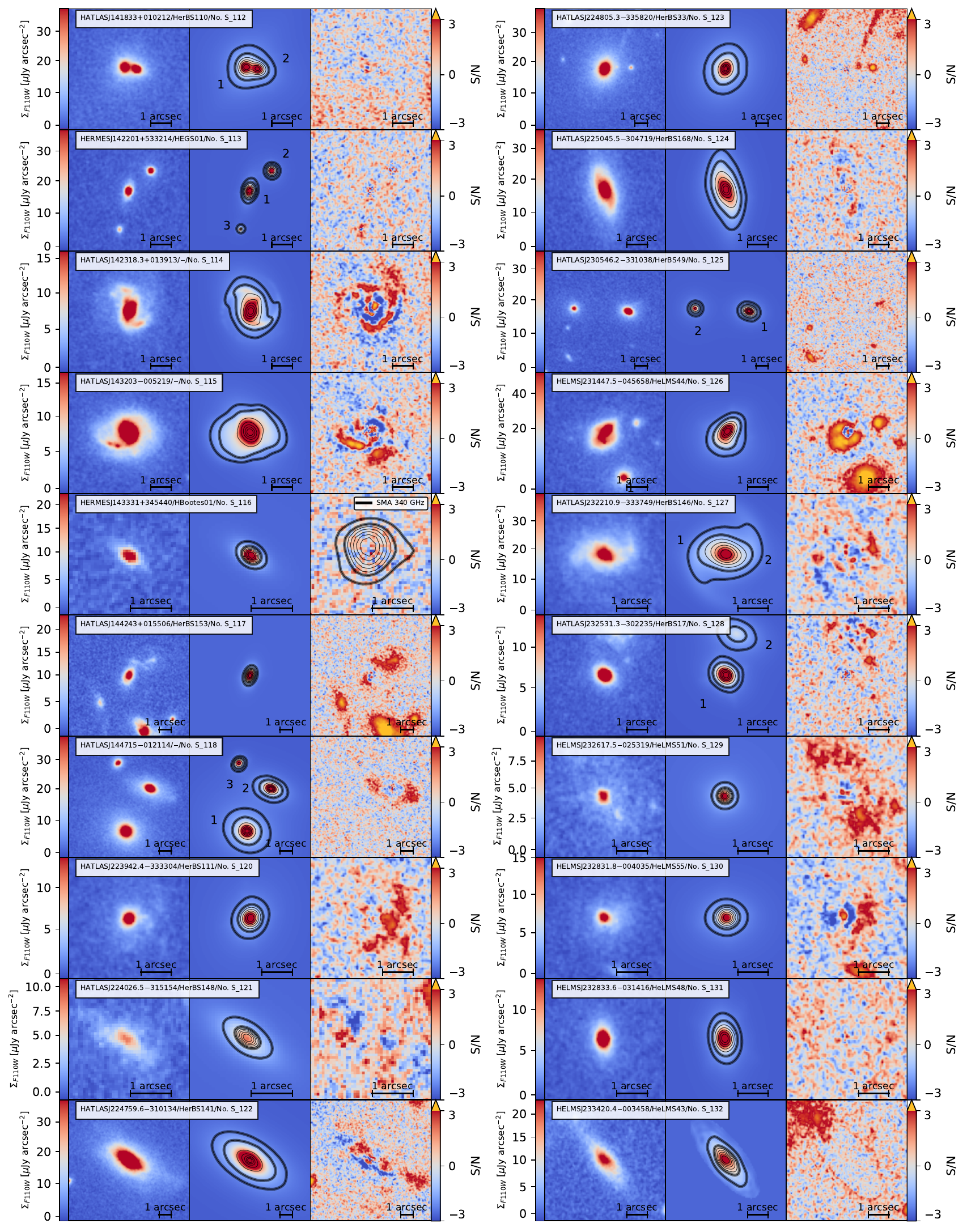}
    \contcaption{}
\end{figure*}

\begin{figure*}
    \centering
    \includegraphics[width=0.45\textwidth]{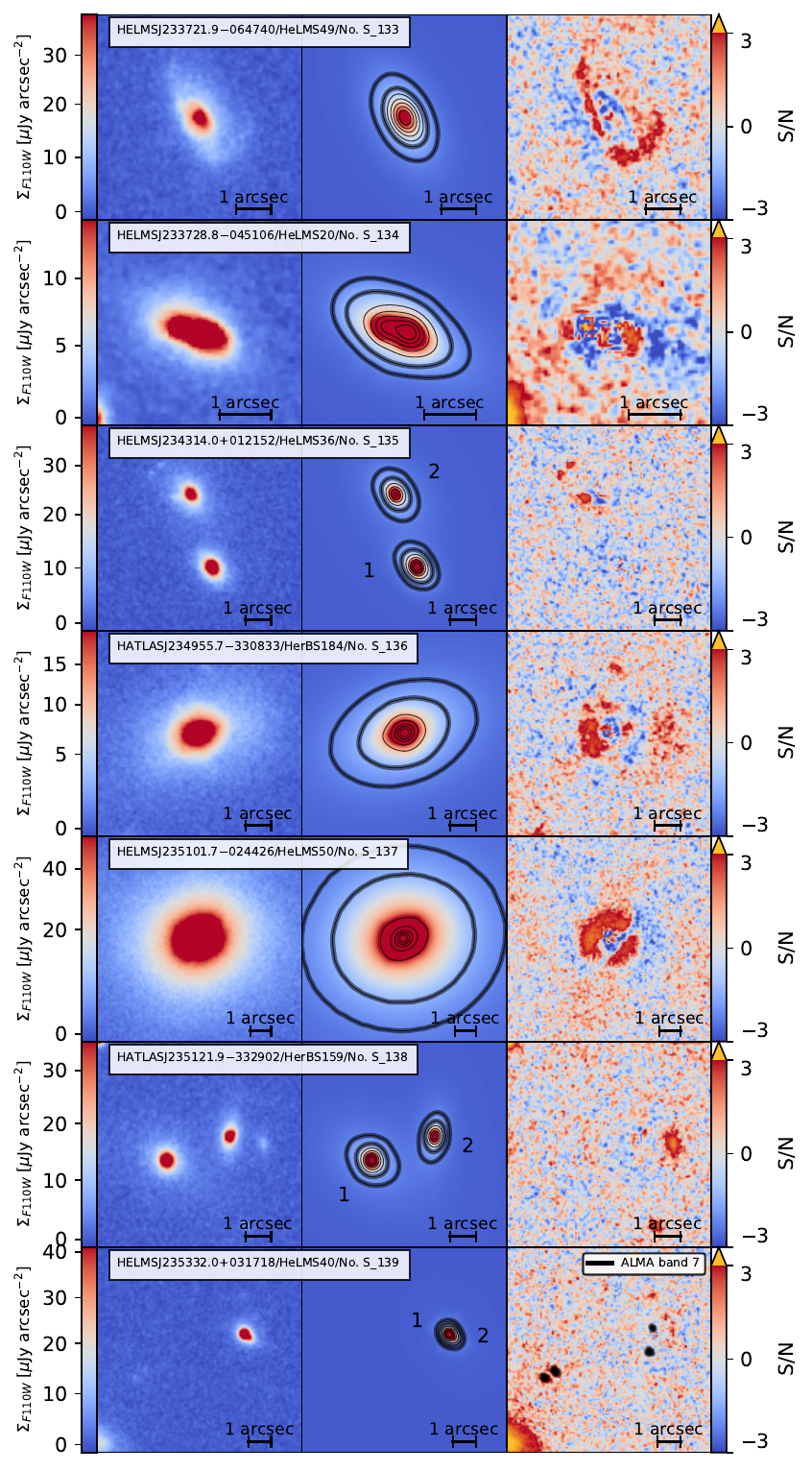}
    \contcaption{}
\end{figure*}
    
\begin{figure*}
    \centering
    \includegraphics[width=0.9\textwidth]{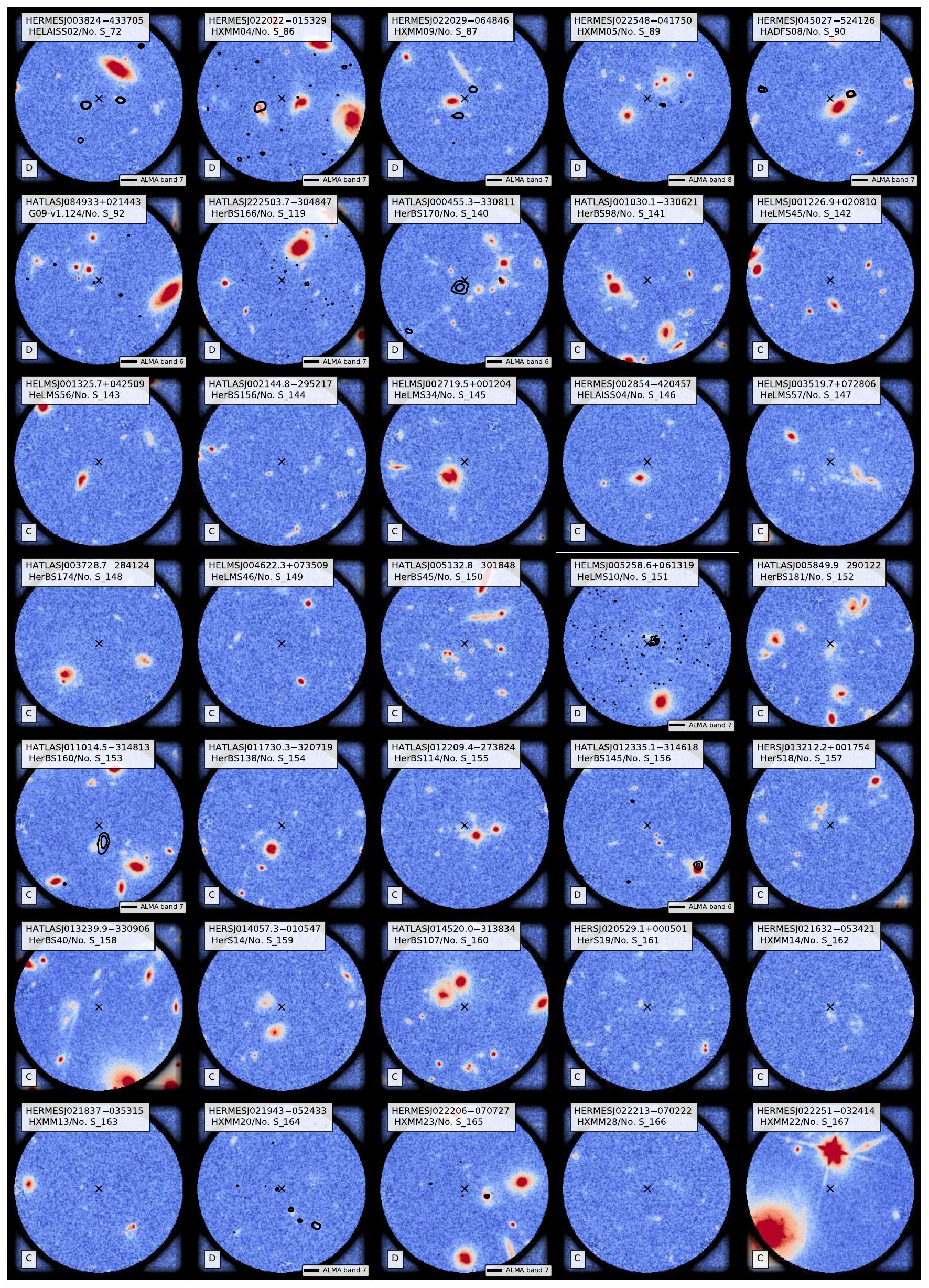}
    \caption{Systems classified as C or D. Each panel show a $\sim\ 20\,{\rm arcsec} \times\, 20\,{\rm arcsec}$} cutout of the \textit{HST} image, the black shaded regions mark the \textit{Herschel}/SPIRE beam at $250\, \mu$m. Lastly, we show the contours of available high-resolution multiwavelength data taken at two levels corresponding to $\textit{SNR} = 5$ and 10.
    \label{fig:CDs}
\end{figure*}

\begin{figure*}
    \centering
    \contcaption{}
    \includegraphics[width=0.9\textwidth]{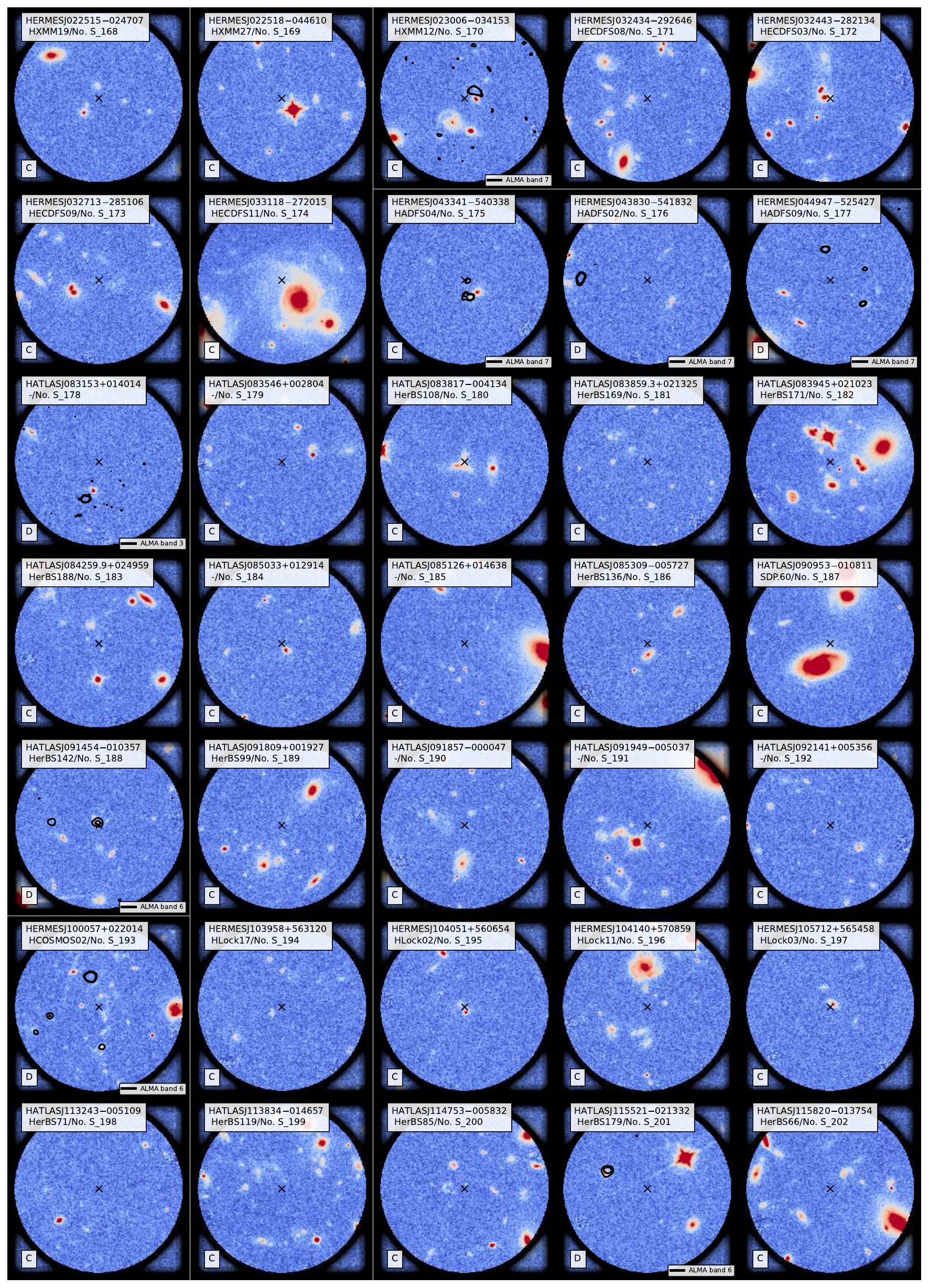}
    
\end{figure*}

\begin{figure*}
    \centering
    \contcaption{}
    \includegraphics[width=0.9\textwidth]{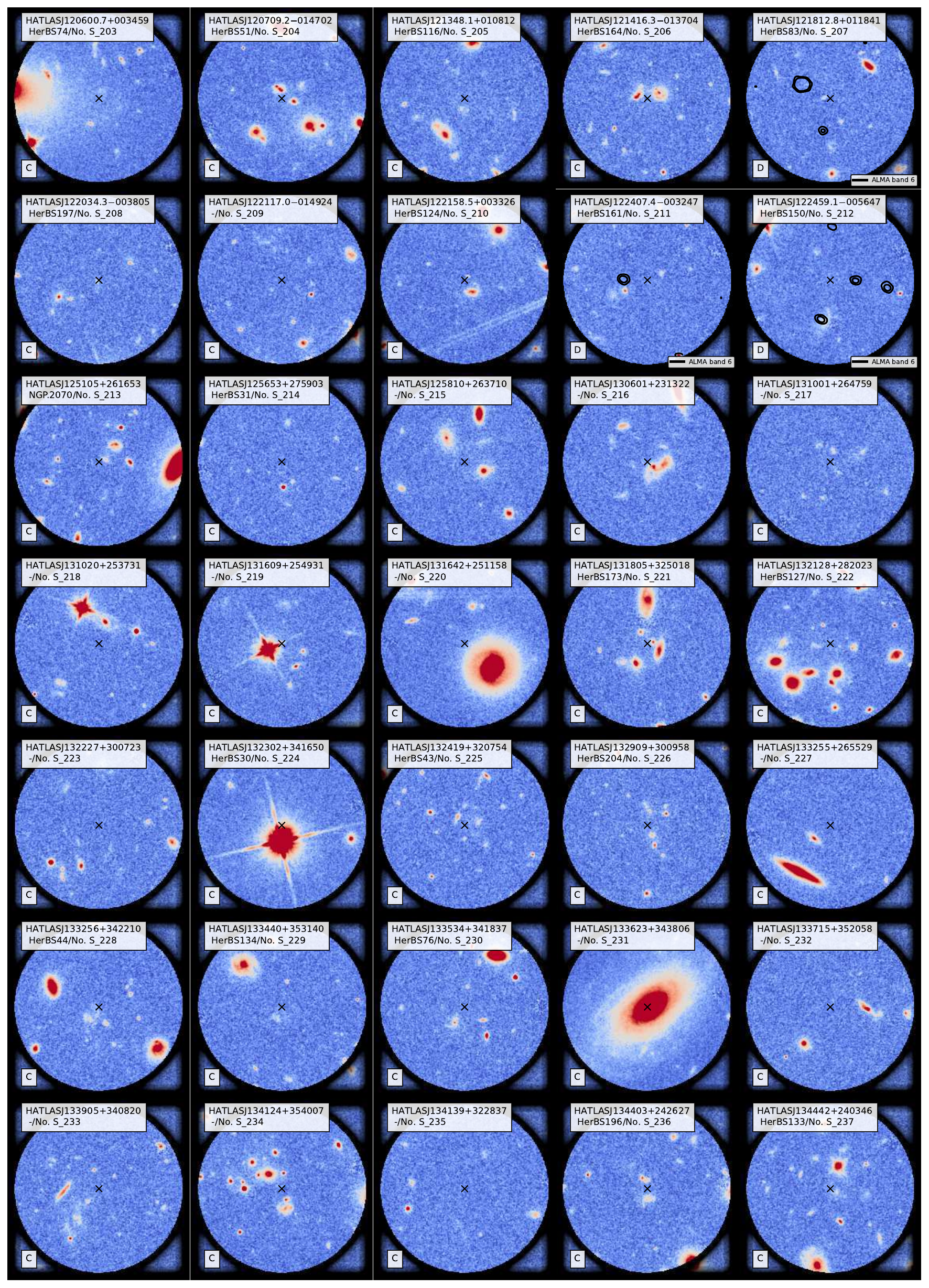}
    
\end{figure*}

\begin{figure*}
    \centering
    \contcaption{}
    \includegraphics[width=0.9\textwidth]{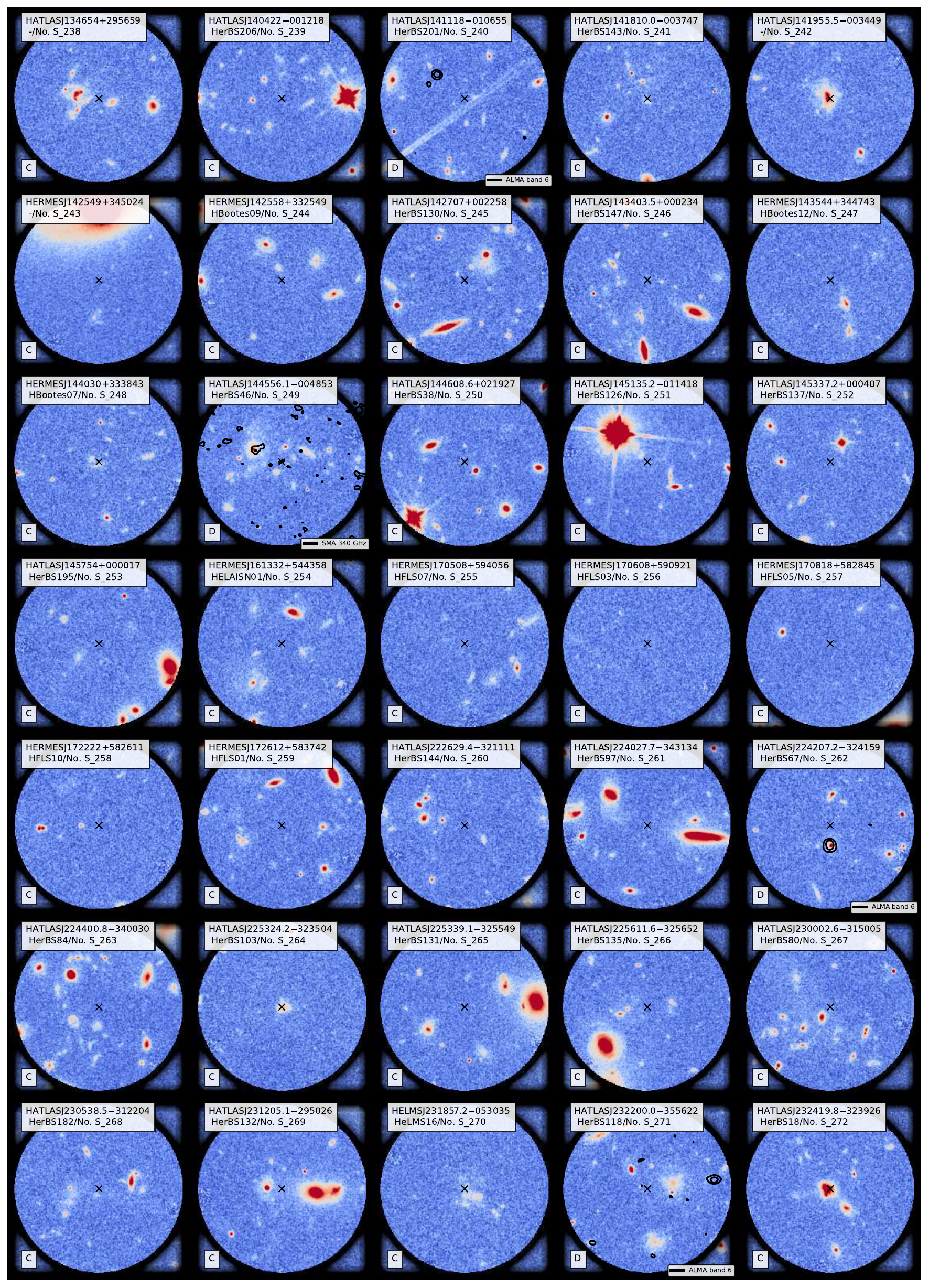}
    
\end{figure*}

\begin{figure*}
    \centering
    \contcaption{}
    \includegraphics[width=0.9\textwidth]{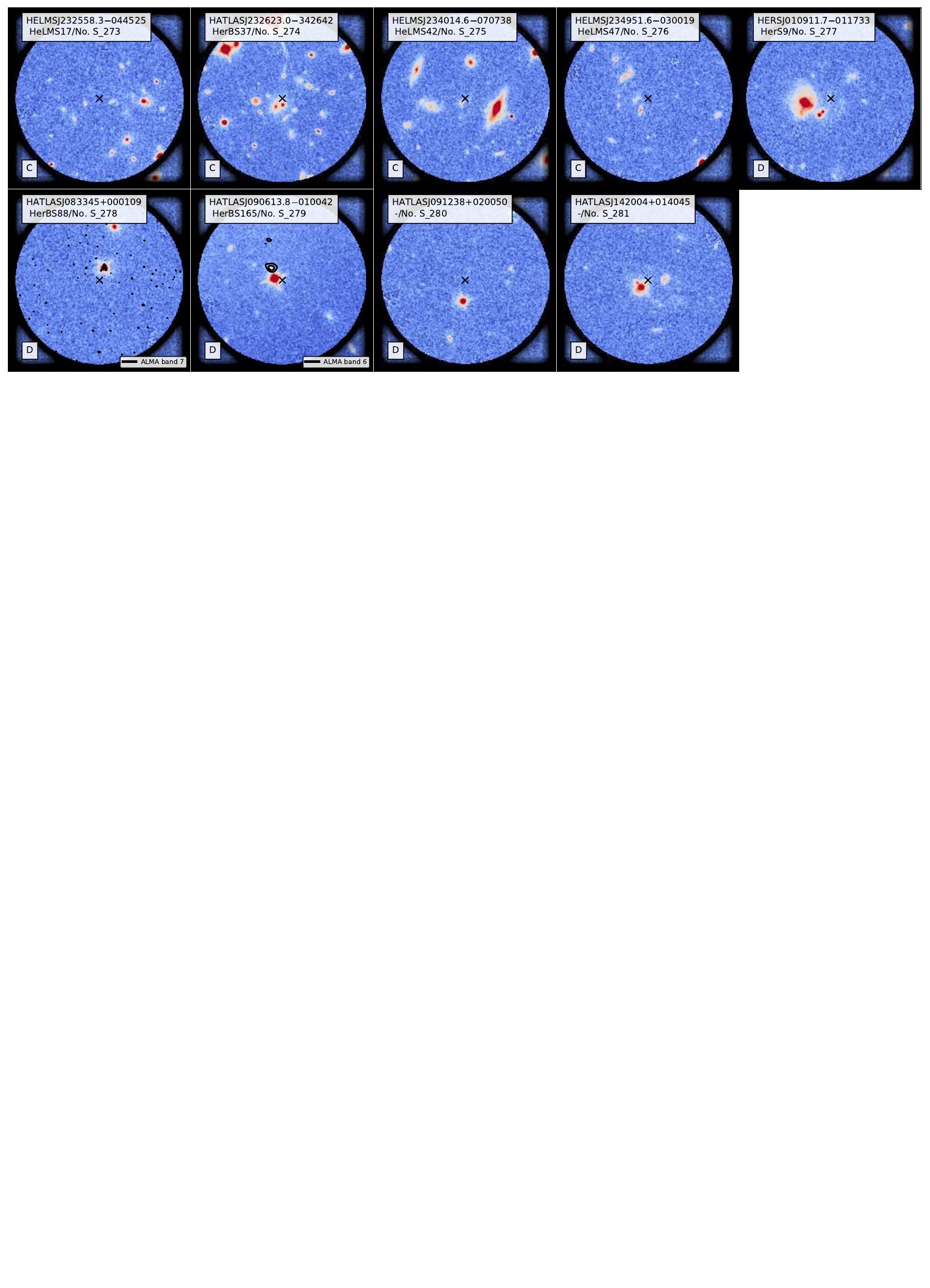}
    
\end{figure*}

\section{Lens modelling results}

\begin{landscape}
    \begin{figure}
        \centering
        \includegraphics[height=0.70\textwidth]{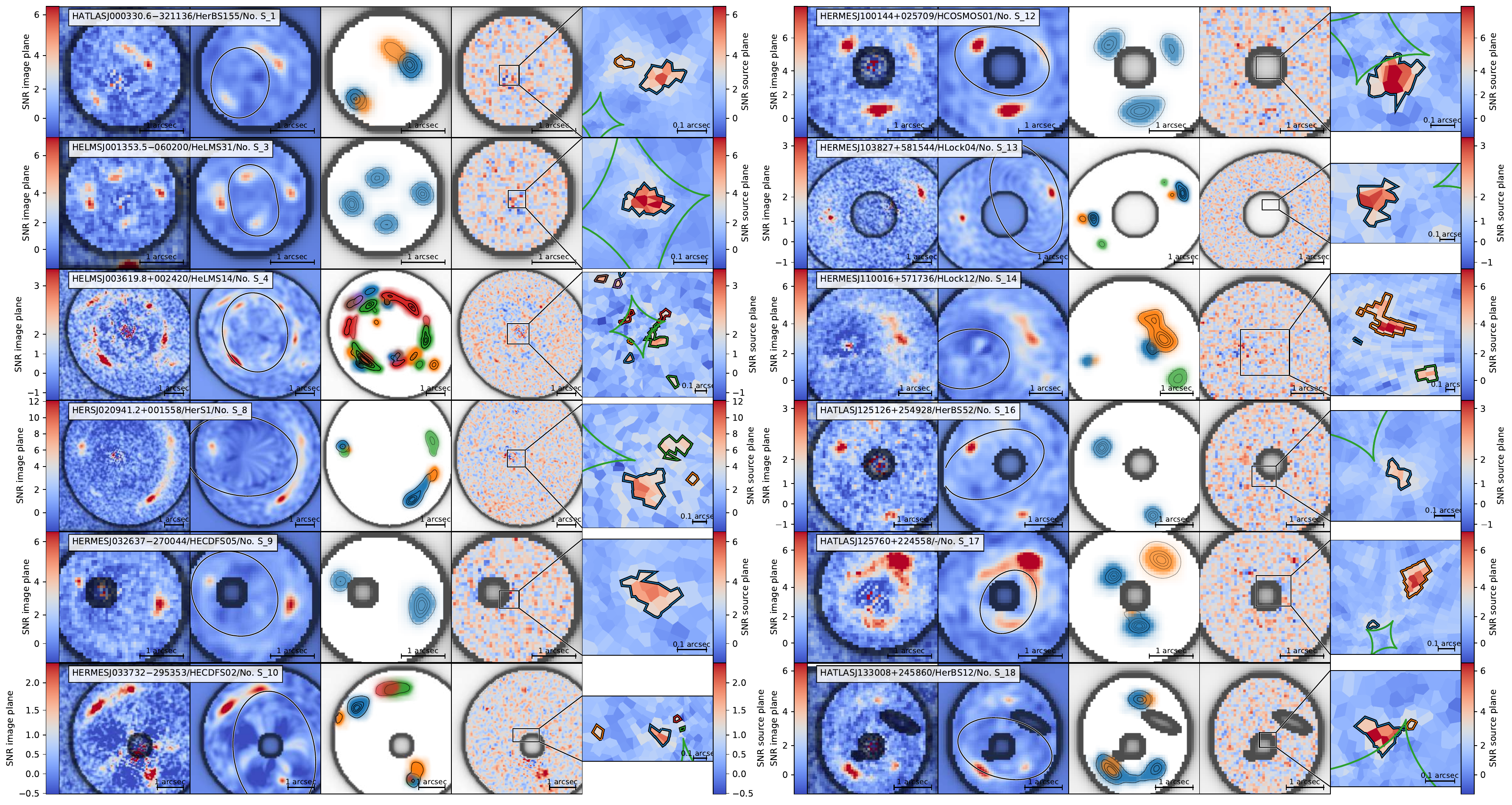}
        \caption{\textit{From left to right panels:\/} Lens subtracted image, best-fitting model of the lensed source, ${\rm SNR}>3$ regions of the reconstructed source plane lensed back to the image plane, SNR map of the residuals obtained after subtracting the model of the lensed source from the input lens-subtracted image, and the reconstructed background source with caustic curves (green line). Each of the first three panels shows the pixel mask (black-shaded region) adopted for the lens modelling. The second panel from the left also shows the critical curves (black curve). In the source plane image, we highlighted with black contours the $\textit{SNR}>3$ region adopted for computing the magnification and source size.}
        \label{fig:LM1}
    \end{figure}
\end{landscape}

\begin{landscape}
    \begin{figure}
        \centering
        \includegraphics[height=0.585\textwidth]{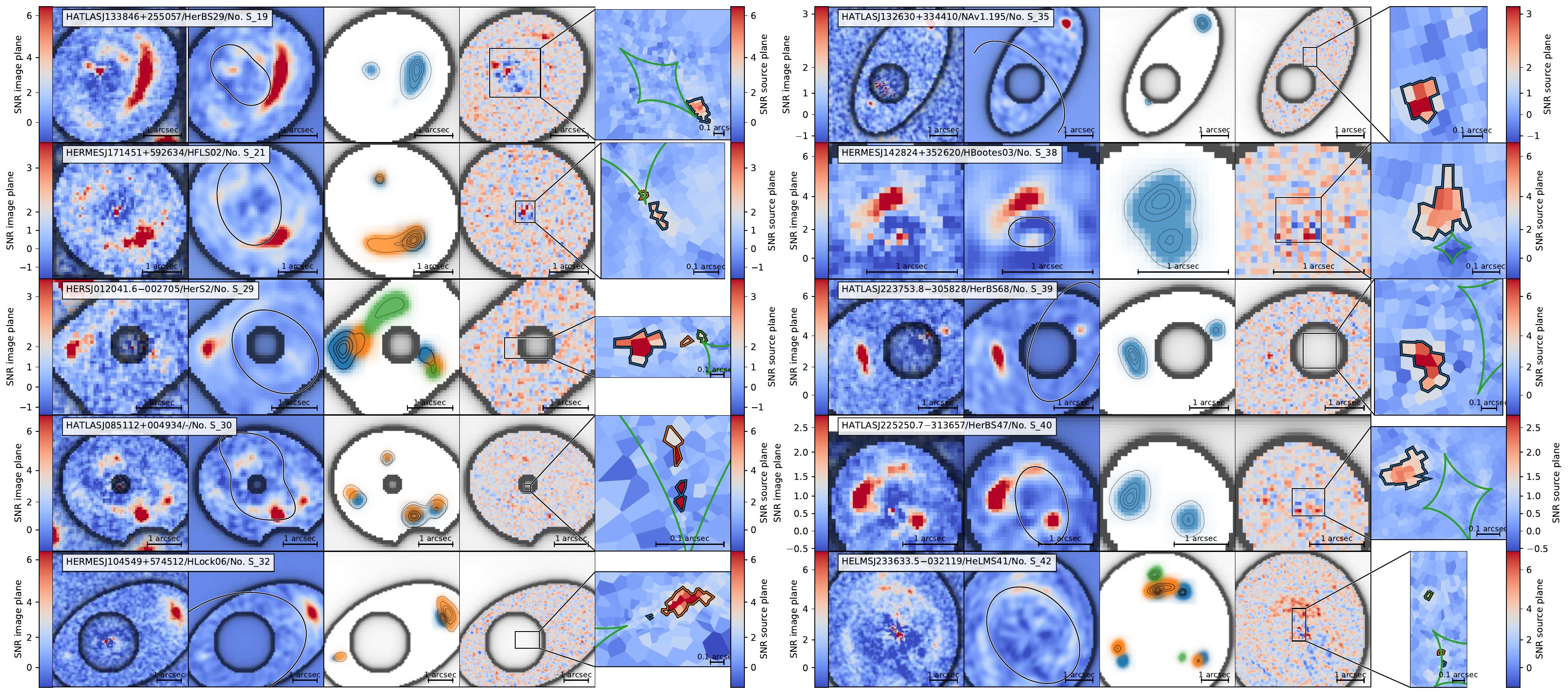}
        \contcaption{}
         
    \end{figure}
\end{landscape}

\section*{Affiliations}
\noindent
{\it
$^{1}$Dipartimento di Fisica e Astronomia "G. Galilei", Universit\`a di Padova, vicolo dell'Osservatorio 3, I-35122 Padova, Italy.\\
$^{2}$Department of Astronomy, University of Cape Town, 7701 Rondebosch, Cape Town, South Africa.\\
$^{3}$INAF - Istituto di Radioastronomia, Via Piero Gobetti 101, 40129, Bologna, Italy.\\
$^{4}$School of Physics and Astronomy, Cardiff University, The Parade, Cardiff, CF24 3AA, UK.\\
$^{5}$INAF - Osservatorio Astronomico di Padova, vicolo dell'Osservatorio 2, I-35122 Padova, Italy.\\
$^{6}$Department of Physics and Astronomy, University of North Carolina Asheville, 1 University Heights, Asheville, NC 28804, USA.\\
$^{7}$Department of Physics, Institute for Computational Cosmology, Durham University, South Road, Durham DH1 3LE, UK.\\
$^{8}$ Department of Physics and Astronomy, Rutgers, the State University of New Jersey, 136 Frelinghuysen Road, Piscataway, NJ 08854-8019, United States.\\
$^{9}$Department of Physics and Astronomy, University of the Western Cape, Robert Sobukwe Road, Bellville 7535, South Africa.\\
$^{10}$Division of Particle and Astrophysical Science, Graduate School of Science, Nagoya University, Aichi 464-8602, Japan.\\
$^{11}$National Astronomical Observatory of Japan, 2-21-1, Osawa, Mitaka, Tokyo 181-8588, Japan.\\
$^{12}$Institut d’Astrophysique Spatiale, B$\hat{a}$t. 121, Université Paris-Sud, 91405 Orsay Cedex, France.\\
$^{13}$Institut de Radioastronomie Millimétrique, 300 rue de la piscine, F-38406 Saint-Martin-d'Hères, France.\\
$^{14}$Blackett Lab, Imperial College, Prince Consort Road, London SW7 2AZ, UK.\\
$^{15}$Department of Physics \& Astronomy, University of California, Irvine, CA 92697, USA.\\
$^{16}$Sorbonne Universit{\'e}, UPMC Universit{\'e} Paris 6 and CNRS, UMR 7095, Institut d'Astrophysique de Paris, 98bis boulevard Arago, 75014 Paris, France.\\
$^{17}$Instituto de Astrofísica de Canarias, C/Vía Láctea, s/n, E-38205 San Cristóbal de La Laguna, Tenerife, Spain.\\
$^{18}$Universidad de La Laguna, Dpto. Astrofísica, E-38206 San Cristóbal de La Laguna, Tenerife, Spain.\\
$^{19}$School of Physics and Astronomy, University of Nottingham, University Park, Nottingham, NG7 2RD, UK.\\
$^{20}$University of Bologna -- Department of Physics and Astronomy “Augusto Righi” (DIFA), Via Gobetti 93/2, I-40129 Bologna, Italy.\\
$^{21}$INAF- Osservatorio di Astrofisica e Scienza dello Spazio, Via Gobetti 93/3, I-40129, Bologna, Italy.\\
$^{22}$Department of Physics and Astronomy, University of Hawai'i, 2505 Correa Road, Honolulu, HI 96822, USA.\\
$^{23}$Institute for Astronomy, 2680 Woodlawn Drive, University of Hawai'i, Honolulu, HI 96822, USA.\\
$^{24}$Departamento de F{\'i}sica, Universidad de Oviedo, C. Federico Garc{\'i}a Lorca 18, 33007 Oviedo, Spain.\\
$^{25}$Instituto Universitario de Ciencias y Tecnolog{\'i}as Espaciales de Asturias (ICTEA), C. Independencia 13, 33004 Oviedo, Spain.\\
$^{26}$Instituto Nacional de Astrofísica, Óptica y Electrónica, Luis Enrique Erro 1, Tonantzintla, Puebla, C.P. 72840, Mexico.\\
$^{27}$Aix-Marseille Universit`e, CNRS and CNES, Laboratoire d’Astrophysique de Marseille, 38, rue Fr`ed`eric Joliot-Curie,
13388 Marseille, France.\\
$^{28}$Cosmic Dawn Center (DAWN).\\
$^{29}$DTU Space, Technical University of Denmark, Elektrovej 327, 2800 Kgs. Lyngby, Denmark.\\
$^{30}$SISSA, Via Bonomea 265, 34136 Trieste, Italy.\\
$^{31}$Université Lyon 1, ENS de Lyon, CNRS UMR5574, Centre de Recherche Astrophysique de Lyon, F-69230 Saint-Genis-Laval, France.\\
$^{32}$Department of Physics and Astronomy, University of British Columbia, Canada.\\
$^{33}$School of Physical Sciences, The Open University, Walton Hall, Milton Keynes, MK7 6AA, UK.\\
$^{34}$Department of Earth and Space Sciences, Chalmers University of Technology, Onsala Observatory, 439 94 Onsala, Sweden.\\
$^{35}$Leiden Observatory, Leiden University, PO Box 9513, 2300 RA, Leiden, The Netherlands.\\
$^{36}$ Inter-university Institute for Data Intensive Astronomy, Department of Astronomy, University of Cape Town, 7701 Rondebosch, Cape Town, South Africa.\\
$^{37}$ Inter-university Institute for Data Intensive Astronomy, Department of Physics and Astronomy, University of the Western Cape, 7535 Bellville, Cape Town, South Africa.\\
$^{38}$SRON Netherlands Institute for Space Research, Landleven 12, 9747, AD Groningen, The Netherlands.\\
$^{39}$Kapteyn Astronomical Institute, University of Groningen, Postbus 800, 9700, AV, Groningen, The Netherlands.}


\bsp	
\label{lastpage}
\end{document}